UNIVERSITÉ DE SHERBROOKE
Faculté de génie
Département de génie électrique et de génie informatique

# Élaboration du Ge mésoporeux et étude de ses propriétés physico-chimiques en vue d'applications photovoltaïques

Thèse de doctorat
Spécialité : génie électrique

Sergii TUTASHKONKO

Jury : Vincent Aimez (directeur)
Richard Arès (directeur)
Mustapha Lemiti (directeur)
Anne Kaminski-Cachopo (directrice)
Denis Morris
Gilles Lerondel
Francois Schiettekatte

À ceux qui vraiment encadrent.

# RÉSUMÉ

Le sujet de cette thèse porte sur l'élaboration du nouveau nanomatériau par la gravure électrochimique bipolaire (BEE) — le Ge mésoporeux et sur l'analyse de ses propriétés physico-chimiques en vue de son utilisation dans des applications photovoltaïques.

La formation du Ge mésoporeux par gravure électrochimique a été précédemment rapportée dans la littérature. Cependant, le verrou technologique important des procédés de fabrication existants consistait à obtenir des couches épaisses (supérieure à 500 nm) du Ge mésoporeux à la morphologie parfaitement contrôlée. En effet, la caractérisation physico-chimique des couches minces est beaucoup plus compliquée et le nombre de leurs applications possibles est fortement limité. Nous avons développé un modèle électrochimique qui décrit les mécanismes principaux de formation des pores ce qui nous a permis de réaliser des structures épaisses du Ge mésoporeux (jusqu'au 10 µm) ayant la porosité ajustable dans une large gamme de 15% à 60%. En plus, la formation des nanostructures poreuses aux morphologies variables et bien contrôlées est désormais devenue possible. Enfin, la maitrise de tous ces paramètres a ouvert la voie extrêmement prometteuse vers la réalisation des structures poreuses à multi-couches à base de Ge pour des nombreuses applications innovantes et multidisciplinaires grâce à la flexibilité technologique actuelle atteinte. En particulier, dans le cadre de cette thèse, les couches du Ge mesoporeux ont été optimisées dans le but de réaliser le procédé de transfert de couches minces d'une cellule solaire à triple jonctions via une couche sacrificielle en Ge poreux.

# REMERCIEMENTS

Je remercie mes directeurs de thèse, Monsieur Vincent Aimez, Monsieur Richard Arès, Monsieur Mustapha Lemiti et Madame Anne Kaminski-Cachopo pour m'avoir offert cet intéressant projet de thèse. Je veux exprimer mes remerciements en particulier à :

- Madame Tetyana Nychyporuk et Monsieur Vladimir Lysenko pour les nombreuses discussions autour de l'analyse des résultats et de l'élaboration du modèle électrochimique. Merci pour leur aide dans les mesures µ-Raman et FTIR, ainsi que pour le soutien moral tout au long de la thèse. Merci aussi pour les conseils et encouragements lors des répétitions de la soutenance de la thèse ;

- Monsieur Olivier Marty pour les mesures par TEM et les discussions fructueuses portant sur l'analyse de la structure Ge mesoporeux ;

- Monsieur Bruno Canut pour les mesures RBS et l'aide avec l'interprétation des résultats ;

- Monsieur Sergii Alekseev pour des discussions inspirantes sur mon projet de recherche, pour la révision de la partie électrochimique de la thèse et pour la générosité avec laquelle il a toujours su à mes questions portant sur la chimie ;

- Monsieur Mustapha Lemiti pour son aide précieuse lors de l'étape de révision et de correction du manuscrit ;

- Madame Mariana Georges, étudiante à la maîtrise, pour l'efficacité de son travail de soutien au niveau de la fabrication de structures de Ge mesoporeux à double couche. Son aide a été précieuse pour l'obtention d'importants résultats de ma thèse.

- Monsieur Denis Morris pour avoir accepté de faire partie de mon jury d'évaluation de thèse et pour les nombreux commentaires et les corrections pertinentes, qui m'ont aidé d'améliorer la lisibilité du manuscrit final ;

- Monsieur Gilles Lerondel et Monsieur Francois Schiettekatte pour leur participation à titre de rapporteurs (examinateurs externes) pour ce jury de thèse. Merci pour leurs commentaires pertinents ;

- La région Rhone-Alpes pour le support financier dans le cadre du programme CMIRA-Explora Doc ;

- Monsieur Romain Mailhes pour la révision du français sur les différentes pages de ma présentation de soutenance de la thèse ;

- L'ensemble des ingénieures, techniciens et professionnels de recherche de la plaforme Nanolyon de l'INL, des salles propres du département de physique de l'Université de Sherbrooke et des salles blanches du 3iT. Merci en particularité à Monsieur Robert Perrin, Monsieur Philippe Girard, Monsieur Khaled Ayadi, Monsieur Denis Pellé et Monsieur Abdelatif Jaouad.

- Doctorants, post-doctorants et stagiaires de l'INL : Monsieur Anton Pastushenko, Monsieur Yury Ryabchikov, Monsieur Anatolii Lukianov et Monsieur Andrii Rogov. Merci pour votre encouragements et pour votre amitié ;
- L'équipe photovoltaïque de L'INL pour la bonne atmosphère de travail ;
- L'ensemble des secrétaires de l'INL et du département GÉGI de l'Université de Sherbrooke dont Madame Virginie Lagarde, Madame Danielle Gagné et Madame Marie-Claire Boisvert pour leur réactivité lors des régulations des questions administratives ;
- Monsieur Michael Jensen et Monsieur Alex Magdzinski pour leur présence amicale, leur soutien moral et pour être toujours prêt d'aider dans les situations difficiles de la vie ;
- Mes parents et mes beaux-parents, Monsieur Volodymyr Tutashkonko, Madame Valentyna Tutashkonko, Monsieur Petro Belelia et Madame Valentyna Shchiebietova pour leur soutien constant dans différents aspects de ma vie et surtout pour leur aide précieuse avec la garde de la petite Veronika lorsque papa travaillait fort à la rédaction de cette thèse.
- Ma petite fille Veronika et ma conjointe Olesia pour la joie et le bonheur qu'elles me procurent tous le jours de ma vie. Leur présence est ma plus grande source de motivation dans la vie.

# TABLE DES MATIÈRES

# LISTE DES FIGURES

# LISTE DES TABLEAUX

# CHAPITRE 1

# Mise en contexte et État de l'art

## 1.1 Photovoltaïque à haut rendement de conversion

### 1.1.1 Cellules solaires à multiple jonctions

Depuis le début du développement des cellules photovoltaïques, la quête permanente est d'augmenter le rendement et de réduire les coûts de conversion (Figure 1.1). Jusqu'à présent, l'approche de réalisation des cellules en tandem ou autrement dit les cellules multi-jonctions démontre le meilleur rendement de conversion d'énergie. Le concept de cellules solaires en tandem est basée sur l'utilisation de plusieurs cellules solaires (ou sous-cellules) de bandes interdites différentes empilées l'une sur l'autre (Figure 1.2). Chaque cellule possède l'énergie de la bande interdite plus élevée que des cellules en-dessous. La lumière incidente est automatiquement filtrée lors de son passage à travers cette empilement [46]. L'utilisation de sous-cellules multiples dans une structure de cellules en tandem permet de diviser le spectre solaire sur de plus petites sections, dont chacune peut être convertie en électricité de façon plus efficace. Le rendement augmente à mesure que le nombre de sous-cellules augmente, le rendement théorique de conversion de la lumière solaire directe est de 86.8% calculé pour une pile infinie de sous-cellules fonctionnant indépendamment [77]. Il serait de 42,5% pour les cellules tandem et 47,5% pour les cellules à 3 jonctions [86], comparativement aux 30% à une cellule solaire sous éclairement AM1.5 (irradiance spectrale de référence standard [46, 86]).

### 1.1.2 Photovoltaïque à concentration (*CPV*)

Le fonctionnement des cellules solaires sous lumière concentrée apporte quelques avantages.

- Amélioration significative du rendement lorsque l'intensité lumineuse augmente (tableau 1.1) grâce à la croissance logarithmique du $V_{oc}$ et du $FF$ [62]. En même temps, étant proportionnels à l'irradiance, les photo-courants $I_{cc}$ des sous-cellules restent accordés [46].

- Moins de perte du rendement dû à l'échauffement. La $V_{oc}$ diminue tandis que le $I_{cc}$ augmente légèrement. Les cellules en Si perdent plus que $0.5\%_{\text{abs}}$ par 1 ℃ en AM1.5

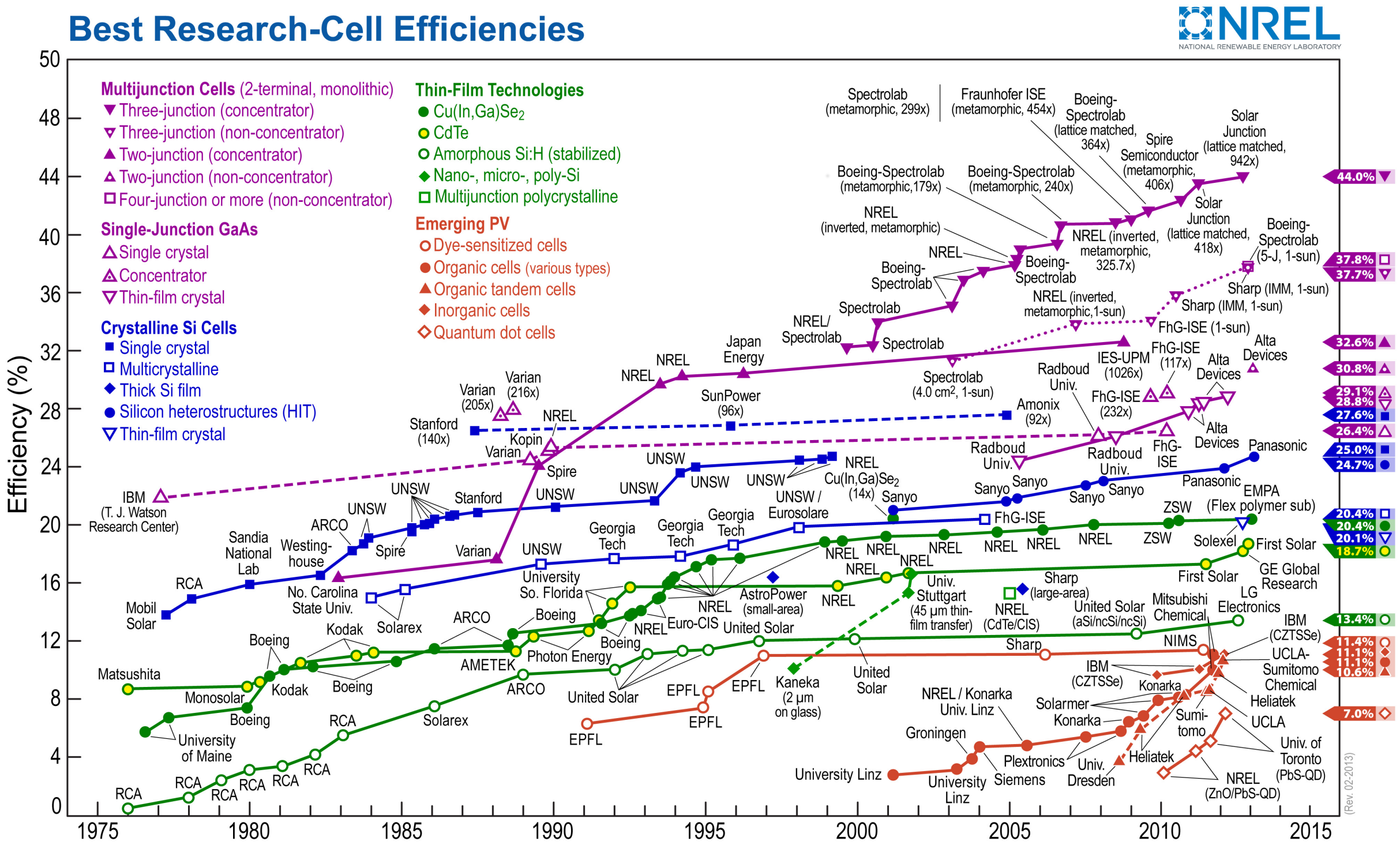

Figure 1.1 Feuille de route du photovoltaïque par NREL.

[60], 0.3–0.4% par 1 °C sous une concentration de 1.5X à 10X [116] et < 0.25%/°C sous une concentration supérieure à 100X [121]. Le coefficient thermique des cellules à triple jonctions est aussi bas que 0.035% [62].

Un système CPV comprend un concentrateur solaire à base de lentilles ou miroirs, un mécanisme de suivi (« *tracker* »), une cellule solaire montée sur un dissipateur de chaleur couplé avec des connecteurs électriques. Afin de maintenir des dimensions raisonnables pour un tel système, le CPV exploite des petites cellules solaires (de 0.25 à 1 cm$^2$). Ceci apporte encore quelques avantages par rapport aux systèmes « plans ».

- La réduction de la taille ou de du nombre de cellules nécessaires permet d'intégrer des cellules multi-jonctions de très haut rendement (facteur économique).

- Une réduction de la quantité des matériaux dont la disponibilité est limitée (par exemple Ge, Ga, In) est nécessaire (facteurs économique et écologique). Cela assure la durabilité dans des scénarios de pénétration très importantes du marché [36].

- Lorsque la production implique des étapes de micro-fabrication (épitaxie, lithographie, etc), la réalisation des cellules de grande surface est compliquée. Le choix de fabriquer des petites cellules assure un meilleur rendement de fabrication (facteurs technologique et économique).

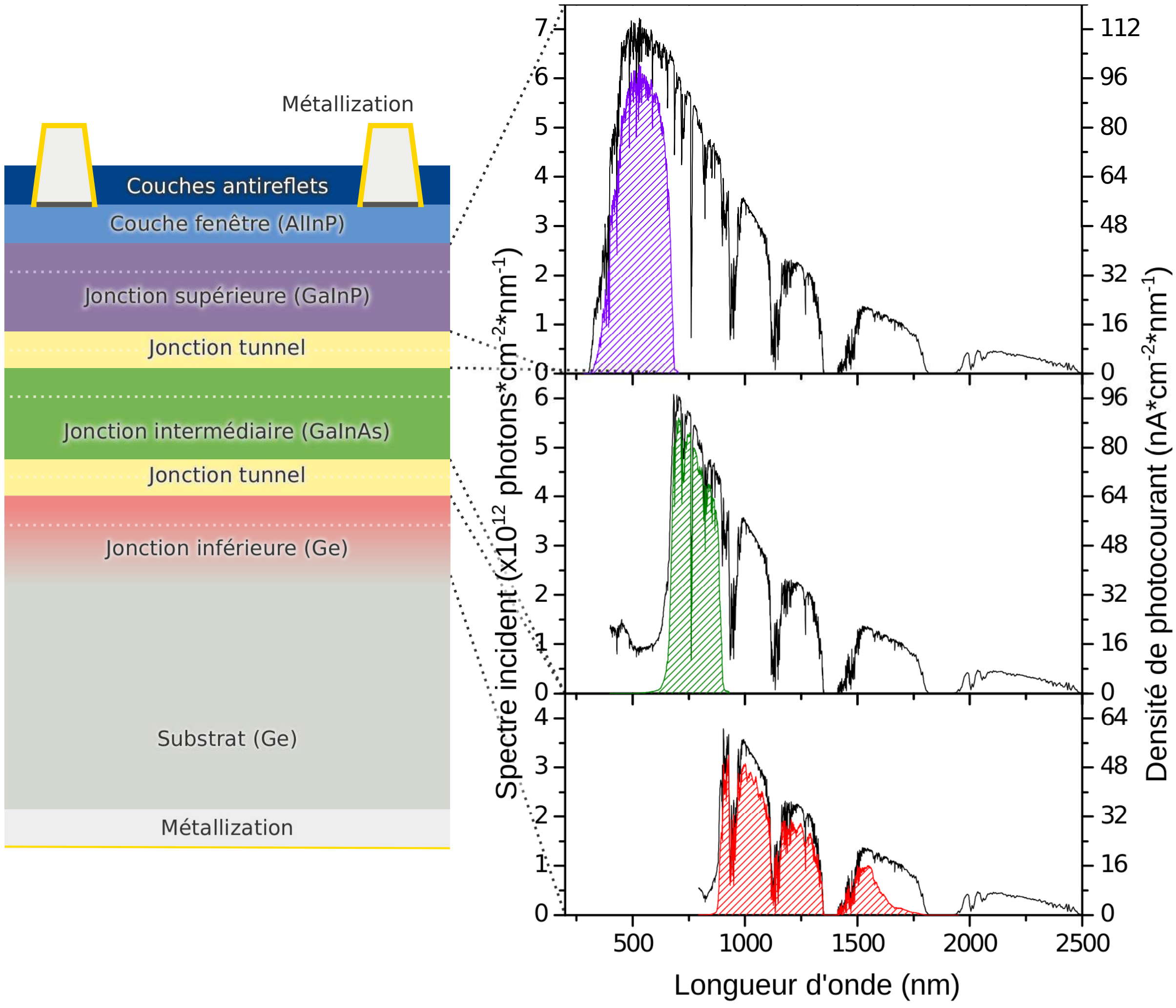

Figure 1.2 Principe de fonctionnement d'une cellule solaire à triple jonctions

- Lorsque un flux thermique augmente par le facteur de concentration, les petites dimensions de module facilitent la gestion de la dissipation de chaleur (facteurs technologique et économique).

| Nombre de jonctions dans une cellule solaire | $\eta$ sous 1 soleil | $\eta$ max. sous concentration |
|---|---|---|
| 1 jonction | 30.8% | 40.8% |
| 2 jonctions | 42.9% | 55.7% |
| 3 jonctions | 49.3% | 63.8% |
| $\infty$ jonctions | 68.2 | 86.8% |

Tableau 1.1 Comparaison du rendement de conversion des cellules solaires sous un éclairement AM1.5 et sous concentration [51]

| N | Eff. (%) | $Eg_1$ (eV) | $Eg_2$ (eV) | $Eg_3$ (eV) | $Eg_4$ (eV) | $Eg_5$ (eV) | $Eg_6$ (eV) | $Eg_7$ (eV) | $Eg_8$ (eV) |
|---|---|---|---|---|---|---|---|---|---|
| 1 | 34.7 | 1.14 | | | | | | | |
| 1 | 34.6 | 1.34 | | | | | | | |
| 2 | 47.2 | 1.57 | 0.93 | | | | | | |
| 3 | 53.0 | 1.86 | 1.34 | 0.93 | | | | | |
| 3 | 53.0 | 1.75 | 1.18 | 0.70 | | | | | |
| 3 | 47.5 | 1.85 | 1.40 | 0.67 | | | | | |
| 3 | 52.5 | 1.75 | 1.18 | 0.67 | | | | | |
| 4 | 56.8 | 1.94 | 1.44 | 1.04 | 0.70 | | | | |
| 4 | 55.8 | 2.04 | 1.57 | 1.21 | 0.92 | | | | |
| 4 | 62.0 | 2.48 | 1.69 | 1.11 | 0.60 | | | | |
| 5 | 65.0 | 2.68 | 1.93 | 1.40 | 0.95 | 0.53 | | | |
| 6 | 67.3 | 2.93 | 2.18 | 1.66 | 1.24 | 0.84 | 0.47 | | |
| 7 | 68.9 | 3.21 | 2.50 | 2.00 | 1.56 | 1.19 | 0.82 | 0.47 | |
| 8 | 70.2 | 3.35 | 2.65 | 2.14 | 1.74 | 1.4 | 1.09 | 0.78 | 0.44 |

Tableau 1.2 Les résultats de simulation du rendement des cellules solaires avec 1–4 [35] et 4–8 [51] jonctions sous une concentration de 500X et une température de 300 K. Étant chauffées à 350K, les cellules à 3 jonctions perdent de 3 à 4.4$\%_{abs}$ d'efficacité

### 1.1.3 CPV : état de l'art et perspectives

En pratique, la voie d'augmentation du rendement des cellules est en contradiction avec une réduction de prix. Il existe alors deux approches majeures de développement : augmenter le rendement malgré le prix ou baisser le coût de fabrication en maintenant un rendement raisonnable. La première approche implique l'utilisation de concentrateurs très complexes avec un facteur de concentration important afin de maximiser le rendement des cellules solaires. Bien que que dans de tels systèmes une cellule solaire n'est pas l'élément le plus coûteux, son rendement est le principal facteur qui détermine la rentabilité économique de l'installation [61].

Les cellules à triple-jonctions « classiques » en accord de maille (angl. *lattice-matched*) à base de GaInP/GaAs/Ge plafonnent actuellement à un rendement maximal de 41.6%. Ainsi, l'augmentation du rendement serait possible en utilisant des structures en désaccord de maille (angl. *lattice-mismatched*) à base de AlGaInAsP, soit par en développant de nouveaux matériaux en accord de maille [35, 61, 73]. Le passage à un rendement supérieur à 40% a été marqué par l'approche utilisant des structures en désaccord de maille. Par exemple, la cellule de Spectrolab $Ga_{.44}In_{.56}P/Ga_{.92}In_{.06}As/Ge$ a été la première à obtenir un rendement de 40.7%. Présentée un an plus tard, la cellule $Ga_{.35}In_{.65}P/Ga_{.83}In_{.17}As/Ge$ du Fraunhofer ISE a montré un rendement de 41.1%. Dans ces deux cellules à haute rendement, des dislocations subsistent néanmoins, générées par les contraintes dues au désaccord de maille. La principale difficulté dans la réalisation de ce type de structure est la croissance de couches tampons entre le Ge et le GaInAs pour réduire les effets de dislocations. Les contraintes apparaissent au début de la croissance. Une autre idée consiste à faire croitre des cellules inversées [35, 61, 73]. Dans ce cas, deux jonctions croissent sous des conditions d'accord de maille. Le Ge est remplacé par le $Ga_{.37}In_{.53}As$ (40.8% NREL). Le passage à des structures ayant quatre jonctions ou plus ne consiste pas simplement à ajouter une jonction comme il a été possible de faire dans le cas de cellules tandem GaInP/GaAs et des cellules en triple-jonctions GaInP/GaAs/Ge [61] (tableau 1.2). Les calculs théorétiques [35] des bandes interdites optimales montrent qu'il n'y a pas de combinaison d'AlGaAsInP idéale pour créer de telles structures [35]. La voie la plus prometteuse est la construction de structures à base de nitrures (InGaN)[56]. Ce matériau dispose d'un coefficient d'absorption plus grand que celui du GaAs et la bande interdite peut atteindre 3.4 eV pour du GaN pur. Cependant, le principal problème est qu'il y a un grand désaccord de maille (10%) entre le GaN et l'InN. La question du choix du substrat avec une maille appropriée reste ouverte.

## 1.2 Report des couches

Lorsque les sous-cellules sont branchées en série, le rendement de la cellule intégrale est limitée par la jonction qui produit le courant le plus faible. Dans des cellules à triple-jonctions « classiques » à deux bornes InGaP/(In)GaAs/Ge, le goulot d'étranglement est la jonction du milieu. La densité de courant produite par chaque jonction est 14.3 mA/cm$^2$, 12.7 mA/cm$^2$ et 20.7 mA/cm$^2$ respectivement (cellule ML-3JSC de Cyrium [50]). Afin de laisser passer plus de photons absorbables par la cellule limitante, l'épaisseur de la cellule supérieure peut être réduite [50]. Une autre approche est d'insérer des boites quantiques dans la jonction milieu pour élargir sa plage d'absorption du coté infra-rouge [123]. Il existe des cellules solaires commerciales avec les boites quantiques démontrant un rendement plus de 40% [52].

Néanmoins, la cellule en Ge produit toujours un courant excédentaire. Les substrats de Ge ont une épaisseur de l'ordre de 140–200 µm nécessaire pour assurer une tenue mécanique lors de la fabrication de la cellule. Cependant, moins de 5 µm d'épaisseur de substrat de Ge sont suffisants pour générer autant de courant que les deux autres jonctions [4]. Le reste du substrat de Ge engendre une résistance en série ainsi que des pertes dues aux recombinaisons volumiques [2]. De plus, le prix du substrat de Ge s'élève à 50% du prix de la cellule complète [24]. L'utilisation des substrats des Ge plus fins (<140 µm) n'est pas une solution intéressante, car lors de la découpe du lingot du Ge en tranches, les pertes en matériau sont trés importantes. L'amincissement chimique ou mécanique d'un substrat améliore le rendement de la cellule mais signifie une perte de matériau coûteux.

L'idée est donc de détacher du substrat de Ge l'épaisseur utilse de la cellule, qui contribue en génération du courant. Cette couche mince peut être transférée vers un substrat à faible coût (par exemple collée directement sur un dissipateur de chaleur d'un concentrateur) tandis que le reste du substrat peut être recyclé pour un nouveau transfert de couche. Ainsi, cette opération doit être réalisée sans affecter les propriétés électriques du matériau constituant la cellule.

Ce procédé en couche mince (porosification, croissance puis transfert) apporte les avantages suivants :

- Économie de Ge et réduction du coût de fabrication.
- Augmentation du rendement.
- Réduction du poids, ce qui est un facteur important pour les applications spatiales.
- Réduction de la résistance thermique.

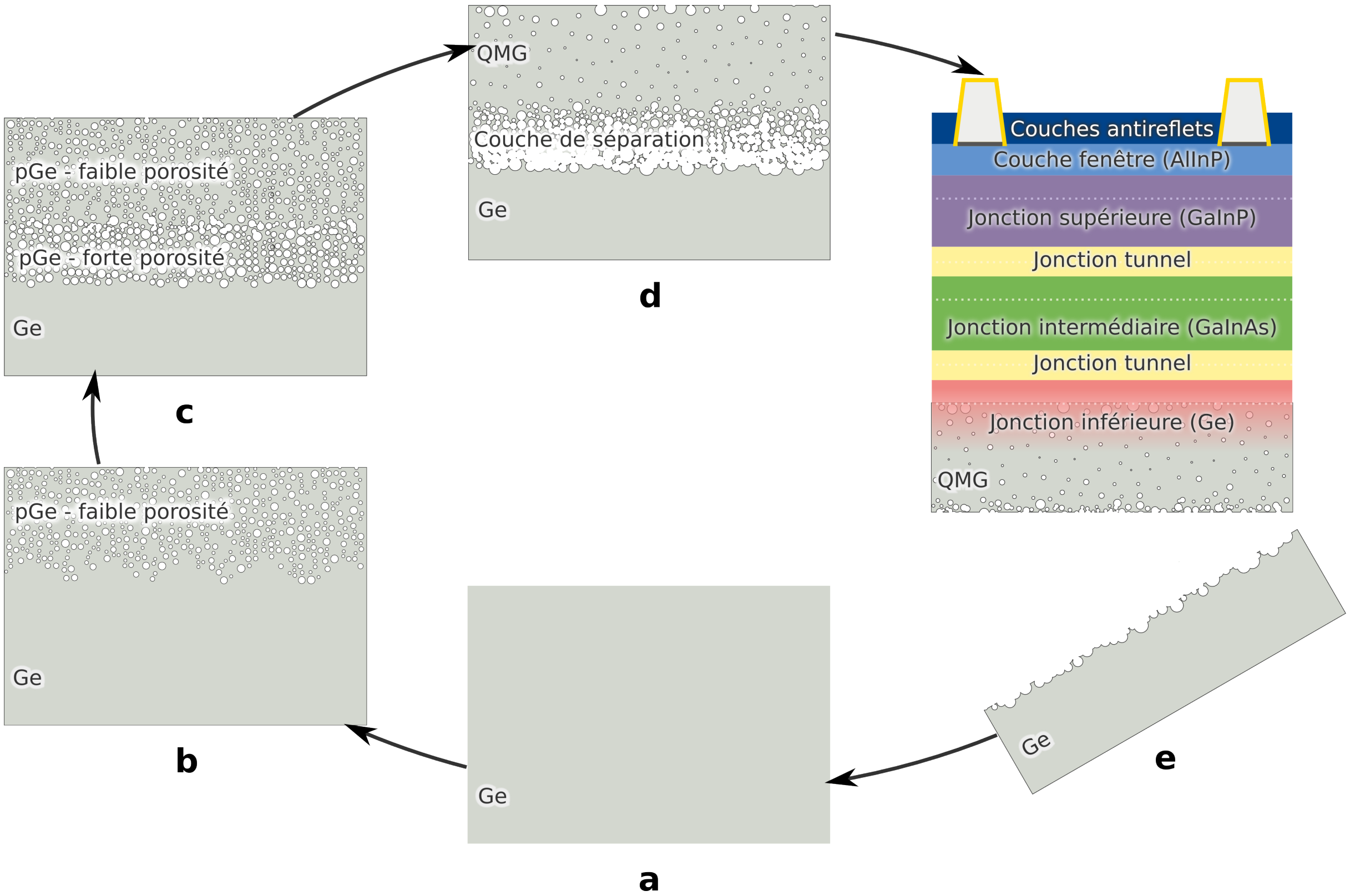

Figure 1.3 Le procédé de transfert de couche mince d'une cellule à 3 jonctions :
(a) le substrat de Ge de départ ;
(b) formation d'une couche poreuse à faible porosité ;
(c) ajout d'une couche poreuse à forte porosité ;
(d) densification de la couche à faible porosité et fragilisation de la couche à forte porosité lors du traitement thermique ;
(e) croissance de la cellule solaire, séparation, transfert et recyclage du substrat.

### 1.2.1 Techniques de report des couches utilisant le Si-poreux

Une méthode de transfert de couches permettant de réutiliser le substrat plusieurs fois est bien développée pour le Si. Bien que l'idée soit assez simple, les réalisations pratiques diffèrent beaucoup entre elles. Les procédés de transfert de couche ayant pour but d'obtenir une couche fine de Si monocristallin sur un substrat étranger pour des applications dans les cellules solaires peuvent être divisés en 2 groupes : ceux qui utilisent le Si poreux et ceux qui ne l'utilisent pas.

Dans le premier cas, la surface du substrat de Si monocristallin est porosifiée. Cette couche poreuse possède une structure à double porosité : une basse porosité au dessus et une porosité élevée en dessous. Ensuite, la structure est recuite à haute température afin de réorganiser les couches poreuses. La couche de faible porosité est de qualité monocris-talline et permet la croissance d'une couche épitaxiale de haute qualité après un recuit (Figure 1.4). La couche épitaxiale formée sur le Si-poreux est détachable du substrat d'origine grâce à la couche de haute porosité, elle est ensuite transférée sur un substrat étranger. La fabrication d'une cellule peut être réalisée avant ou après ce transfert ce qui modifie l'enchaînement des opérations.

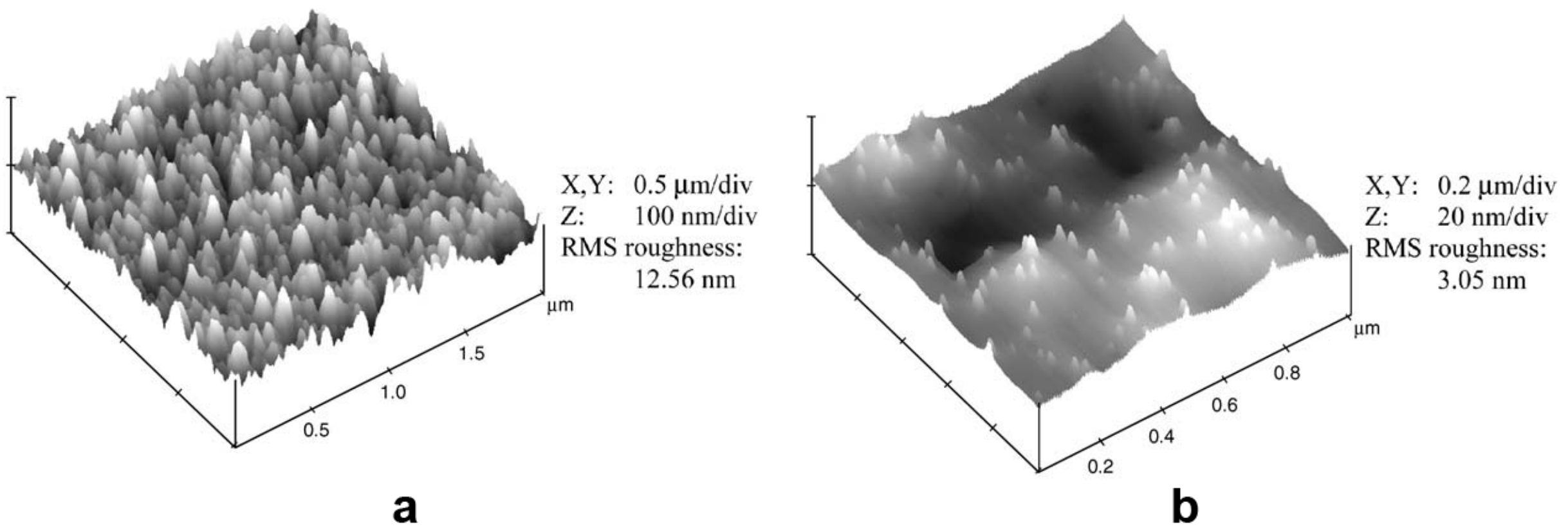

Figure 1.4 Deux images AFM de la surface de Si poreux après la préparation (a) et après le recuit sous une température de 1050°C durant 30 minutes sous une atmosphère d'hydrogène. [106]

Les procédés comme le « Smart-cut », « Ion-cut », « Via-hole etching for the separation of thin film (VEST) » et « Epilift » se situent dans le deuxième groupe où aucune porosification n'est utilisé. Les procédés « Smart-cut » et « Ion-cut » sont basés sur l'implantation d'ions. Le processus « VEST » utilise l'oxydation du Si et sa recristallisation par fusion de zone (ZMR). Le processus « Epilift » utilise des masques d'oxyde de Si pour l'épitaxie à phase liquide (LPE). Dans toutes les méthodes mentionnées, la croissance de Si est effectuée par CVD, LPE ou par dépôt assisté par ions (IAD), sauf dans les cas du « Smartcut » et de l'Ion-cut où une croissance n'est pas nécessaire. L'ensemble des méthodes de transfert de couches qui utilisent la porosification, classifiées selon la séquence des étapes de fabrication, est présenté à la Figure 1.5.

Le procédé de transfert de couches de Si pour la fabrication des cellules solaires a été aussi développé à l'INL dans le cadre des thèses de Sévak Amtablian [6] et de Jed Kraiem [63]. Des transferts ont été réalisés avec succès à sept reprises en utilisant le même substrat de départ, de diamètre 50 mm [63]. La consommation de Si est estimée à 5 µm par cycle et la rugosité du substrat recyclé plafonne à 8 nm. L'adaptation de la même procédure aux substrats de diamètre 100 mm a pu être également démontrée [6].

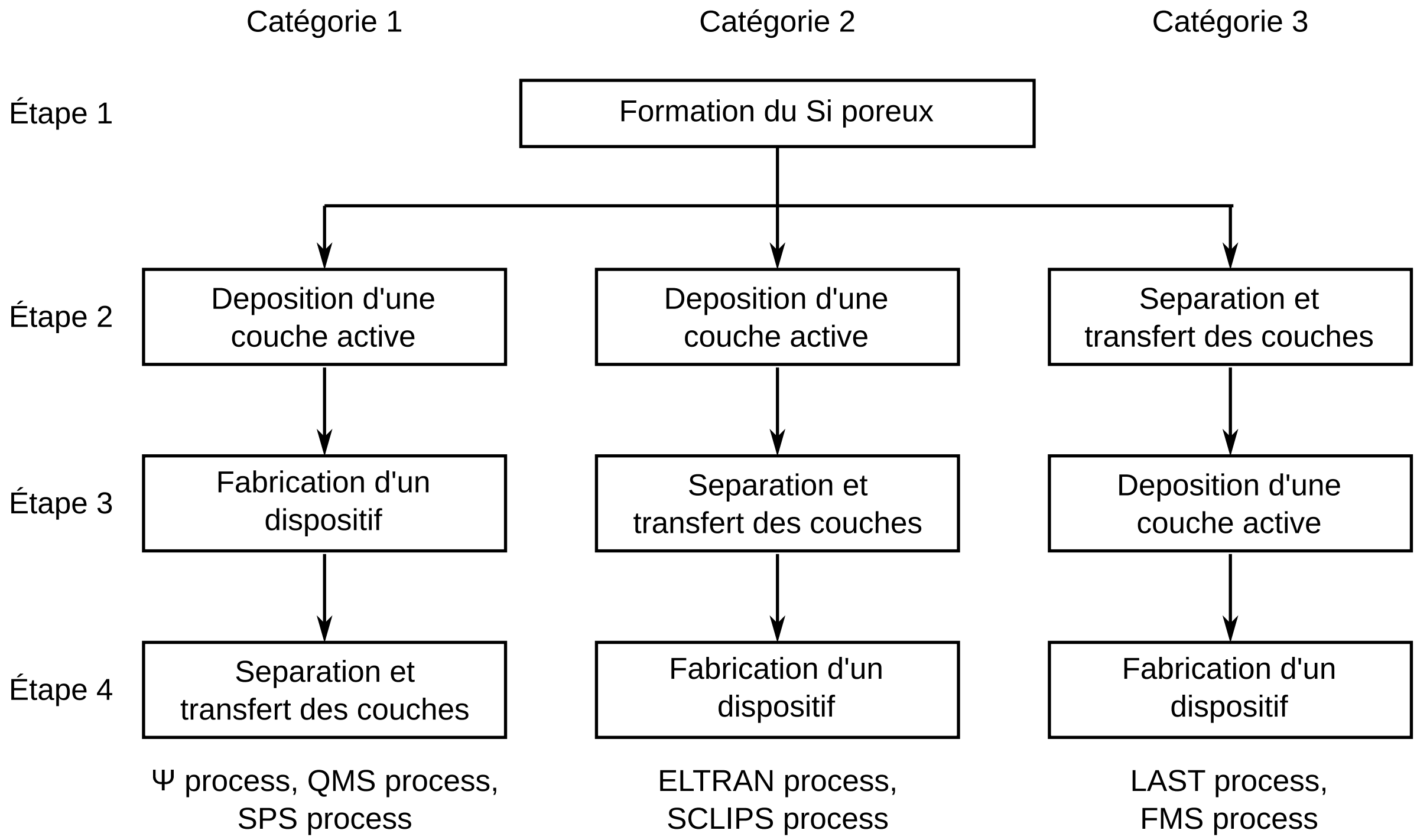

Figure 1.5 Classification de méthodes de transfert de couches de Si. Chaque processus comprend 4 étapes principales. [106]

La séquence des opérations utilisées pour ce procédé est présentée sur la Figure 1.6. Cette méthode fait partie de la catégorie 1 de la classification présentée sur la Figure 1.5.

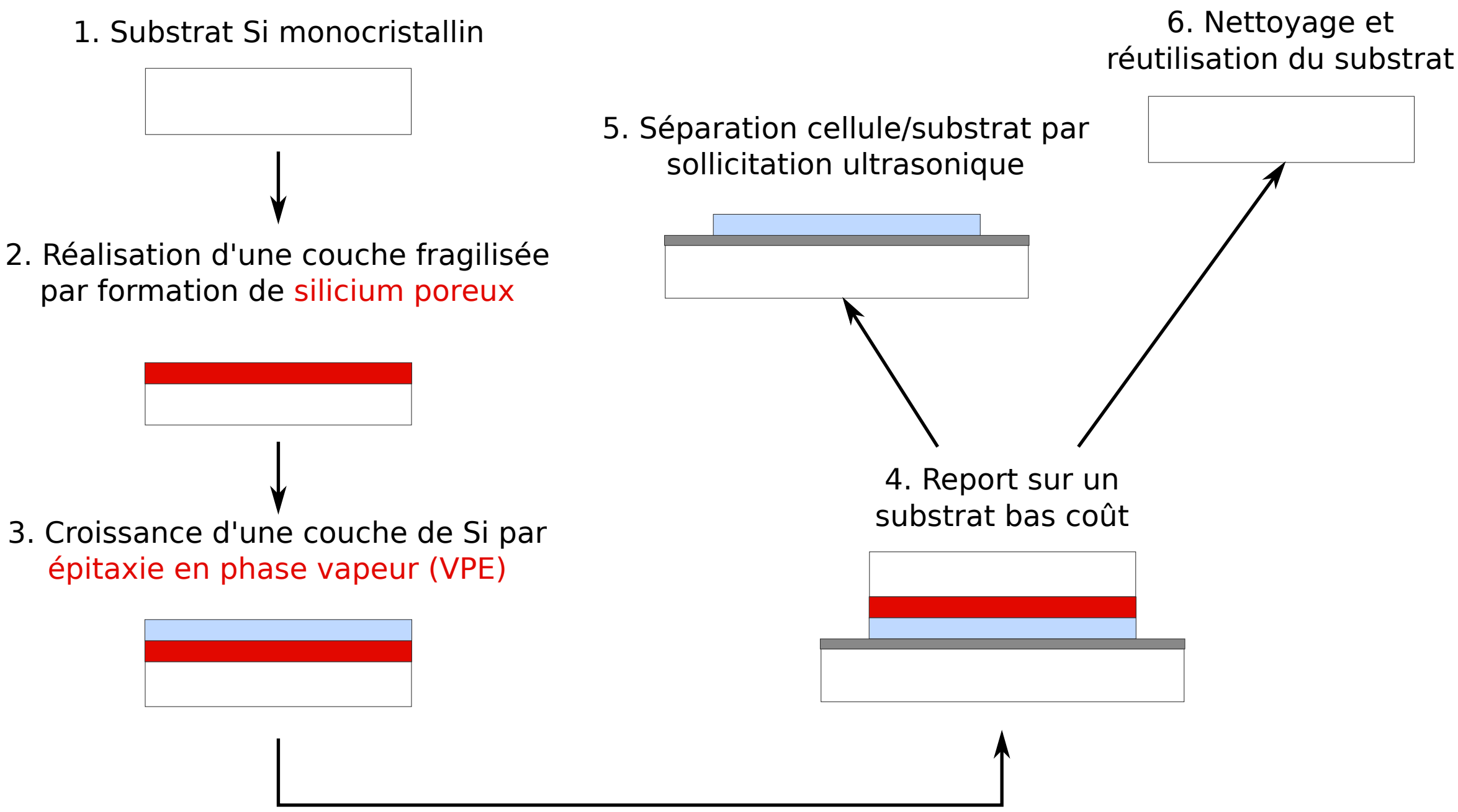

Figure 1.6 Procédé de transfert de couche mince de Si monocristallin développé à l'INL [6],[63]

Les caractéristiques de la bicouche formée sont :

- Une couche de faible porosité présentant une porosité voisine de 25% et une épaisseur d'environ 1.1 µm.
- Une couche de forte porosité présentant une porosité voisine de 70% et une épaisseur d'environ 700 nm.

Le tableau 1.3 synthétise les techniques de transfert de couche utilisant le Si poreux.

| | ELTRAN -SCLIPS | SPS | PSI | FMS | QMS | INL |
|---|---|---|---|---|---|---|
| Structure poreuse | Multicouche 10 µm | Tricouche 10 µm | Bicouche 1 µm | Monocouche 20–50 µm | Bicouche >25 µm | Bicouche 2 µm |
| Épitaxie | Oxydation + LPE | CVD | CVD | CVD | - | CVD |
| Séparation | Jet d'eau après l'épitaxie | Action mécanique | | Électro -polissage | Recuit haute T°C | Action mécanique (ultrasons) |

Tableau 1.3 Récapitulatif des techniques de fabrication de couches minces utilisant le Si poreux [6]

### 1.2.2 Report des couches utilisant le Ge-poreux

Une possibilité de séparation d'une couche active de Ge en vue d'application photovoltaïque a été démontrée par E Garralaga Rojas et al. [40, 97]. Contrairement à une approche classique impliquant une formation de deux couches de porosité différente, une seule couche poreuse a été formée sur la surface de Ge (Figure 1.7). Lorsque le temps de gravure est assez long (>10h), l'épaisseur de la couche poreuse atteint 900 nm. Lorsqu'une inhomogénéité de porosité apparait, le haut et le bas de la couche se recristallisent différemment lors d'un recuit. On peut observer une faible rugosité de la surface après le recuit (<2 nm), ce qui est favorable pour l'épitaxie subséquente. En revanche, les auteurs ne précisent pas si la cellule solaire est réalisée avant ou après la séparation, ce qui ne permet pas de classer ce procédé sur le diagramme 1.5. La température de recuit étant assez modéré (400–500°C) inférieure aux températures d'épitaxie typiques (600–700°C). Suite à des résultats de recuit présentés dans le travail [40], il y a un risque d'effondrement de la couche à forte porosité et donc, de la structure entière lorsqu'elle est exposée à des températures au-delà de 500°C. Dans ce procédé, le détachement de la couche poreuse est de nature spontanée, tandis qu'un procédé de transfert des couches nécessite un détachement contrôlé.

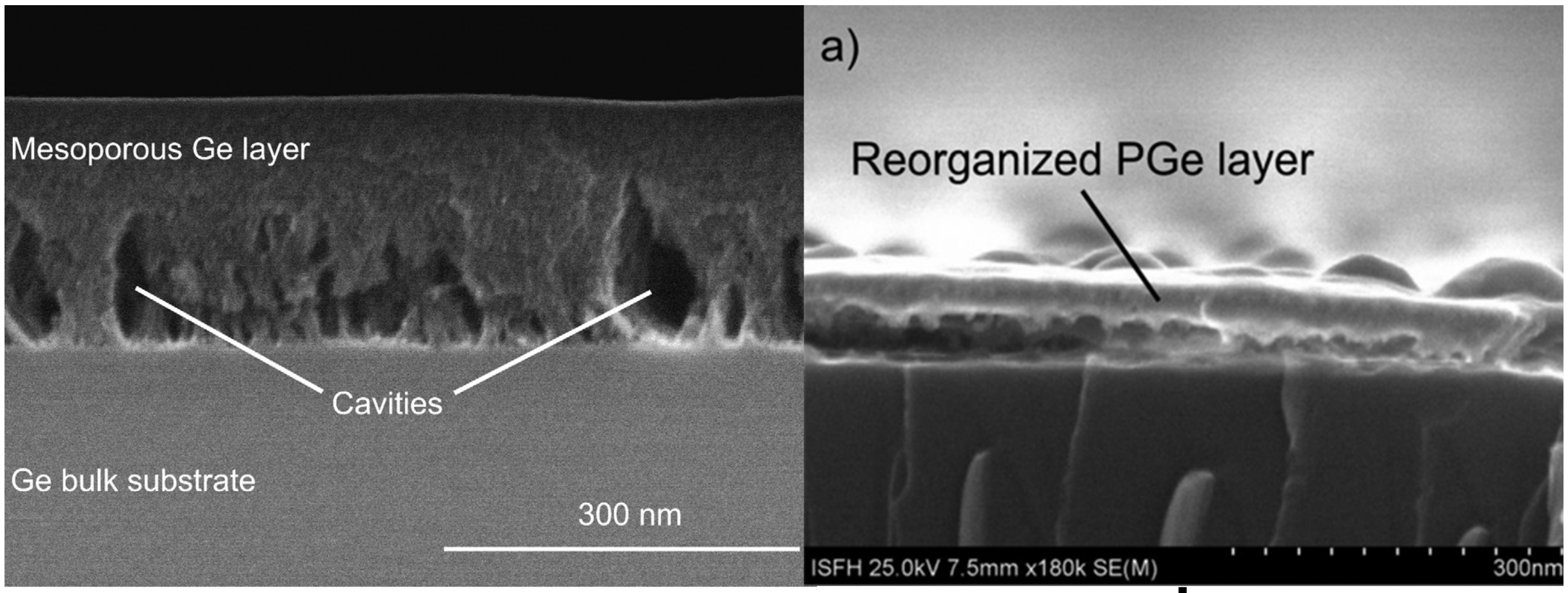

Figure 1.7 (a) Image MEB transversale de Ge mesoporeux gravé pendant 22 h dans $HF_{50\%}$ avec une densité de courant de 0.5 $mA/cm^2$ ; (b) Image MEB de la couche poreuse détachée spontanément après recuit [97].

Il apparaît ainsi que le procédé de porosification du germanium présente plusieurs limitations, c'est pourquoi ce travail de thèse s'est donné pour objectif de développer un nouveau procédé de porosification.

## 1.3 Germanium poreux

Le silicium poreux a été découvert par hasard en 1956 par Arthur Uhlir Jr. et Ingeborg Uhlir au Bell Labs aux Etats-Unis lors de développement d'une technique de polissage et de texturisation des surfaces de silicium et de germanium. Cependant, il a été constaté que, sous plusieurs conditions un produit brut sous la forme d'un film noir, rouge ou brun était formé sur la surface du matériau.

Après la découverte en 1990 d'une photoluminescence intense de Si poreux nanostructurés à la température ambiante [18, 28], les semi-conducteurs poreux ont suscité un grand intérêt grâce à leur potentiel d'applications en nanoélectronique, nanophotonique, photocatalyse, ainsi que dans les domaines de la conversion d'énergie et de la détection biologique [34].

Une grande attention a également été accordée au Ge. En effet, le Ge comme le Si, est un semi-conducteur avec une large gamme de propriétés qui pourrait susciter de nouvelles applications. Il possède l'une des constantes diélectriques les plus élevées ($\varepsilon(\text{Ge}) \approx 16.2$) parmi les semi-conducteurs classiques, un indice de réfraction très élevé ($n(\text{Ge}) \approx 4.14$ dans le proche infrarouge), et est donc d'un intérêt notable pour des applications dans les cristaux photoniques [30]. L'absence d'un oxyde stable permet une fonctionnalisation de sa surface assez facile [72]. La réduction dimensionnelle de Ge à une échelle nanométrique induit l'apparition de nouvelles fonctionnalités, telle qu'un confinement quantique [85]. Les couches poreuses avec des pores cristallographiques (par exemple, avec une morphologie colonnaire) sont d'un intérêt particulier pour leurs propriétés de biréfringence [44].

Dans le cadre de cette thèse, le Ge mésoporeux est développé en vue de la réalisation de transfert de couche pour des applications photovoltaïques, mais l'élaboration et la maîtrise de ce nouveau matériau contribuerait largement à la science des matériaux et pourrait servir pour les nombreuses applications mentionnées ci-dessus.

### 1.3.1 Caractérisation d'un solide poreux

Tout matériau solide qui contient des cavités, des canaux ou des interstices peut être considéré comme poreux, mais dans un contexte particulier, une définition plus restrictive peut être appropriée. Avant de faire le point sur l'état de l'art du développement de Ge poreux, il est nécessaire d'introduire une terminologie et les paramètres qui caractérisent une couche poreuse.

#### Paramètres quantitatifs.

Les matériaux poreux sont caractérisés par plusieurs paramètres quantitatifs [100].
La **porosité** $P$ est un rapport du volume total de pores $V_p$ au volume de la couche poreuse $V$. S'il s'agit d'une poudre de particules poreuses, $V$ est le volume apparent de particules excluant les vides interparticulaires.

$$P = \frac{V_p}{V} \tag{1.1}$$

Dans certains cas, on peut distinguer entre la porosité ouverte (c'est à dire le volume des pores accessibles à une sonde donnée) et une porosité fermée. Les méthodes utilisées pour mesurer $V_p$ et $V$ doivent être indiquées. Le volume $V$ d'une couche poreuse homogène est un produit simple de la **surface porosifiée** avec **l'épaisseur de la couche**.

3 paramètres différents se réfèrent à la **densité** $\rho$ :

$\rho_{vraie}$ : la densité de matière hors pores et vides interparticulaires ;
$\rho_{apparente}$ : la densité de matière, y compris les pores fermés et inaccessibles ;
$\rho_{volumique}$ : la densité de matière, y compris les pores et les vides interparticulaires.

La porosité peut être alors exprimée en termes de $\rho$ :

$$P = 1 - \frac{\rho_{volumique}}{\rho_{vrai}} \tag{1.2}$$

La **surface totale** du matériau mesurable par expérience est celle des pores ouverts ($A_p$) et de la surface libre (hors pores fermés et inaccessibles).

La **taille des pores** $L_p$ (en général, la largeur de pores) est une distance entre deux parois opposées d'un pore. Une valeur $L_p$ a un sens précis quand la forme géométrique des pores est bien définie et connue (par exemple cylindrique, sphérique, en forme de fente, etc). Néanmoins, la taille limite d'un pore est celle de sa plus petite dimension (la largeur d'un pore en forme de fente, le diamètre d'un pore cylindrique, le diamètre d'un pore sphérique). Par analogie, $L_{cr}$ — est la **taille d'un cristallite** formé entre les pores.

Finalement, la densité de tassement de pores est donnée par le paramètre $L_{ip}$ — **l'espacement entre les pores**, qui est défini comme une distance entre les centres de deux pores voisins.

Selon la classification par UICPA (Union internationale de chimie pure et appliquée), il y a 3 types des pores [104] :

- Micropores ($L_p < 2$ nm) ;
- Mésopores ($L_p = 2$–50 nm) ;
- Macropores ($L_p > 50nm$).

Un matériau poreux homogène, dont la plupart des pores est de même classe, est appelé matériau *micro/nanoporeux*, *mésoporeux* ou *macroporeux*.

La **distribution de la taille des pores** $\chi(L_p)$ est représentée par les dérivés $\frac{dA_p}{dLp}$ ou $\frac{dV_p}{dLp}$ en fonction de $L_p$. La **distribution de la taille des cristallites** $\varphi(L_{cr})$ est determinée d'une manière similaire.
Les **valeurs moyennes** de $L_p$, $L_{cr}$ et $L_{ip}$ à l'intérieur d'une couche poreuse uniforme sont données par les paramètres $d_p$, $d_{cr}$ et $d_{ip}$ respectivement.

Les cristallites, étant séparés par des pores, sont interconnectés par des isthmes cristallins de largeur moyenne $d_c$. Ce dernier paramètre est particulièrement important, car il détermine la capacité de transfert du courant et de l'énergie thermique entre des cristallites et alors au travers de la couche poreuse.

Les caractéristiques présentées ci-dessus, cependant, ne portent pas d'information sur la morphologie d'une couche poreuse.

**Description qualitative — Morphologie.**

La forme des pores individuels et leur arrangement dans un solide détermine une morphologie de matériau.

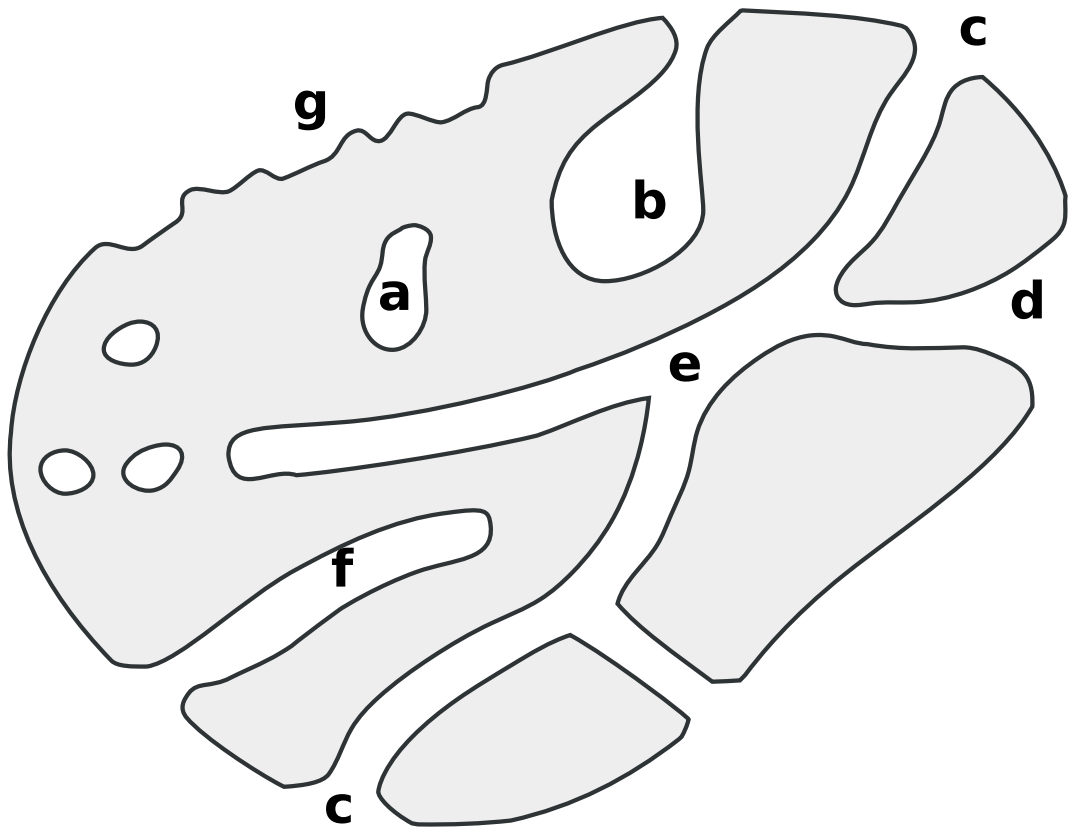

Figure 1.8 Section transversale schématique d'un solide poreux [100].

À l'aide de la Figure 1.8, les pores peuvent être classés en fonction de leur disponibilité à une fluide externe [100]. De ce point de vue, une première catégorie de pores corresponde à ceux totalement isolés de leurs voisins, comme dans la région (a) ; ce sont des pores fermés. Ils influencent les propriétés macroscopiques comme la densité apparente, la résistance mécanique et la conductivité thermique, mais sont inactifs dans des processus tels que l'écoulement du fluide et l'adsorption de gaz. D'autre part, les pores qui ont un canal de communication continu avec la surface externe du corps, comme (b), (c) (d) (e) et (f), correspondent à des pores ouverts. Certains peuvent être ouverts à une seule extrémité (comme b) et (f) ; ils sont alors désignés de pores aveugles (i.e. sans issue, ou sacciformes). D'autres peuvent être ouverts aux deux extrémités (à travers les pores), comme autour de (e).

Les pores peuvent également être classés selon leur forme : ils peuvent être de forme cylindrique (soit ouverte (c) ou sans issue (f)), en forme de bouteille d'encre (b), en forme d'entonnoir (d), ou en forme de fente. Une surface rugueuse (g) ne peut être considérée comme poreuse qu'à condition que les pores soient plus profonds que larges.

Lorsque la gravure est faite à l'aide d'un courant électrique (gravure électrochimique), les pores sont répartis en deux groupes majeurs :

- Les **pores cristallographiques** se forment de manière indépendante de l'orientation planaire et de la direction actuelle. La forme de prisme permet d'exposer les « plans d'arrêt ».

- Les **pores de lignes de courant** se forment en suivant la direction des flux de courant.

Les pores cristallographiques sont souvent influencés par le courant. Par exemple, parmi 8 directions cristallographiques d'une famille <133>, les pores se propagent le long de ceux qui forment un produit scalaire négatif avec le vecteur du champ électrique.

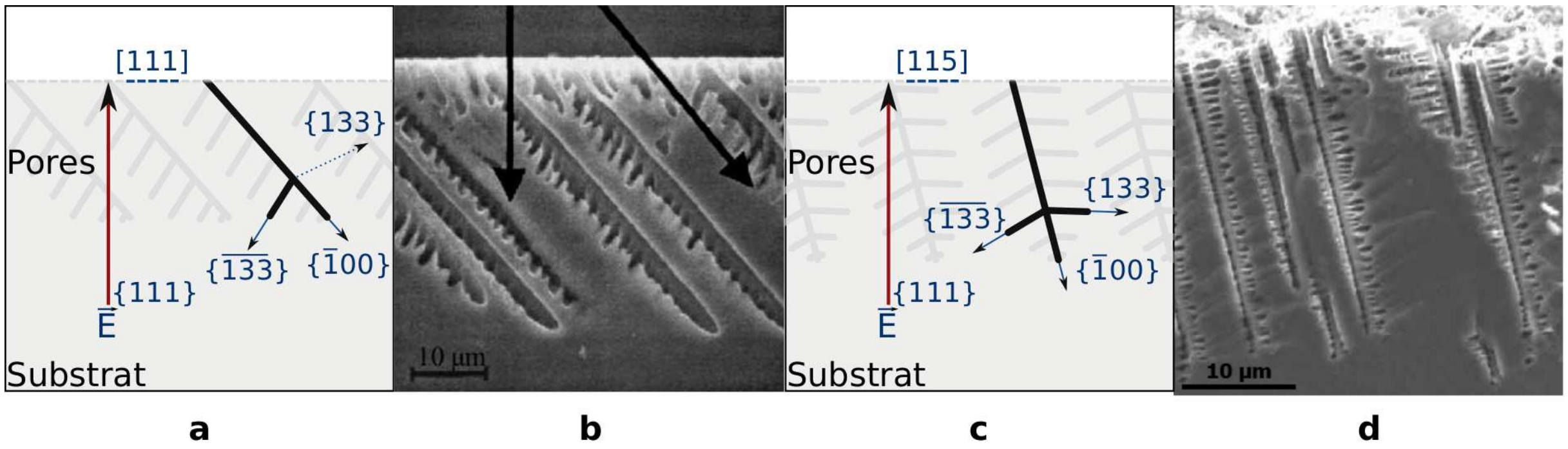

Figure 1.9 (a,c) : illustration de la règle de sélection des directions de propagation de pores cristallographiques ; (b,d) : images MEB en coupe transversale de macropores cristallographiques dans le Si type n [34].

Dans le Ge, des pores qui suivent les directions <100> et <111> en fonction de l'orientation du substrat ont été rapportés auparavant. Les plans d'arrêt sont {110}, mais sont aussi parfois {100}. Cependant, aucun pore cristallographique n'a été rapporté à ce jour pour le Ge.

Désormais, dans le texte, la nomination de pores fera référence à leur aspect visuel (« pyramides », « éponge », « arbres », etc.) accompagnée par une explication en termes de caractéristiques présentées ci-dessus.

## 1.3.2 Les propriétés spécifiques du Ge

- Le Ge possède la constante diélectrique la plus élevée parmi les semi-conducteurs conventionnels : $\varepsilon(\text{Ge}) = 16.2$, $\varepsilon(\text{Si}) = 11.7$, $\varepsilon(\text{GaAs}) = 12.9$ ; $\varepsilon(\text{InP}) = 12.5$ ;

- Sa bande interdite est peu élevée (0.66 eV).

- La concentration de porteurs de charge intrinsèques est 1000 fois plus élevée que dans le Si ($2 \cdot 10^{13}$ cm$^{-3}$ vs $1 \cdot 10^{10}$ cm$^{-3}$).

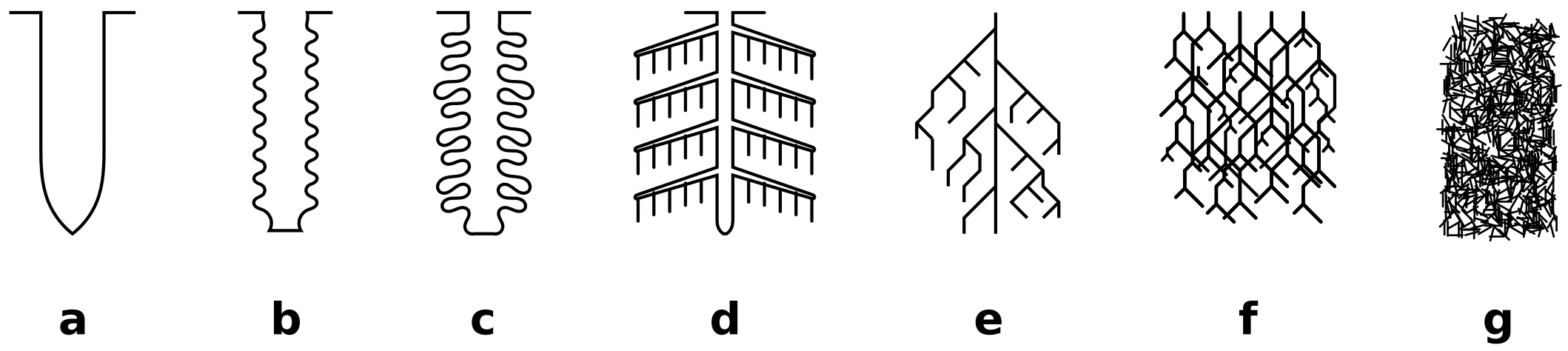

Figure 1.10 Classification des pores en fonction du degré de ramification, inspirée de la référence [122] :
(a) Parois de pores lisses ;
(b) Branches plus courtes que le diamètre ;
(c) Branches de deuxième niveau seulement ;
(d) Pores principaux avec les branches de deuxième et de troisième niveaux ;
(e) Branches dendritiques : ramification récursive des branches latérales ;
(f) Pores dendritique lorsqu'un tronc principal n'est pas prononcé ;
(f) Branches denses, aléatoires et courts. Souvent, un cas extrême de pores dendritiques fortement interconnectés.

- Électrocapillarité : la surface peut passer de l'état hydrophile (probablement dû à la présence de liaisons $-$H) à l'état hydrophobe (probablement dû à la présence de liaisons $-$OH) grâce à un changement du potentiel électrique appliqué [21]
- Le Ge n'est pas polaire (comme le Si) et son oxyde est instable (comme pour les matériaux III-V) Ceci complique la gravure, mais facilite la fonctionnalisation de surface [72].
- Le Ge est insoluble dans les acides et bases dilués, mais se dissout lentement dans l'acide sulfurique concentré.
- Les anions sont adsorbés par la surface de Ge dans l'ordre suivant :
  $I^- \longrightarrow Br^- \longrightarrow Cl^- \longrightarrow F^- \longrightarrow SO_4^{2-}$ [49].
- Le plan de clivage n'est pas toujours bien défini : normalement $\{110\}$.
- La dissolution uniforme et rapide du matériau, même durant sa croissance, est stable [31, 33, 98].
- La germination et la croissance de pores se comportent de manières différentes sur les surfaces polies, rugueuses et mécaniquement endommagées [31].
- Les pores sont souvent remplis de $GeO_2$ [31].
- Il n'y a pas d'auto-ordonnance des pores, même s'il y a des oscillations de voltage et de courant [31].

### 1.3.3 Gravure électrochimique du Ge

La gravure électrochimique est une méthode de porosification la plus fréquemment utilisée. Ce procédé est basé sur l'oxydation d'atomes superficiels et une dissolution de produits dans une solution électrolytique. Il convient de rappeler que, par « oxydation », on sous-entend une augmentation de l'état d'oxydation de l'atome lorsqu'il réagit avec un agent oxydant. Dans le cas de la gravure d'un semi-conducteur, les atomes sont oxydés lors de l'injection de trous, qui se localisent sur des liaisons covalentes des atomes superficiels.

Du point de vue de l'électrochimie, une porosification électrochimique ne diffère pas de l'électropolissage. Le mode effectif dépend des conditions du procédé, telles que la densité de courant anodique et la composition de l'électrolyte. Trois régions (porosification, transition et électropolissage) peuvent être identifiées sur le plan log(j)-log|HF| (Figure 1.11 (b)). Dans la réference [105], les limites de la région de transition ont été déterminées depuis la position du pic de courant de la courbe 1.11 (a), tandis que dans la référence [89], l'état de la surface gravée est déterminée par des mesures AFM.

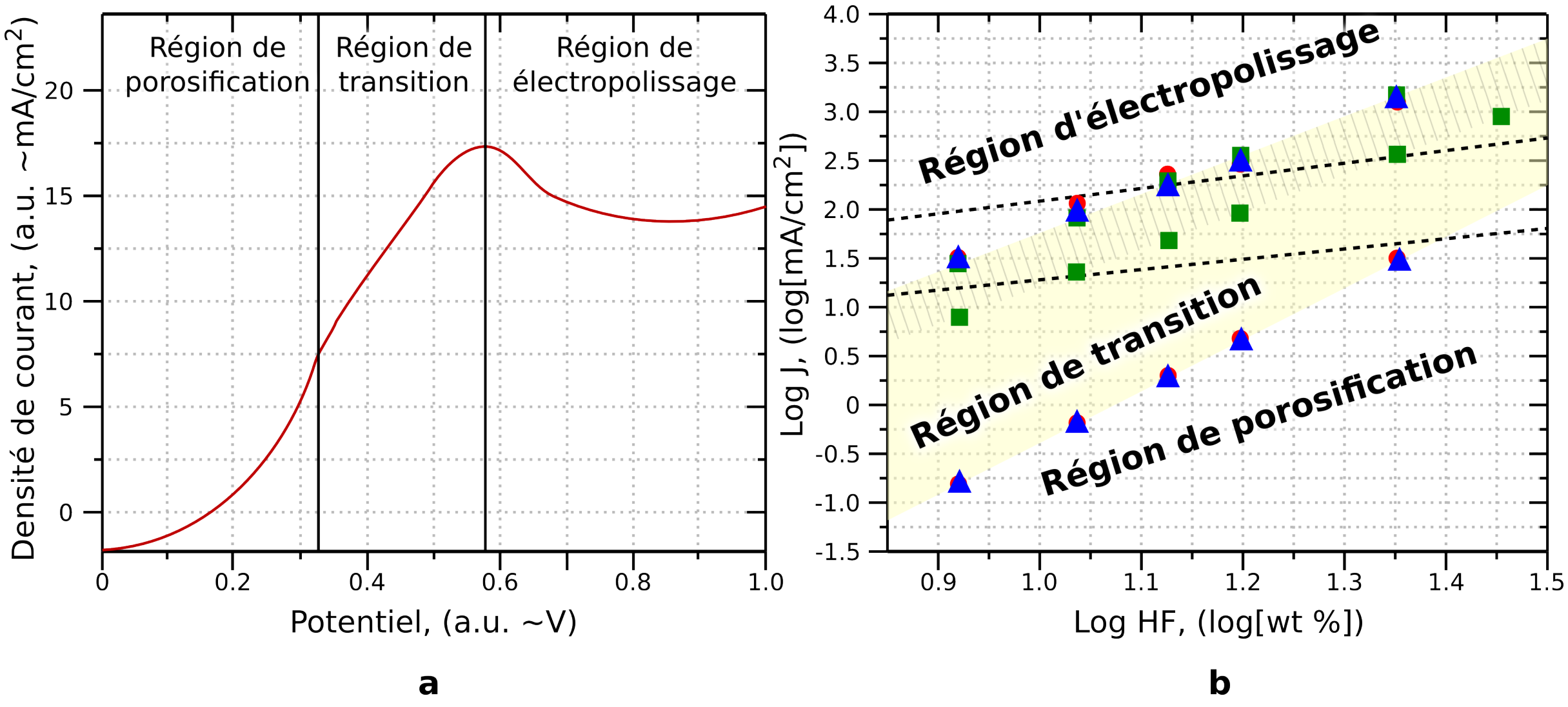

Figure 1.11 (a) Courbe j(V) typique de Si dans l'HF. Les unités exactes des échelles dépendent du dopage du substrat et de la composition de l'électrolyte. (b) Cartographie topologique log(j)-log|HF| des régions de porosification, transition et électropolissage [89]. Les données de [105] sont indiquées par les lignes en pointillés.

Un mode de transition signifie un régime de formation de fosses en surface avec une profondeur à peu près égale à la largeur.

Plusieurs effets sont responsables de la formation de pores [34, 105] :

- la germination sur des défauts volumiques ou de surface ;
- le transfert de porteurs de charge dans les cristallites lorsqu'une partie du cristallite est occupé par la zone de charge d'espace (ZCE) ;
- la distribution inhomogène du champ électrique autour des extrémités des pores dont le rayon de courbure de la surface est plus petit ;
- des effets quantiques dans les cristallites lorsque leurs dimensions sont de l'ordre nanométrique ;
- claquage locale de la ZCE, etc.

Différents modèles expliquent la formation des différents types de pores. Cependant, une théorie générale qui explique tous les phénomènes observés n'est pas encore établie.

Une réaction de dissolution anodique ne se déroule que si les produits de réaction attendus sont solubles. Ainsi, l'anodisation du Si nécessite la présence d'anions de fluorine dans l'électrolyte. Consécutivement, les électrolytes sont classifiés par rapport à la présence d'HF :

- Les électrolytes aqueux ont pour base un mélange d'acide fluorhydrique (HF) ou fluorure d'ammonium ($NH_4F$) et d'eau. La concentration de $F^-$ (sous n'importe quelle forme) peut varier de 0.001 à 49%. Les électrolytes peuvent contenir des additifs d'éthanol ($C_2H_5OH$), d'acide acétique ($CH_3COOH$) ou d'autres agents servant à réduire la tension superficielle, à ajuster le niveau de PH, à ajuster la viscosité ou encore à aider à obtenir les résultats désirés pour d'autres raisons. Les électrolytes aqueux ont un fort pouvoir oxydant.
- Les électrolytes organiques ont pour base un mélange de HF avec un solvant organique. Ce type d'électrolyte contient toujours de l'eau qui vient du HF dilué à 49%. Une grande variété de solvants peut être utilisée. Parmi les plus utilisés, l'acétonitrile (MeCN), diméthylformamide (DMF), et diméthylsulfoxyde (DMSO) peuvent être soulignés. Même si la différence entre les électrolytes aqueux et les électrolytes organiques n'est pas évidente dans cette classification, les électrolytes de chaque type démontrent des propriétés électrochimiques très différentes. Le pouvoir oxydant des électrolytes organiques est beaucoup plus modéré que celui les électrolytes aqueux.
- Les électrolytes pour l'oxydation anodique ne contiennent jamais d'ions $F^-$, mais contiennent des agents oxydants. Leur pouvoir d'oxydation dépend de la composition exacte d'électrolyte. Ceci peut notamment être mesuré en appliquant un courant continu. Lorsque l'oxyde est formé (sans toutefois être dissout), le voltage doit être augmenté pour maintenir le courant fixe. Au début de cette expérience, les courbes $V(t)$ sont toujours linéaires et leur pente $dV/dt$ peut être utilisée comme mesure du pouvoir oxydant. Les électrolytes pour l'oxydation ont des applications limitées, mais ils ont un rôle clé dans la compréhension de l'électrochimie des semiconducteurs.

- Le dernier groupe d'électrolytes est une combinaison des 3 précédents. Par exemple, le $H_3PO_4$ devient un électrolyte de type oxydant avec l'ajout de HF. Les électrolytes qui ne contiennent pas d'eau [94] peuvent être d'un grand intérêt pour la gravure du Ge.

Le germanium est un semi-conducteur pour lequel l'électrochimie est bien étudiée. En dépit de ce fait, la réalisation de couches de Ge mésoporeux avec des caractéristiques prédéfinies est toujours un défi. Afin de mettre en évidence les points problématiques de porosification du Ge et pour justifier la stratégie de recherche choisie, le point sur l'état de l'art de réalisation des couches de Ge poreux sera fait dans les sections suivantes de ce chapitre.

### Gravure par courant continu

Les premières tentatives pour réaliser du Ge-poreux ont utilisé une gravure anodique sous polarisation continue identique au procédé utilisé pour produire du Si poreux. Une bonne revue de la littérature ainsi qu'une étude approfondie des conditions de porosification a été faite par Fang, Foll et Carstensen [31]. Les paramètres analysés : type de substrat (« p » et « n »), orientation de substrat ((100), (110), (110), (115)), composition d'électrolyte (aqueux HCl, HF et organique HCl/DMSO), concentration d'électrolyte, état de surface (polie ou rugueuse).

Une formation de macropores (Figure 1.12 (a)) avec des problèmes spécifiques pour le Ge, telle que la gérmination non uniforme et tendant à former des domaines (Figure 1.13) a été démontrée. L'utilisation de cette approche n'a pas conduit à la formation du Ge mésoporeux [33]. La passivation de surface insuffisante a été désignée comme le principal enjeu dans la formation du Ge poreux (PGe). En effet, le dioxyde de Germanium, étant instable, il est soluble dans le plupart des solvants, y compris l'eau. La surface de Ge après une désoxydation est couverte par des liaisons pendants et s'oxyde rapidement sous une polarisation anodique. Cela signifie une grande vitesse de dissolution anodique dans tout électrolyte contenant de l'eau. Il a été démontré, que la formation d'une couche poreuse est toujours accompagnée par une gravure uniforme du substrat (Figure 1.12 (b)).

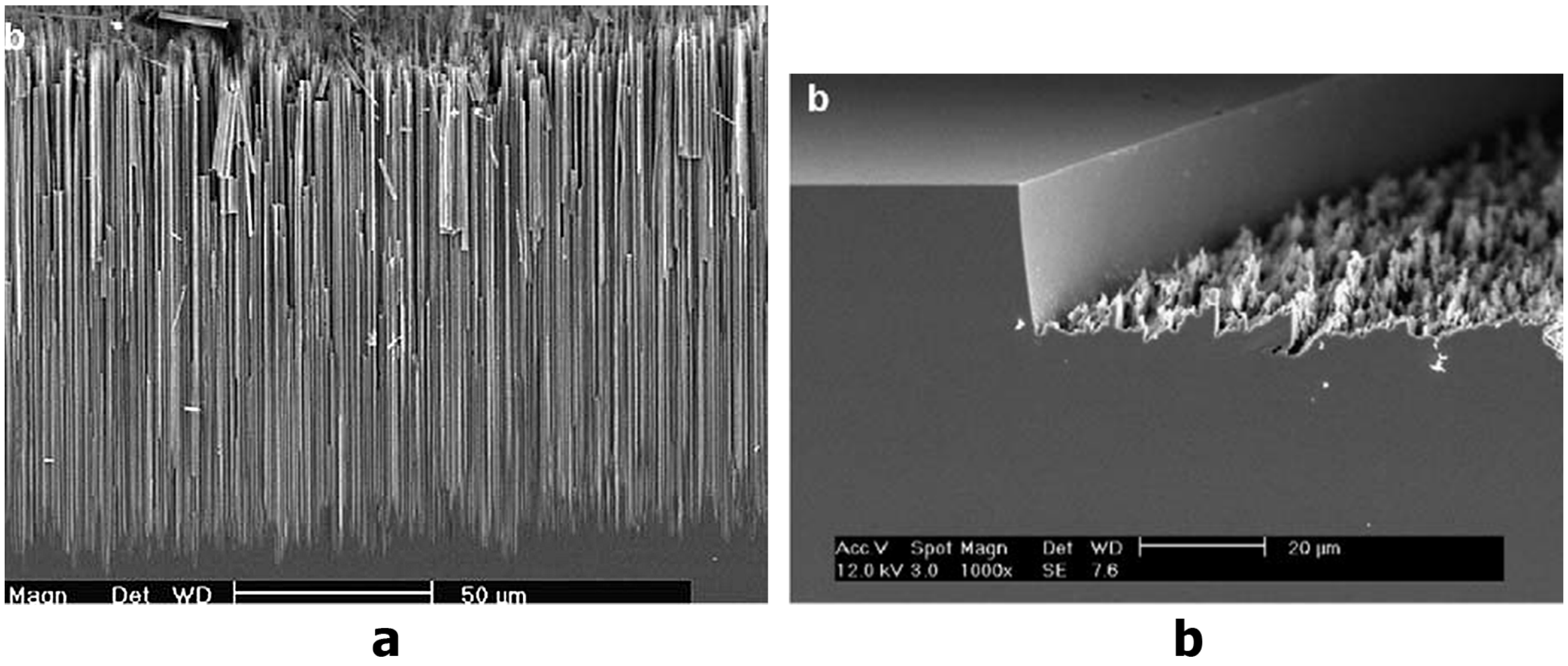

Figure 1.12 (a) image MEB en coupe transversale du Ge macroporeux en colonnes réalisé sur un substrat n-Ge (111) avec $HCl_{2.7\%}$ T=14 °C, t=1050 min. [31]
(b)image MEB de bord d'un échantillon démontrant la baisse du niveau de la surface suite à la dissolution uniforme qui agit en parallèle à la porosification.[31]

Il faut noter, qu'une possibilité de formation de macropores cristallographiques a été rapportée pour le Ge de type n indépendamment du niveau de dopage. Pour les substrats de type p fortement dopés, l'électro-polissage a lieu dans toute la gamme des courants continus. Dans le cas des substrats de type p faiblement dopés, une couche spongieuse rouge-brune fragile, dont le diamètre de pores $d_p$ = 0.5–1 µm, est formée (Figure 1.14).

La difficulté de réalisation d'une couche poreuse sur un substrat de type p est en cohérence avec nos propres résultats. Par analogie avec le Si (Figure 1.11), le Ge est attendu pour passer en mode de porosification lorsqu la concentration d'HF est élevée et que le courant est relativement faible. Cependant, une texturisation suivant les plans cristallographiques (Figure 1.15) ou même une formation d'une structure irrégulière (Figure 1.16) a été observée.

### Gravure bipolaire

Une autre technique de gravure électrochimique (Figure 1.17 (a)) permettant obtenir des couches épaisses de Ge macroporeux (1–15 µm) (Figure 1.17 (b)) a été proposée par H. C. Choi et J. Buriak [15, 23]. Dans ce procédé nommé par ses auteurs « gravure électrochimique bipolaire » (*Bipolar Electrochemical Etching*, BEE), l'étape de la gravure anodique a été suivi d'une étape de cathodisation. Sous polarisation anodique, la gérmination des pores a lieu, tandis que sous polarisation cathodique, les pores se propagent et leur surface est hydrogénée. La formation d'une couche poreuse sous une polarisation cathodique peut être probablement expliquée par protonation des liaisons covalentes entre les atomes de la surface et du volume du Ge. Comme l'énergie de liaison Ge-Ge est assez élevée, ce processus requiert l'application d'une tension de polarisation importante, ce qui est en cohérence avec la valeur du courant utilisé (350 mA/cm$^2$). Ainsi, il sera assez difficile d'utiliser cette méthode pour réaliser une couche de Ge mesoporeux.

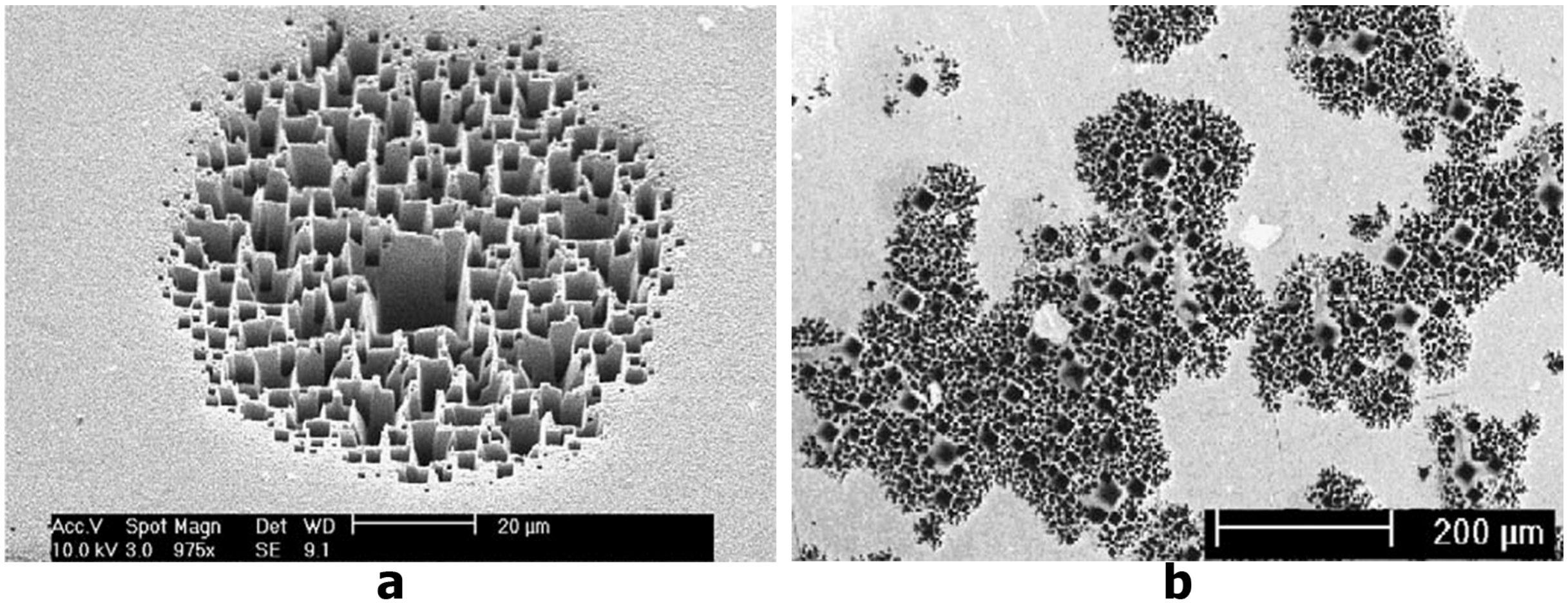

Figure 1.13 Images MEB en vue plane d'un domaine de pores (a) et d'une agglomération de domaines formés sur la surface polie de n-Ge ($1 \cdot 10^{16}$) polie dans $HCl_{5\%}$ aprés 2 h de gravure [31]

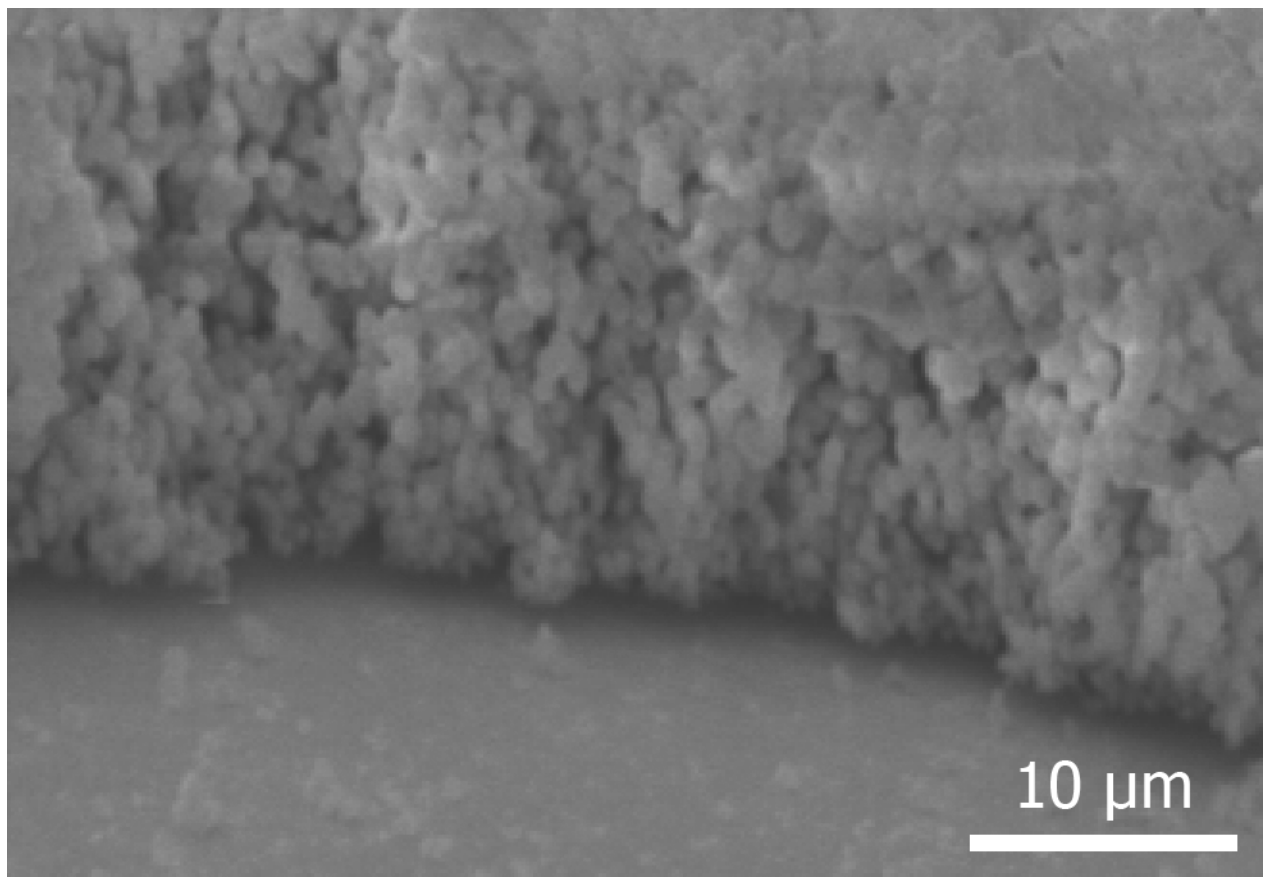

Figure 1.14 Image MEB en coupe transversale d'une couche macroporeuse de Ge en éponge réalisé sur un substrat de type p de forte résistivité [33]

L'idée d'alternance du courant a été par la suite développée par E. Garralaga Rojas et al. pour réaliser des couches de Ge mesoporeux (Figure 1.18 a). Dans ce procédé, la tension appliquée sur le substrat de Ge est commutée d'une polarisation anodique à une polarisation cathodique d'une manière cyclique (désormais, le terme « gravure bipolaire » désignera une telle commutation cyclique). Durant les impulsions anodiques les pores

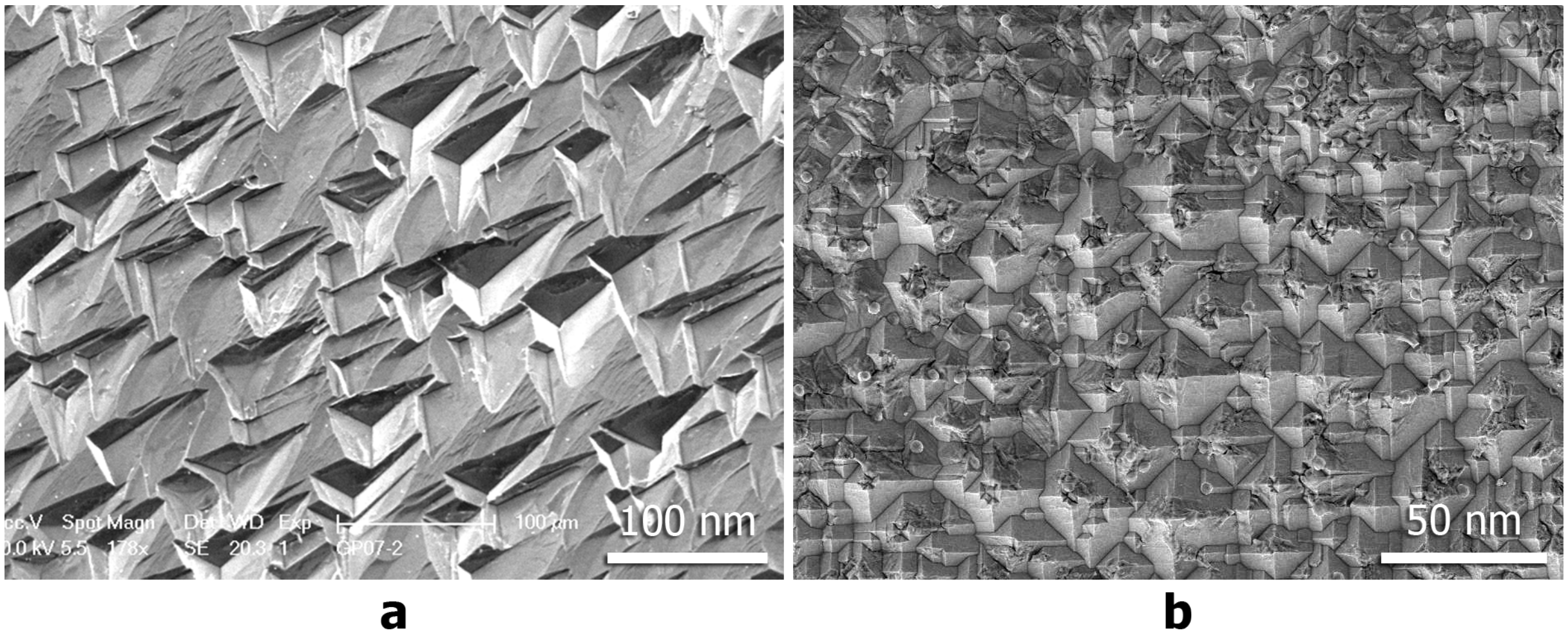

Figure 1.15 Images MEB en vue plane de surfaces texturisées de Ge en résultat de gravure dans l'$HF_{49\%}$ sous une densité de courant de 5 $mA/cm^2$ d'un substrat (111) type p à forte résistivité $\rho = 1–10\ \Omega \cdot cm$ (a) et d'un substrat type p fortement dopé $\rho = 0.06–0.015\ \Omega \cdot cm$ (b).

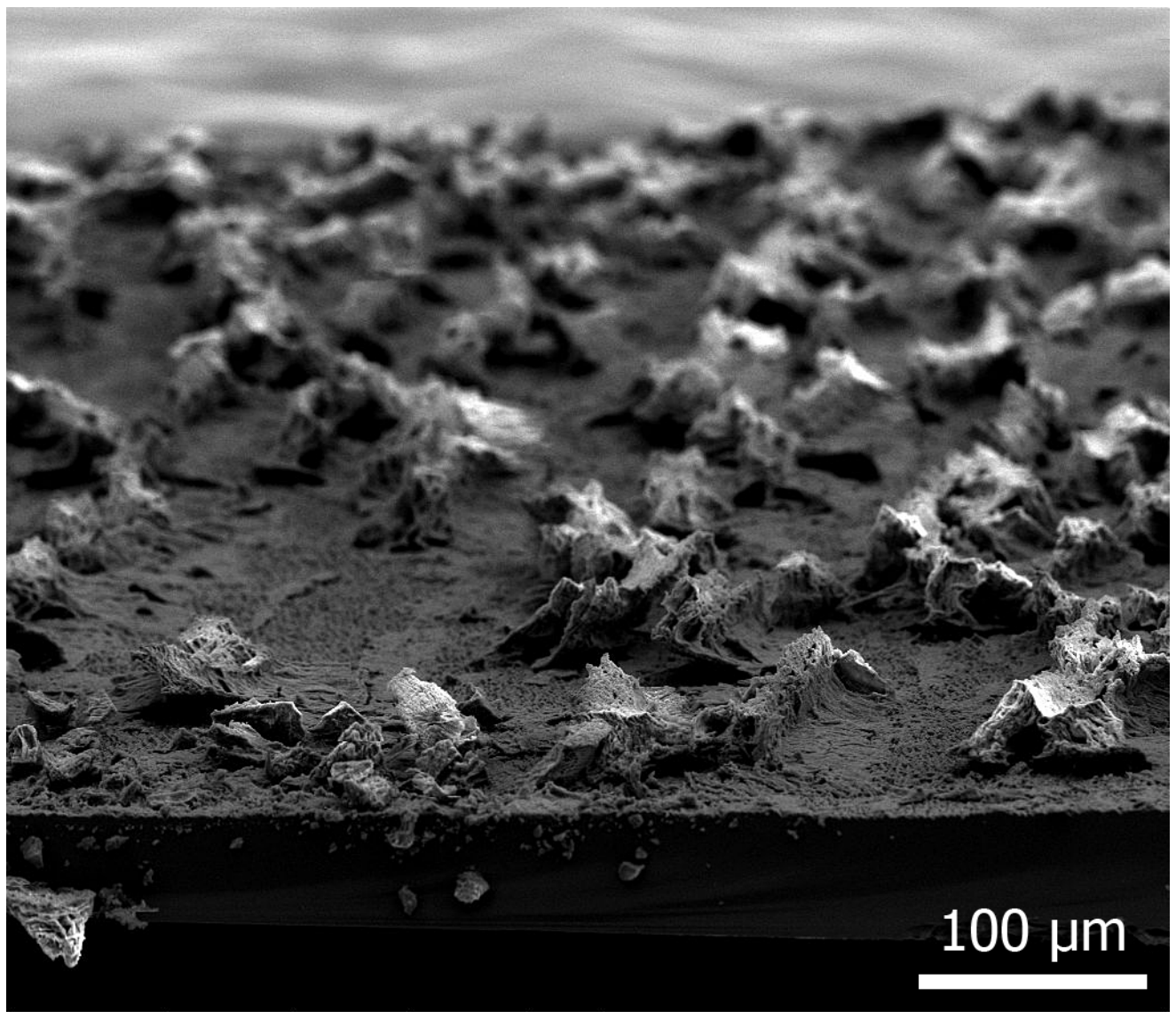

Figure 1.16 Image MEB d'un substrat incliné de Ge démontrant une destruction de sa surface en résultat de gravure dans l'$HF_{49\%}$ sous une densité de courant continu 10 $mA_{/}cm^2$ pendant 2 heures.

sont formés tandis que les impulsions cathodiques servent à passiver la surface interne des pores avec l'hydrogène. La passivation des pores avec $H^+$, assurée de cette manière, fournit une barrière énergétique pour le transfert des charges et, par conséquent, inhibe la dissolution de la couche poreuse formée. Contrairement au procédé de H. C. Choi et J. Buriak, l'amplitude des impulsions est assez faible, alors aucune gravure du matériau durant les impulsions négatives n'a lieu. Par ailleurs, il a été démontré que la porosité

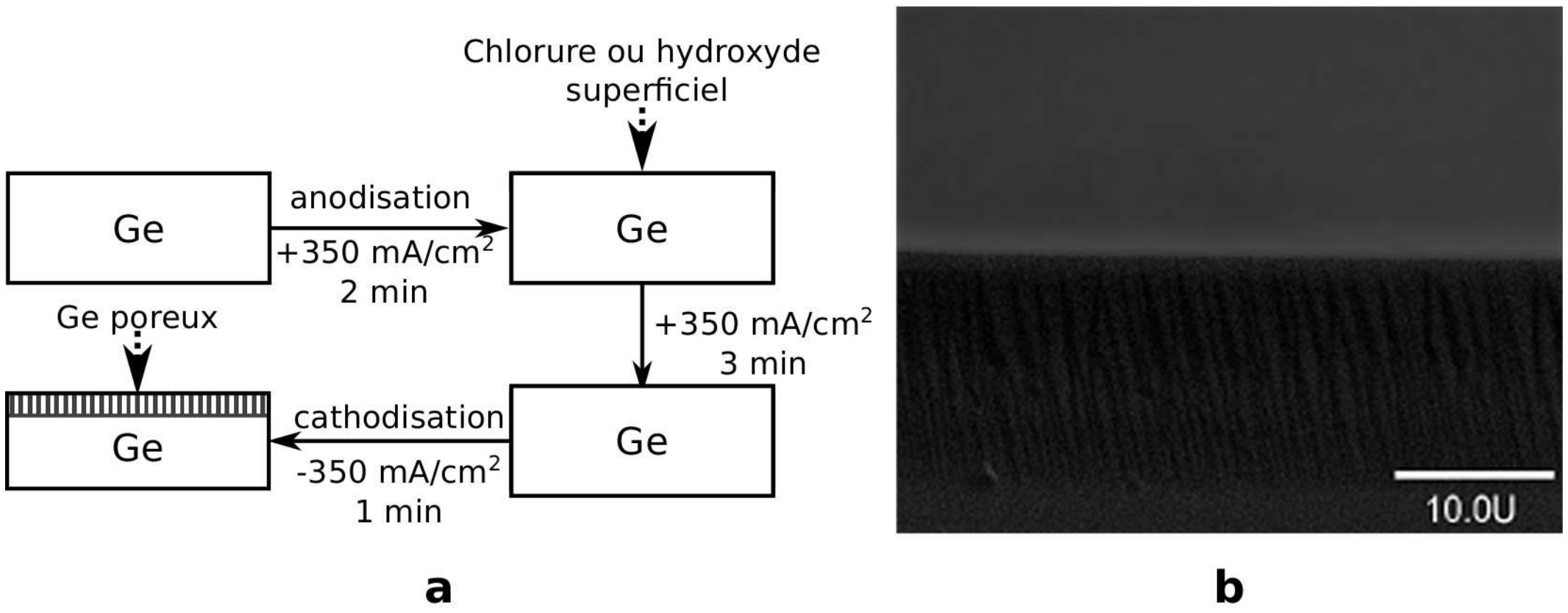

Figure 1.17 (a) Représentation schématique du procédé de gravure proposé par H. C. Choi et J. Buriak [15] ; (b) Une couche de Ge macroporeuse en colonnes a été réalisée par ce procédé.

ainsi que le taux de dissolution uniforme du substrat augmente rapidement lorsqu'une concentration d'HF diminue. Ainsi, l'$HF_{50\%}$ a été utilisé. Des couches poreuses réalisées par cette technique ont été utilisées pour le transfert de couches décrit dans la section 1.2.2. Une formation de bi- et de multi-couches est aussi possible (Figure 1.18 b).

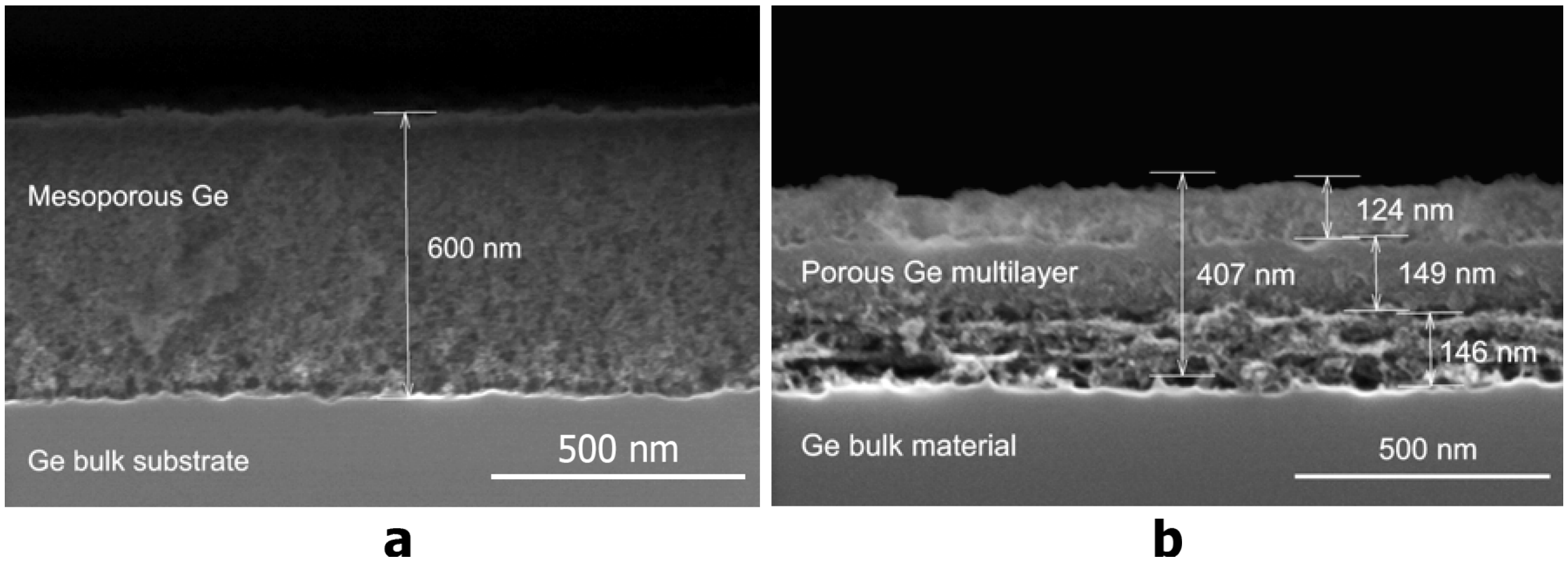

Figure 1.18 (a) couche spongieuse uniforme et (b) structure multi-couches de Ge mesoporeux formées sur un substrat de Ge type p ($10^{15}$), (100)

En même temps, l'aspect et les propriétés de ces couches sont loin de celles du Si mesoporeux typique. Les couches sont de la même morphologie — éponge isotrope. Cependant, la vitesse de gravure de Ge décroit rapidement lors de porosification. Le temps de gravure doit être supérieur à 10 h pour réaliser une couche d'épaisseur de plus que 500 nm. Le fait que les structures formées sont extrêmement minces (<1 µm) rend impossible l'étude approfondie des morphologies optimales de PGe pour assurer le cahier des charges du transfert des couches. Ainsi, après une longue gravure, un gradient de porosité apparait, ce qui indique que le problème de destruction de la surface des pores n'est pas résolu complètement.

En dépit des limitations mentionnées, le procédé BEE demeure le plus prometteur. Ce procédé a été choisi pour le développement ultérieure tout en reconsidérant le choix des conditions de gravure.

### 1.3.4 Autres méthodes de gravure de Ge

Il convient de mentionner d'autres méthodes de porosification qui ont été considérés mais que ce sont avérées mal adaptées à la réalisation de couches de Ge mésoporeux.

#### Gravure chimique — « stain-etching »

Bien que la méthode « stain-etching » n'utilise pas de source électrique, elle est électrochimique par nature. Les processus d'oxydoréduction sont décrits par les réactions anodiques et cathodiques locales (équation 1.5)[112].

$$\textit{Anode} : \begin{cases} \mathrm{Ge} + 2\mathrm{H_2O} + m\mathrm{h^+} \longrightarrow \mathrm{GeO_2} + 4\mathrm{H^+} + (4-m)\mathrm{e^-} \\ \mathrm{GeO_2} + 6\mathrm{HF} \longrightarrow \mathrm{H_2GeF_6} + 2\mathrm{H_2O} \end{cases} \tag{1.3}$$

$$\textit{Cathode} : \begin{cases} \mathrm{HNO_3} + 3\mathrm{H^+} \longrightarrow \mathrm{NO} + 2\mathrm{H_2O} + 3\mathrm{h^+} \end{cases} \tag{1.4}$$

$$\textit{Total} : \begin{cases} 3\mathrm{Ge} + 4\mathrm{HNO_3} + 18\mathrm{HF} \longrightarrow 3\mathrm{H_2GeF_6} + 4\mathrm{NO} + 8\mathrm{H_2O} + \\ + 3(4-m)\mathrm{h^+} + 3(4-m)\mathrm{e^-} \end{cases} \tag{1.5}$$

Lors de la réduction d'un agent oxydant (par ex. $HNO_3$), des trous sont injectés dans le volume du Ge (1.4). Ces trous sont par la suite utilisés pour oxyder le Ge (1.3). Similairement à une gravure électrochimique, une surface peut être polie ou texturisée dépendement de la concentration de trous (qui est proportionnelle à une concentration d'agent oxydant) et de la concentration d'HF.

Il existe un grand nombre d'articles sur la gravure chimique du Ge, mais ils ne font référence qu'aux propriétés optiques. Les rares images prises par MEB démontrent qu'il ne s'agit pas vraiment de couches poreuses.

Après la gravure, une surface très rugueuse avec beaucoup de cristaux cubiques, pyramidaux et sphériques est formée (Figure 1.19). Les cristaux possèdent une très grande dispersion en taille. Aucun ordre n'est observé. Le but de la plupart des articles est de mesurer les propriétés luminescentes des surfaces gravées. Les cristaux d'une taille de 1 µm (de 0.1 µm à 10 µm dépendamment de la méthode de gravure), avec une largeur de la bande interdite $E_g$ plus petite que celle du matériau volumique, sont formés sur la surface. Les cristaux sont isolés par du dioxyde de Ge. Il n'est donc pas surprenant que les spectres de PL soient décalés vers la région visible. Cependant, une telle surface est absolument inutilisable pour l'épitaxie.

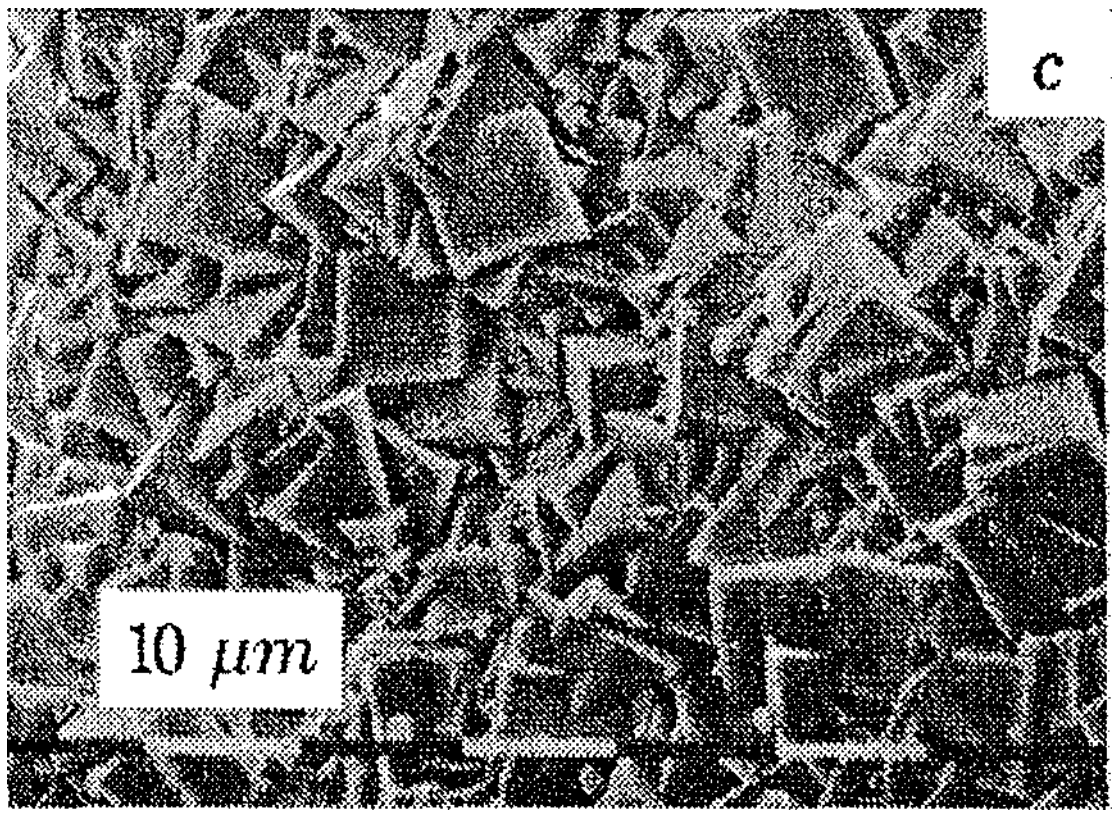

Figure 1.19 La surface de Ge gravé dans l'$HF_{40\%}$. [29]

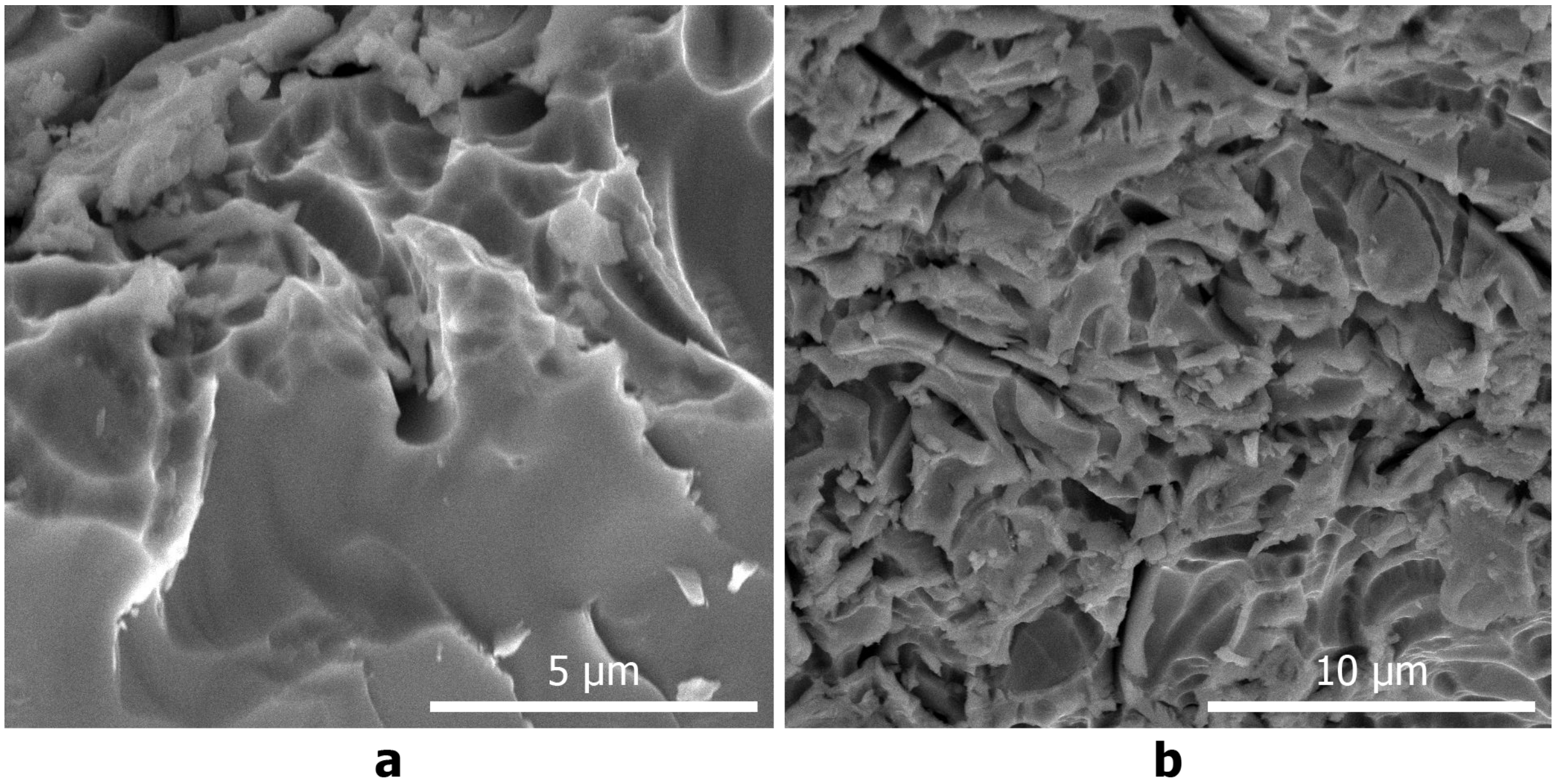

Figure 1.20 Une couche de Ge poreux réalisée par gravure chimique dans la solution HF : $H_3PO_4$ : $H_2O_2$ = 34 : 17 : 1 sur Ge (111) type p 1–10 $\Omega \cdot$ cm pendant 2 heures : (a) vue en coupe et (b) vue plane

La meilleure couche de Ge poreux réalisée dans le cadre de cette thèse par gravure chimique est présentée sur la Figure 1.20. Cette méthode est considérée comme une voie sans issue et n'a été plus développée dans le cadre de cette thèse.

### Gravure assistée par des métaux

La gravure assistée par métaux (angl. « Metal-assisted etching », MAE) consiste à graver le substrat localement à l'aide de nanoparticules d'un métal catalytique. Cete gravure est purement chimique. Grâce à la différence des travaux d'extraction des électrons d'un métal noble (Ag, Pt, Au, etc) et un semi-conducteur [22], des électrons sont injectés dans une particule et créant des lacunes (trous) dans des liaisons covalentes d'un semi-conducteur. Par la suite, ces trous participent à l'oxydation des atomes d'un semi-conducteur selon une réaction similaire à 1.3.

Le procédé est fait en deux étapes :

1. D'abord, les particules sont déposées sur la surface d'un semi-conducteur (par exemple depuis une solution aqueuse de $AgNO_3$ grâce à une réaction de déplacement galvanique [5, 19], mais l'utilisation d'une solution contenant des particules colloïdales est aussi possible).

2. Par la suite, l'échantillon est immergé dans une solution de gravure contenant du HF.

La formation des couches mésoporeuses avec des pores ordonnés en colonnes a été rapportée pour le Si (Figure 1.21).

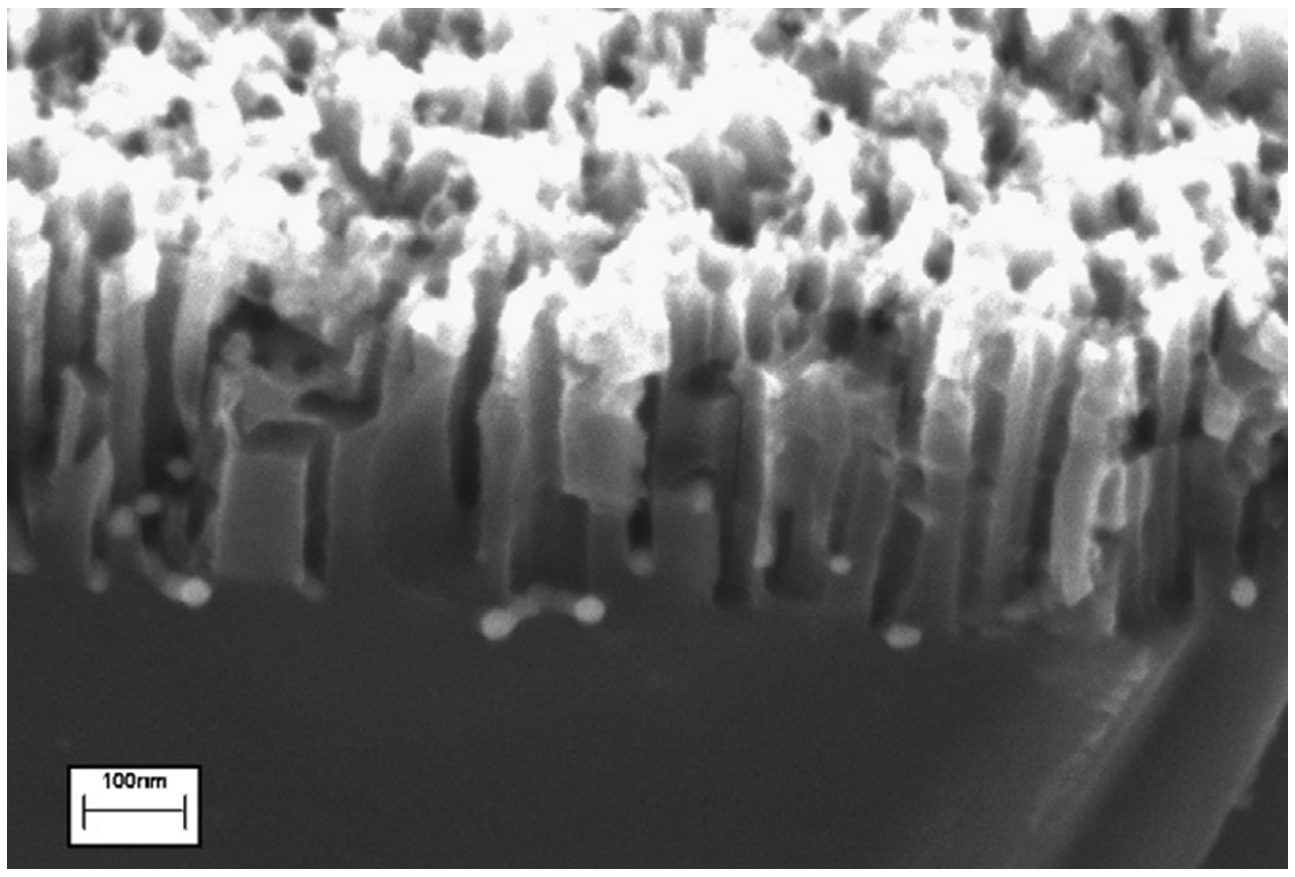

Figure 1.21 Image MEB en coupe d'un échantillon de Si (100) type p chargé par des particules en Ag et gravé dans HF : $H_2O_2$ : $H_2O$ = 25 : 10 : 75 pendant 30 s.[20]

Dans le cas du Ge, la situation est plus difficile. Dans une solution à base de HF, la surface de Ge n'est pas passivée. De plus, la longueur de diffusion est très importante et peut aller jusqu'au 300 µm dans les substrats utilisés. Suite à quoi, les trous générés peuvent donc être transportés hors de l'interface avec une particule. La dissolution du Ge a lieu toute au long de la surface, et pas seulement à proximité d'un catalyseur métallique.

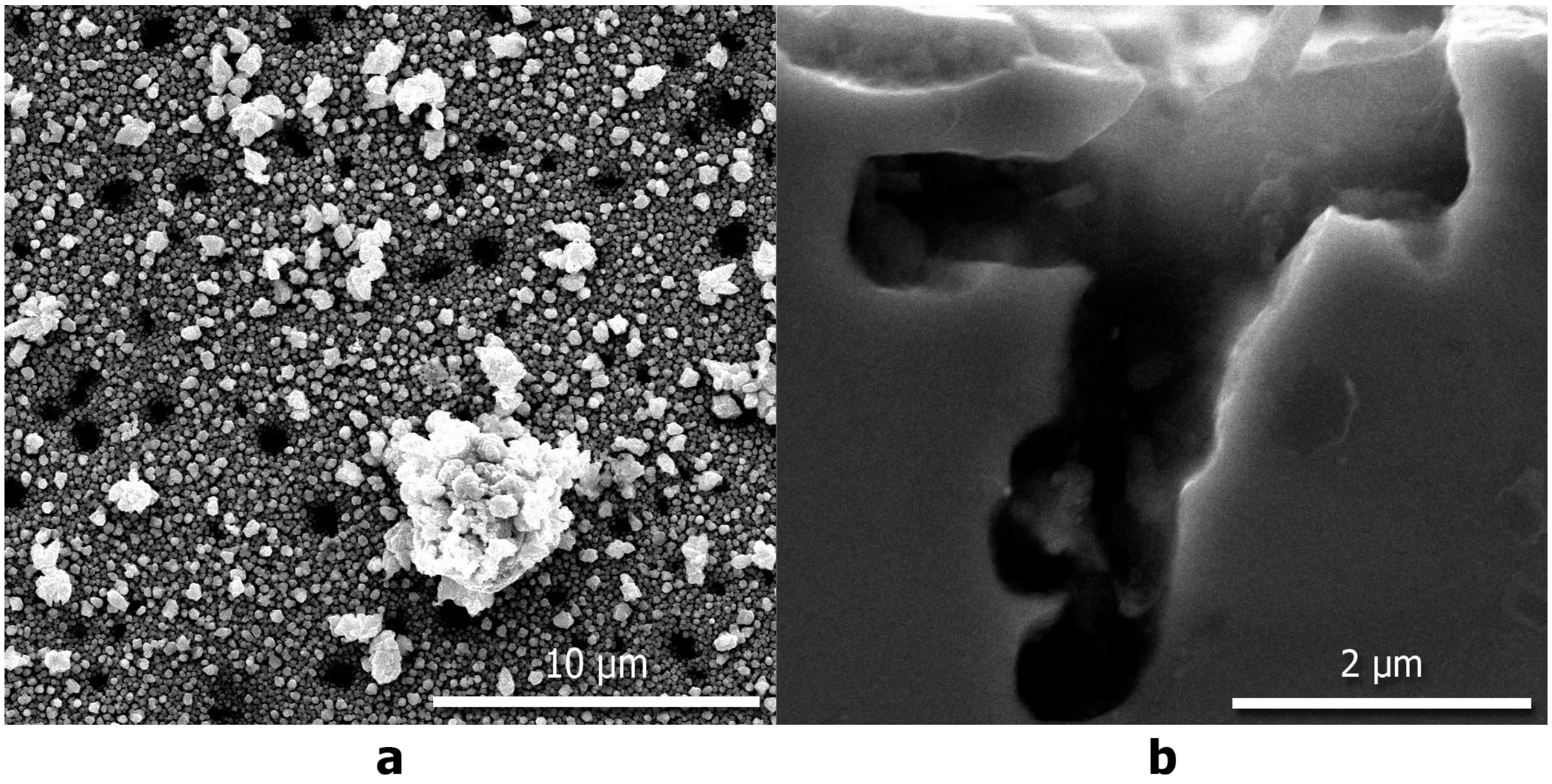

Figure 1.22 Images MEB en vue plane (a) et en coupe (b) de la surface du Ge avec des macropores formés dans une solution $H_2O$ : $H_2O_2$ : HF = 75 : 10 : 25 pendant 30 min par des particules d'Ag déposées depuis une solution de $AgNO_3$ (2 mM) pendant 30 min

Figure 1.22 montre des pores macroscopiques réalisés sur le Ge type p à forte résistivité (1–10 $\Omega \cdot$ cm) après une optimisation des conditions de gravure.

## 1.4 Conclusions du chapitre

Le mise en œuvre du transfert de couches d'une cellule solaire, exige le développement des procédés de porosification et de recuit fiables. Bien qu'il existe déjà des travaux publiés sur le sujet, de nombreux problèmes demeurent non résolus.

Les exigences définies pour des couches poreuses :

- Absence de la dissolution chimique lors de la gravure électrochimique.
- Épaisseur des couches : 100 nm–5 µm ou plus.
- Taille des pores : <10 nm.
- Uniformité des couches (normale et latérale).
- Contrôle sur la porosité dans la gamme 10–50% ou plus.
- Possiblité de faire des bi- ou multi-couches de Ge poreux de porosités différentes avec des interfaces abruptes.
- Contrôle de la morphologie. Réalisation de pores cristallographiques similaires à ceux du Si mesoporeux.

Les exigences définies pour des couches recristallisées :

- Très bonne qualité cristalline ;
- Épaisseur d'une couche recristallisée : 100 nm–2 µm ;
- Minimisation de la densité des cavités dans une couche recristallisée ;
- Faible rugosité de surface ;
- Décapage contrôlé (et pas spontané) ;
- Uniformité latérale de la couche ;

Comme il sera décrit dans les chapitres suivantes, la gravure électrochimique bipolaire avec les impulsions anodiques/cathodique optimisées permet de pallier aux limitations liées à la fabrications des filmes poreux adaptées au procédé de transfert des couches. Cette méthode est la technique principale employée dans le cadre de ce travaille de thèse.

# CHAPITRE 2

# Méthodes

Dans ce chapitre nous présenterons l'appareillage d'élaboration du Ge mesoporeux et les techniques de caractérisation des échantillons utilisés durant ce travail de thèse.

## 2.1 Technique de fabrication

### 2.1.1 Anodisation électrochimique

L'anodisation électrochimique est la méthode principale utilisée dans le cadre de cette thèse pour élaborer les nanostructures du Ge-poreux. L'appareillage utilisé pour le procédé d'anodisation est décrit ci-dessous.

#### Cellule d'anodisation

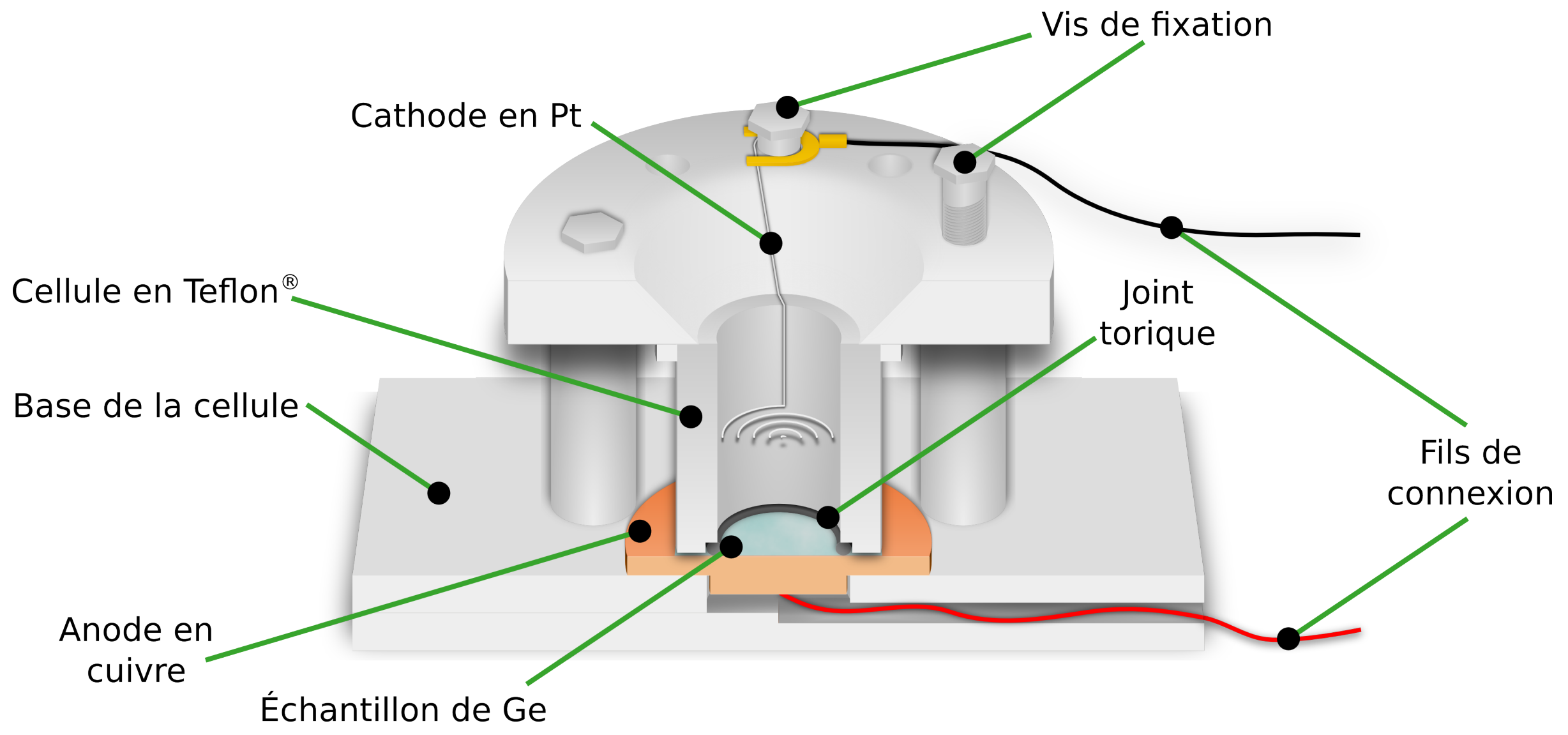

Figure 2.1 Représentation schématique de la cellule d'anodisation, vue en coupe.

La structure de la cellule est représentée schématiquement sur la Figure 2.1, elle est constituée de Teflon™, matériau chimiquement inerte. L'échantillon de Ge est mis en contact direct avec l'électrode en cuivre par sa face arrière. La partie de la surface du Ge à porosifier est délimitée par le joint torique en viton résistant aux acides. Dessus se trouve une cellule cylindrique contenant l'électrolyte. À mi-hauteur de la cellule se trouve une contre-électrode constituée d'un fil en alliage Pt/Rh de 0.5 mm de diamètre. La forme de la contre-électrode doit assurer l'uniformité des lignes du champ électrique entre les deux électrodes. En même temps, lorsque l'électrolyte est agité dans la partie supérieure de la cellule, la contre-électrode ne doit pas affecter la circulation du liquide dans la cellule. La contre-électrode est de la forme d'une spirale d'Archimède avec un pas $\approx 2$ mm. L'étanchéité de l'ensemble est assuré par un couvercle visé à la base de la cellule.

La gravure du Ge conduite sous faibles courants est fortement affectée par la géométrie de la cellule. Seules les cellules cylindriques (dont le diamètre de l'orifice est le même que celui du joint torique) ont démontré une bonne uniformité superficielle de la gravure. La cellule de diamètre $d = 11$ mm et une surface $S = 1$ cm$^2$ a été principalement utilisée. Certains échantillons ont été fabriqués à l'aide des cellules de diamètre égal à 7 ou 18 mm.

### Source d'alimentation

L'anodisation est effectuée dans des conditions galvanostatiques utilisant des sources programmables Keithley 220 ou Keithley 6221. Le courant est appliqué sous la forme d'onde rectangulaire. L'amplitude des pulsations anodiques et cathodiques $J_+$ et $J_-$ varie dans le plage de 0.125 mA/cm$^2$ à 8 mA/cm$^2$, tandis que la durée des pulsations $t_{on}$ et $t_{off}$ varie entre 0.125 s et 60 s. La durée totale du procédé varie de 2 min et 20 h. La gravure commence toujours à partir d'une impulsion positive et se termine en appliquant un niveau «0».

### Traitement des échantillons

La gravure électrochimique est conduite sur des substrats de Ge type p (0.005–0.050 $\Omega\cdot$cm) dopés au Ga fournis par Umicore ou AXT. Les substrats sont orientés (100) avec une désorientation de 6° vers une direction (111). Après le clivage en morceaux carrés, les échantillons sont nettoyés pendant 5 minutes en 3 étapes : (1) eau déionisée, (2) acétone, (3) éthanol, puis immédiatement séchés sous le jet d'azote. Par la suite, un échantillon est chargé dans la cellule d'anodisation et la gravure a lieu. Après la vidange de la cellule, l'échantillon est rincé dans l'éthanol puis séché à l'air pendant 5 minutes afin d'assurer une évacuation lente de l'éthanol des pores. Afin d'éviter la destruction des pores due aux effets de capillarité, le séchage doit se faire assez lentement. Le rinçage de certains échantillons épais a été effectué dans l'hexane qui possède le coefficient de la tension de surface inférieur à celui d'éthanol et de l'eau.

## 2.2 Moyens de caractérisation

### 2.2.1 Microscope électronique à balayage (MEB)

Un microscope électronique à balayage (MEB ou *SEM* pour *Scanning Electron Microscopy* en anglais) fournit des informations sous forme d'images lumineuses, résultant de l'interaction d'un faisceau d'électrons avec un volume microscopique de l'échantillon étudié. La microscopie électronique à balayage est une méthode principale de caractérisation du Ge-poreux, permettant de mesurer l'épaisseur, la porosité, la morphologie des couches poreuses ainsi que la rugosité de sa surface.

#### Principe de fonctionnement

Le principe général de fonctionnement d'un MEB consiste à balayer successivement, ligne par ligne, la surface de l'échantillon avec un faisceau d'électrons puis à transmettre le signal du détecteur à un écran cathodique dont le balayage est exactement synchronisé avec celui du faisceau électronique incident.

Le système du MEB est constitué d'une enceinte sous vide secondaire avec une source d'électrons, un système de focalisation du faisceau d'électrons et la platine d'échantillons mobile à 3 axes. La chambre avec la platine est séparée de la colonne d'émission pour maintenir la pression basse constante dans la colonne lors de l'ouverture de la chambre pour charger les échantillons ou lors du travail sous pression élevée.

Le principe du canon à électrons est d'extraire les électrons d'un matériau conducteur vers le vide où ils sont accélérés par un champ électrique. Il existe deux familles de canon à électrons selon le principe utilisé pour extraire les électrons :

- l'émission thermoïonique, avec les filaments de tungstène et pointes $LaB_6$ ;
- l'émission par effet de champ.

La qualité des images et la précision analytique que l'on peut obtenir avec un MEB requièrent que la tache électronique sur l'échantillon soit à la fois fine, intense et stable. Longtemps limité à une résolution juste submicrométique, le MEB atteint actuellement des résolutions nanométriques (<10 nm) grâce à l'utilisation d'émetteurs d'électrons de type Schottky (émission de champ). Dans ce canon les électrons sont accélérés par un potentiel de l'ordre de 0.5 à 30 keV.

Le flux d'électrons est ensuite limité par les diaphragmes et focalisé sur la surface de l'échantillon à l'aide de différentes lentilles électromagnétiques sous forme d'une tache.

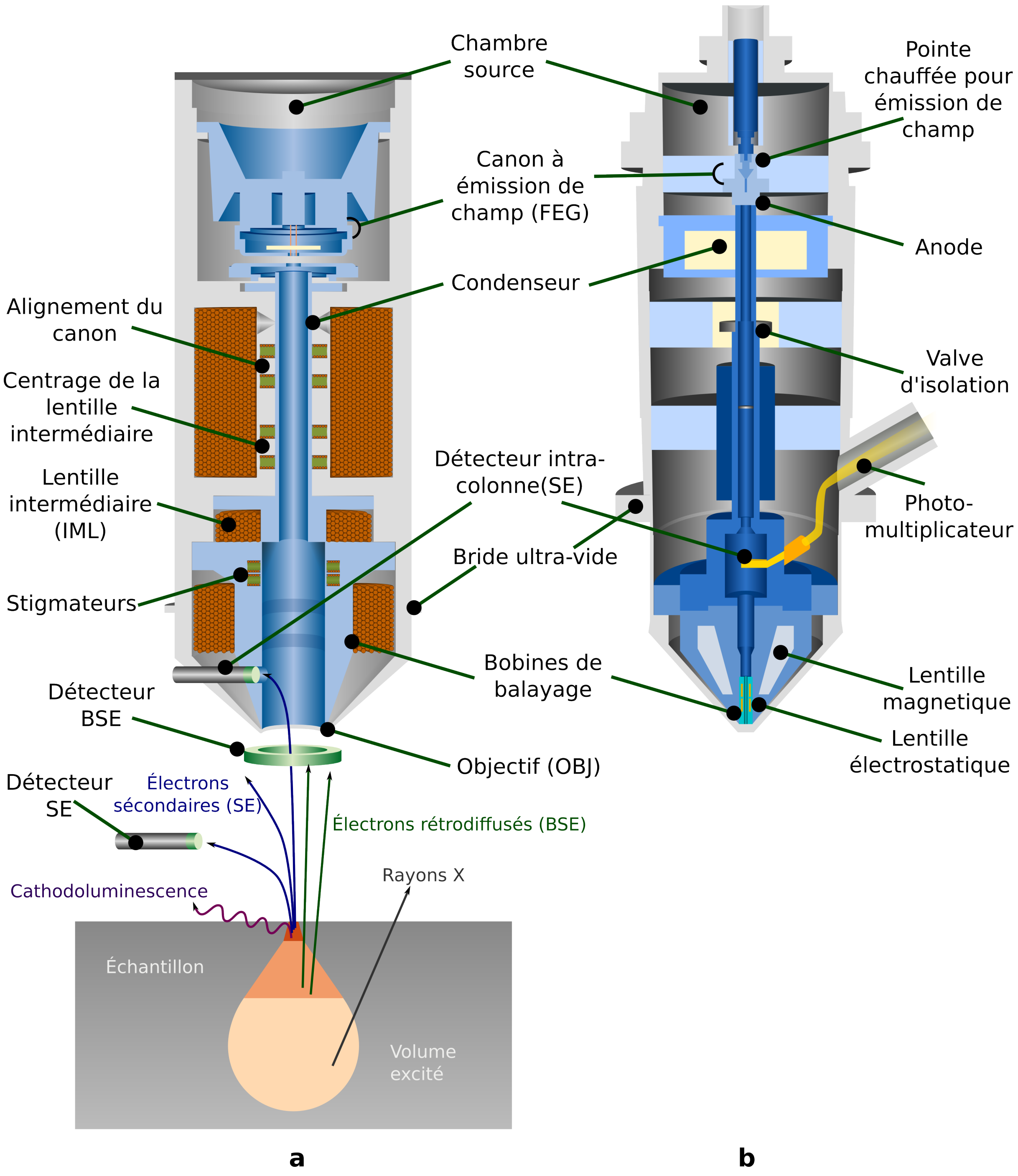

Figure 2.2 (a) Schema d'un MEB Tescan MIRA 3 ; (b) Schema d'une colonne Gemini d'un MEB Zeiss LEO 1530 VP

Des bobines déflectrices en X et en Y permettent de déplacer le pinceau électronique à la surface de l'échantillon.

Les électrons primaires, issus du canon à électrons, frappent la surface de l'échantillon. Certains électrons sont diffusés de manière élastique, ceux-ci sont les électrons dits "rétrodiffusés". Les autres, au cours du choc, cèdent une partie de leur énergie cinétique aux atomes, provoquant l'ionisation de l'atome par éjection d'un électron dit «secondaire ». L'énergie de ces derniers est faible (typiquement quelques dizaines d'eV), donc seuls les électrons venant des couches superficielles ressortent de la matière. L'atome ainsi ionisé se désexcite. Un électron d'une couche supérieure descend occuper la place laissée vide, ce qui provoque soit l'émission d'un photon X (émission secondaire), soit d'un électron Auger. Des électrons rétrodiffusés et des électrons secondaires émis par l'échantillon sont recueillis sélectivement par des détecteurs qui transmettent un signal à un écran cathodique dont le balayage est synchronisé avec le balayage de l'objet. Le taux d'électrons arrivant au détecteur dépend de la morphologie de la surface. La détection des électrons secondaires est le mode classique d'observation de la morphologie de la surface.

### Équipement

**L'équipement Tescan Mira 3 LMH** est un MEB à très haute résolution disponible à l'INL sur le site de l'INSA. La construction de la colonne ainsi que le placement des détecteurs et des analyseurs auxiliaires sont illustrés sur la Figure 2.2 a. Les caractéristiques principales du MEB Tescan Mira 3 sont données dans le Tableau 2.1.

Le microscope peut fonctionner dans l'un des cinq modes.

Le **mode «Résolution»** (*«Resolution Mode»*) est un mode de base fait pour les observations à haute résolution latérale. Les lentilles IML sont éteintes, la lentille OBJ excitée focalise le faisceau sortant d'électrons.

- Haute résolution ;
- Petite profondeur du champ.

Le **mode «Profondeur»** (*«Depth Mode»*) diffère du mode précédent par la lentille auxiliaires IML qui est activée. Ce mode est utilisé pour augmenter la profondeur de champ.

- Bonne résolution ;
- Profondeur de champ accrue.

**Optique électronique**

| | | | |
|---|---|---|---|
| Filament | source Schottky à effet de champ | | |
| Tension d'accélération | de 200 V à 30 kV | | |
| Taille de sonde | SE : | 1.2 nm | 30 kV |
| | | 1.5 nm | 15 kV |
| | | 2.5 nm | 3 kV |
| | | 4.5 nm | 1 kV |
| | in-Beam SE : | 1.0 nm | 30 kV |
| | | 1.2 nm | 15 kV |
| | | 2.0 nm | 3 kV |
| | | 3.5 nm | 1 kV |
| Courant de faisceau | de 2 pA à 200 nA | | |
| Grossissement | de 3.5X à 1,000,000X | | |
| Détecteur / distance de travail | Intra-colonne *«In-Beam»* : | 1 mm | |
| | Électrons secondaires *«SE»* : | 5 mm | |
| | Électrons retrodiffusés *«BSE»* : | 8 mm | |
| Champ de vision maximal | mode High-Vacuum : | 78 mm | |

**Platine**

| | | |
|---|---|---|
| Déplacement de l'échantillon | X=80 mm : | |
| | Y=60 mm : | |
| | Z=47 mm : | |
| | Rotation=360° : | continue |
| | Inclinaison : | de -30° à 90° |
| Taille d'échantillon | Z<80 mm : | |

Tableau 2.1 Caractéristiques du microscope Tescan MIRA 3 LMH installé à l'INL site INSA

Le **mode «Champ»** (*«Field Mode»*) utilise la lentille intermédiaire IML pour la focalisation du faisceau d'électrons alors que l'objectif OBJ est éteint. Ce mode est utilisé pour se rendre sur sur l'échantillon (ou la partie de l'échantillon) à analyser. La désavantage de ce mode est la taille du spot importante ; le grossissement max. est quelques milliers fois.

- Large champ de vision ;
- Grande profondeur de champ ;
- Résolution moins bonne.

Le **mode «Large Champ»** (*«Wide Field Mode»*) utilise la lentille intermédiaire IML pour la focalisation du faisceau d'électrons alors que l'objectif OBJ est excité à haut niveau. Ce mode est utilisé pour localiser les échantillons sur la platine. Une mise au point précise est compliquée par la profondeur de champ élevée. Les mesures d'objets ne donnent pas des chiffres véritables (image déformée ; valeur du grossissement dépendant de la mise au point).

- Champ de vision extra-large ;
- Grande profondeur de champ ;
- Très faible grossissement ;
- Déformation d'images (effet minimisé par le logiciel).

Le **mode de «Canalisation»** (*«Channeling Mode»*) est conçu pour l'acquisition de motif de canalisation électronique (*Electron Channeling Pattern, ECP*). Le balayage et la focalisation des lentilles sont contrôlés de telle manière que le faisceau d'électrons soit en contact avec le même point de la surface de l'échantillon tout le temps ; seulement l'angle d'incidence du faisceau d'électrons change. En tenant compte des capacités optiques de la colonne, la taille minimale des cristaux est de 100–150 µm. Les motifs ECP sont faits par les électrons rétrodiffusés ; le mode requiert l'utilisation d'un détecteur des électrons retrodiffusés («BSE»).

- Champ de vision extra-large ;
- La position du point spécifié dans l'image dépend de l'inclinaison du faisceau.

Lors d'observations des couches du Ge-poreux l'échantillon a été centré sur l'écran en mode «Large Champ», puis le MEB a été mis en mode «Résolution». La topographie de la surface du Ge-poreux avec la différence d'altitude de quelques nm fait que les modes à profondeur de champ sont inutiles. Les échantillons inclinés ou ceux avec une couche poreuse décollée ont étés focalisés par la mise au point dynamique.

**Zeiss LEO 1530 / Supra 55 VP** Le 3IT de l'Université de Sherbrooke possède 4 microscopes électroniques à balayage :

- JEOL 6300 équipé de NPGS
- LEO 1540XB CrossBeam FIB équipé de NPGS
- LEO 1530
- LEO Supra 55 VP équipé de cathodoluminescence, de microcaractérisation et de NPGS

LEO-VP est un appareil multi-fonctions pouvant être utilisé pour : la microscopie électronique à balayage, la caractérisation par les techniques de cathodoluminescence (CL), la technique du courant induit par faisceau d'électrons (EBIC), la spectroscopie de rayons-X (EDX), la lithographie par faisceau d'électrons à haute résolution (à l'aide du logiciel NPGS). Ce microscope possède les meilleures caractéristiques d'imagerie. C'est pourquoi il a été choisi pour les observations.

La construction de la colonne est illustrée sur la Figure 2.2 b. Les caractéristiques principales du MEB LEO Supra 55 VP sont données dans le Tableau 2.2.

| | | |
|---|---|---|
| **Optique électronique** | | |
| Filament | source Schottky à effet de champ | |
| Tension d'accélération | de 200 V à 30 kV | |
| Taille de sonde | In-Beam : | 1.0 nm 20 kV<br>3.0 nm 1 kV |
| Courant de faisceau | de 4 pA à 20 nA | |
| Détecteurs | Intra-colonne *«In-Lens»*<br>Électrons secondaires *«SE»*<br>Électrons retrodiffusés *«BSE»* | |
| Grossissement | de 20X à 900,000X | |
| **Platine** | | |
| Déplacement de l'échantillon | X=75 mm : | motorisé |
| | Y=75 mm : | motorisé |
| | Z=55 mm : | dont 25 mm motorisé |
| | Rotation=360° : | continue |
| | Inclinaison : | de -15° à 65° |
| Taille d'échantillon | X<100 mm<br>Y<100 mm<br>Z< 15 mm | |
| Distance de travail | 1<Z<50 mm | |

Tableau 2.2 Caractéristiques du microscope Zeiss LEO 1530 / Supra 55 VP installé au 3IT, Université de Sherbrooke

### Préparation des échantillons

Normalement, aucune préparation spéciale des échantillons n'est requise. Pour les observations en coupe, les échantillons sont clivés à l'aide d'une pince de clivage TED PELLA 8295. Dû à une désorientation des substrats de 6 ° vers la direction <111>, les plans de clivage ne sont pas parfaitement perpendiculaires à la surface.

Cependant, lors des observations, des effets de charge dans le volume de la couche poreuse sont constatés. Ce phénomène provient de la faible conductance électrique du Ge-poreux. L'effet est encore plus important pour des échantillons avec une surface libre fortement oxydée. Pour faciliter les observations en vue pleine, la surface observable est délimitée par une bande adhésive conductrice.

Étant serrées entre 2 barrettes conductrices, la plupart des échantillons ne pose aucun problème lors d'observations en coupe. Sinon, le porte-échantillon peut être incliné. Les électrons qui traversent la couche poreuse sont évacués par le substrat conducteur, ce qui évite de charger le Ge-poreux derrière le plan d'observation.

### 2.2.2 microscopie électronique en transmission (MET)

La microscopie électronique en transmission (MET ou *TEM* pour *Transmission electron microscopy* en anglais) est une technique qui permet d'observer et de caractériser les défauts structuraux présents dans les échantillons minces par leur influence sur le trajet des électrons traversant le matériau. L'interaction des électrons incidents avec les atomes de l'échantillon se traduit par des diffusions élastiques et inélastiques. La diffusion élastique est à l'origine de la diffraction électronique, elle-même à l'origine des contrastes des images utilisées pour la caractérisation structurale. Un système de lentilles magnétiques projette l'image de l'échantillon sur un écran phosphorescent qui transforme l'image électronique en image optique.

Les effets d'interaction entre les électrons et l'échantillon donnent naissance à une image, dont la résolution peut atteindre 0.5 Å=50 pm. Les images obtenues ne sont généralement pas explicites, et doivent être interprétées à l'aide d'un support théorique. L'intérêt principal de ce microscope est de pouvoir combiner cette grande résolution avec les informations de l'espace de Fourier, c'est-à-dire la diffraction. Il est aussi possible d'étudier la composition chimique de l'échantillon en étudiant le rayonnement X provoqué par le faisceau électronique. Contrairement aux microscopes optiques, la résolution n'est pas limitée par la longueur d'onde des électrons ($E = 200$ keV, $\lambda = 2.5$ pm), mais par les aberrations dues aux lentilles magnétiques. Une résolution sub-Ångström est réalisable grâce à utilisation des correcteurs d'aberrations teles que CEOS Gmbh.

Le microscope électronique en transmission a deux principaux modes de fonctionnement :

**mode image :** Le faisceau électronique interagit avec l'échantillon suivant l'épaisseur, la densité ou la nature chimique de celui-ci, ce qui conduit à la formation d'une image contrastée dans le plan image. En plaçant le détecteur dans le plan image, on peut observer une image par transparence de la zone observée.

**mode diffraction :** Ce mode utilise le comportement ondulatoire des électrons. Lorsque le faisceau traverse un échantillon cristallographique, il donne lieu au phénomène de diffraction. Le faisceau est diffracté en plusieurs petits faisceaux, et ceux-ci se recombinent pour former le cliché de diffraction, grâce aux lentilles magnétiques.

#### Analyse des clichés de diffraction

L'analyse des clichés de diffraction d'électrons par l'échantillon observé est essentielle pour compléter les observations TEM. Cette analyse comprend une évaluation qualitative d'une cristallinité de structure et un étiquetage des anneaux et tes taches de diffraction avec des indices de Miller (h,k,l) appropriés.

Ayant les taches identifiées, plusieurs paramètres structuraux peuvent être déterminés :

- espacement inter-planaire, $d$ ;
- constante de réseau, $a$ ;
- orientation cristalline ;
- nature des défauts.

Ainsi, l'identification d'une phase ou l'analyse de défauts n'est pas possible sans l'analyse du cliché de diffraction.

Rappelons, qu'une notation (hkl) désigne un plan, {hkl} — famille des plans, [hkl] — une direction et <hkl> — une famille des directions.

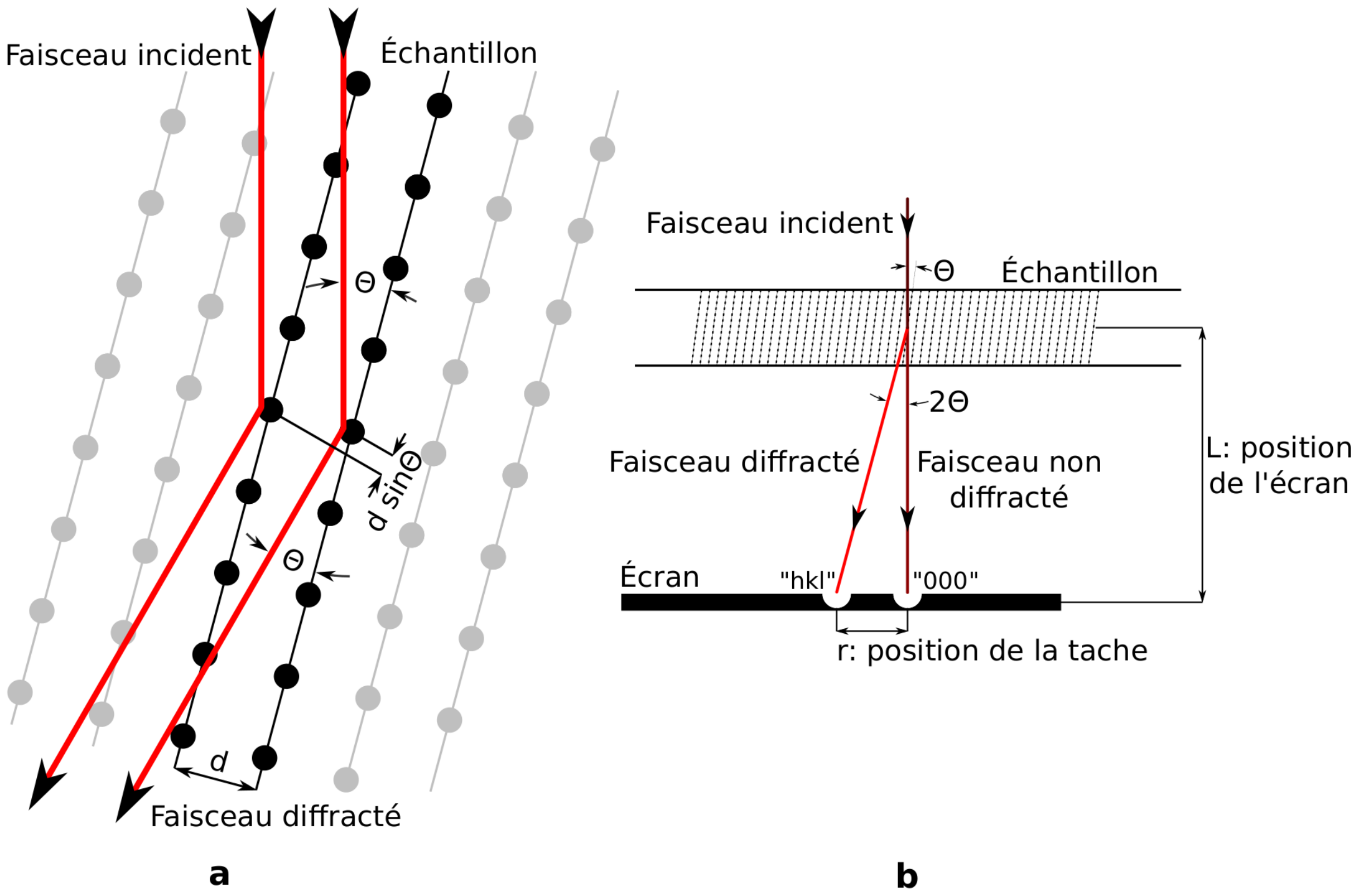

Figure 2.3 (a) Illustation de la loi de diffraction de Bragg, (b) Schéma de la position du faisceau diffracté en fonction des paramètres de système.

Grâce à la dualité onde-corpuscule, les électrons subissent une diffraction lorsqu'ils traversent un échantillon fin cristallin (Figure 2.3 a). La condition d'interférence constructive s'exprime par la loi de Bragg, initialement introduit pour les rayons-X :

$$2d \sin\theta = n\lambda \tag{2.1}$$

où $\lambda$ est la longueur d'onde de De Broglie d'électrons incidents, $\theta$ - angle entre un plan cristallin et le faisceau incident, $n$ — ordre de diffraction. Dans le plupart des cas, seulement le premièr maximum de diffraction ($n = 1$) est observé.

Les électrons, étant chargés négativement, sont dispersés dans un matériau par les électrons et les noyaux. Ainsi, contrairement aux rayons-X, il s'agit d'interaction avec des champs locaux de l'échantillon et pas d'une excitation et une relaxation d'un atome [118].

Lorsque l'angle d'incidence est suffisamment petit ($\theta \approx \sin\theta \approx \tan\theta = r/2L$)(Figure 2.3 (b)), la condition d'interférence constructive prend la forme suivante :

$$rd = L\lambda \tag{2.2}$$

L'espacement $d$ dépende de la constante de réseau et des indices de Miller :

$$d = \frac{a}{\sqrt{h^2 + k^2 + l^2}} \tag{2.3}$$

Finalement, une combinaison des équations 2.3 et 2.2 donne une relation entre la mesure ($r$) et les paramètres de la structure ($a$, *(hkl)*) :

$$r = \frac{L\lambda}{a}\sqrt{h^2 + k^2 + l^2} \tag{2.4}$$

Une section 2-D d'un réseau réciproque peut être définie par deux vecteurs. Il est alors nécessaire d'indexer seulement 2 taches pour l'identifier (Figure 2.4). Toutes les autres peuvent être déduites par l'addition de vecteurs.

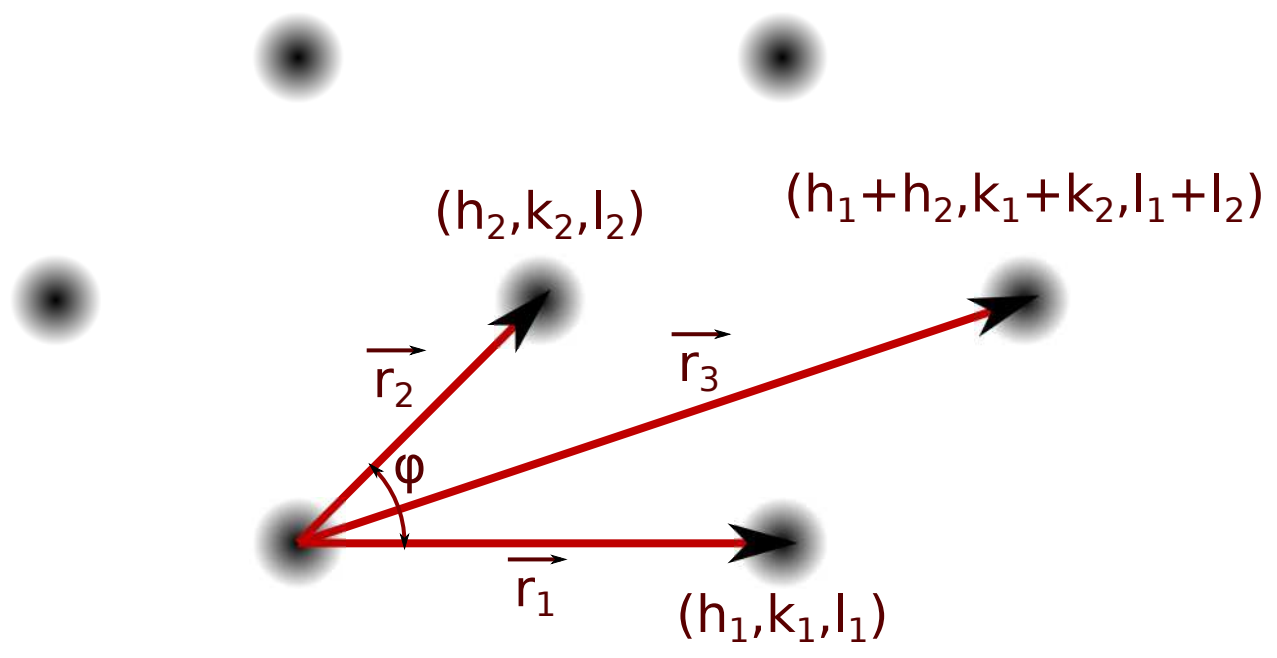

Figure 2.4 Dessin d'un cliché de diffraction et ses mesures principales

L'angle $\varphi$ est calculé comme suit :

$$\cos\varphi = \frac{h_1h_2 + k_1k_2 + l_1l_2}{\sqrt{h_1^2 + k_1^2 + l_1^2}\sqrt{h_2^2 + k_2^2 + l_2^2}} \tag{2.5}$$

Le problème principal — une identification des taches sur un cliché de diffraction, est simplifié par le fait que les plans cristallins, qui forment les taches de diffraction, obéissent à des règles de sélection selon leurs indices de Miller (tableau 2.3).

| Réseau de Bravais | Exemples | Réflexions permises |
|---|---|---|
| Cubique primitif | Po | Tous h, k, l |
| Cubique centré | Fe, W, Ta, Cr | h+k+l=pair |
| Cubique à faces centrées | Cu, Al, Ni, NaCl, LiH, PbS | h,k,l tous pairs ou tous impairs |
| Cubique à faces centrées : « diamant » | Si, Ge | tous impairs ou tous pairs avec $h + k + l = 4m, m \in \mathbb{Z}$ |
| Triangulaire | Ti, Zr, Cd, Be | $l$ impair, $h + 2k \neq 3m, m \in \mathbb{Z}$ |

Tableau 2.3 Règles de sélection pour les indices de Miller [118]

Le germanium, ayant un réseau cristallin de type « Diamant », possède les réflexions permises suivantes : {111}, {022}, {113}, {133}, {224}, {115}, {044}, {135}, {335}, {155}, {117} etc.

Le cliché de diffraction d'un matériau monocristallin est une matrice des taches, une représentation du cristal dans l' espace réciproque. Ainsi, si la structure cristalline est connue, la procédure du rapport pour l'indexation est la suivante [74] :

1. Choisir l'origine. Un endroit différent de la tache (000) peut être également choisi.

2. Mesurer la distance à une tache marquante, $r_1$. Pour une meilleure précision, mesurer l'espacement entre plusieurs points le long d'une ligne et établir une moyenne.

3. Mesurer l'espacement d'un deuxième point, $r_2$. La deuxième tache ne doit pas être alignée avec la première et l'origine. La meilleure façon est de choisir $r_2$ perpendiculaire à $r_1$, si possible.

4. Mesurer l'angle $\varphi$ entre $\vec{r_1}$ et $\vec{r_2}$.

5. Préparer un tableau donnant les rapports des distances des plans de diffraction admis dans la structure connue (Équation 2.4, Tableau 2.3). Commencer par les plans les plus espacés (petits $r$) ; à faire une fois pour chaque structure.

6. Localiser une valeur de rapport $r_1/r_2$ dans le tableau et repérer les indices de Miller qui lui correspondent. Préférer les cellules dont les indices de Miller sont les plus petits.

7. Calculer l'angle entre deux plans des familles de plans déterminées à l'étape précédente à l'aide de l'équation 2.5 et comparer le avec $\varphi$ mesuré. Si deux valeurs coïncident — accepter l'indexage. Sinon, revoir le tableau et sélectionnez une autre paire de plans possibles.

8. Indexer toutes les taches du cliché utilisant une combinaison linéaire des vecteurs $\vec{r_1}$ et $\vec{r_2}$.

| {hkl} | $r_2 \downarrow$ | {111} | {022} | {113} | {133} | {224} | {115} | {044} | {135} |
|---|---|---|---|---|---|---|---|---|---|
| $r_1 \rightarrow$ | $\sqrt{h^2+k^2+l^2}$ | 1,7321 | 2,8284 | 3,3166 | 4,3589 | 4,8990 | 5,1962 | 5,6569 | 5,9161 |
| {111} | 1,7321 | 1 | 1,6330 | 1,9149 | 2,5166 | 2,8284 | 3,0000 | 3,2660 | 3,4157 |
| {022} | 2,8284 | 0,6124 | 1 | 1,1726 | 1,5411 | 1,7321 | 1,8371 | 2,0000 | 2,0917 |
| {113} | 3,3166 | 0,5222 | 0,8528 | 1 | 1,3143 | 1,4771 | 1,5667 | 1,7056 | 1,7838 |
| {133} | 4,3589 | 0,3974 | 0,6489 | 0,7609 | 1 | 1,1239 | 1,1921 | 1,2978 | 1,3572 |
| {224} | 4,8990 | 0,3536 | 0,5774 | 0,6770 | 0,8898 | 1 | 1,0607 | 1,1547 | 1,2076 |
| {115} | 5,1962 | 0,3333 | 0,5443 | 0,6383 | 0,8389 | 0,9428 | 1 | 1,0887 | 1,1386 |
| {044} | 5,6569 | 0,3062 | 0,5000 | 0,5863 | 0,7706 | 0,8660 | 0,9186 | 1 | 1,0458 |
| {135} | 5,9161 | 0,2928 | 0,4781 | 0,5606 | 0,7368 | 0,8281 | 0,8783 | 0,9562 | 1 |

Tableau 2.4 Tableau de rapports des espacements inter-planaires de Ge tenant en compte les règles de sélection sur les indices de Miller des plans (Tableau 2.4)

Lorsqu'un matériau est polycrystallin, son cliché de diffraction contient des anneaux concentriques. Dans ce cas, l'analyse du cliché consiste à mesurer les rayons des anneaux $r_1$, $r_2$, $r_3$, ... etc.

- Mesurer les rayons des anneaux $r_1$, $r_2$, $r_3$, ... etc.

- Calculer les espacements $d_1$, $d_2$, $d_3$, ... etc. des plans qui donnent naissance aux anneaux de diffraction (Équation 2.2)

- Identifier le matériau et/ou ses plans de diffractions depuis un tableau référentiel [118]

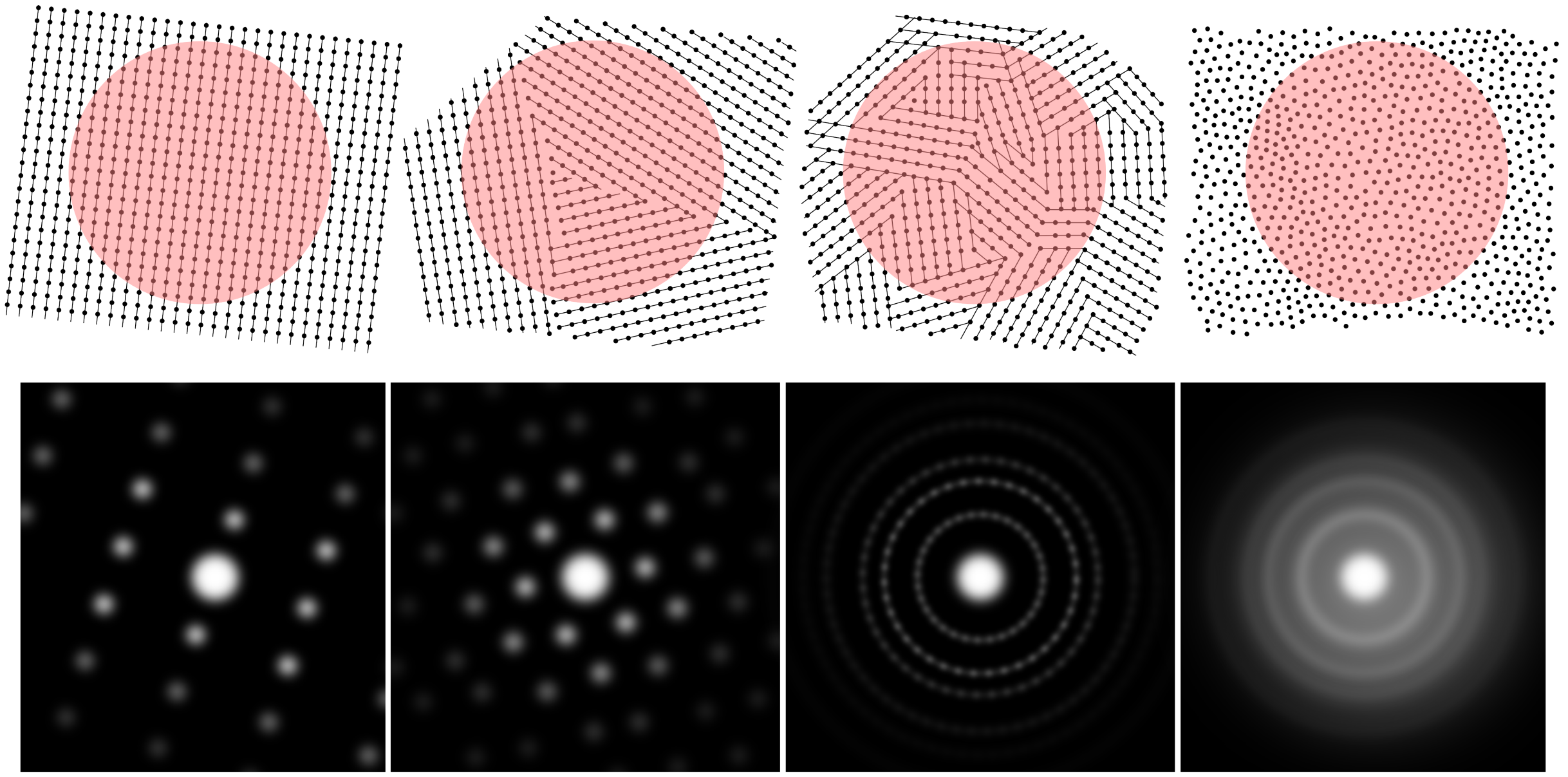

Figure 2.5 Dessins schématiques d'un matériau monocristallin (a), à grands grains (b), polycristallin (c) et amorphe (e) et des clichés de diffraction qui leur correspondent.

### 2.2.3 La spectroscopie Raman

#### Principe de base

Lorsqu'on éclaire un cristal par un faisceau de lumière monochromatique, cette radiation excitatrice peut être transmise, réfléchie, absorbée ou diffusée par le milieu. Il peut y avoir alors une diffusion élastique (diffusion Rayleigh, Figure 2.6 (a)) quand l'énergie électromagnétique est simplement redistribuée spatialement selon une direction particulière (réflexion spéculaire sur une surface plane) ou dans toutes les directions (diffusion par des fluctuations d'indice optique, par une surface rugueuse). Dans ce cas la fréquence de la lumière diffusée est égale à celle de la lumière incidente. Il existe aussi la diffusion inélastique qui se traduit par un échange d'énergie entre le rayonnement et la matière. Cet effet est très faible, approximativement 1 photon sur 1 million sera émis avec une fréquence légèrement différente de la fréquence d'onde incidente : c'est l'effet Raman. Dans cet effet, la différence d'énergie entre le photon incident et le photon diffusé correspond à l'émission ou l'absorption d'un mode de vibration de la matière.

L'absorption ou l'émission d'un mode de vibration ne se fait pas directement par le champ électromagnétique. Comme les fréquences des photons incidents sont bien supérieures à celles des phonons, la probabilité d'interaction directe photon-phonon est très faible. Cette interaction s'effectue alors par l'intermédiaire des électrons. Ainsi, le processus de diffusion peut être décrit par un mécanisme en trois étapes :

1. Interaction photon incident - électron avec la création d'une paire électron-trou ;

2. Interaction d'un membre de celle-ci avec un phonon mis en jeu ;

3. Recombinaison radiative de la paire électron-trou par émission d'un photon.

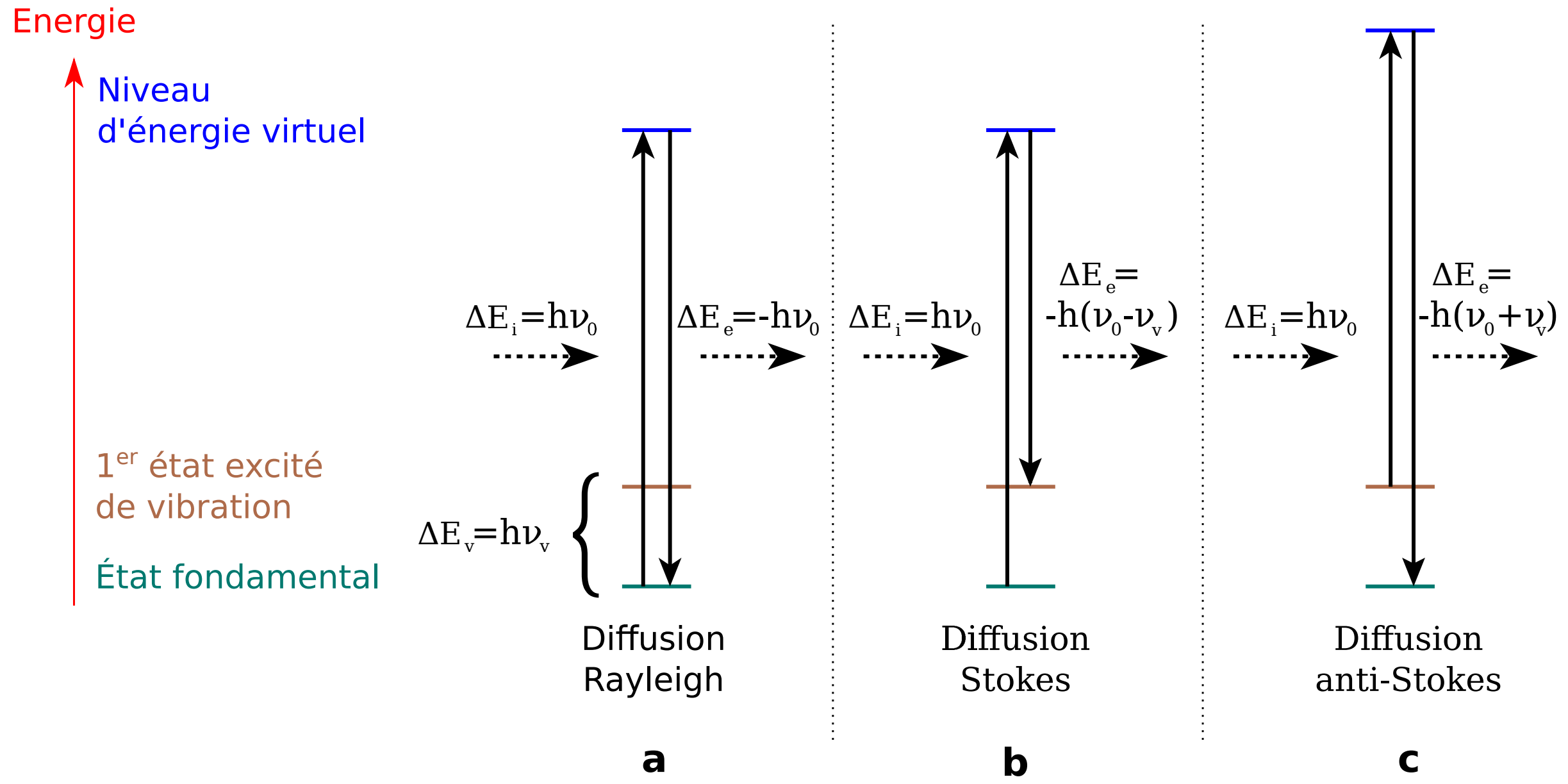

Figure 2.6 Schéma énergétique de la diffusion de la lumière : diffusion Rayleigh (a), Stokes (b) et anti-Stokes (c)

Il existe deux types de diffusion Raman : Stokes et anti-Stokes. Lorsque l'état vibrationnel final possède une énergie supérieure à celle de l'état initial, le système pendant l'interaction gagne un quantum d'énergie vibrationnelle, autrement dit il y a une création d'un phonon (Figure 2.6 (b)). Cette énergie a été cédée par la lumière incidente, qui est diffusée avec une fréquence inférieure. C'est la diffusion Stokes. Autrement, si l'état vibrationnel final possède une énergie inférieure à celle de l'état initial (Figure 2.6 (c)), le système perd un phonon. Cette énergie a été gagnée par la lumière qui est diffusée avec une fréquence supérieure. Dans ce cas on parle de la diffusion anti-Stokes.

Le processus de diffusion Raman obéit à des règles de sélection imposées par la conservation de l'énergie et de la quantité de mouvement dans le processus d'interaction et la symétrie moléculaire ou cristalline du système diffusant.

Le premier point fait intervenir les fréquences ($\nu$) et les vecteurs d'onde ($\vec{k}$) des rayonnements incident (i) et diffusé (d) ainsi que ceux des phonons (p) créés («+ » — Stokes) ou perdus («− » — anti-Stokes). Les égalités suivantes doivent donc être satisfaites :

$$h\nu_i = h\nu_d \pm h\nu + p \tag{2.6}$$

$$\vec{k}_i = \vec{k}_d \pm \vec{k}_p \tag{2.7}$$

Or, dans le cas de la spectroscopie Raman, les longueurs d'onde des rayonnements incident et diffusé se situent dans le domaine du visible ou de proche infrarouge. Statistiquement et thermodynamiquement, la population du niveau énergétique le plus bas est la plus grande. Ainsi, les raies Stokes sont celles utilisées dans la détection Raman puisqu'elles sont plus intenses et donc plus faciles à détecter.

### Ge-poreux

L'exploitation des spectres Raman peut conduire à la détermination de la taille des nanocristallites constituant une couche poreuse. Sur la base des travaux sur le Si-poreux [7, 32, 54, 55], le SiC-poreux [87], ainsi que sur le nano-Ge [38, 75, 120] il est connu que si la forme du spectre Raman est asymétrique vers des énergies plus basses, ce fait souligne l'existence de nanocristaux, même s'ils ne sont pas observables par le MEB.

### Modèle de confinement des phonons

Afin d'interpréter les spectres, un modèle phénoménologique du confinement des phonons est très largement utilisé [53]. Ce modèle, proposé à l'origine par Richter et al. [95] a été ensuite étendu par Campbell et Fauchet [16]. Lors d'un processus de diffusion Stokes dans un cristal infini la différence du vecteur d'onde $\vec{k} = \vec{k}_L - \vec{k}_S$ est transféré à un phonon avec le vecteur d'onde $\vec{q_0}$. La fonction de Bloch d'un phonon avec un vecteur d'onde $\vec{q_0}$ dans un cristal infini est suivante :

$$\Phi(\vec{q_0}, \vec{r}) = u(\vec{q_0}, \vec{r})e^{-i\vec{q_0}\vec{r}} \tag{2.8}$$

où $u(\vec{q_0}, \vec{r})$ est une fonction avec la périodicité du réseau, $\vec{r}$ - le vecteur spatial. Pour décrire un phonon dans un nanocristal, une fonction de localisation du phonon (angl. *« phonon-weighting function »*) $W(\vec{r}, L)$ est utilisée :

$$\Psi(\vec{q_0}, \vec{r}) = W(\vec{r}, L)\Phi(\vec{q_0}, \vec{r}) = \Psi'(\vec{q_0}, \vec{r})u(\vec{q_0}, \vec{r}) \tag{2.9}$$

Le choix typique pour le $W(\vec{r}, L)$ est une Gaussianne $W(\vec{r}) = \exp(-\alpha^2 r^2/L^2)$, où $\alpha$ est un paramètre arbitraire. $\alpha$ est fixé de manière à obtenir une petite valeur du $W$ à la limite d'un cristallite sphérique de diamètre $L$, qui est une caractéristique du système. La fonction de gaussienne a été choisie pour simplifier le calcul. Cependant, il n'y a pas de raison physique de choisir telle fonction $W$ ou de supposer qu'elle est différente de zéro aux limites. Une fonction de localisation carrée et une fonction de localisation Gaussienne donnent les mêmes résultats si $\alpha$ est choisi correctement [1, 99]. Les auteurs rapportent différentes valeurs de $\alpha$ pour traiter les spectres des semiconducteurs nanostructurés : $\sqrt{2}$ [95], $\sqrt{8}\pi$ [16], 6.3 [3], $\sqrt{6}$ [99]. En général, les valeurs de $\alpha$ trouvées dans la littérature sont dans l'intervalle de 1.4 à 10.4 [3].

Confinée dans un nanocristal, $\Psi'(\vec{q_0}, \vec{r})$ n'est plus une onde progressive, mais un paquet d'onde, qui peut être développé en série de Fourier

$$\Psi'(\vec{q_0}, \vec{r}) = \int C(\vec{q_0}, \vec{q}) e^{i\vec{q}\vec{r}} d^3q \tag{2.10}$$

avec les coefficients de Fourier suivants :

$$C(\vec{q_0}, \vec{q}) = \frac{1}{2\pi^3} \int \Psi'(\vec{q_0}, \vec{r}) e^{-i\vec{q}\vec{r}} d^3r \tag{2.11}$$

L'intensité de diffusion Raman est alors décrite par une superposition continue des courbes lorentziennes de largeur de bande $\Gamma$ centrées sur les nombres d'onde $\omega(q)$ de la courbe de dispersion des phonons et pondérées par le facteur $|C(\vec{q_0}, \vec{q})|^2$ En supposant une zone de Brillouin sphérique, une courbe de dispersion isotrope ainsi que $\vec{q_0} = \vec{k}_L - \vec{k}_s \approx 0$ pour une diffusion à 1 photon, l'intensité Raman peut être écrite comme suit :

$$I'(\omega) \propto \int \frac{|C(0, \vec{q})|^2}{[\omega - \omega(q)]^2 + (\Gamma_0/2)^2} d^3q \tag{2.12}$$

Alors que la périodicité du réseau est brisée par l'introduction de la fonction de localisation, l'intégration doit être fait de $-\infty$ à $+\infty$ [99] et pas juste dans la $1^{re}$ zone de Brillouin [16]. Dans le cas d'une fonction de localisation Gaussienne avec le facteur $\alpha = \sqrt{6}$ les coefficients Fourrier prennent la forme :

$$|C(0, \vec{q})|^2 \cong |e^{-q^2L^2/2\alpha^2}|^2 = |e^{-q^2L^2/12}|^2 \tag{2.13}$$

$I'(\omega)$ représente l'intensité d'un cristallite sphérique de diamètre L ou d'un échantillon composé de cristallites identiques. En y introduisant la distribution des cristallites par taille $\varphi(L)$, l'intensité de l'ensemble des nanocristallites s'écrit comme suit :

$$I(\omega) = \int \varphi(L) I'(\omega, L) dL \tag{2.14}$$

La distribution $\varphi(L)$ peut être approximée par une Gaussienne :

$$\varphi(L) \approx \frac{1}{\sqrt{2\pi\sigma^2}} \exp\left(-\frac{1}{2}\frac{(L-L_0)^2}{\sigma^2}\right) \tag{2.15}$$

où le diamètre moyen $L_0$ et l'écart type $\sigma$ sont les caractéristiques de la distribution.

Si $\sigma \leq L_0/3$ l'intensité Raman prend la forme suivante [54] :

$$I(\omega) \propto \int f(q) \exp\left(-\frac{f(q) L_0^2 q^2}{2\alpha^2}\right) \frac{d^3 q}{[\omega - \omega(q)]^2 + (\Gamma_0/2)^2} \tag{2.16}$$

avec $f(q) = [1 + (\sigma q)^2/\alpha^2]^{-1/2}$

Finalement, le problème est de choisir une bonne expression décrivant $\omega(q)$, en raison de la forte anisotropie des courbes de dispersion des phonons du Ge [114]. En se basant sur la règle de la somme de Brout [14], Paillard et al. [92] ont calculé la fonction moyenne de dispersion des phonons optiques pour des nanocristaux du Si. Ayant utilisé les résultats de la spectroscopie de diffusion des neutrons thermiques par Nilsson et Nelin [88], Wellner et al. ont adapté ce modèle au Ge [115] :

$$\omega(q) = \sqrt{\omega_0^2 - \frac{43565 q_r^2}{|q_r| + 0.566}} \tag{2.17}$$

où $q_r = (a/2\pi)q$ est le vecteur d'onde réduit, $a = 0.566$ nm — le paramètre de maille cristalline du Ge et $\omega_0 = 300.2$ cm$^1$ — la fréquence de phonons optique au centre de la zone de Brillouin.

Les Figures 2.7 et 2.8 montrent des résultats de simulation des spectres Raman du Ge nano-structuré. Les spectres ont été calculés à partir des équations générales (2.14) et (2.15). Pour des structures avec une taille moyenne de cristallites supérieure à 10 nm, les spectres Raman sont indiscernables de celui du Ge volumique. Lorsque $L_0$ diminue, le pic Raman s'élargit asymétriquement et se décale vers le rouge (Figure 2.7 (a)). La position du pic Raman tracée en fonction de $L_0$ (Figure 2.7 (b)) permet de faire une première estimation du diamètre moyen des cristallites dans la structure, sans traiter le spectre.

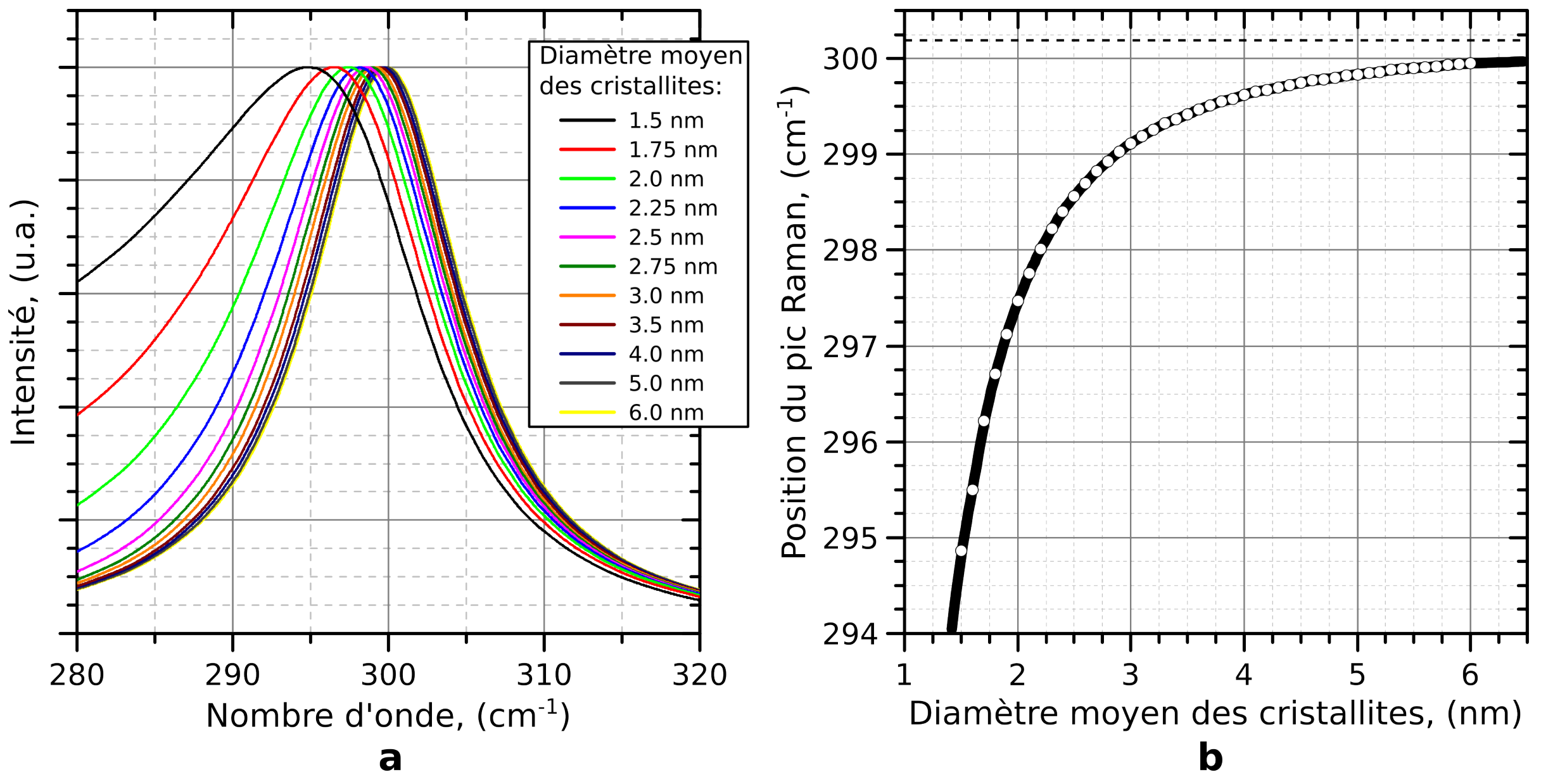

Figure 2.7 (a) Simulation des spectres de diffusion Raman pour différentes valeurs du diamètre moyen des cristallites $L_0$ lorsque $\sigma$ est fixé à 0.5 nm ; (b) décalage du pic Raman en fonction du diamètre moyen des cristallites $L_0$. Le trait pointillé indique la position du pic pour le Ge volumique.

La diminution de $\sigma$, lorsque $L_0$ est fixé, mène à un résultat similaire (Figure 2.8 (a)). Un phénomène notable est la formation d'une « marche » sur la partie gauche du spectre lorsque $\sigma$ est grand. L'intensité du signal dans cette partie de spectre augmente jusqu'à ce qu'elle atteigne l'amplitude du pic Raman principal. Au-delà, l'amplitude de ce nouveau pic anormal reste inchangée tandis que le pic principal disparait graduellement.

Ces trois régimes d'élargissement ont été observés pour tous les $L_0$ analysés (Figure 2.8 (b)) :

1. allongement de l'épaulement gauche (régime « normal », $\sigma \lessapprox L_0/4$) ; la largeur du spectre dépend peu du sigma et approche asymptotiquement au $\Gamma_0 = 12$ cm$^{-1}$ ;

2. formation d'une marche d'amplitude plus petite que l'amplitude à mi-hauteur du pic principal (régime « supranormal », $L_0/4 \lessapprox \sigma \lessapprox L_0/3.5$) ;

3. croisement rapide de la largeur de la distribution due à la marche qui dépasse la mi-hauteur du pic principal (régime « anormal », $L_0/3.5 \lessapprox \sigma$) ;

Ainsi, $L$ et $\sigma$ sont les paramètres du système responsables du décalage et de l'élargissement asymétrique du pic Raman vers le rouge.

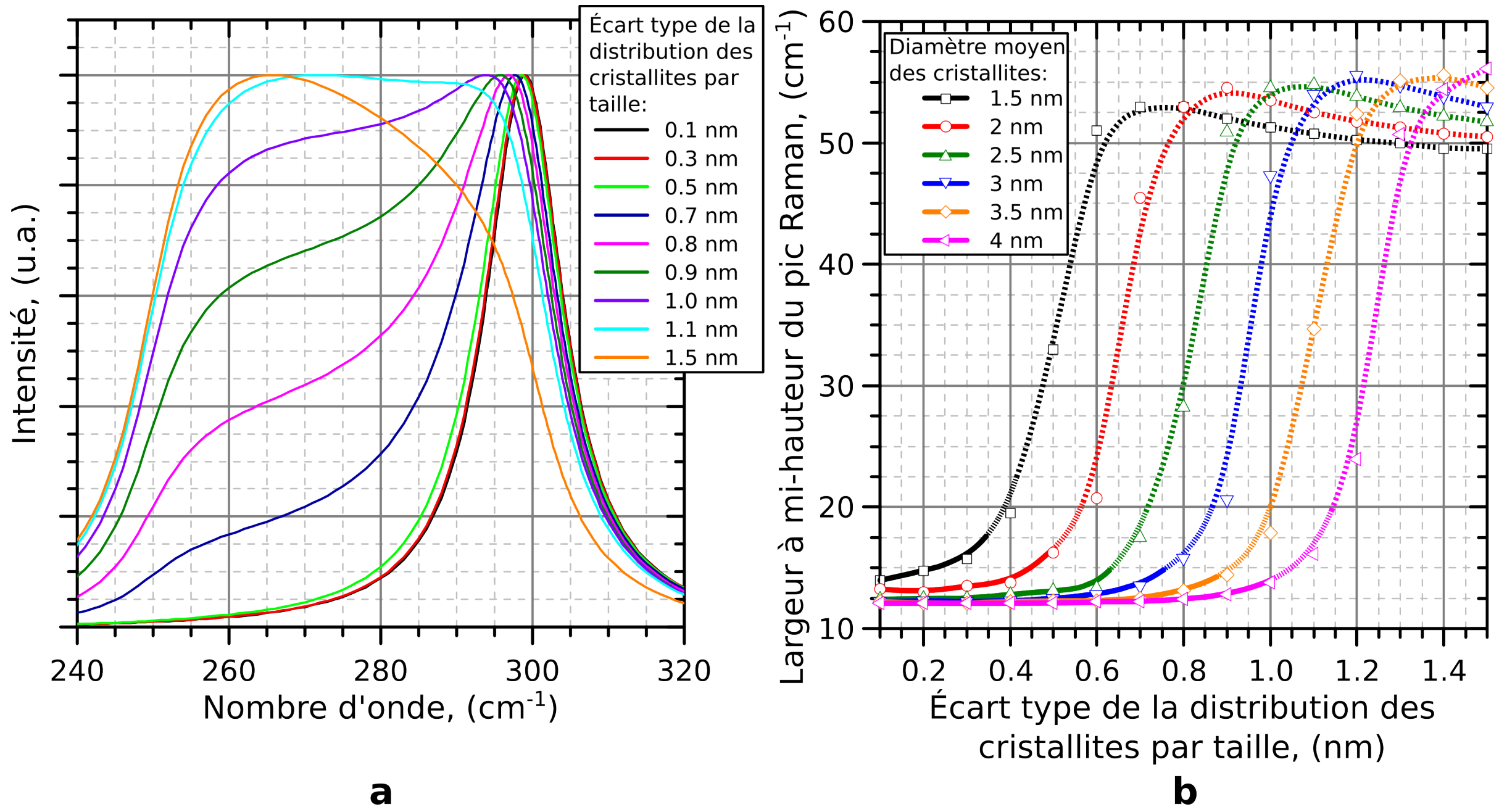

Figure 2.8 (a) Simulation des spectres de diffusion Raman pour différentes valeurs de $\sigma$ lorsque le diamètre moyen des cristallites $L_0$ est de 2.5 nm ; (b) largeur du spectre simulé mesurée à mi-hauteur du pic Raman en fonction de $\sigma$

Remarquons, que l'expression simplifiée 2.16 ne contient pas de $\sigma$. Elle décrit, ainsi, le régime « normale ». Cependant, l'approximation $\sigma < L_0/3$ utilisée est trop large ; elle mène au résultat incorrect lorsque $L_0/4 \lessapprox \sigma \lessapprox L_0/3$.

Une large distribution granulométrique des nanocristaux donne des profils Raman similaires à un mélange de nanocristallites et une petite quantité d'une phase amorphe [54]. Dans ce cas, un modèle de particules avec un noyau cristallin et une coquille amorphe peut être appliqué [53]. Par conséquent, la présence d'une queue plus forte et plus longue dans des spectres Raman laisse penser qu'une phase amorphe existe, ce qui peut ne pas être le cas. Alors, pour traiter les spectres Raman correctement, les couches poreuses doivent être analysées par d'autres méthodes, telles que le TEM (Microscopie Électronique en Transmission) et la RBS (Spectroscopie de rétrodiffusion de Rutherford). En pratique, le décalage et élargissement des spectres des nanocristaux du Ge sont moins prononcés que ceux du nano-Si. Le phénomène s'explique par une plus faible dispersion des phonons optiques[55].

### Montage expérimentale

Les spectres Raman ont été réalisés à l'aide du micro-spectromètre Renishaw RM1000, équipé d'un laser à Argon comme source excitatrice à la longueur d'onde 514.5 nm. Le schéma de cet appareil est présenté sur la Figure 2.9.

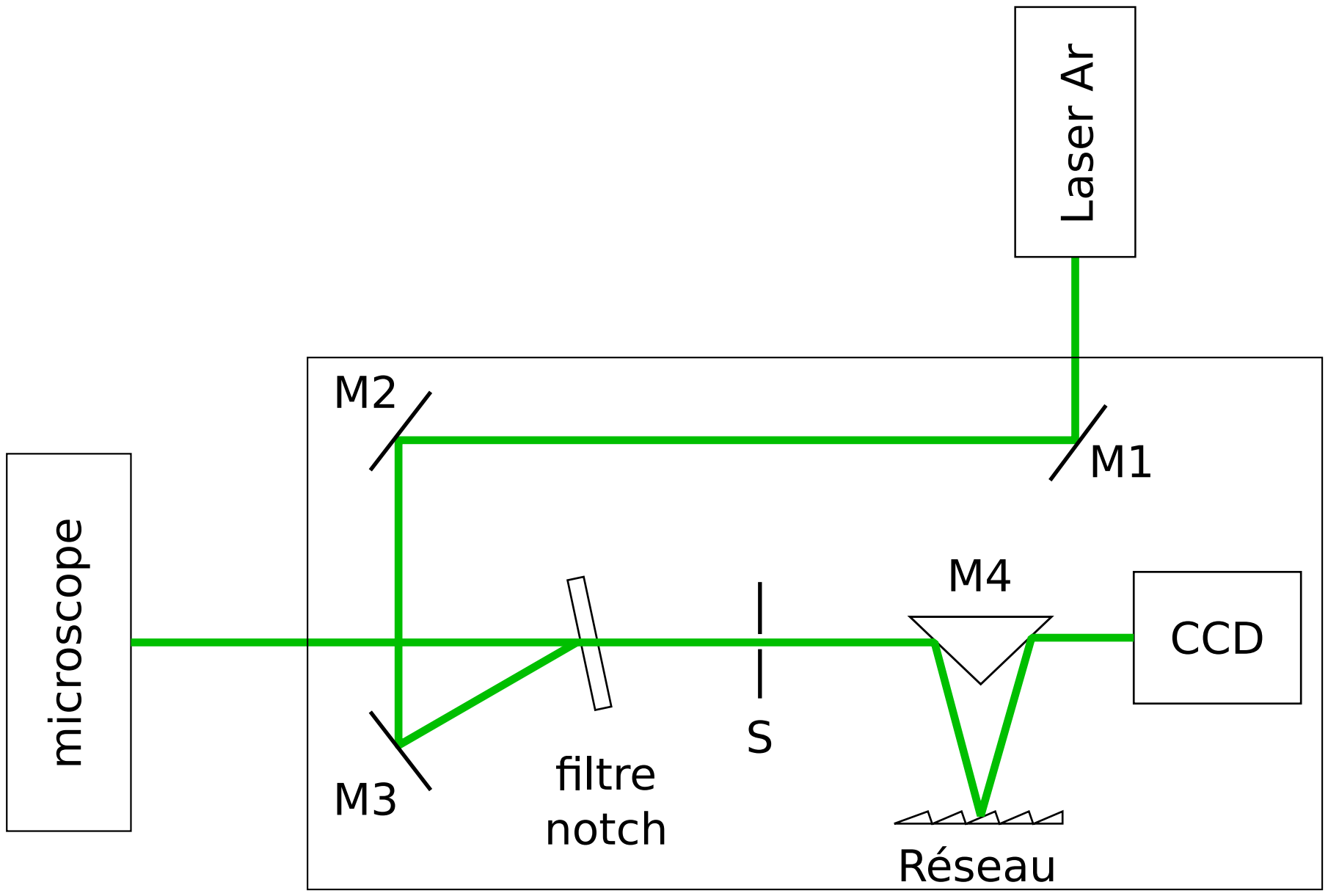

Figure 2.9 Schéma du spectromètre Renishaw RM1000 [27].

Les miroirs M1 et M2 sont mobiles et sont utilisés pour aligner le faisceau avec les optiques. Le miroir M3 est fixe et permet de renvoyer le faisceau sur un filtre « notch » qui joue aussi le rôle de miroir. Ce filtre coupe les fréquences comprises entre $-100$ et 100 $cm^{-1}$ par rapport à la longueur d'onde excitatrice. Le faisceau incident, polarisé ou non est envoyé vers un microscope Olympus. Après interaction avec l'échantillon, la lumière diffusée est renvoyée sur la fente S qui est la fente d'entrée du spectromètre 1. Un miroir M4 de forme triangulaire renvoie le faisceau vers le réseau ayant 1800 traits/mm. La lumière dispersée est finalement détectée par une caméra CCD refroidie par effet Pelletier. Le faisceau laser peut être focalisé sur un diamètre de 5 µm) sur l'échantillon à l'aide d'un objectif 50X. Les spectres peuvent être enregistrés dans une fenêtre allant de 200 à 1200 $cm^{-1}$ [27].

La profondeur d'absorption de la lumière incidente dans le Ge poreux est inférieure à l'épaisseur typique des couches poreuses. Afin d'éviter l'influence de la couche amorphisée qui peut se trouver sur la surface des échantillons du Ge poreux, la couche poreuse a été détachée mécaniquement du substrat et a été reposée la couche amorphe en bas. Lors d'analyse de la morphologie des couches poreuses en fonction de l'épaisseur des couches (Section 3.2.2), les échantillons ont été mesurés en coupe transversale.

### 2.2.4 La spectroscopie Infrarouge à Transformée de Fourier (*FTIR*)

La Spectroscopie Infrarouge (IR) est basée sur l'absorption d'un rayonnement infrarouge par le matériau analysé. Elle permet, via la détection des vibrations caractéristiques des complexes chimiques, d'effectuer l'analyse chimique d'un matériau.

Le principe de la spectroscopie IR est le suivant : lorsque l'énergie du faisceau lumineux est voisine de l'énergie de vibration d'une liaison chimique, cette dernière absorbe le rayonnement et on enregistre une diminution de l'intensité réfléchie ou transmise à cette longueur d'onde. Le domaine spectral entre 4000 $cm^{-1}$ et 400 $cm^{-1}$ (2.5–25 µm) correspond au domaine de l'énergie de vibration de diverses molécules.

Toutes les vibrations ne donnent pas lieu à une seule bande d'absorption, cela va dépendre aussi de la géométrie de la molécule et en particulier de sa symétrie. Pour une géométrie donnée on peut déterminer les modes de vibration actifs en infrarouge grâce à la Théorie des Groupes. Par conséquent, à un matériau de composition chimique et de structure données va correspondre un ensemble de bandes d'absorption caractéristiques permettant d'identifier le matériau.

Il existe deux sortes de spectromètres infrarouge : les spectromètres à balayage et les spectromètres à transformée de Fourier. Dans le cadre de cette thèse le spectromètre infrarouge Perkin Elmer GSX-2 à transformée de Fourier (interféromètre de Michelson) a été utilisé.

Le principe de la spectroscopie FTIR est le suivant. Le faisceau IR provenant de la source infrarouge (dans notre cas, c'est une lampe halogène en tungstène produisant le rayonnement IR proche ou un filament chauffé à 1350 K fournissant le rayonnement dans l'IR moyen et lointain) est dirigé vers l'interféromètre de Michelson qui va moduler chaque longueur d'onde du faisceau à une fréquence différente. Le faisceau modulé est alors réfléchi vers l'échantillon où des absorptions interviennent et arrive ensuite sur le détecteur pour être transformé en signal électrique. Le signal du détecteur apparaît comme un interférogramme, c'est à dire une signature de l'intensité en fonction de la position du miroir mobile de l'interféromètre Michelson. Cet interférogramme est ensuite converti en un spectre infrarouge (intensité de l'absorption en fonction de la longueur d'onde) par transformée de Fourier. L'intensité intégrale d'une bande d'absorption peut être reliée à la concentration des liaisons chimiques responsables de l'absorption.

Il existe une grande diversité de modes opératoires en spectroscopie FTIR. Dans le cadre de ce travail deux modes principaux ont été utilisés : (i) Réflexion et (ii) Réflexion Totale

Atténuée. Le premier mode permet de déterminer la porosité d'une couche. Son principe est présenté sur la Figure 2.10 a. Des interférences constructives ou destructives entre le faisceau incident puis réfléchi à l'interface air / Ge poreux et le faisceau réfléchi à l'interface Ge poreux / Ge monocristallin peuvent être clairement observées sur les couches dont l'épaisseur $d \leq \lambda_{reflechi}$ (Figure 2.10 b). L'indice optique de la couche ($n$) est déterminé à partir du spectre de réflectivité, en considérant la différence de chemin optique entre les deux faisceaux réfléchis. Les conditions pour les maximums d'interférence d'ordres $k$ et

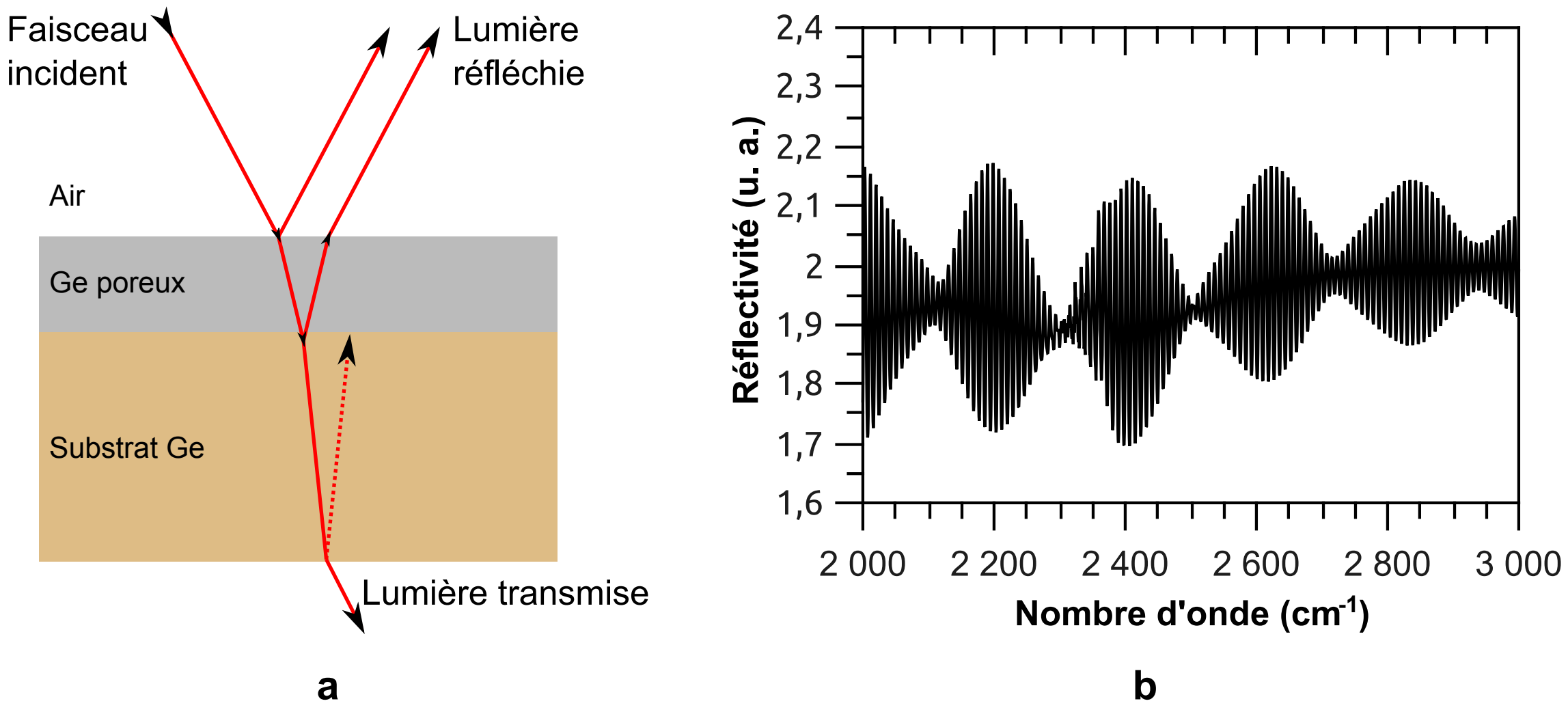

Figure 2.10 (a) L'interférence d'un rayon infrarouge sur une couche du Ge poreux ; (b) spectre de réflectivité d'une couche du Ge poreux (éponge, $P = 75\%$, $d = 3$ µm). Les oscillations fines avec la période $\approx 10$ cm$^{-1}$ correspondent à l'interférence sur le substrat

$k + N$ peuvent être ainsi écrites :

$$2dn = k\lambda_k \tag{2.18}$$

$$2dn = (k+N)\lambda_{k+N} \tag{2.19}$$

où $\lambda_x$ est la longueur d'onde correspondante au maximum de l'ordre $x$ ; $k$ et $N$ sont des nombres entiers ; $N$ — le nombre des maximums d'interférence entre $\lambda_k$ et $\lambda_{k+N}$. Ainsi, on obtient :

$$n = \frac{N}{2d}\left(\frac{1}{\lambda_k} - \frac{1}{\lambda_{k+N}}\right)^{-1} \tag{2.20}$$

La porosité moyenne peut être estimée à partir de ces mesures de l'indice de réflexion en utilisant les modèles du milieu effectif de Brouggeman ou de Maxwell-Garnett [25]. Selon le modèle de milieu de Brouggeman, par exemple, l'équation mettant en corrélation l'indice de réflexion et la porosité d'une couche est suivante :

$$(1-P)\frac{n_{\mathrm{Ge}}^2-n^2}{n_{\mathrm{Ge}}^2+2n^2}+P\frac{1-n^2}{1+2n^2}=0 \tag{2.21}$$

où $P$ est la porosité moyenne d'une couche du Ge poreux, $n_{\mathrm{Ge}}$ — l'indice de réfraction du Ge monocristallin.

Le principe du deuxième mode appelé ATR (*Attenuated Total Reflection*) est expliqué sur la Figure 2.11. L'échantillon à analyser est placé sur la surface d'un cristal à fort indice de réfraction (typiquement le Ge, comme son indice de réfraction est le plus élevé parmi les semi-conducteurs conventionnels ($n$(Ge) $\approx 4.14$ dans le proche infrarouge). L'analyse des échantillons de Ge poreux est toujours possible grâce à son indice de réfraction inférieur à celui du Ge monocristallin. Le faisceau IR guidé dans le cristal initial traverse le Ge et subissant une réflexion totale à l'interface Ge-échantillon est dirigé vers le détecteur. Le faisceau guidé est légèrement perturbé par l'existence d'ondes progressives transversales appelées ondes évanescentes. Celles-ci pénètrent dans une certaine profondeur de l'échantillon se trouvant en contact direct avec le Ge. Une partie de l'énergie des ondes évanescentes est absorbée par des liaisons chimiques et la réflexion totale du faisceau principal est dite atténuée. Etant très sensible, cette technique permet d'identifier et de quantifier les liaisons chimiques présentes sur la surface interne du Ge poreux en évitant les saturations des spectres en régime de transmission sur les couches auto-portées ou des pics parasites dûs à l'interférence en régime de réflexion sur les couches fines de Ge poreux.

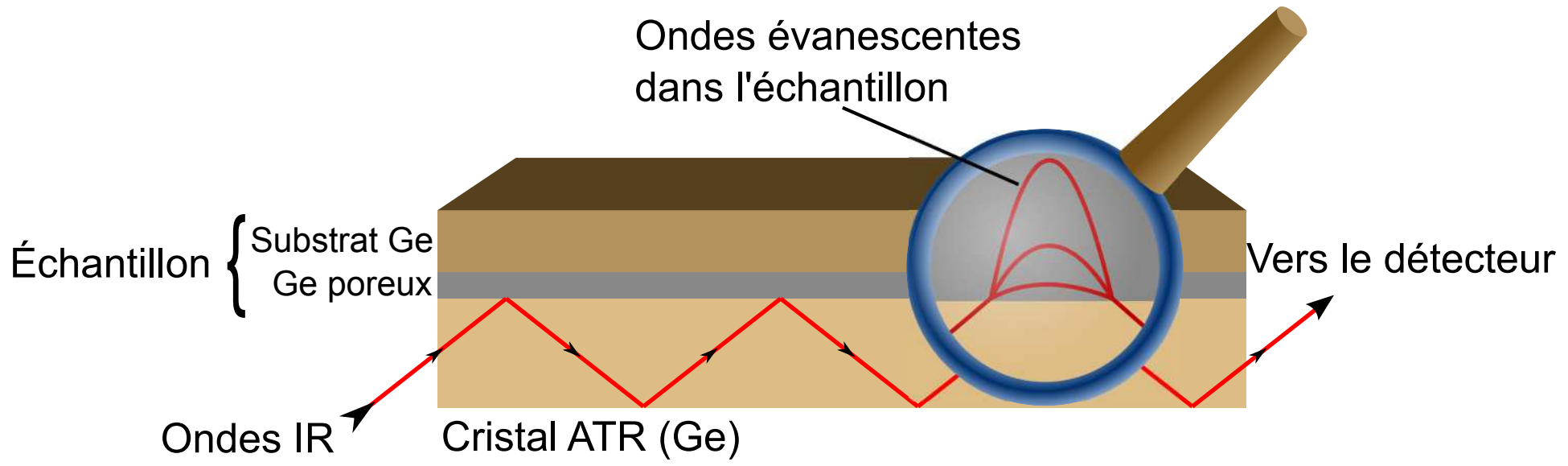

Figure 2.11  Principe de l'ATR

### 2.2.5 RBS

La spectroscopie de rétrodiffusion de Rutherford (*Rutherford Backscattering Spectroscopy*, «*RBS*») est une technique utilisée pour déterminer la structure et la composition de matériaux par l'analyse de la rétrodiffusion d'un faisceau d'ions, en l'occurrence des particules $\alpha$ de quelques MeV émises par une source radioactive. Ces $\alpha$ sont interceptées après diffusion sur l'échantillon par un détecteur solide. La mesure de leurs énergies résiduelles permet la mesure de la masse de l'atome diffuseur et par conséquent l'identification de ce dernier.

Les points principaux de l'analyse RBS :

- La pratique de l'analyse RBS nécessite une infrastructure lourde : accélérateur et équipements associés, personnel qualifié, normes de protection à respecter, etc ;
- L'acquisition des données est rapide (15 min. par échantillon) et leur exploitation quantitative est facile à effectuer en général ;
- L'analyse RBS ne nécessite pas de préparation particulière de l'échantillon et elle est non destructive ;
- La sensibilité de la méthode augmente fortement avec le numéro atomique $Z$ (soit la charge du noyau). Le cas favorable est celui d'une impureté lourde présente dans une matrice légère. Cependant la sélectivité décroît pour les fortes masses atomiques.
- L'analyse RBS permet dans de nombreux cas de déterminer la composition d'un mélange de divers constituants. Par contre, elle ne fournit aucune information sur l'état chimique (états de charge, nature des liaisons) des espèces atomiques impliquées dans ce mélange.
- Les profils de répartition sont mesurables sur des profondeurs de 10 nm à quelques µm

#### Principe de base

En première approximation, la rétrodiffusion de Rutherford peut être considérée comme une collision élastique entre sphères dures. En assumant la conservation de l'énergie et de la quantité de mouvement, l'énergie de l'élément diffusé peut être déterminée. Le facteur cinématique $K$ est le rapport entre l'énergie $E$ de l'ion incident après diffusion sur un atome en surface de la cible et son énergie initiale $E_0$ (Figure 2.12 a).

$$K = \left( \frac{m_1 \cos\theta + \sqrt{m_2^2 - m_1^2 \sin^2\theta}}{m_1 + m_2} \right)^2 \tag{2.22}$$

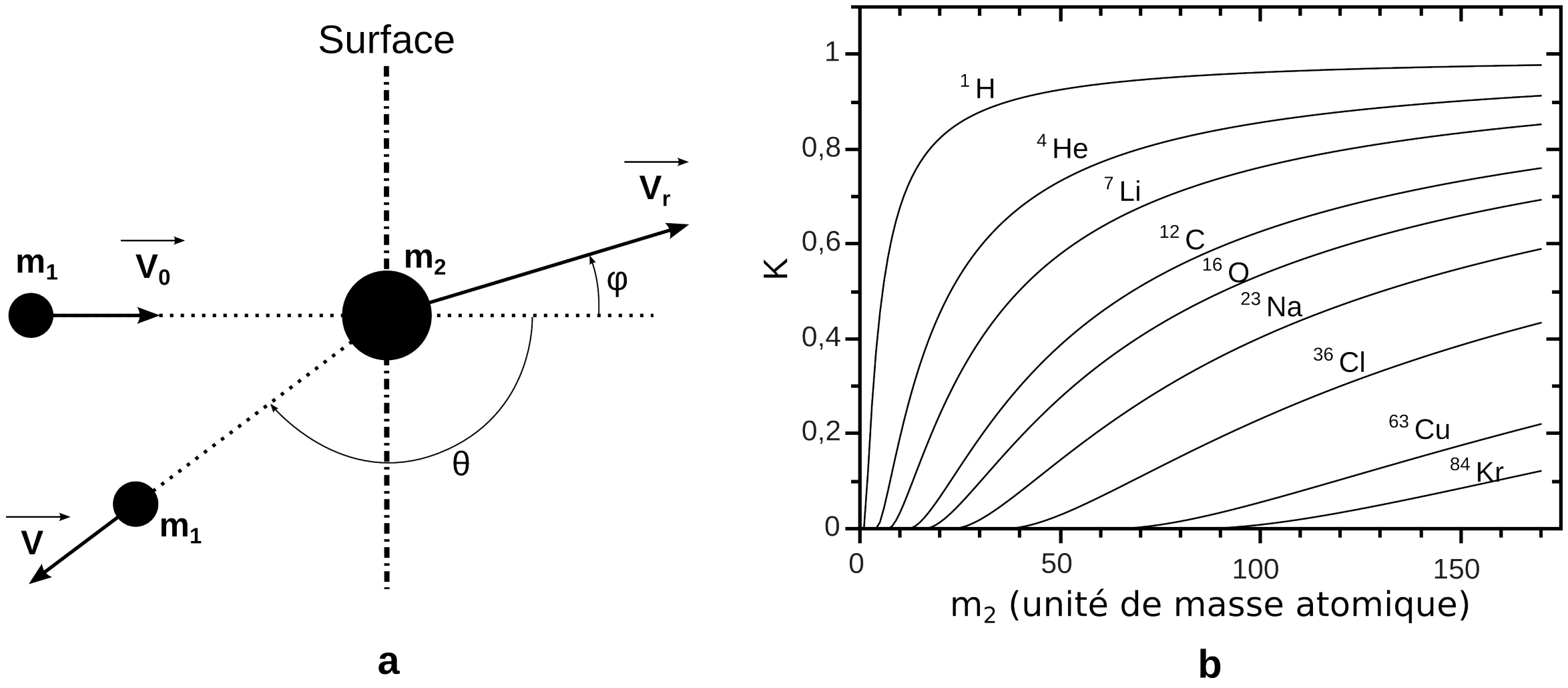

Figure 2.12 (a) Représentation d'une diffusion à grand angle. L'interaction coulombienne entre l'ion incident et le noyau cible est schématisée par un choc entre deux sphères dures ; (b) Évolution du facteur cinématique K en fonction de la masse atomique de l'atome cible pour divers projectiles incidents. Angle de diffusion : $\theta = 172°$

Dans des conditions expérimentales données (paramètres $m_1$, $E_0$ et $\theta$ fixés), $E$ est une fonction univoque de $m_2$. En d'autres termes, la mesure de l'énergie des ions ayant rétrodiffusé en surface de la cible permet de remonter à la masse atomique des atomes qui la constituent (Figure 2.12 b).

Afin d'obtenir la distribution angulaire des particules rétrodiffusées, il est nécessaire de déterminer la section efficace différentielle : En se basant sur un potentiel coulombien entre le projectile (charge $Z_1 q$) et le noyau cible (charge $Z_2 q$), Rutherford et Darwin ont établi la formule suivante :

$$\sigma = \left(\frac{Z_1 Z_2 q^2}{16\pi\varepsilon_0 E_0}\right)^2 \frac{4}{\sin^4\theta} \frac{\left\{\left[1-((m_1/m_2)\sin\theta)^2\right]^{1/2}+\cos\theta\right\}^2}{\left[1-((m_1/m_2)\sin\theta)^2\right]^{1/2}} \tag{2.23}$$

Les équations 2.22 et 2.23 permettent de déterminer l'énergie et la probabilité qu'une particule soit détectée à un angle solide pour une seule collision. La rétrodiffusion élastique est un événement rare. En fait, l'immense majorité des ions incidents va progressivement ralentir dans la cible et s'y implanter. Dans le domaine d'énergie considéré ($\approx$ 1 MeV par unité de masse du projectile), le ralentissement est essentiellement gouverné par les processus dits « électroniques », correspondant à l'interaction entre l'ion incident et les électrons du matériau cible. Compte tenu de l'énorme rapport de masse entre un ion (même léger) et un électron, le ralentissement électronique s'effectue sans déviation de trajectoire du projectile et par conséquent agit de façon analogue à une force de frottement. La section efficace d'arrêt $\varepsilon$ (unité : eVcm$^2$) chiffre la perte d'énergie $dE$ subie par le projectile lorsqu'il traverse une couche de matière de densité atomique $N_{at}$ (en cm$^{-3}$) et d'épaisseur $dx$ (Figure 2.13) :

$$dE = -\varepsilon(E)N_{at}dx \tag{2.24}$$

Considèrerons le mouvement d'un ion à travers la matière. Émis avec une énergie $E_0$

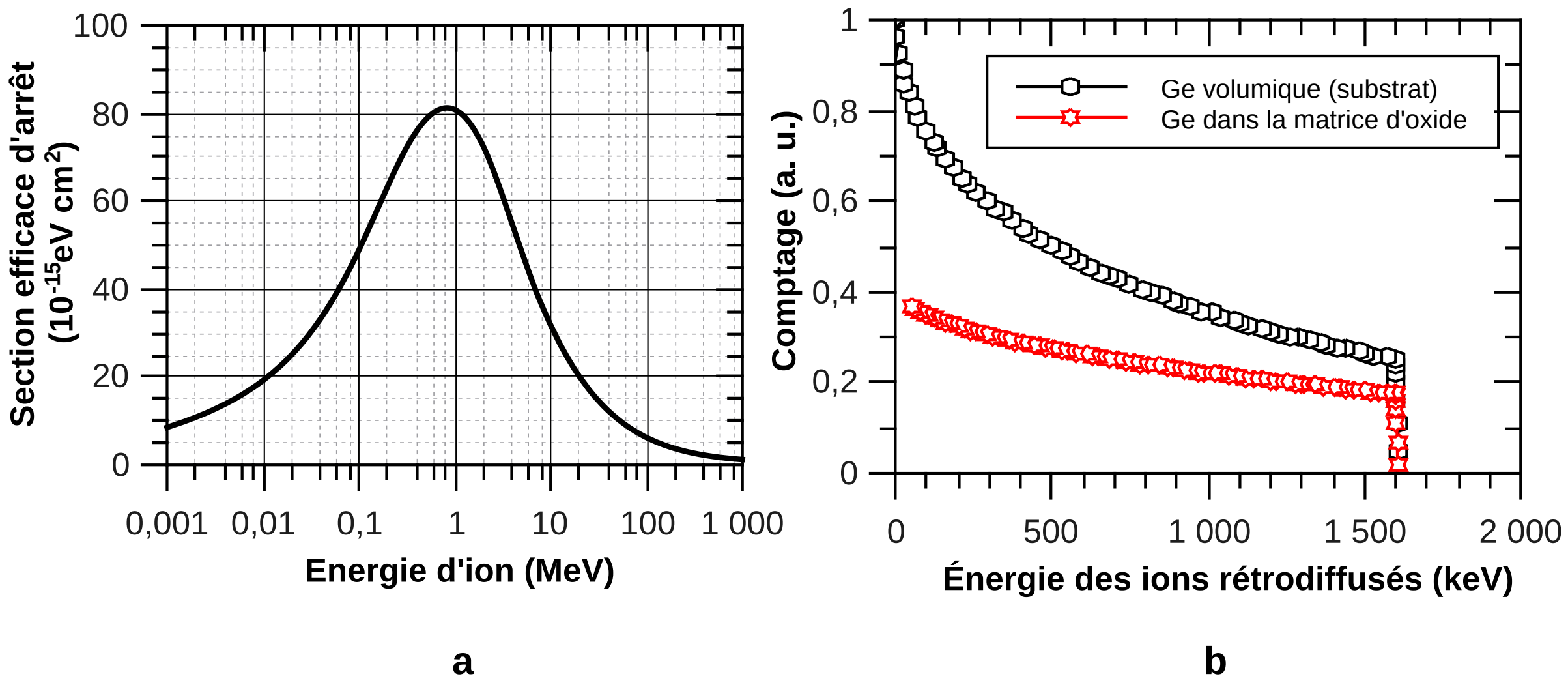

Figure 2.13 (a) Section efficace d'arrêt $\epsilon$ d'ions $^4$He$^+$ dans une cible du Ge [10, 70, 71] ; (b) Simulation des spectres RBS du Ge identique pour tous les ions, il est ralenti par une couche de matière d'épaisseur $l$ à l'énergie $E'(E_0, l)$. L'énergie $E'$ est calculée par intégration numérique de l'équation 2.24 connaissant la dépendance $\varepsilon(E)$. Étant rétrodiffusé avec une probabilité $\propto \sigma(E')$, l'ion perd de l'énergie par le facteur cinématique $K$. Finalement, l'ion traverse la matière en sens inverse en obtenant l'énergie $E = E'(K \cdot E'(E_0, l), l)$, qui est mesurée par le détecteur. Seuls les ions d'énergie $E > 0$ sort du matériau. Le spectre RBS est donc un graphique paramétrique $y \propto \sigma(E'(E_0, l))$, $x = E$ pour tous $l \in [0, \infty)$ (Figure 2.13). Les valeurs de $m_2$ et $N_{at}$ peuvent être extraites par ajustement des spectres mesurés sur les courbes simulées.

Les expériences ont été menée à l'aide d'ions $^4He^+$ de 2 MeV délivrés par l'accélérateur Van de Graaff de 4 MV de l'Institut de Physique Nucléaire de Lyon (IPNL). Les particules rétrodiffusées sont détectées avec une jonction implantée et fixée à un angle de 172° par rapport à l'axe du faisceau dont la résolution de détection est de 13 keV.

# CHAPITRE 3

# Ge mesoporeux par gravure électrochimique

Dans ce chapitre, nous montrerons :

(a) comment en choisissant les conditions d'anodisation adéquates nous pouvons fabriquer des couches épaisses et homogènes de Ge mésoporeux en utilisant la technique BEE (gravure électrochimique bipolaire) ;

(b) l'influence des différents paramètres d'anodisation sur la morphologie des couches formées ;

(c) enfin, nous terminerons par l'analyse conceptuelle des mécanismes de formation du Ge poreux en s'appuyant sur la base des données disponibles.

## 3.1 Vers la compréhension du processus BEE

Nos premières expériences ont été basées sur les paramètres de la gravure BEE proposés par E. Garralaga Rojas et al. [39, 41, 98]. La gravure électrochimique du Ge a été réalisée dans $HF_{49\%}$ sur des substrats de Ge de type p (0.005–0.04 $cm^{-1}$) de 100 µm d'épaisseur. Les amplitudes des impulsions étant les mêmes ($J_+ = J_-$) elles ont été variées dans la gamme de 0.25 $mA/cm^2$ à 7.2 $mA/cm^2$, avec la durée des impulsions $t_{on} = 60$ s et $t_{off} = 5$ s.

L'image typique du Ge mesoporeux formé sous ces conditions expérimentales est présentée sur la Figure 3.1 (a). Une fine couche nanoporeuse d'épaisseur de $\approx$ 290 nm sans structure de pores clairement définie est observée. En effet, comme il a été déjà mentionné par E. Garralaga Rojas et al.[39, 98], ces paramètres de gravure électrochimique n'assurent pas le contrôle suffisant de la vitesse de gravure et par conséquent, de l'épaisseur de la couche poreuse, en raison de la dissolution importante du PGe déjà formé.

Nous avons également observé qu'en utilisant ce protocole, il est assez difficile d'ajuster la porosité des couches du PGe. Seules des couches poreuses d'épaisseur inférieure à 1 µm et de porosité supérieure à 50% peuvent être obtenues indépendamment de la durée de la gravure (>5 h) et de la densité de courant d'anodisation appliquée. De plus, nous avons observé que la densité de courant d'anodisation n'avait pas un effet apparent sur la morphologie des couches poreuses, contrairement au cas des autres semi-conducteurs poreux comme Si ou SiC. Seules les couches nano-spongieuses semblables à celle présentées sur la Figure 3.1 (a), ont été obtenues (Figure 3.2).

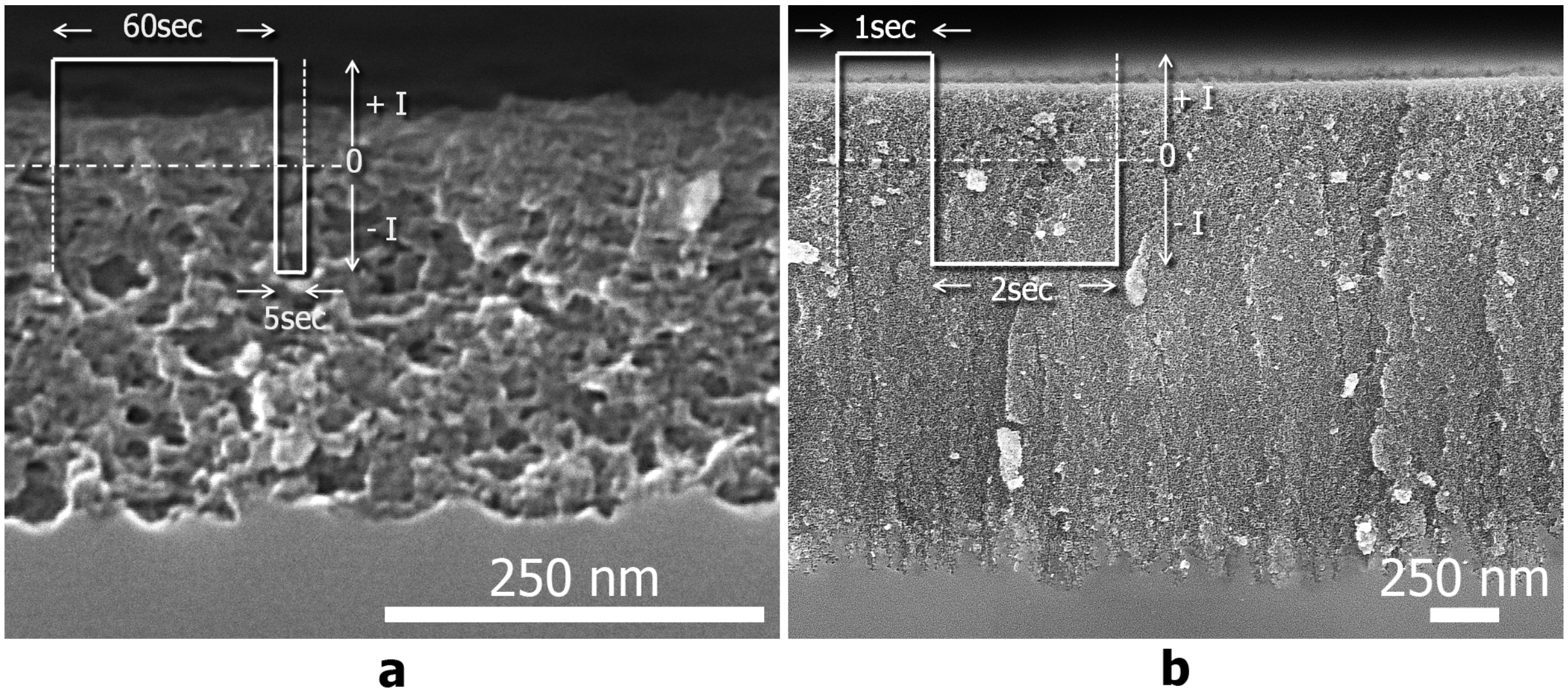

Figure 3.1 Images MEB en coupe transversale des couches de Ge poreux formées par la technique *BEE* dans l'électrolyte $HF_{49\%}$ pur pendant 3 h à 1.7 mA/cm$^2$ avec des durées d'impulsions suivantes :
(a) $t_{on} = 60$s, $t_{off} = 5$ s ; (b) $t_{on} = 1$ s, $t_{off} = 2$ s.

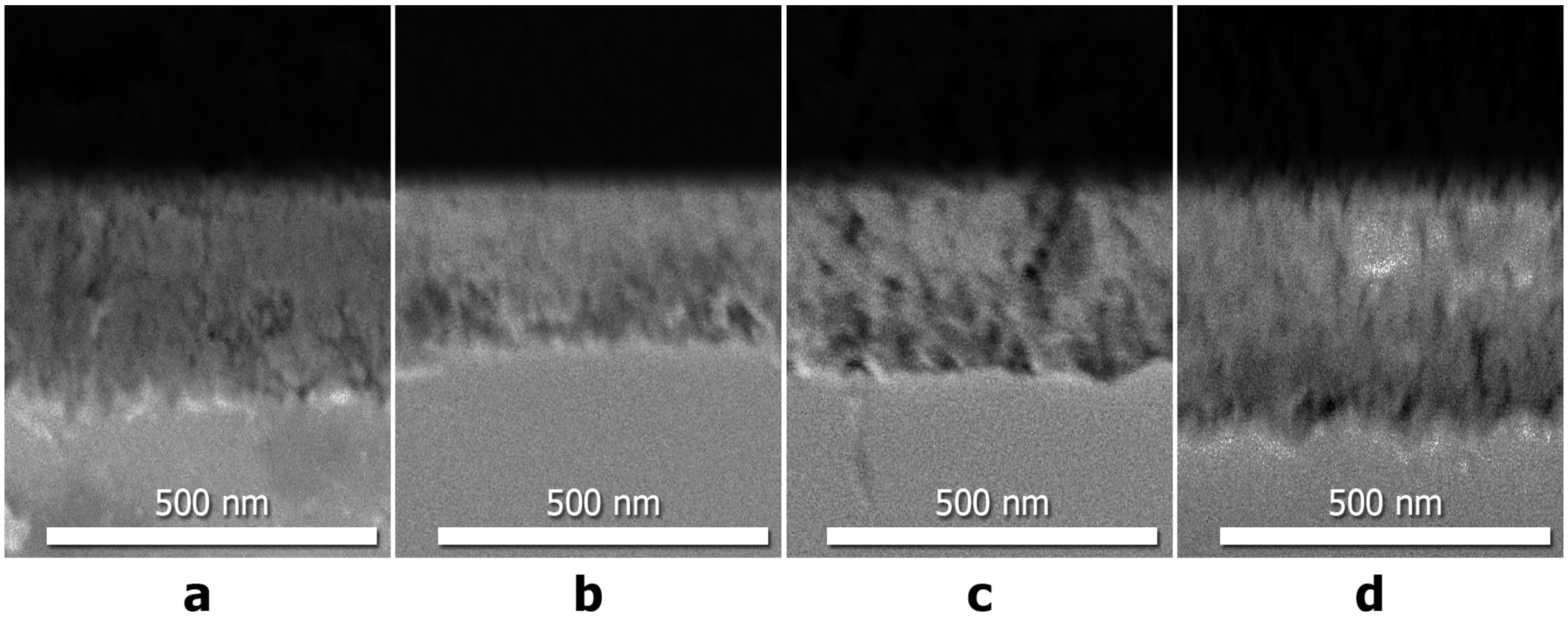

Figure 3.2 Images MEB en coupe transversale des couches de Ge poreux formées par la technique *BEE* dans l'électrolyte $HF_{49\%}$ pur pendant 3 h avec $t_{on} = 60$ s, $t_{off} = 5$ s et la densité de courant suivante :
(a) 0.45 mA/cm$^2$ ; (b) 0.9 mA/cm$^2$ ; (c) 1.8 mA/cm$^2$ ; (d) 3.6 mA/cm$^2$ ;

La faible épaisseur des couches du PGe formées et l'importante détérioration de leur structure poreuse nous suggère l'hypothèse que pour ces conditions expérimentales un endommagement notable des parois des pores a lieu pendant les étapes anodiques de la gravure BEE. Ceci est en raison d'un temps de persistance de la couche de passivation des pores avec $H^+$. En effet, si ce temps de persistance de la couche passée est trop courte par rapport à la durée de l'étape anodique ($t_{on}$), la dissolution de la couche poreuse déjà formée va fortement perturber la croissance des pores suivants. Cela signifie que sans barrière énergétique assurée par la passivation induite par l'apport d'ions $H^+$, le courant ne passera pas exclusivement par les extrémités des pores, mais également à travers les parois latérales des pores, ce qui entrainera la destruction de la structure poreuse déjà formée.

Il existe un autre phénomène associé à cette observation, soit la dissolution purement chimique. En effet, les pores n'ayant pas de barrière de la protection subissent la gravure purement chimique. Par conséquent, afin d'éviter la dissolution de la couche poreuse (quel que soit le mécanisme de dissolution chimique et/ou électrochimique) et de favoriser la croissance des pores, la durée de chaque impulsion anodique doit être inférieure ou comparable au temps caractéristique de persistance de la couche de passivation des parois des pores avec $H^+$ ($\tau_p$). Le temps caractéristique $\tau_p$ définit la durée de résidence des ions $H^+$ pourraient rester à la surface du Ge sous une polarisation anodique. La valeur de $\tau_p$ donne donc une indication de la période pendant laquelle la surface interne du Ge poreux déjà formé sera protégée contre la dissolution (chimique et/ou électrochimique).

Quelques indications concernant ce temps de persistance caractéristique peuvent être trouvées dans la littérature [31, 39, 98], pourtant aucun valeur typique n'a été donnée. Il a été mentionné que $\tau_p$ dépend principalement de la densité de courant de l'étape anodique subséquente et de la concentration de l'électrolyte. On s'attend à ce que la valeur de $\tau_p$ diminue avec une augmentation de la concentration de l'électrolyte et de la densité de courant appliquée. Dans nos expériences, ce temps de persistance caractéristique de la passivation avec $H^+$ a été déterminée expérimentalement. Pour nos conditions expérimentales : $HF_{49\%}$ et $|J_+| = |J_-| = 0.5$–$3.0$ mA/cm$^2$, $\tau_p$ a été estimé à environ de 1 s. D'ailleurs, la phase de cathodisation doit être terminée aussitôt que la surface est complètement passivée. Pour la même gamme de courants, la durée de la phase cathodique $t_{off}$ qui corresponde à une passivation compète de la surface a été estimé à environ de 2 s. Sur la base de ces résultats et pour les expériences suivantes, la durée de chaque impulsion anodique ($t_{on}$) a été diminuée de 60 s à 1 s et la durée de chaque impulsion cathodique ($t_{off}$) a été diminuée de 5 s à 2 s.

La structure poreuse réalisée avec le nouvel ensemble de paramètres est montrée à la Figure 3.1 (b). Une différence explicite entre les images (a) et (b) peut être notée. Bien que les deux structures soient obtenues après 3 h de gravure, la seconde est 7 fois plus

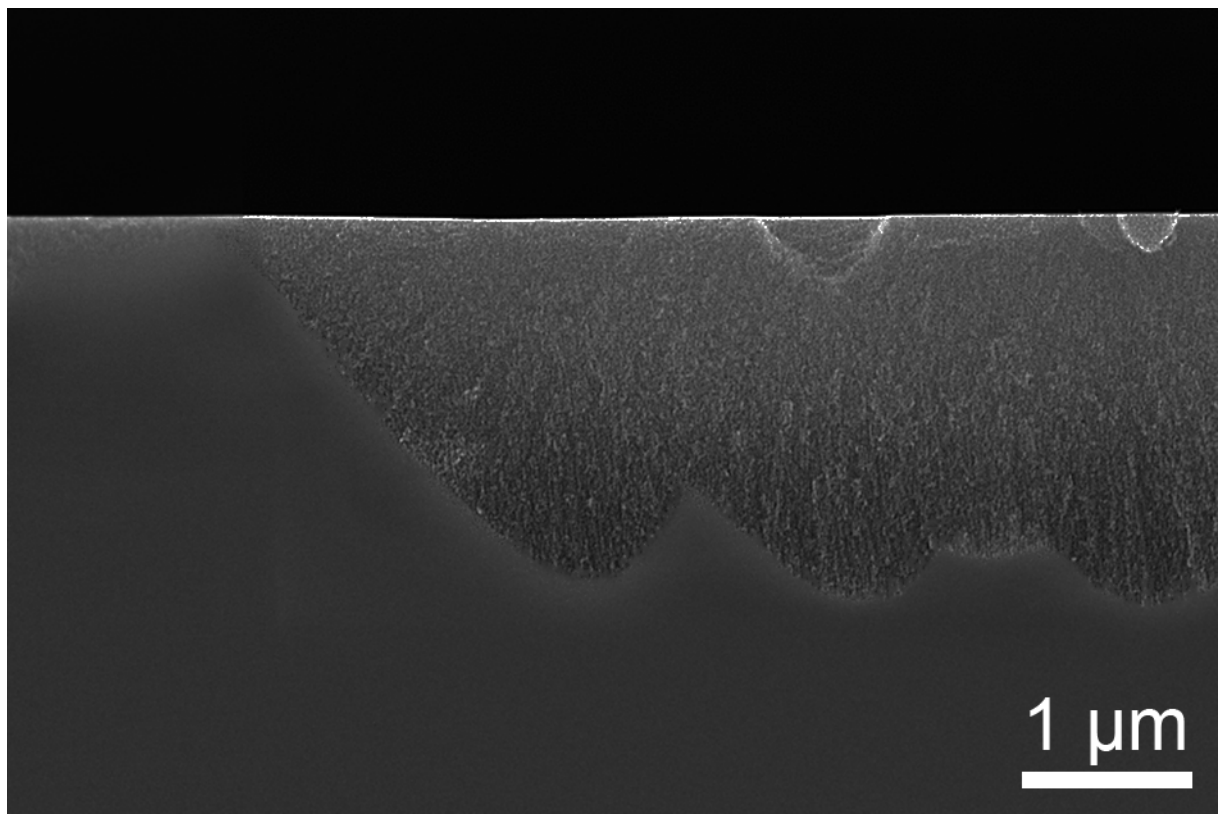

Figure 3.3 Image MEB en coupe transversale de la couche de Ge poreux (BEE, $HF_{49\%}$, 2 mA/cm$^2$, $t_{on}/t_{off} = 1$ s/2 s, 2 h) capturée au bord de la zone porosifiée

épaisse (2 µm par rapport à 290 nm de la structure (a)), homogène et montre un squelette poreux bien défini, similaire à celui du Si poreux typique. Comme on peut le voir sur la Figure 3.3, la surface de la partie porosifiée est au même niveau que celle du substrat de Ge, ce qui indique qu'elle ne se déplace pas en cours de porosification. Ainsi, l'inhibition de la dissolution de la surface de l'échantillon rend possible la réalisation de nouvelles structures poreuses et l'analyse de leurs propriétés.

## 3.2 Influence des paramètres du procédé sur les propriétés des structures

En général, toutes les propriétés structurelles et morphologiques des semi-conducteurs poreux, telles que l'épaisseur, la porosité, le diamètre des pores, la taille des cristallites et la surface spécifique, sont contrôlées par :

- la densité de courant anodique ($J_+$) ;
- le temps d'anodisation ($T$) ;
- la composition et la concentration de l'électrolyte ($HF_{\%}$) ;
- le type et le niveau de dopage du substrat ;
- le niveau d'illumination.

Dans le cas de la gravure bipolaire du Ge, les paramètres de gravure à ajouter sont :

- la durée des impulsions anodiques et cathodiques ($t_{on}$, $t_{off}$ respectivement) ;
- la densité de courant de cathodisation ($J_-$).

L'influence de ces paramètres sur la morphologie et la structure des couches de PGe sera étudiée en détail dans les paragraphes suivants.

### 3.2.1 Influence de la densité de courant anodique et du temps d'anodisation

Parmi les paramètres du procédé de gravure électrochimique, la densité de courant et le temps d'anodisation semblent avoir l'effet le plus direct sur la morphologie [17]. C'est la raison pour laquelle nous avons choisi en premier ces paramètres pour étudier leurs influences sur la morphologie du Ge poreux. Dans nos expériences la densité de courant variait entre 0.25 et 5.0 mA/cm$^2$ et le temps de gravure total de 5 min à 20 h. Il est à noter également que pour des densités de courant plus importantes, l'effet d'électropolissage commence à jouer un rôle majeur, ce qui conduit à la destruction de la couche poreuse.

La Figure 3.4 présentent les images MEB de l'interface entre le Ge monocristallin et la région poreuse pour des densités de courant anodique différentes et deux temps de gravure : (i) 3 h et (ii) >6 h.

Pour les densités de courant supérieures à 1.8 mA/cm$^2$, les pores sont bien séparés et présentent une orientation bien définie. Lorsque le temps de gravure total est assez court ($\approx$ 3 h), les pores individuels se propagent de façon rectiligne dans la direction préférentielle <100> (Figure 3.5), formant ainsi une morphologie colonnaire avec des branches courtes de second niveau seulement (Figure 3.4 a). La formation de nombreux pores latéraux a été observée uniquement pour les temps de gravure assez long (>6 h) (Figure 3.4 b). Dans ce cas les pores latéraux ont des branches de deuxième et de troisième niveau et même de niveaux supérieurs. Contrairement aux pores principaux, les branches ne suivent pas les directions cristallines. La colonne droite des images révèle une transformation de morphologie dans une couche de transition. Lorsque le procédé est suffisamment long (et les structures sont suffisamment épaisses), dans les trois cas, il se forme des canaux dendritiques souples, autrement appelés des « fils ». Le fait que la densité et la morphologie des « fils » peut dépendre de la morphologie de la couche supérieure signifie une similarité de mécanismes de formation des pores.

Indépendamment du temps de gravure total, l'application des densités de courant faibles (< 1.8 mA/cm$^2$) conduit à la formation d'un réseau de pores fortement interconnectés sans une direction de croissance bien définie, comme nous pouvons le voir sur les Figures 3.4 c-f. La morphologie dendritique peut être obtenue uniquement pour les densités de courant inférieures à 1.6 mA/cm$^2$ (Figure 3.4 e, f), alors que pour les densités de courant de

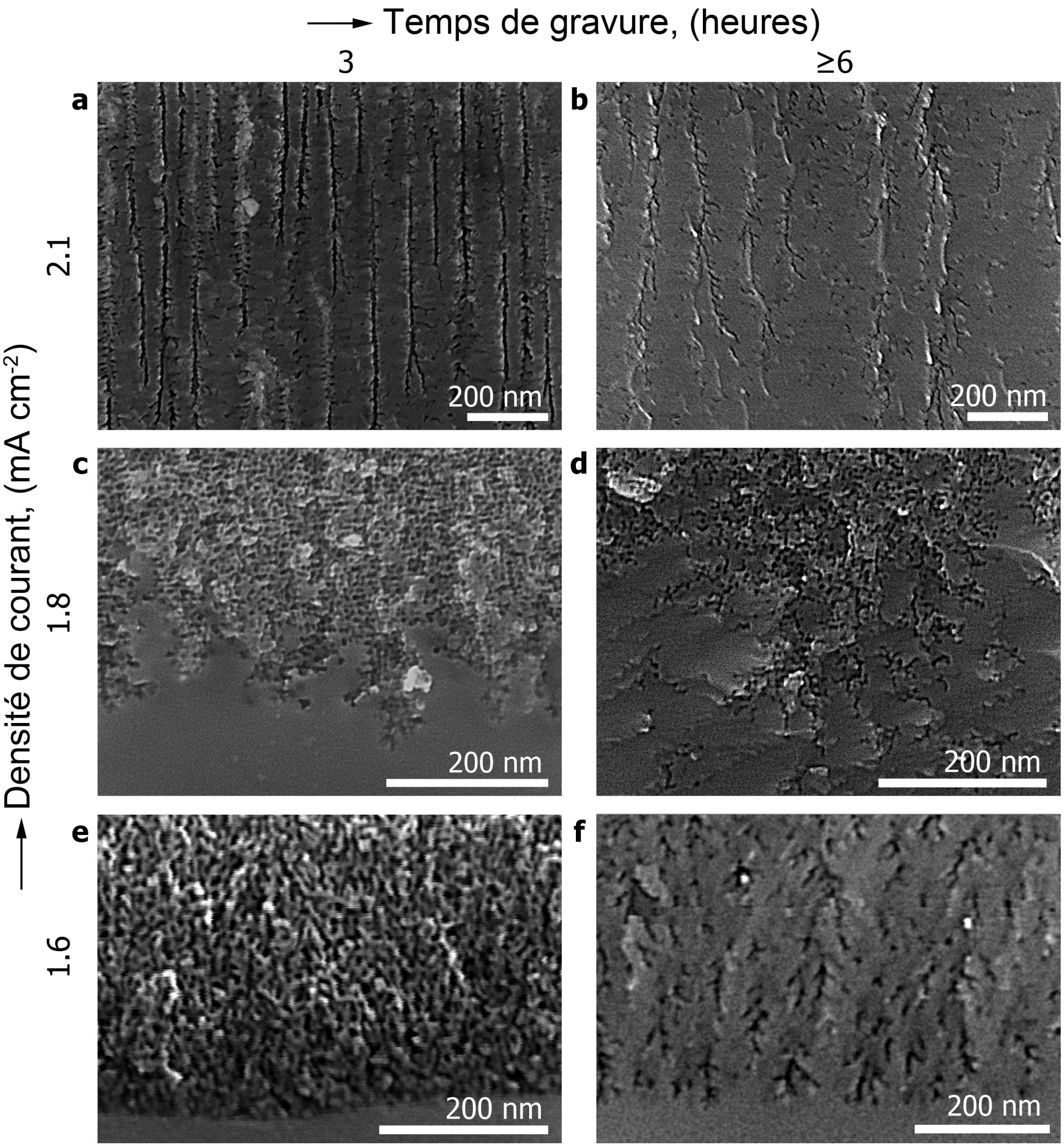

Figure 3.4 Images MEB de l'interface entre le Ge volumique et le Ge poreux pour différentes densités de courant et deux temps de gravure : (i) 3 h, (ii)>6 h. Six morphologies peuvent être identifiées :
(a) « colonnes » avec des courtes branches de second niveau seulement,
(b) « colonnes » avec des branches de deuxième et troisième niveaux,
(c) « éponge » à haute porosité avec des branches étroites, aléatoires et courtes,
(d) « éponge » à faible porosité,
(e) « dendrites » à faible porosité et (f) « dendrites » à haute porosité.

1.6 mA/cm$^2$ à 1.8 mA/cm$^2$ il se forme des couches de Ge mesoporeux avec une morphologie de type « éponge » (Figure 3.4 c, d).

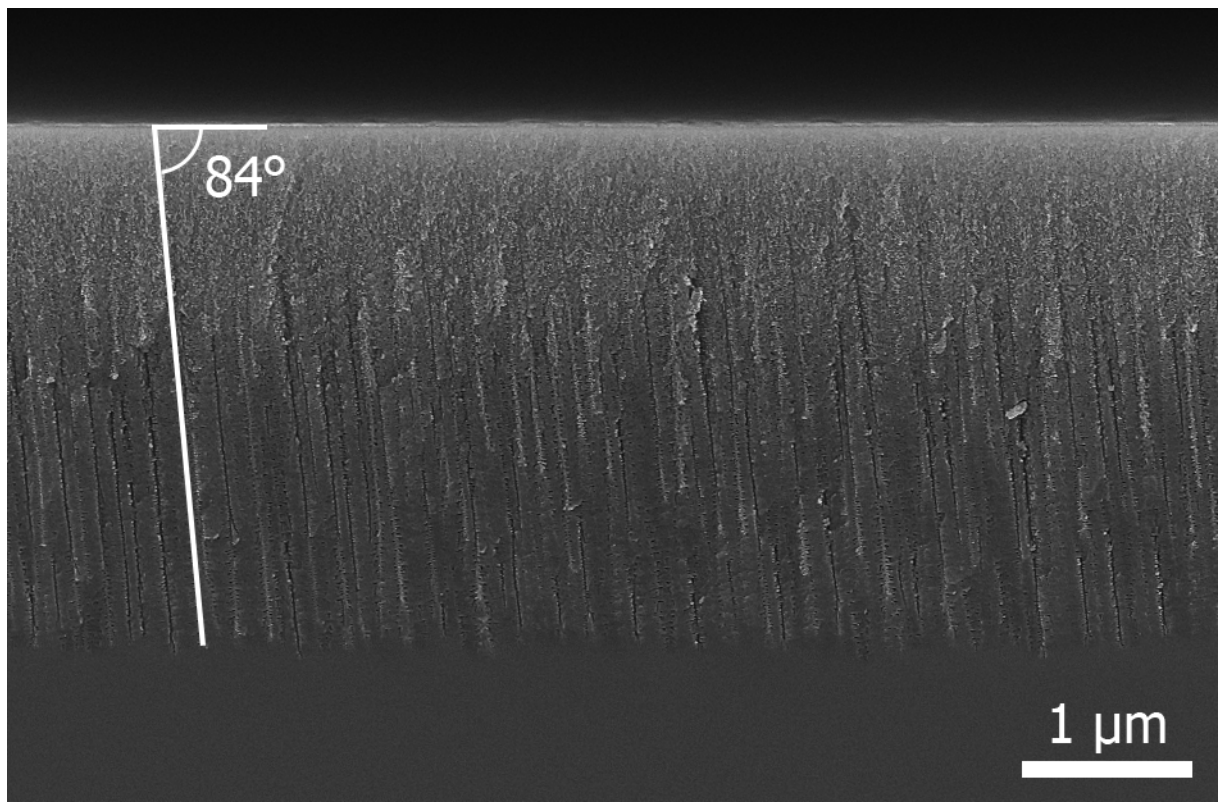

Figure 3.5 Image MEB en coupe transversale de la couche du Ge poreux avec les pores en colonnes. La direction de propagation des pores forme un angle de 84°avec la surface, ce qui correspond à la direction <100>.

Le Tableau 3.1 présente les valeurs du diamètre moyen des pores des couches de Ge poreux présentées sur la Figure 3.4. Quelques tendances générales peuvent être observées. Tout d'abord le diamètre moyen des pores augmente progressivement de 4 à 15 nm avec la densité de courant appliquée. Ainsi, selon l'UICPA [104], les couches poreuses obtenues peuvent être classées comme mésoporeuses (2–50 nm). Deuxièmement, toutes les couches de PGe formées aux faibles densités de courant ($< 1.8$ mA/cm$^2$) (Figure 3.4 (c-f)) sont caractérisées par de petits pores ayant la distribution en taille bien étroite (4–5 nm). Dans le cas des densités de courant plus importantes ($> 1.8$ mA/cm$^2$), le diamètre des pores dépend fortement du temps de gravure. Pour un temps de gravure $\approx 3$ h, le diamètre des branches de second niveau (4–6 nm) est environ deux fois plus petit que celui-ci des pores principaux (8–15 nm) (Figure 3.4 (a)). Il est à noter également que le diamètre des pores principaux diminue avec la profondeur et devient comparable à celui-ci des branches (Figure 3.4 (b)).

| Densité de courant anodique (mA/cm$^2$) | Temps de gravure total (h) | Diamètre des nanopores (nm) |
|---|---|---|
| 1.6 mA/cm$^2$ | 3 | 5–8 |
| 1.6 mA/cm$^2$ | 8 | 5–9 |
| 1.8 mA/cm$^2$ | 3 | 4–10 |
| 1.8 mA/cm$^2$ | 6 | 4–10 |
| 2.1 mA/cm$^2$ | 3 | 8–15 (troncs)<br>4–6 (branches) |
| 2.1 mA/cm$^2$ | 20 | 5–8 |

Tableau 3.1 Paramètres de gravure et diamètre des nano-pores des échantillons présentés sur la Figure 3.4

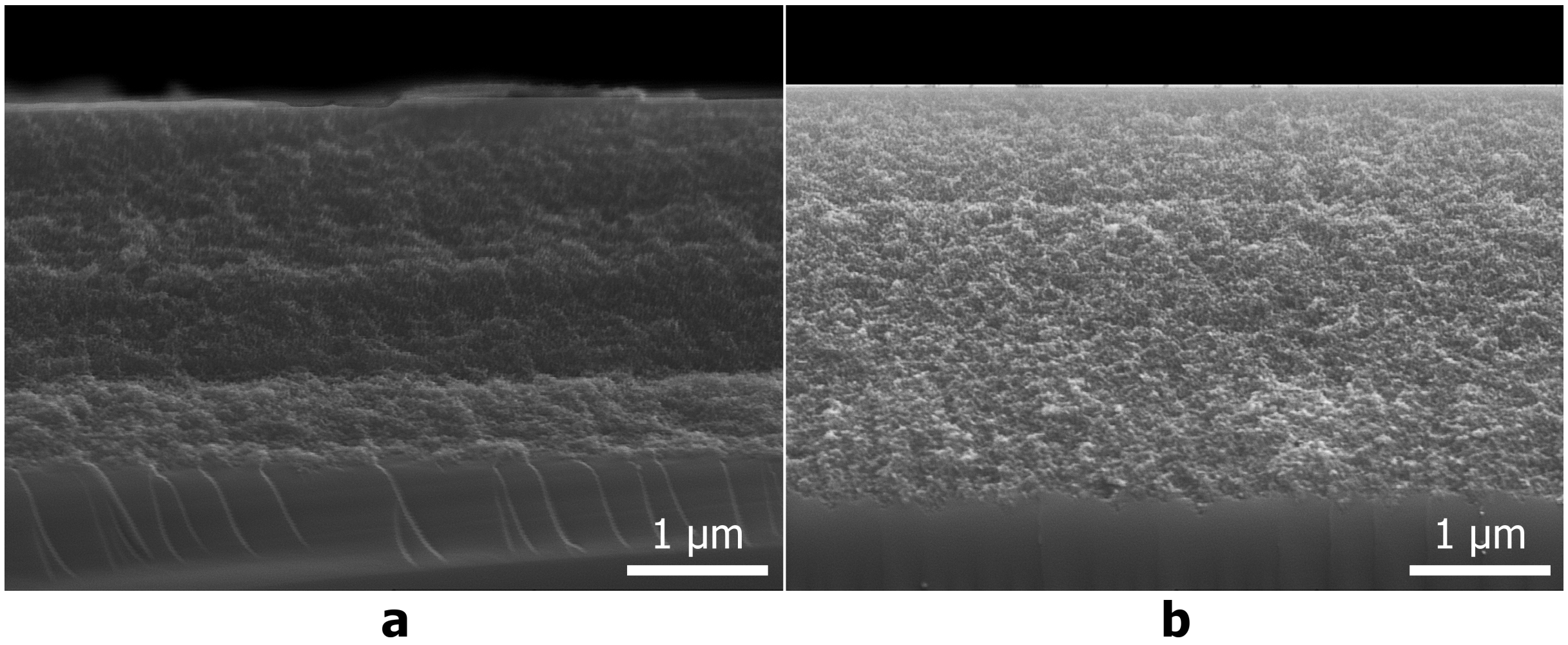

Figure 3.6 Images MEB en coupe transversale des structures de Ge poreux de type «éponge » formées sur un substrat AXT ($t_{on}$ = 1 s, $t_{off}$ = 2 s, $HF_{49\%}$, $t = 2$ h) sous la densité de courant de 4 mA/cm$^2$ (a) et 8 mA/cm$^2$.

Le phénomène de changement de morphologie lorsque l'amplitude de courant change est lié, probablement, aux propriétés des substrats UMICORE utilisés dans cette série d'expériences. Lorsque la résistivité des substrats AXT est la même que celles des substrats d'UMICORE, seule la morphologie en éponge est observée pour toute la gamme de courants explorée (Figure 3.6). La Figure 3.7 montre la porosité et la vitesse de croissance d'une couche de Ge poreux réalisée sur un substrat AXT en fonction de la densité de courant appliquée.

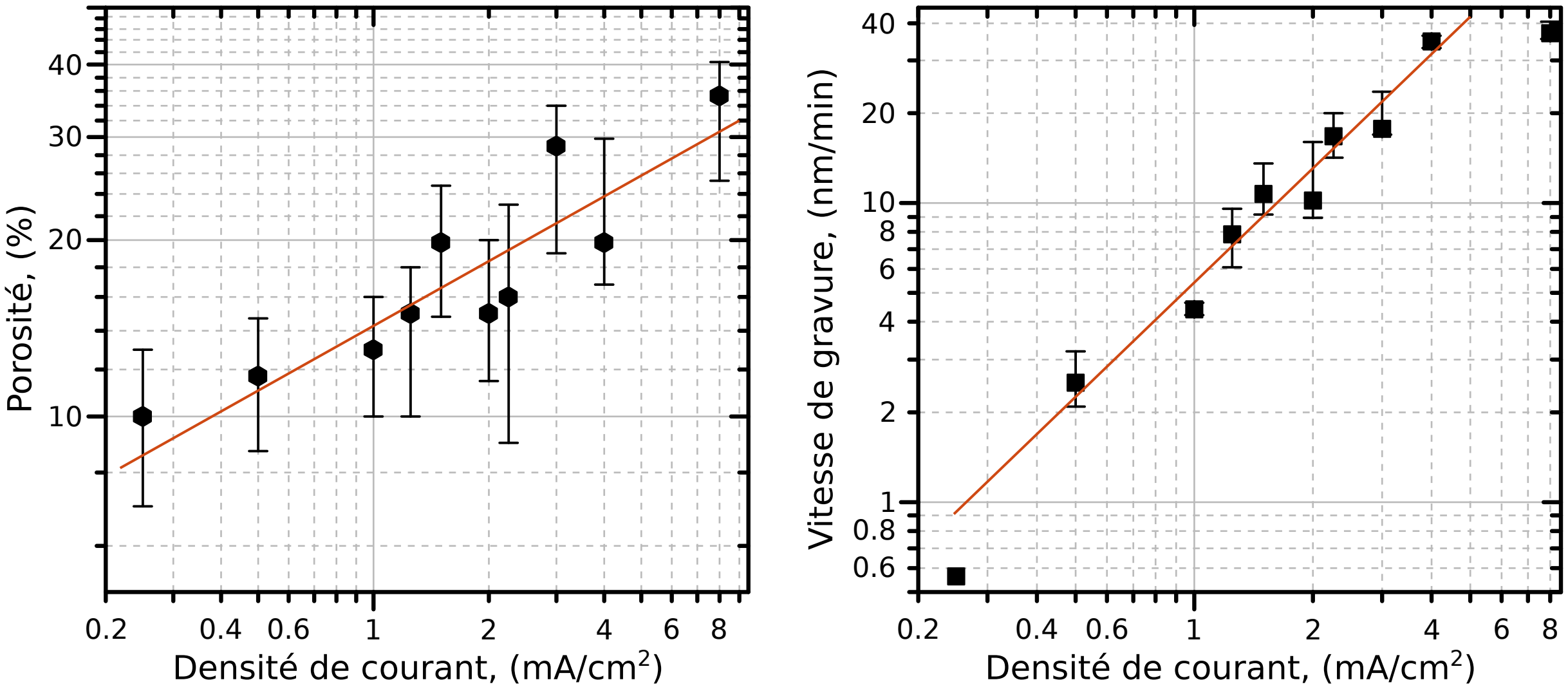

Figure 3.7 Évolution de la vitesse de propagation des pores (a) et évolution de la porosité (b) d'une couche poreuse en éponge formée sur un substrat AXT avec $t_{on}$ = 1 s, $t_{off}$ = 2 s, $HF_{49\%}$ en fonction de la densité de courant $J_+ = |J_-| = 0.25$–8 mA/cm$^2$

La vitesse de gravure augmente rapidement avec la densité de courant et atteint des valeurs élevées de 40 nm/min (ce qui correspond à une épaisseur de la couche $\approx$ 5 µm). Ici la vitesse moyenne de gravure est calculée comme le rapport de l'épaisseur d'une couche et la durée totale du procédé $T$. La courbe tend à saturer aux grandes valeurs de $J$. La réalisation de couches sous des courants encore plus élevés est compliquée par l'instabilité mécanique des couches très épaisses et ce, malgré un traitement spécial post-fabrication de la couche par Hexane.

La porosité des couches spongieuses peut être ajustée de 15 à 50%. Tant que le diamètre des pores ne change pas avec le courant, le changement de la porosité est dû à la formation de nouvelles cavités dans l'espace du cristal entre les pores existants, ce qui signifie l'augmentation de la densité volumique des pores. Une fois la largeur des isthmes entre les cristallites $d_c$ devient inférieure à une largeur de la zone de charge d'espace(ZCE), la courbe de la porosité sature et une dissolution ultérieure du matériau n'apporte qu'un épaississement de la couche poreuse.

Il est possible de réaliser des couches de Ge poreux de morphologies différentes en éponge par un ajustement des temps $t_{on}$ et $t_{off}$ ou par une variation des courants $J_+$ et $J_-$ de manière indépendante. Désormais, le fournisseur du substrat sera indiqué si les résultats obtenus sur les deux types de substrats sont différents.

### 3.2.2 Variation de la morphologie du Ge poreux avec la profondeur

La Figure 3.8 (a) présente une vue en coupe du Ge mesoporeux d'épaisseur environ 7 µm formée à 2.1 mA/cm$^2$. Trois régions caractéristiques marquées comme $A$, $B$ et $C$ peuvent être identifiées. Elles correspondent à trois morphologies différentes. La surface de la couche de Ge poreux (la partie supérieure de la région A) est caractérisée par la morphologie de type éponge. À une profondeur environ 500 nm sous la surface, la formation des colonnes avec des branches de second niveau a lieu (partie inférieure de la région A). L'espace entre les colonnes est remplie avec des pores de type éponge. Leur taux de remplissage diminue progressivement avec la profondeur et disparait complètement à une profondeur de 1.5 µm sous la surface, ce qui conduit à la mise en place d'une morphologie purement « colonnaire » (région B). Cette région colonnaire s'étend jusqu'à une profondeur de 3 µm au-delà de laquelle les pores avec des gros troncs (les pores principaux de cette région) sont petit à petit transformés en pores de type fil avec de nombreuses branches de deuxième et troisième niveau (région C). La Figure 3.8 (b) montre la variation de la porosité du Ge poreux en fonction de la profondeur de la couche présentée sur la Figure 3.8 (a). Étant extrêmement élevée à la surface (70–80%), la porosité diminue progressivement avec la profondeur. Ensuite, elle se stabilise à l'interface avec la région B au niveau 20–25%. Il est à noter également qu'il n'y a pas de différence dans la porosité entre les régions B et C. Une analyse structurale supplémentaire peut être obtenue en utilisant la spectroscopie Raman. L'analyse Raman de l'échantillon en coupe transversale a été effectuée grâce à la taille de spot très petite (1–2 µm). Les spectres Raman de trois types de morphologies qui coexistent dans la structure à multicouches (Figure 3.8 (a)) ainsi que du Ge monocristallin sont présentés sur la Figure 3.8 (c).

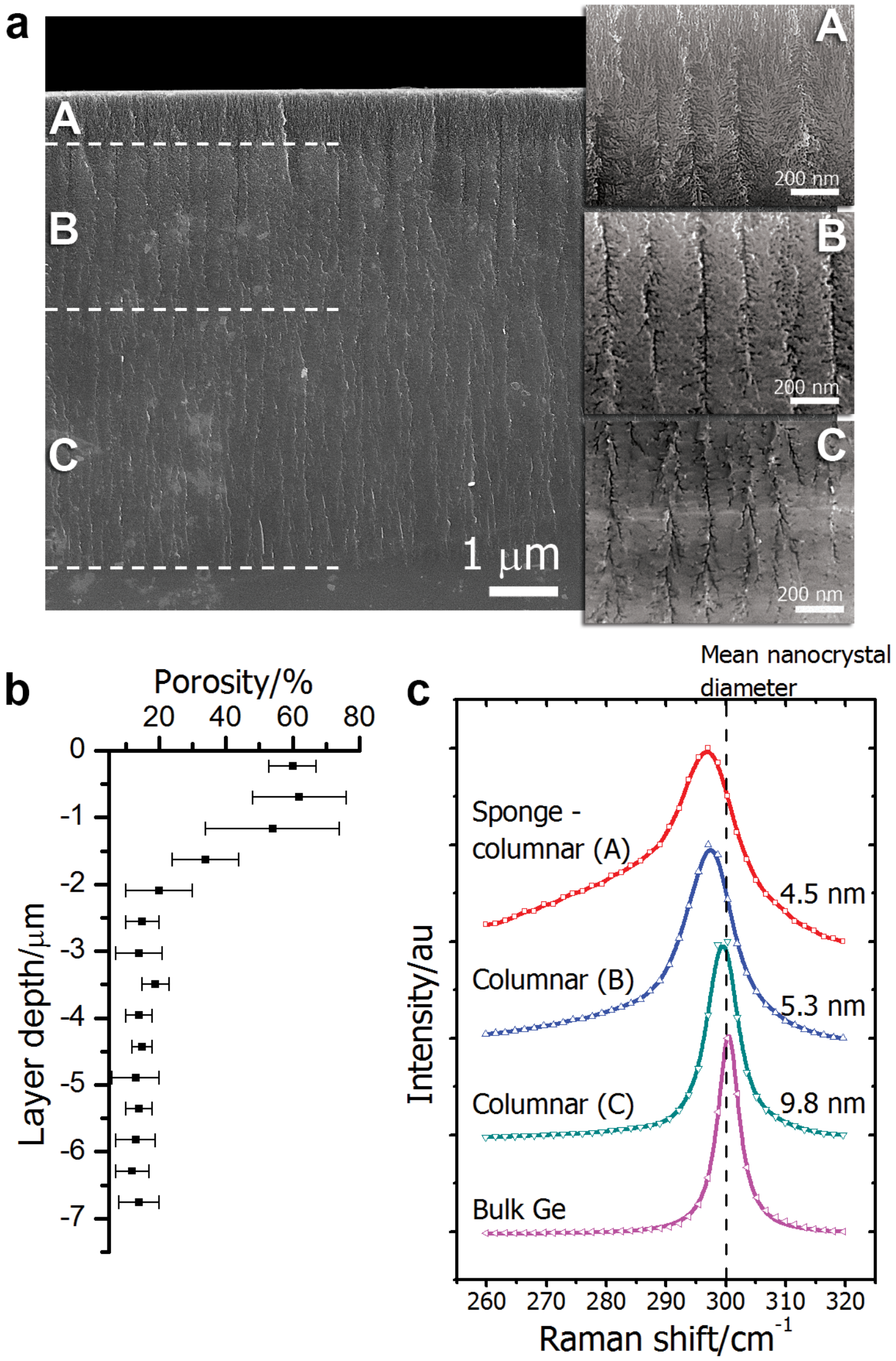

Figure 3.8 Variation des paramètres d'une couche de Ge poreux (morphologie, porosité, taille des cristallites) avec la profondeur

Ces spectres affichent clairement les éléments suivants :

- Toutes les morphologies présentent un pic autour de 300 $cm^{-1}$ ce qui indique la nature cristalline du Ge mesoporeux.

- La largeur, la forme et la position spectrale de ces pics sont fortement différentes de celles du Ge monocristallin.

- Tous les spectres sont décalés vers le rouge et sont asymétriques et élargis vers les basses fréquences par rapport au Ge monocristallin.

- Une large bande, située autour de 295 $cm^{-1}$, peut être observée dans le spectre de la couche superficielle du Ge poreux, dont la morphologie est de type « éponge-colonnaire ». Cette bande est liée à une fraction du Ge amorphe [54].

Pour estimer le diamètre moyen des nanocristallites de Ge, constituant une couche poreuse, la modèle phénoménologique du confinement des phonons, décrite dans la section 2.2.3 a été adopté. Les résultats de calcul sont présentés sur la Figure 3.8 (c). Le diamètre moyen des nanocristallites augmente progressivement de 4.5 à 9.8 nm en profondeur. L'analyse de la variation de la morphologie des couches de Ge poreux en profondeur, formées dans des conditions différentes, a révélé les mêmes tendances. Tout d'abord, l'existence de la région superficielle A, ayant une morphologie éponge-colonnaire, a été observée pour tous les échantillons de Ge poreux formés à une densité de courant élevée ($> 1.8$ mA/$cm^2$). De plus, son épaisseur relative est sensiblement la même pour tous les échantillons ( 20%). La formation de la région C dépend fortement de l'épaisseur de la couche. Pour les couches fines, ayant une épaisseur inférieure à 3 µm, le Ge poreux présente une structure à deux couches avec une région A sur la surface (morphologie de type « éponge-colonnaire ») et une région B au-dessous (morphologie « purement colonnaire »). Les échantillons ayant une épaisseur supérieure à 3 µm ont une structure à trois couches, similaire à celle présentée sur la Figure 3.8 (a).

Dans le cas des échantillons de Ge poreux, formés à des densités de courant faibles ($<$ 1.8 mA/$cm^2$), une diminution de la densité des pores et du degré de leur interconnection a été constatée (Figure 3.9), ce qui résulte de la mise en place avec le temps, d'une région de faible porosité. Ce phénomène semble être exclusivement dépendant de l'épaisseur de l'échantillon. En effet, les couches de Ge poreux de type «éponge » ou «dendrites » sont bien homogènes jusqu'à une épaisseur $L_u$ d'environ 2 µm. Au-delà, la couche de transition commence à se former par dessous.

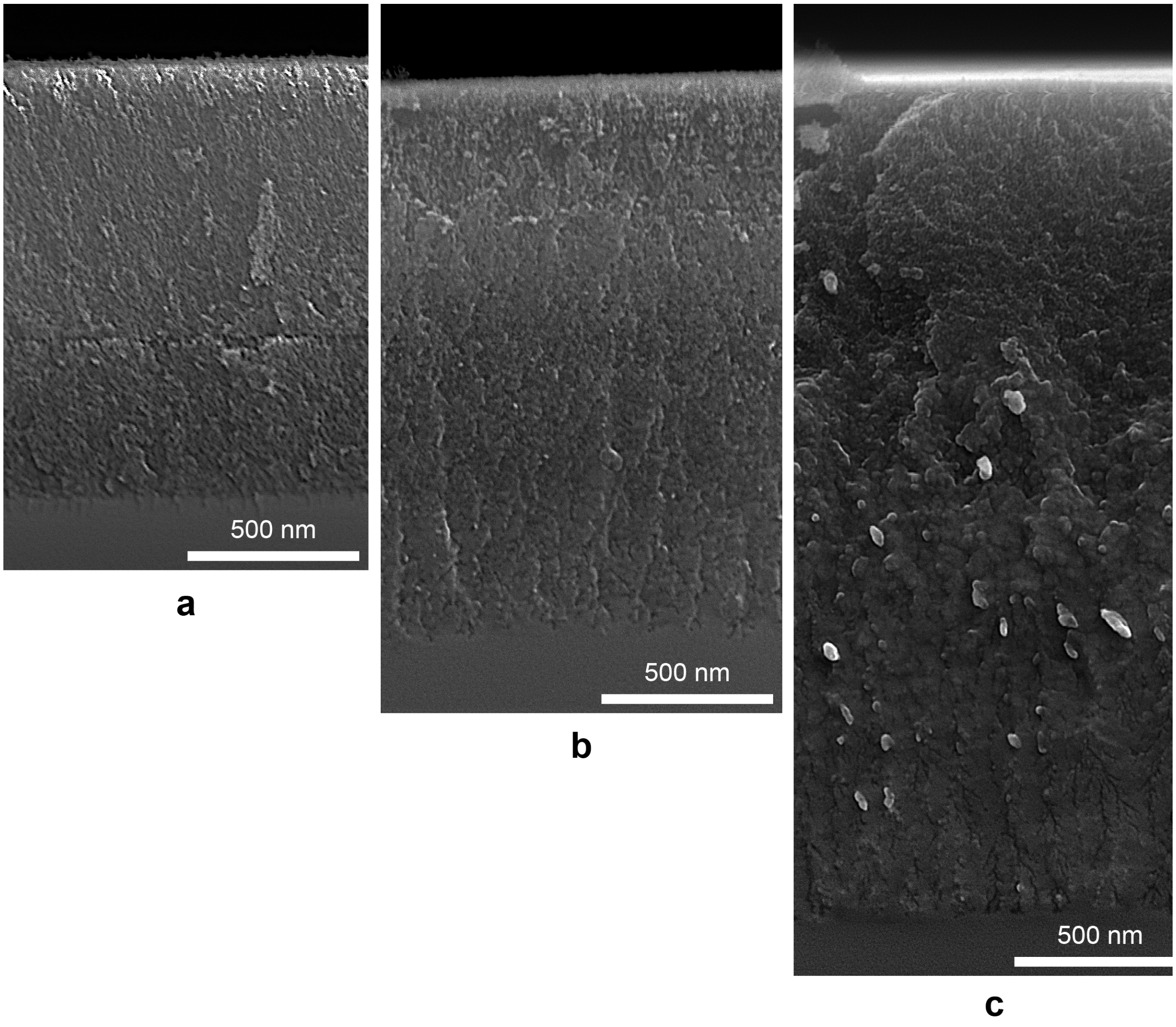

Figure 3.9 Images MEB en coupe transversale d'une structure ($J_+ = J_- = 1.6$ mA/cm$^2$, $t_{on} = 1$ s, $t_{off} = 2$ s, $HF_{49\%}$) composée d'une couche supérieure en éponge et une couche inférieure de morphologie dendritique après une gravure pendant 1 h(a), 2 h(b) et 3 h(c).

Les Figures 3.9 et 3.10 montrent l'évolution de la structure de Ge poreux lors d'une gravure avec $J_+ = J_- = 1.6$ mA/cm$^2$, au fil du temps. D'abord une couche de type éponge à forte porosité est formée. Ensuite, lorsque l'épaisseur de la structure dépasse une valeur $L_u$, la croissance d'éponge ralentit et les pores qui s'ajoutent au bas de la structure prennent progressivement une forme de fils.

Le phénomène de changement de porosité et/ou de morphologie est probablement lié à une diminution d'électrolyte en profondeur de pores. Autrement dit, les réactions anodiques sur le Ge sont limitées par la diffusion des réactifs, ce qui diminue l'ordre partiel de la réaction par rapport à l'HF (section 3.3.2).

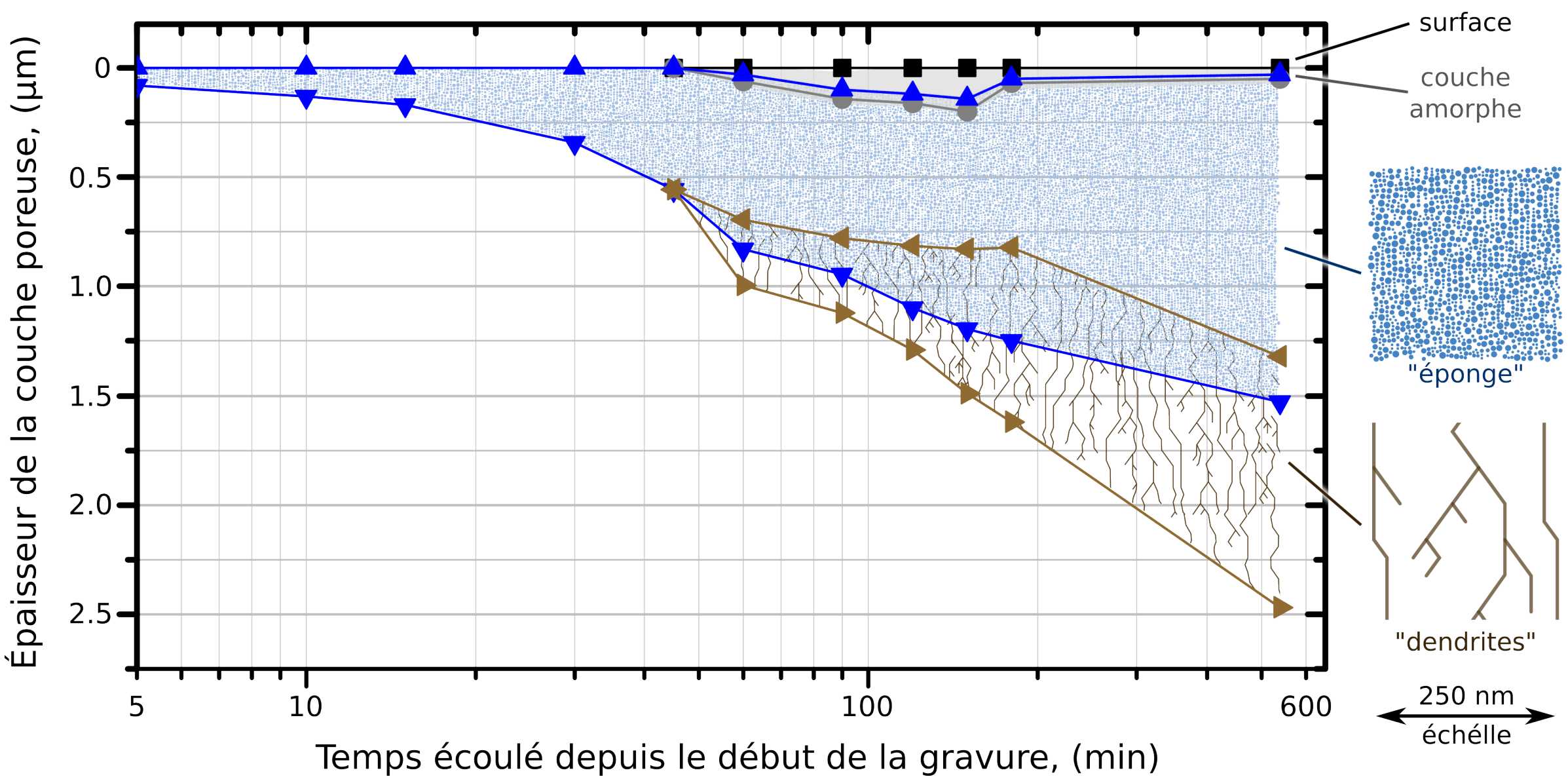

Figure 3.10 Schéma de composition structurale d'une couche de Ge poreux réalisée sous les conditions $J_+ = J_- = 1.6$ mA/cm$^2$, $t_{on} = 1$ s, $t_{off} = 2$ s, $HF_{49\%}$ en fonction de $T$

### 3.2.3 Influence de la durée des impulsions cathodiques ($t_{off}$)

Les autres paramètres qui influent sur la formation du Ge mesoporeux sont les paramètres des impulsions de cathodisation. Dans cette section nous considérons le cas de la durée d'impulsions cathodiques inférieures à deux secondes. Comme il a été déjà mentionné dans le paragraphe précédent, le temps caractéristique de 2 s correspond à la formation d'une monocouche d'hydrogène sur la surface développée du Ge poreux.

Pour cette étude tous les autres paramètres ont été fixés. Les densités de courant anodique $J_+$ et cathodique $J_-$ ont été fixées à 2 mA/cm$^2$. La concentration du HF reste inchangée soit 49%. La durée des impulsions cathodiques $t_c$ est égale à 1 s.

**Morphologie « éponge ».** La Figure 3.11 présente une image d'un échantillon de Ge mesoporeux formée avec $t_{off} = 2$s. La morphologie de ces couches est de type l'éponge : la structure est composée de cavités elliptiques interconnectées, séparées par une carcasse cristalline. La représentation schématique ainsi que les images MEB d'une structure spongieuse sont données sur la Figure 3.11. Les pores ne sont pas ordonnés à petite distance, cependant un certain ordre à longue distance est observé. La taille des pores $d_p$ et la distance entre les pores $d_{ip}$ varient entre 5 et 10 nm. Une morphologie similaire a été obtenue pour tous les échantillons formés avec $t_{off}$ supérieur à 1 s. Une telle morphologie est induite par le dégrée de passivation, qui est considérée comme « complet ». Si la durée des impulsions cathodiques est plus de deux secondes (pour nos conditions expérimentales), l'hydrogène moléculaire se libère lors d'une réaction chimique (Équation 3.21), ce qui conduit à la destruction progressive de la couche poreuse.

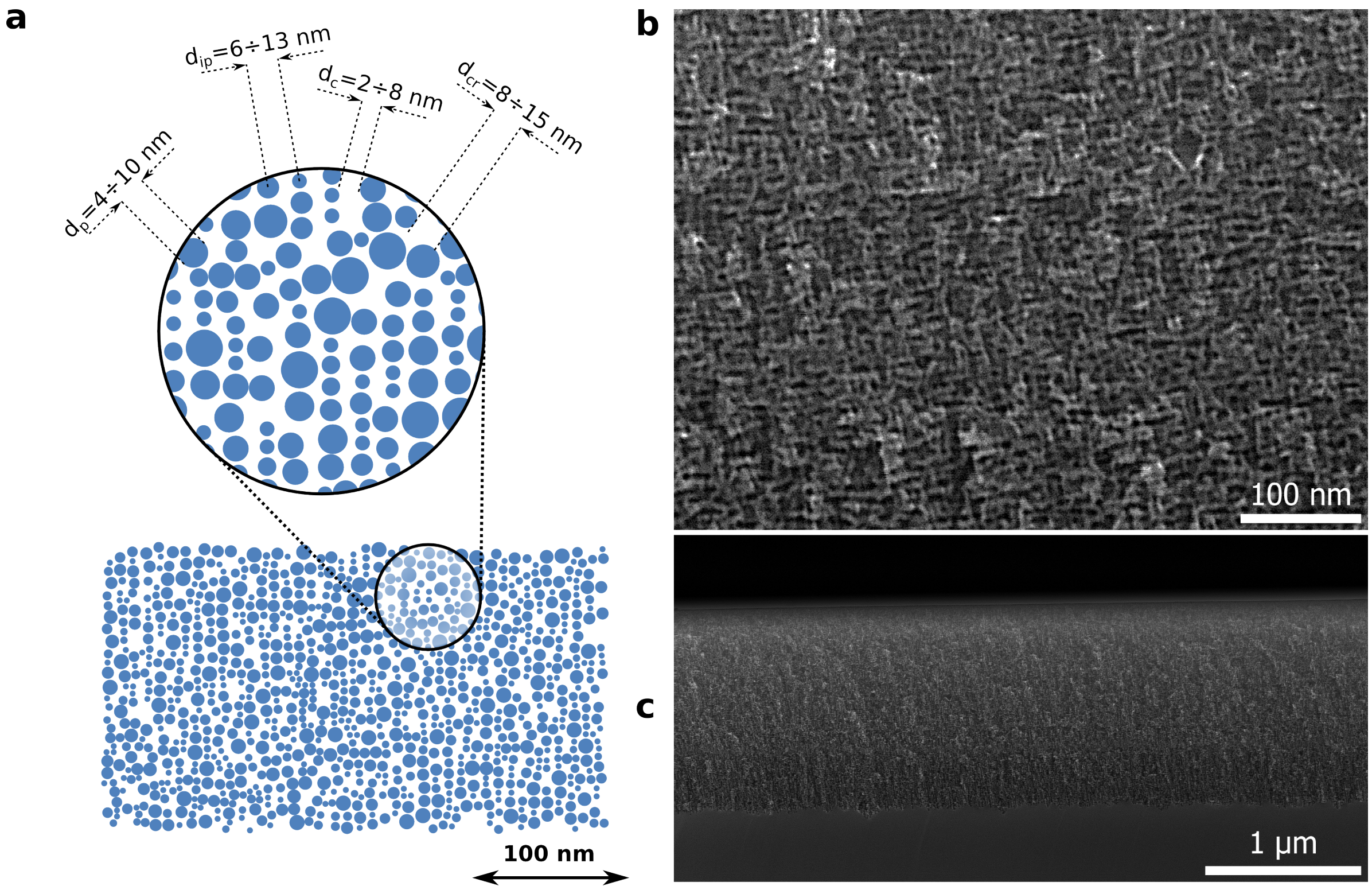

Figure 3.11 (a) Schéma structurale de pores type « éponge » (la désignation des paramètres $d_p$, $d_{ip}$, $d_c$ et $d_{cr}$ est donnée dans la section 1.3.1) ; images MEB en coupe transversale d'une structure ($J_+ = J_- = 2$ mA/cm$^2$, $t_{on} = 1$ s, $t_{off} = 1.0$ s, $HF_{49\%}$) composée entièrement de Ge poreux avec une morphologie « éponge » : fort (b) et faible (c) agrandissement.

**Morphologie « sapin ».** L'autre cas extrême est l'échantillon fabriqué avec $t_{off} = 0.25$ s dont l'image MEB est présentée sur la Figure 3.12. Il est à noter que tous les autres paramètres expérimentaux sont identiques. La formation de la nouvelle morphologie, qui était nommé « arbre «ou « sapin », a eu lieu. La structure poreuse avec cette morphologie est constituée de pores verticaux qui suivent la direction [100] et de pores latéraux le long des directions [111], [11-1], [1-11], [1-1-1]. Cette morphologie est parfaitement périodique. Le diamètre des pores principaux est de 4 à 11 nm. L'épaisseur des pores latéraux varie de 4 à 11 nm. La distance verticale entre les pores latéraux est entre 20 et 25 nm. Enfin, la distance entre les pores principaux varie entre 80 et 110 nm. Cette dernière distance ne dépend presque pas des conditions de gravure lorsque des « arbres » sont observés, contrairement au diamètre de pores et à l'espacement entre les branches. Ainsi, une ger-mination des pores se fait hypothétiquement sur des défauts périodiques de la surface d'un cristal, tandis qu'une croissance régulière des pores est contrôlée par le champ électrique appliquée. Le mécanisme de formation des « arbres » est expliqué dans la section 3.3.5, alors que le phénomène de la germination ordonnée est un sujet à étudier.

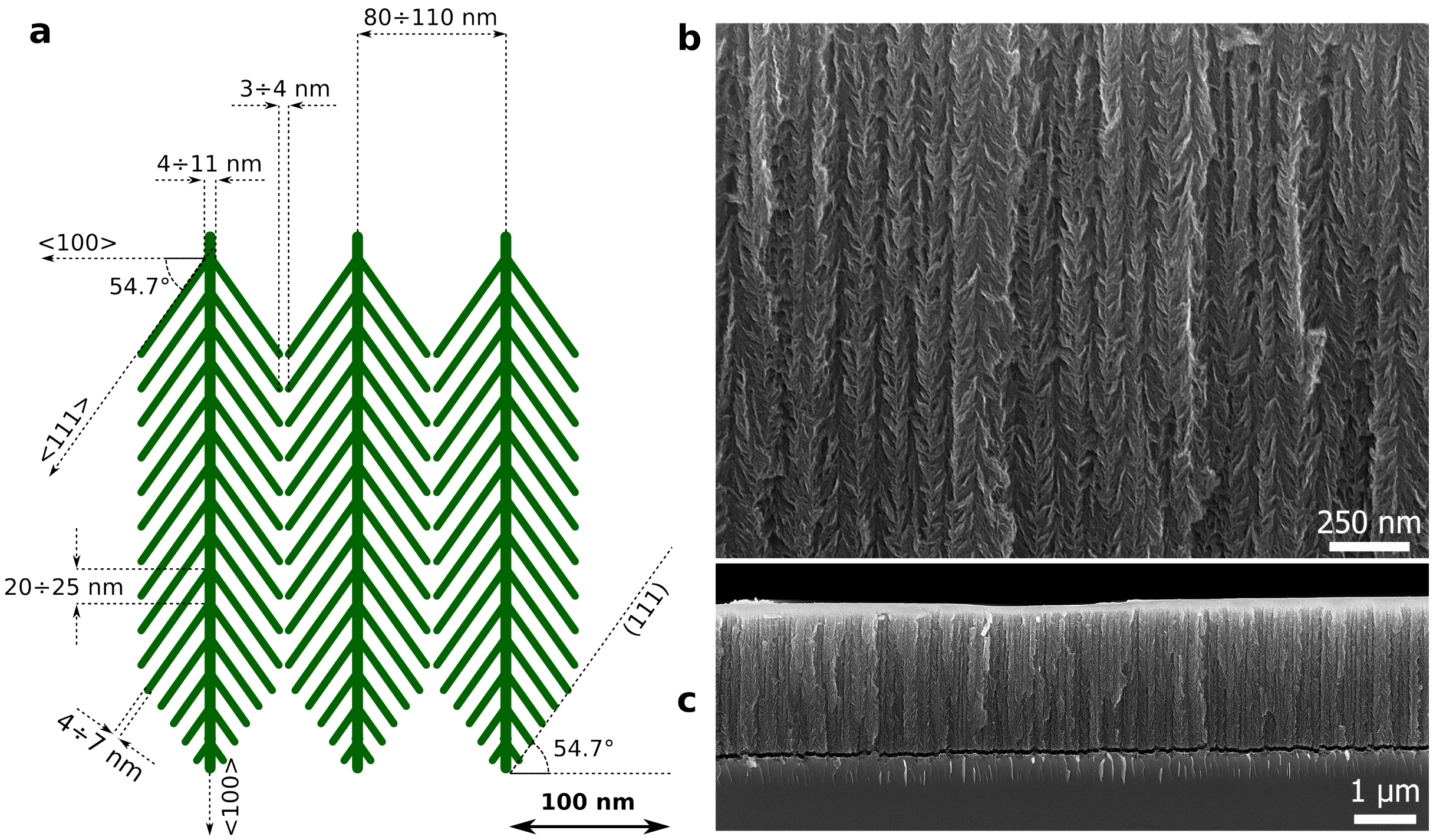

Figure 3.12 (a) Schéma structurale de pores type « sapin » ; images MEB en coupe transversale d'une structure ($J_+ = J_- = 2$ mA/cm$^2$, $t_{on} = 1$ s, $t_{off} = 0.375$ s, $HF_{49\%}$) composée entièrement de Ge poreux de morphologie « sapin » : fort (b) et faible (c) agrandissement.

La formation de cette morphologie était observée dans tous les échantillons formés avec $t_{off}$ variant entre 0.1 et 0.375 s. Il convient également de noter qu'avec la valeur limite $t_{off} = 0$, qui correspond à un courant continu, aucune porosification n'est observée (section 1.3.3). La surface est texturée en pyramides convexes dont les facettes sont les plans <111>. Cependant, l'exploration du régime de transition entre une porosification et une texturisation ($t_{off}$ = 0–0.1 s) reste assez difficile. En effet, une gravure avec $t_{off}$ inférieur à 0.25 s est accompagnée par l'apparition de contraintes mécaniques ce qui induit la destruction de la couche poreuse. (Figure 3.13)

**Morphologie « arête de poisson ».** Lorsque $t_{off}$ est dans la gamme de 0.375 s à 1 s, la formation d'un empilement de couches des morphologies différentes est observée. « L'éponge » est formée sur la surface. La couche intermédiaire présente une nouvelle morphologie, nommée par la suite « arête de poisson ». Enfin, la morphologie « sapin » termine l'empilement. Décrivons la morphologie « arête de poisson ». Cette structure est composée de pores principaux dont la direction est [100] et de pores latéraux suivant les directions <113>. Plusieurs paramètres structuraux de cette morphologie sont similaires à ceux de la morphologie « sapin ». En particulier, la distance entre les pores principaux est de 80 à 100 nm, la distance verticale entre les pores latéraux varie entre 20 et 25 nm, leur largeur est autour de 4–7 nm et enfin, les pores latéraux cessent de croitre dès qu'ils s'effondrent, et la distance entre eux devient de l'ordre de 3–4 nm. Cependant, contrairement au cas de la morphologie « sapin », la largeur des troncs principaux est environs deux fois plus grande autour de 12–16 nm. Il est à noter, que dans une telle structure, l'emplacement des troncs du « sapin» et de « l'arête de poisson » coïncide. À l'interface de ces deux morphologies, les branches changent brusquement d'orientation passant de <113> à <111> et les troncs s'amincissent sans changer la direction de propagation <100>.

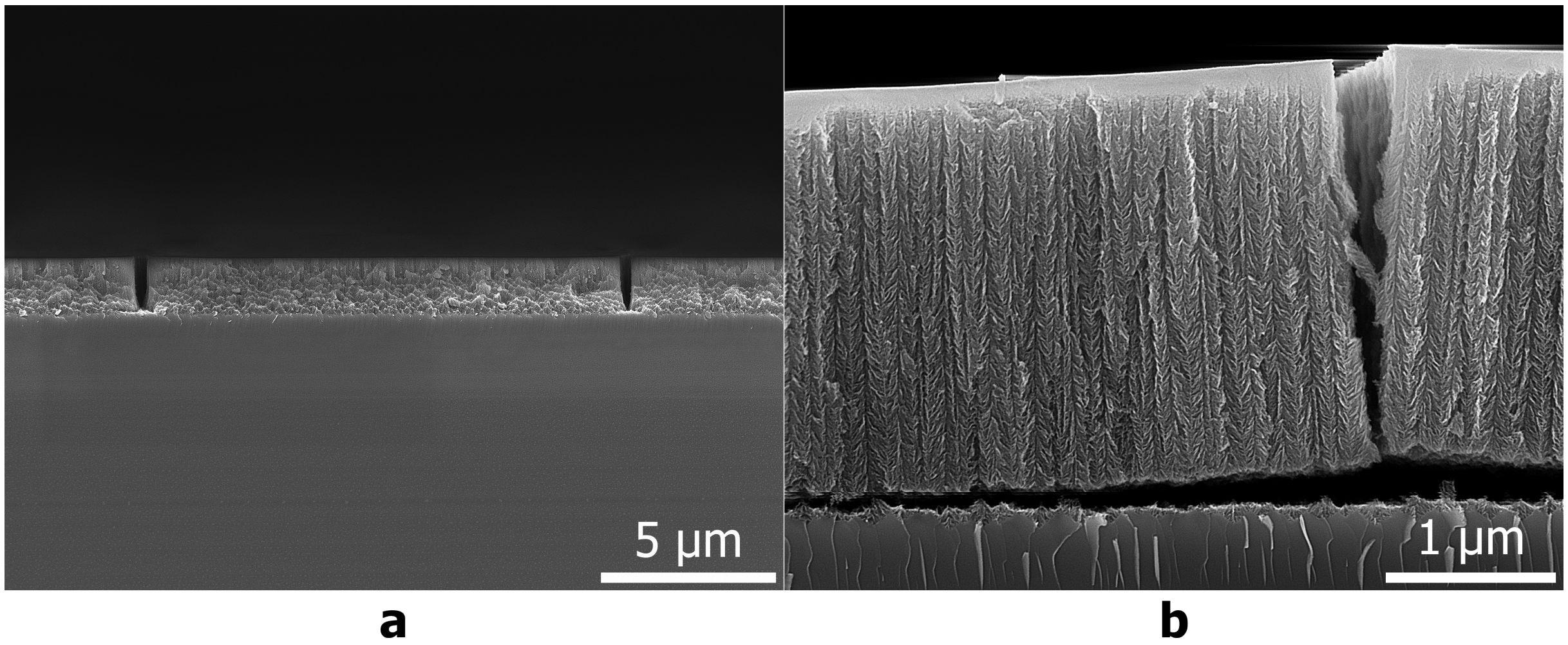

Figure 3.13 Processus destructeurs dans le Ge poreux lorsque $t_{off} \to 0$ : craquage(a) et décollement(b)

La formation de macropores <113> a été rapportée pour le Si orienté (100) [34]. La structure démontrée ci-dessus parait d'être unique et propre au Ge.

La généralisation des résultats est présentée sur le diagramme de la Figure 3.15. Elle montre la variation de la morphologie en profondeur (section transversale) en fonction du temps $t_{off}$. Il est à noter que pour tous les échantillons, utilisés pour produire ce diagramme, le temps de dissolution $T_A$ est fixé à une heure. Autrement dit, il s'agit du temps qui prend en compte uniquement la durée des impulsions positives, puisque la gravure a lieu

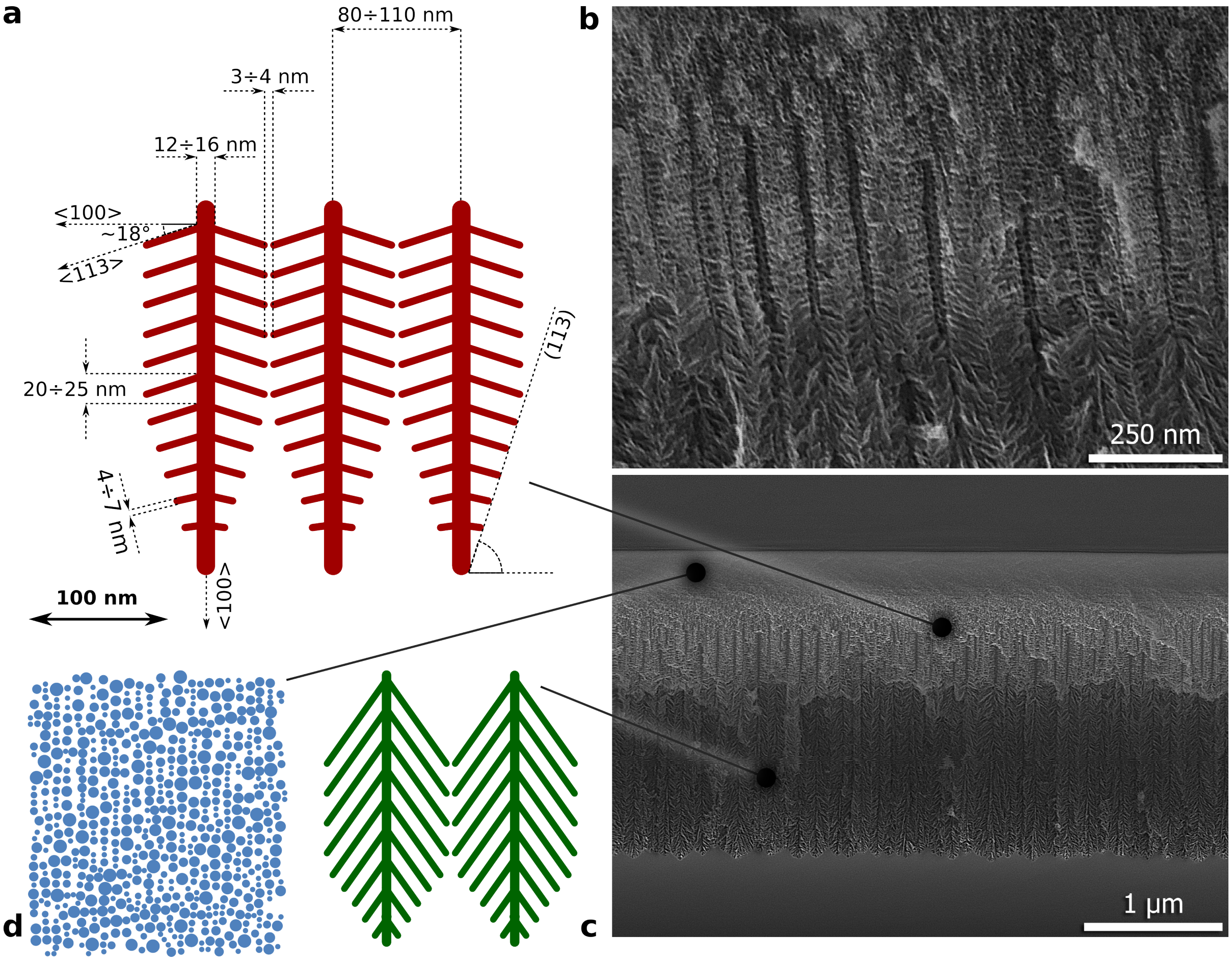

Figure 3.14 (a) Schéma structural de pores de type « arête de poisson » ; images MEB en coupe transversale d'une structure ($J_+ = J_- = 2$ mA/cm$^2$, $t_{on} = 1$ s, $t_{off} = 0.5$ s, $HF_{49\%}$) composée du Ge poreux de 3 morphologies, dont celle ayant le milieu en « arête de poisson » : faible(b) et fort(c) agrandissement ; une « éponge » et un « sapin » sont les deux autres morphologies (d) uniquement sous un régime anodique (équation 3.1).

$$T_A = \sum_{i=1}^{\frac{T}{t_{on}+t_{off}}} t_{on} = T\frac{t_{on}}{t_{on}+t_{off}} \tag{3.1}$$

Les zones bleue, verte et rouge de ce diagramme correspondent aux morphologies de type « éponge », « sapin » et « arête de poisson », respectivement. L'ordre des couches est toujours respecté quelles que soient les conditions du procédé. Les échantillons ayant à la fois trois, deux ou une morphologie unique peuvent être obtenus en ajustant la durée de l'impulsion cathodique ou la durée du procédé. En effet, de nouvelles couches apparaissent progressivement, l'une sous l'autre.

La Figure 3.16 montre le front de propagation des pores lorsque $T_A$ est inférieure à 2 h. Les images MEB révèlent la germination des « arbres » (a) ou des « arêtes de poisson » (b) sous une couche de Ge poreux existante. Une couche homogène composée uniquement de pores type « arête de poisson » peut être réalisée si $t_{off} = 0.375$–$0.75$ s et si le procédé est de courte durée : $T_A \leq 30$ min

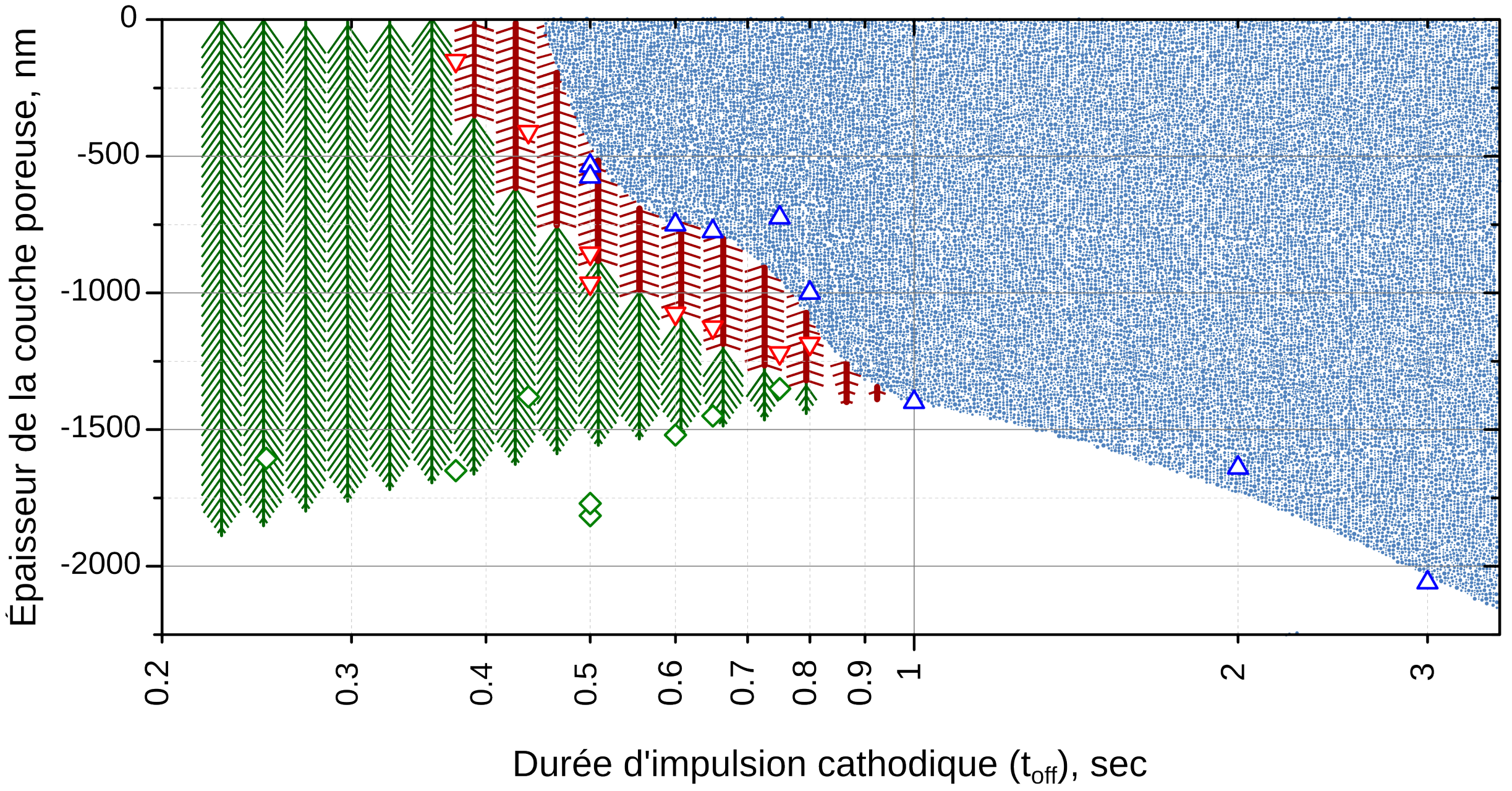

Figure 3.15 Schéma de composition structurale d'une couche de Ge poreux réalisée sous les conditions $J_+ = J_- = 2$ mA/cm$^2$, $t_{on} = 1$ s, $HF_{49\%}$, $T_A = 2$ h en fonction de $t_{off}$

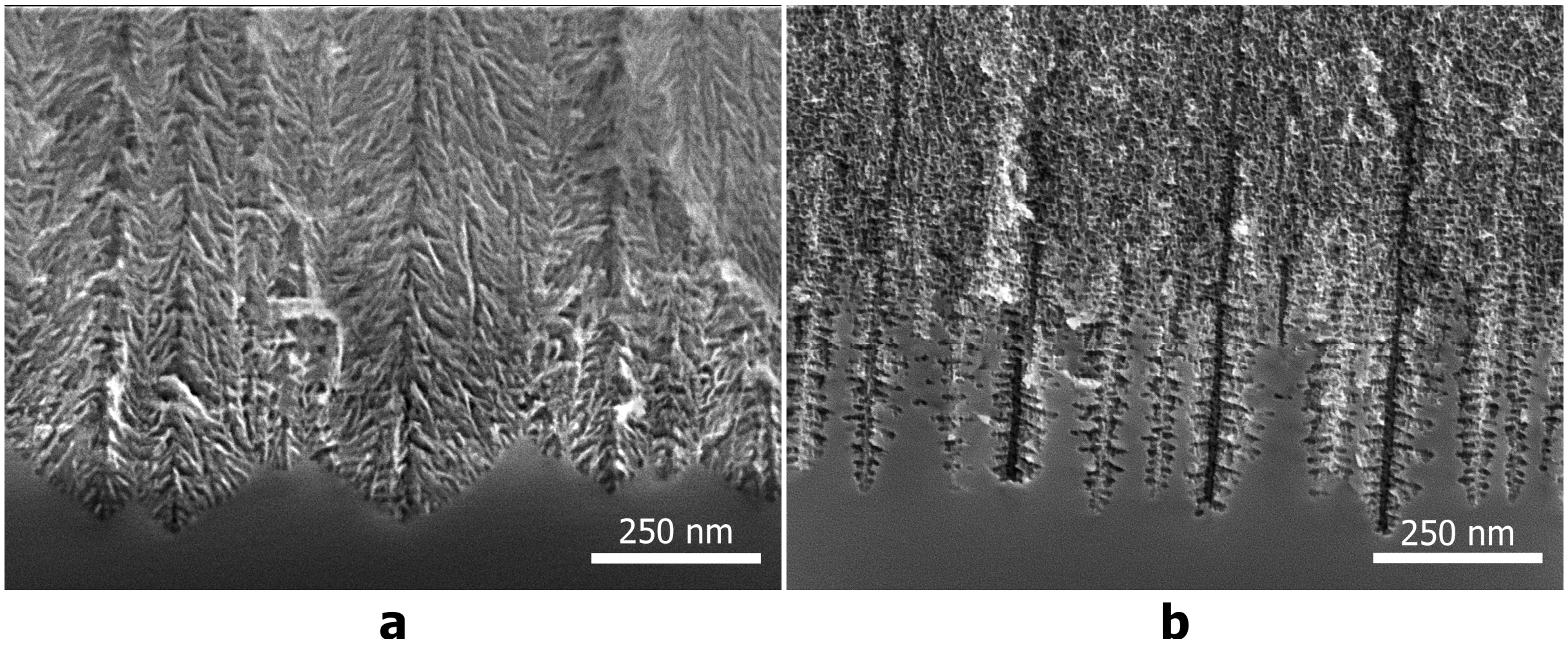

Figure 3.16 Images MEB en coupe transversale des interfaces entre le Ge volumique et le Ge poreux réalisé sous les conditions $J_+ = J_- = 2$ mA/cm$^2$, $t_{on} = 1$ s, $t_{off} = 0.625$ s, $HF_{49\%}$ et une durée du procédé $T = 1$ h (a) et $T = 1$ h 30 (b)

### 3.2.4 Influence simultanée de la durée des impulsions et de l'amplitude de courant cathodique

La durée des impulsions cathodiques n'est pas le paramètre unique qui détermine le degré de passivation de la surface développée du Ge poreux. En effet, la charge totale négative injectée durant une étape de cathodisation dépend en même temps de la durée des impulsions et de l'amplitude de courant cathodique. Ainsi, la qualité de la passivation est déterminée par l'aire sous la courbe d'une impulsion cathodique. Pour confirmer cette hypothèse, des échantillons ont été fabriqués avec différentes durées d'impulsions et d'amplitudes du courant cathodique, mais avec la même charge totale négative injectée (aire sous la courbe). La Figure 3.17 présente une paire d'échantillons de ce type : (a) $-2$ mA/cm$^2 \cdot 0.5$ s et (b) $-10$ mA/cm$^2 \cdot 0.1$ s. Les paramètres des impulsions anodiques ont été les mêmes. Dans les deux cas la charge totale des électrons injectée durant un cycle de cathodisation était de 1 mC. Comme nous pouvons le voir sur la Figure 3.17 (a) et (b), les deux structures fabriquées ont la même morphologie et la même vitesse de gravure. Cependant, il est à noter que pour les amplitudes de courant cathodique supérieures à 20 mA, les couches poreuses se détruisent. L'explication de ce phénomène sera discutée dans la section 3.3.4.

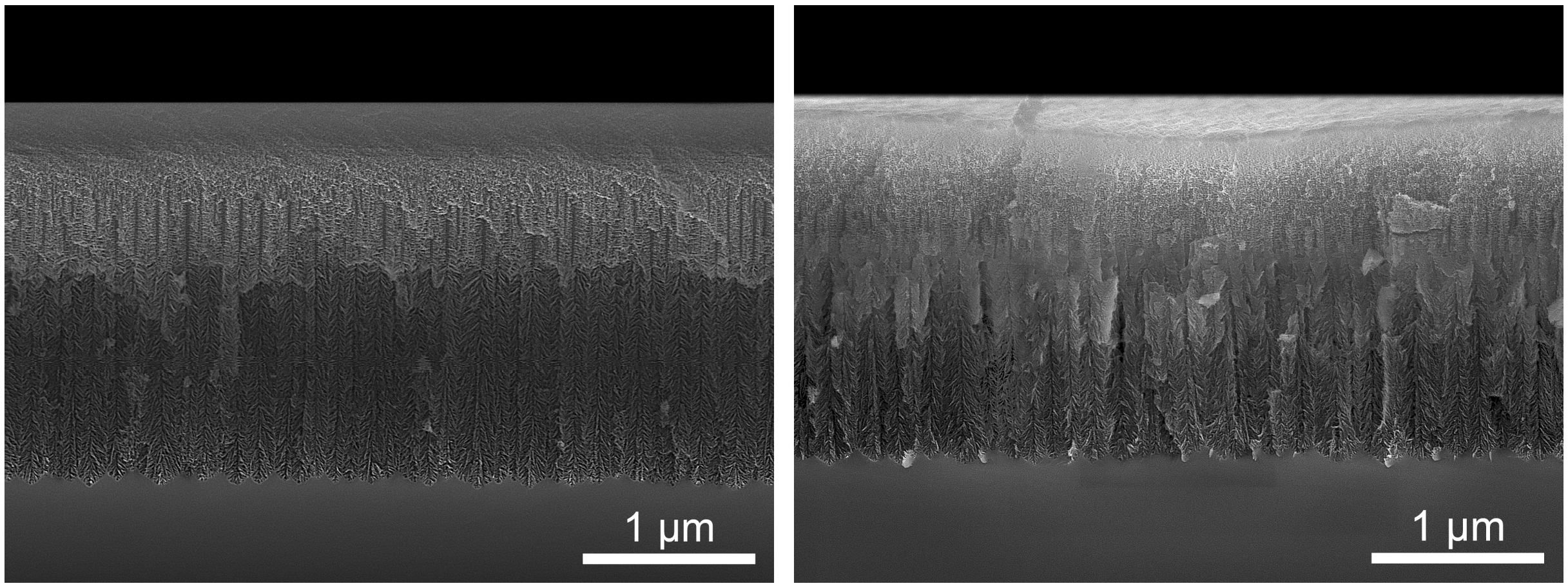

a b

Figure 3.17 Images MEB en coupe transversale des couches du Ge poreux fabriquées avec $J_+ = 2$ mA/cm$^2$, $t_{on} = 1$ s, $HF_{49\%}$, $T_A = 2$ h et des paramètres des parties cathodiques suivantes :
(a) $J_- = -2$ mA/cm$^2$ ; $t_{off} = 0.5$ s et (b)$J_- = -10$ mA/cm$^2$ ; $t_{off} = 0.1$ s

### 3.2.5 Influence de la durée des impulsions anodiques ($t_{on}$)

L'autre paramètre important qui influe la formation du Ge mesoporeux est la durée des impulsions anodiques ($t_{on}$). Comme il a été déjà indiqué dans la section 3.1 de ce chapitre, si la durée $t_{on}$ est supérieure au temps de persistance caractéristique de la passivation des parois des pores avec $H^+$ ($\tau_p$), la surface interne du Ge mesoporeux perd sa protection contre la dissolution chimique et l'endommagement notable des parois des pores a lieu. Pour nos conditions expérimentales, la durée $\tau_p$ trouvée est autour d'une seconde. D'autre part, pour toutes les durées des impulsions anodiques inférieures à 1 s, la morphologie de la couche poreuse doit rester indépendante du temps $t_{on}$. La Figure 3.18 présente une couples d'échantillons ayant la morphologie de type « sapin ». Dans les deux cas, la porosification se déroule sous une passivation partielle. Comme nous pouvons le constater, la durée des impulsions anodiques n'est pas suffisante pour créer des dommages à la couche poreuse, la formation de deux morphologies est indépendante du temps $t_{on}$.

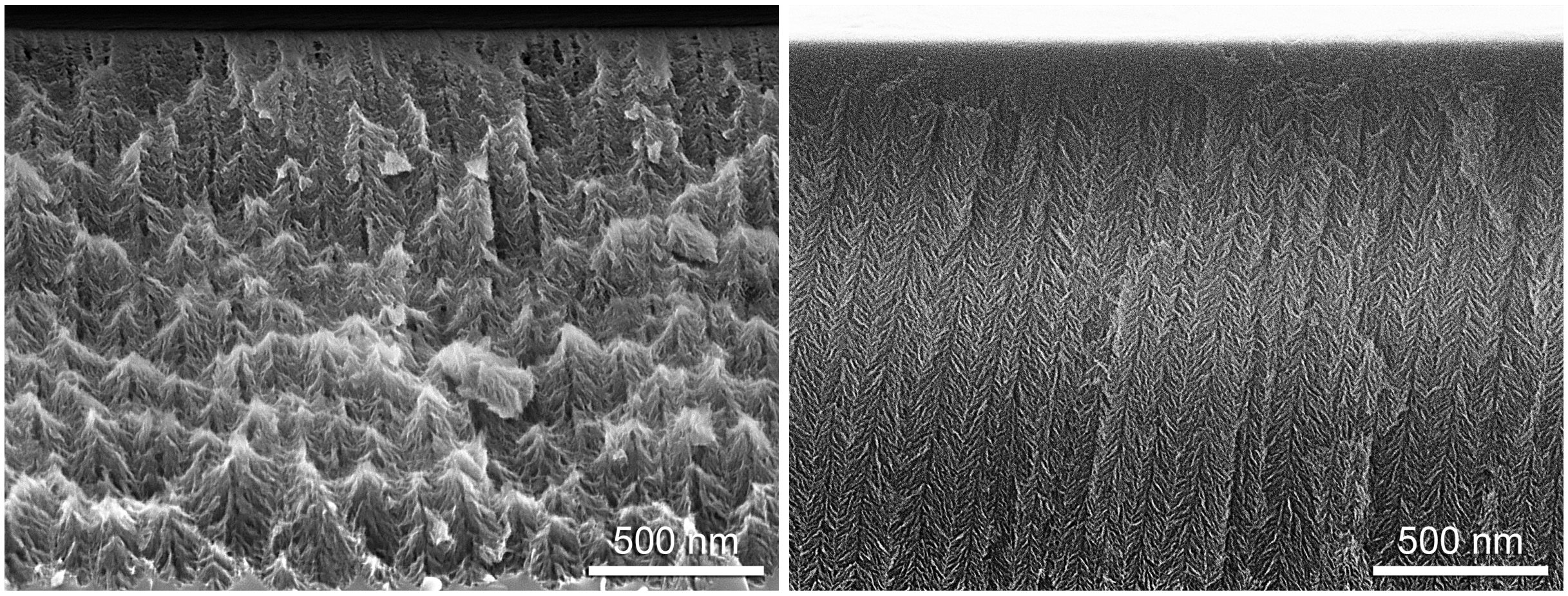

a b

Figure 3.18 Images MEB en coupe transversale des couches du Ge poreux fabriquées avec $J_+ = J_- = 2$ mA/cm$^2$, $t_{off} = 0.25$ s, $HF_{49\%}$, $T_A = 2$ h et des paramètres des parties anodiques suivantes :
(a) $t_{on} = 1.0$ s et (b) $t_{on} = 0.5$ s

## 3.3 Analyse des mécanismes de formation du Ge mesoporeux

Afin de décrire la formation de pores d'une manière qualitative, un modèle électrochimique des réactions ayant lieu lors de la gravure doit être établi.

Le procédé *BEE* est basé sur l'alternance des régimes cathodiques et anodiques. D'abord, le comportement électrochimique du Ge sera considéré pour les deux régimes anodique et cathodique. En se basant sur des résultats expérimentaux, les réactions impliquées lors de la formation des pores et leurs facteurs cinématiques seront déterminés. Par la suite, une explication qualitative de formation des différentes morphologies sera donnée. Finalement, une conclusion générale sera faite sous la forme d'un algorithme pratique permettant d'obtenir une couche du Ge poreux avec les paramètres prédéfinis.

### 3.3.1 Régime anodique

#### Dissolution tétravalente

Pour la première fois, le mécanisme de la dissolution anodique du Germanium a été proposé par D. R. Turner [111] en se basant sur l'analyse voltamétrique de l'interface Ge-électrolyte et des résultats expérimentaux antérieurs [13]. Il a été démontré, qu'une oxydation anodique du Ge implique presque 2 fois moins de charge ($2 \cdot 10^{-4}$ C/cm$^2$) qu'une réduction cathodique du Ge(IV) [111]. Ceci signifie que lors de la dissolution anodique continue, la couche superficielle de Ge contient toujours une quantité définie de matériau réductible (dont une mono-couche d'oxyde ou d'hydroxyde de Ge). Mais comme aucune évolution d'$O_2$ n'est observée, il est suggéré que les molécules d'eau ou des ions $OH^-$ par voie électrochimique réagissent avec les atomes superficiels du Ge. Ces atomes possèdent des électrons non liés (soit 2 dans le plan (100)) qui passent dans la bande de conduction suite à la réaction suivante :

$$=\mathrm{Ge}\colon + 2\mathrm{OH}^- \longrightarrow =\mathrm{Ge(OH)}_2 + 2\mathrm{e}^- \tag{3.2}$$

où (=) signifie deux liaisons covalentes d'un atome superficiel, (:) deux électrons non liés d'un atome superficiel.

Un atome de Ge associé avec les radicaux hydroxyle passe dans la solution. Des liaisons covalentes entre des atomes superficiels et le volume sont rompues avec la participation de 2 trous suite à la réaction suivante :

$$=\mathrm{Ge(OH)}_2 + 2\mathrm{e}^+ \longrightarrow \mathrm{Ge(OH)}_2{}^{2+} \tag{3.3}$$

Cette réaction est suivie par l'apparition de nouveaux atomes superficiels de Ge avec deux électrons non liés qui réagissent avec des ions $OH^-$ et des molécules d'eau selon la réaction (3.2). L'ion de Ge réagit avec des ions $OH^-$ comme suit :

$$Ge(OH)_2{}^{2+} + 2OH^- \longrightarrow H_2GeO_3 + H_2O \tag{3.4}$$

Finalement, la réaction globale :

$$Ge + 3H_2O + 2e^+ \longrightarrow H_2GeO_3 + 4H^+ + 2e^- \tag{3.5}$$

La vitesse de réaction est déterminée par l'étape (3.3) qui dépend du transport de trous du volume vers la surface.

L'équation (3.6) présente le modèle de dissolution anodique à l'état tétravalent du Ge d'orientation (100) proposé par Beck et Gerischer [9] modifié pour le cas des solutions acides [12]. En se basant sur le modèle de Turner [111], les étapes intermédiaires sont expliqués en introduisant les coefficients variables $\gamma$ et $\lambda$ de participation des trous et des électrons dans différentes étapes de la dissolution anodique.

La réaction (3.6a) correspond à la réaction (3.2) du modèle de Turner et peut se dérouler sous des potentiels négatifs. La rupture consécutive de l'une des deux autres liaisons covalentes d'atomes superficiels de Ge nécessite une énergie d'activation importante. Cette rupture est facilitée par la dissociation des groupes hydroxyles de surface (étape 3.6b avec la participation des molécules d'eau. Un radical négatif d'Oxygène contribue à l'apparition d'un trou sur l'une des deux liaisons covalentes de l'atome superficiel (étape 3.6c), plus précisément, le départ d'un des deux électrons de valence entraine l'apparition d'un état libre (trou) dans la bande de valence de semi-conducteur. La localisation d'un trou $e^+$ sur l'une des liaisons covalentes de l'atome de Ge, conduit finalement à une rupture de cette liaison grâce aux vibrations thermiques du réseau (étape 3.6d). Cette étape, qui est considérée comme la plus difficile, détermine la vitesse du processus entier de dissolution anodique du Ge. Tandis que le trou est localisé sur la liaison (étape 3.6c), la rupture de liaison est réversible jusqu'à ce que le groupe $Ge^+O^-OH$ pivote loin de la surface (étape 3.6d). Dans la dernière réaction le Ge(III) est instable, la réaction (3.6e) s'effectue facilement.

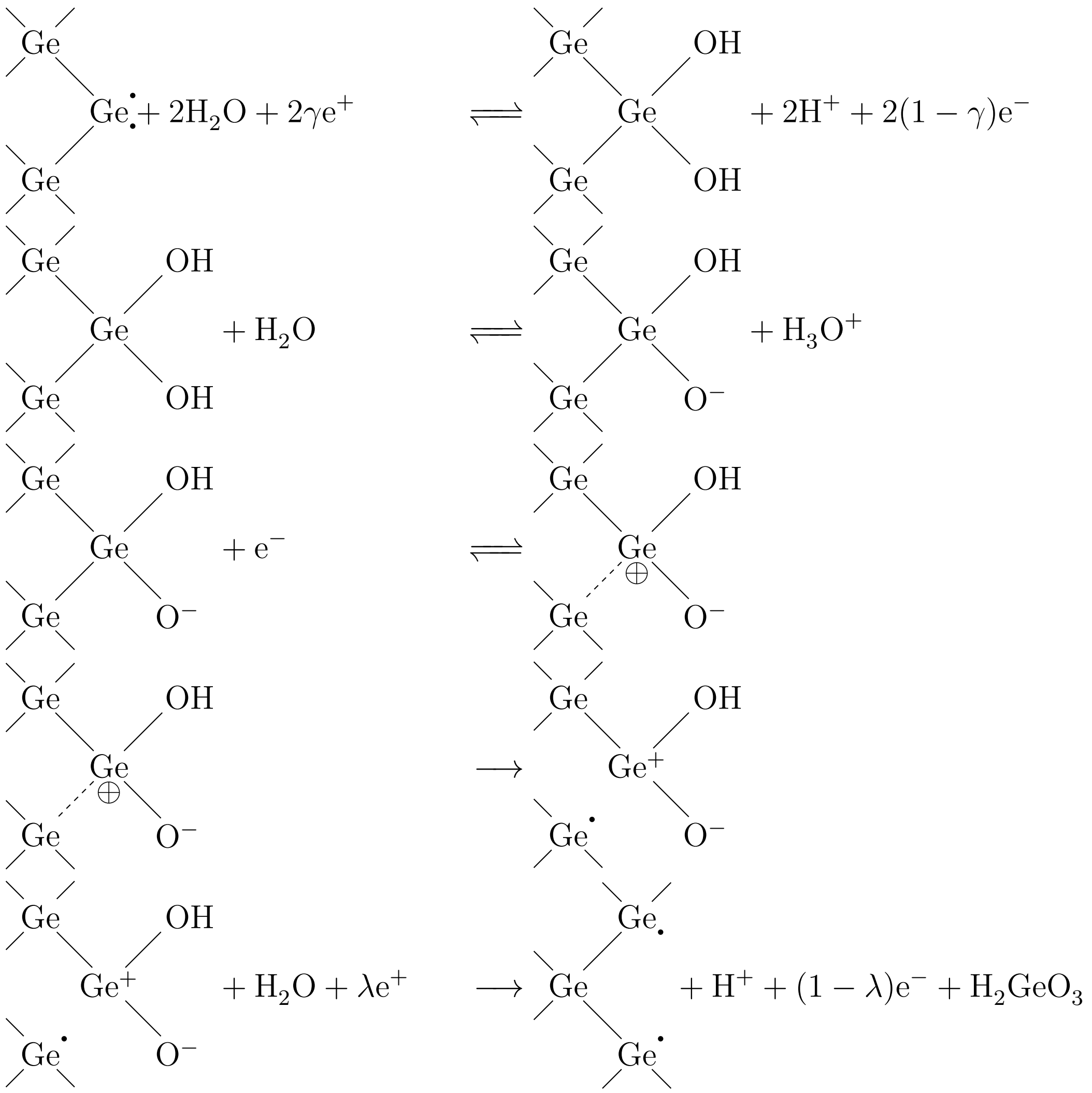

(3.6a)

(3.6b)

(3.6c)

(3.6d)

(3.6e)

La réaction globale de dissolution anodique à l'état tétravalent du Ge de Beck et Gerischer s'écrit :

$$\mathrm{Ge} + 3\mathrm{H_2O} + m\mathrm{e}^+ \longrightarrow \mathrm{H_2GeO_3} + 4\mathrm{H}^+ + (4-m)\mathrm{e}^- \qquad (3.7)$$

Le taux de dissolution mesuré pour une surface orientée (110) est très similaire à celui d'une surface (100). Les atomes superficiels d'une surface (110) sont attachés avec 3 liaisons, ce qui devrait compliquer le détachement des atomes lors de la dissolution. Cependant, la surface (110) peut être facilement nucléée par un détachement d'un atome avec deux liaisons. Ce processus peut commencer sur un défaut et se propager le long d'une ligne d'atomes (équation 3.8). Les réactions chimiques impliquées sont les mêmes que celles correspondant au cas d'une surface (100).

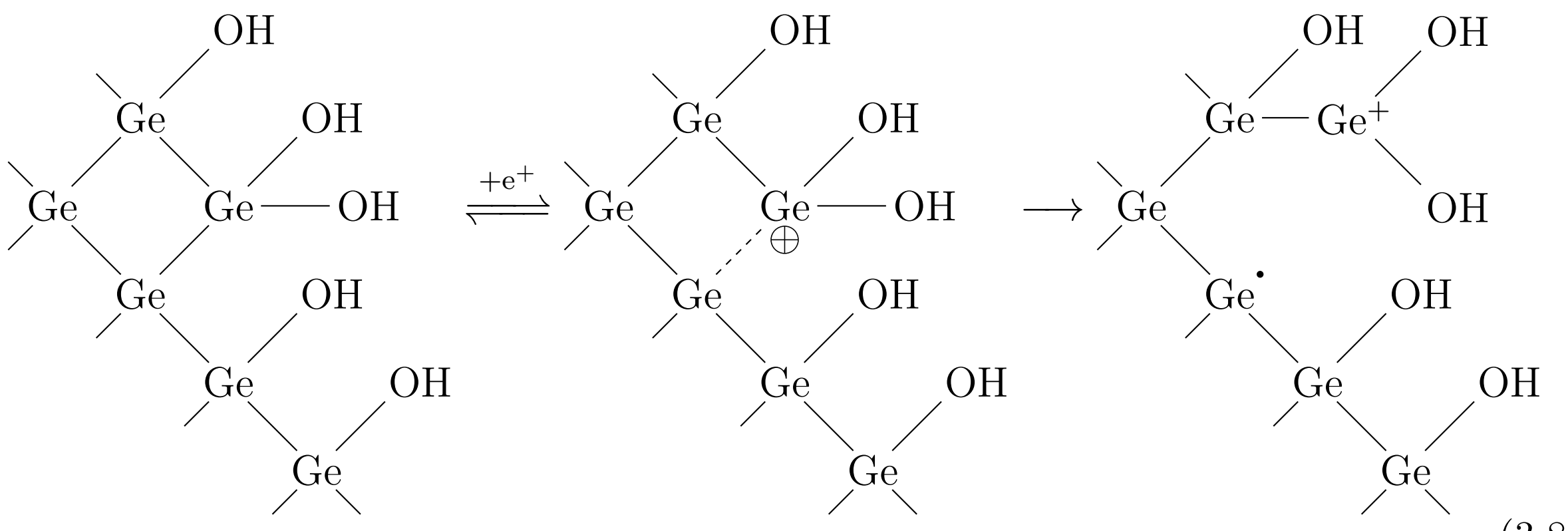

(3.8)

## Dissolution bivalente

Batenkov [8] propose le schéma de la dissolution anodique du Ge à l'état bivalent dans des solutions acides et neutres (équation 3.9), qui se produit en parallèle avec sa dissolution à l'état tétravalent. L'orientation de l'anode (le plan (011)) a été choisie parallèle à l'une des quatre liaisons covalentes tétraédriques de deux atomes de surface d'un semi-conducteur. Les trois autres liaisons de chacun des deux atomes de la surface, et de tous les atomes analogiques sont réparties de la façon suivante :

- une liaison — avec un atome de Ge volumique ;
- la deuxième liaison — avec un autre atome superficiel de Ge ;
- la troisième liaison — tournée vers la solution. Cette dernière liaison est répartie dans la réaction précédente et porte un électron non lié.

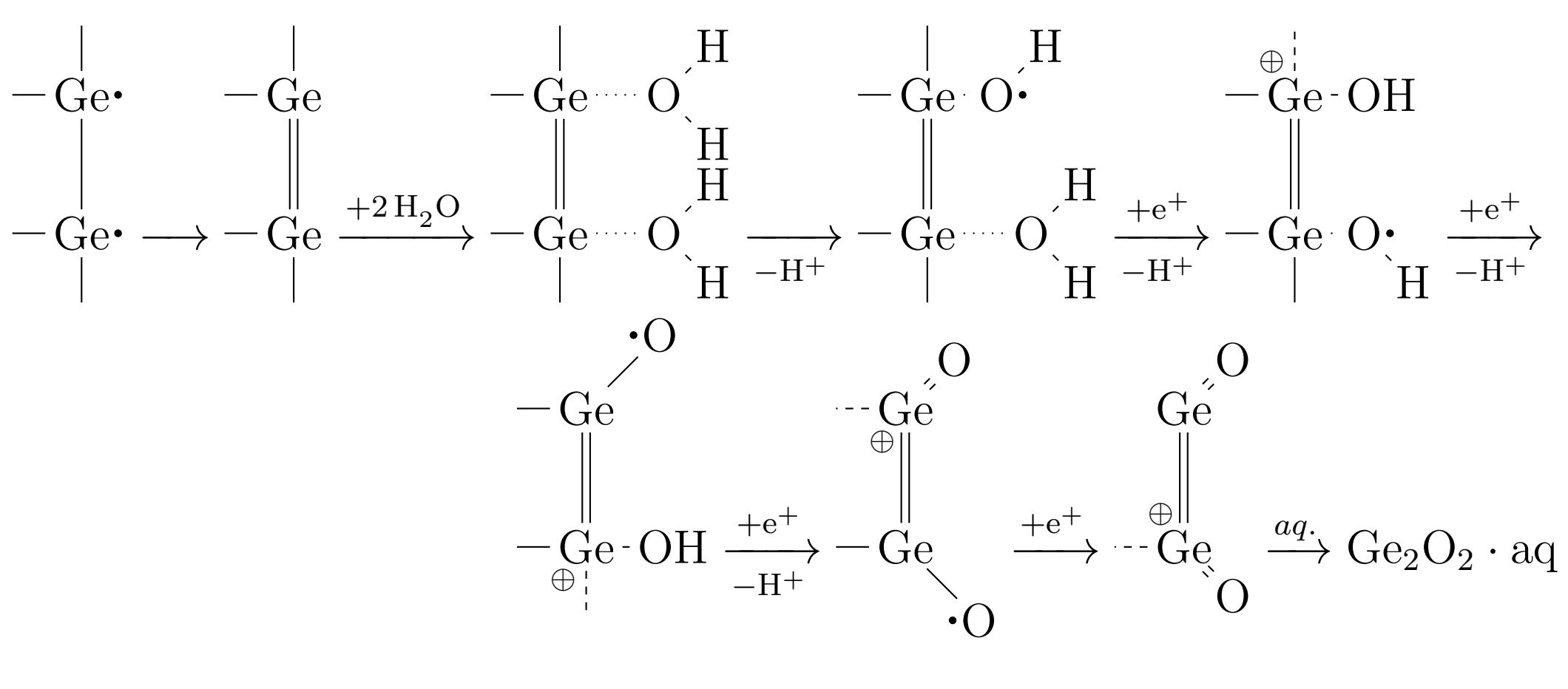

(3.9)

E. Garralaga Rojas propose [39] une réaction (3.10) de dissolution bivalente au Ge(IV) par analogie avec le Si [66]

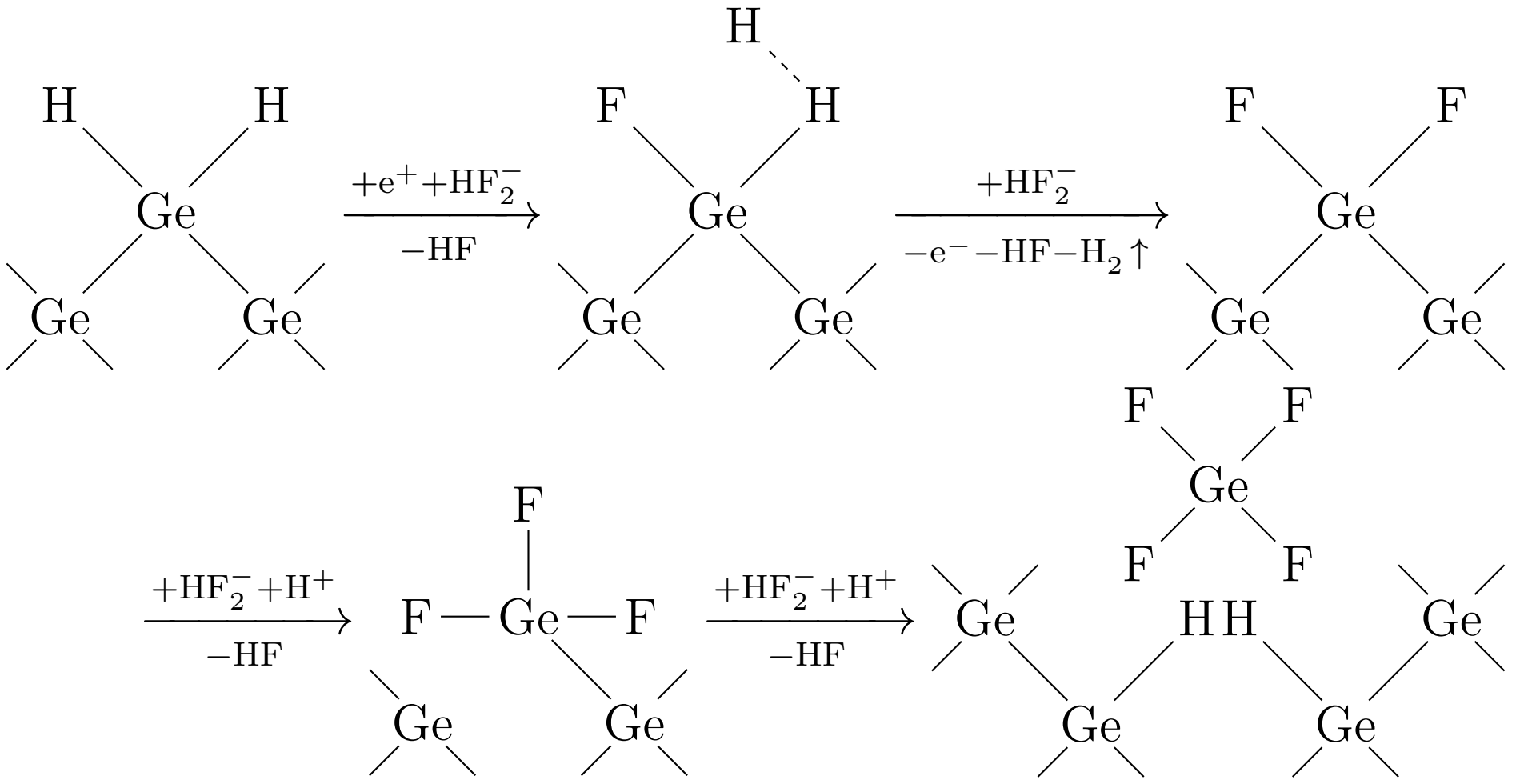

(3.10)

Cependant, l'occurrence de la réaction (3.10) est peu probable. Suite à cette réaction, la surface devrait être hydrogénée, ce qui n'a pas été rapporté et ne se confirme pas par nos propres mesures FTIR (section 4.2.1). Après l'anodisation sous faibles courants, la surface de Ge est couverte par du GeO (souvent — polymérisée). Étant anodisé sous plus forts courants, le $GeO_2$ peut être trouvé sur la surface. Suite à la réaction chimique avec l'HF et l'eau, le $GeO_2$ se transforme en $GeF_4$ et puis en $H_2GeF_6$. La formation du $GeF_4$ via HF seul n'a jamais été rapportée dans la littérature. Ainsi, la dissolution bivalente au Ge(II) avec l'HF n'est pas possible parce que le $GeF_2$ n'existe pas en solution aqueuse.

### 3.3.2 Cinétique des réactions électrochimiques

Le courant qui traverse l'interface réactif est décrit par la Relation de Butler-Volmer :

$$j = j^0[(j_a/j_a^0) + (-j_k/j_k^0)] = j^0[\exp(\alpha\eta_{0,a}nF/RT) - \exp(-(1-\alpha)\eta_{0,k}nF/RT)] \quad (3.11)$$

où :

$j$ — densité de courant (en $A \cdot m^{-2}$) ;
$j^0$ — densité de courant d'échange (incluant la constante de vitesse) ;
$\eta_{0,a}$ et $\eta_{0,k}$ — les sur-potentiels d'électrode sous polarisation anodique ou cathodique respectivement ;
$T$ — température (en K) ;
$n$ — nombre d'électrons intervenant dans l'étape déterminant la vitesse de réaction ;
$F$ — constante de Faraday (en $C \cdot mol^{-1}$) ;
$R$ — constante des gaz parfaits (en $J \cdot K^{-1} \cdot mol^{-1}$) ;
$\alpha$ — coefficient de transfert de charge ;

Lorsque le courant anodique est supérieur au courant d'échange ($j_a > J^0$), l'équation (3.11) se transforme en équation de Tafel :

$$\eta_a = (RT/\alpha nF)\ln(j_a/j^0) = a + b \lg j_a \quad (3.12)$$

avec

$$a = -(RT/\alpha nF)\ln j^0 \quad (3.13)$$

La présence d'une section droite sur une courbe de polarisation tracée en échelle logarithmique signifie une prédominance du courant anodique sur le courant d'échange et la nature électrochimique du sur-potentiel. La pente de cette droite $a$ (aussi appelée une pente de Tafel) permet de déterminer les paramètres électrochimiques $\alpha$, $n$ et $j^0$ et donc, de juger sur les réactions ayant lieu sur l'électrode.

Il a été rapporté auparavant, que la pente de Tafel pour le Ge en régime anodique peut être ajustée de manière continue par variation d'une composition d'électrolyte. La pente prend sa valeur maximale de 0.12 V pour des solutions diluées ($HF_{\approx 0\%}$) et diminue à 0.06 V pour des solutions concentrées ($HF_{21.6\ mol/l}$) [79]. Batenkov confirme ces résultats [8] mais rapporte aussi l'existence de la deuxième droite de Tafel observable sous polarisations anodiques inférieures au 0.15 V (Figure 3.19). La pente de Tafel de la deuxième droite dépend peu de la concentration d'HF et reste dans le plage de 0.10 à 0.12 V. L'existence de deux droites de Tafel sur une même courbe est liée apparemment à deux réactions chimiques dominantes différentes. Dans les solutions aqueuses d'HF de n'importe quelle concentration, lorsque le sur-potentiel est inférieur à 0.15 V, l'étape qui contrôle la dissolution anodique du Ge est la réaction de son oxydation au Ge(II) :

$$=\mathrm{Ge_2OOH^-} + \mathrm{e^+} \longrightarrow -\mathrm{Ge_2O_2^-} + \mathrm{H^+} \tag{3.14}$$

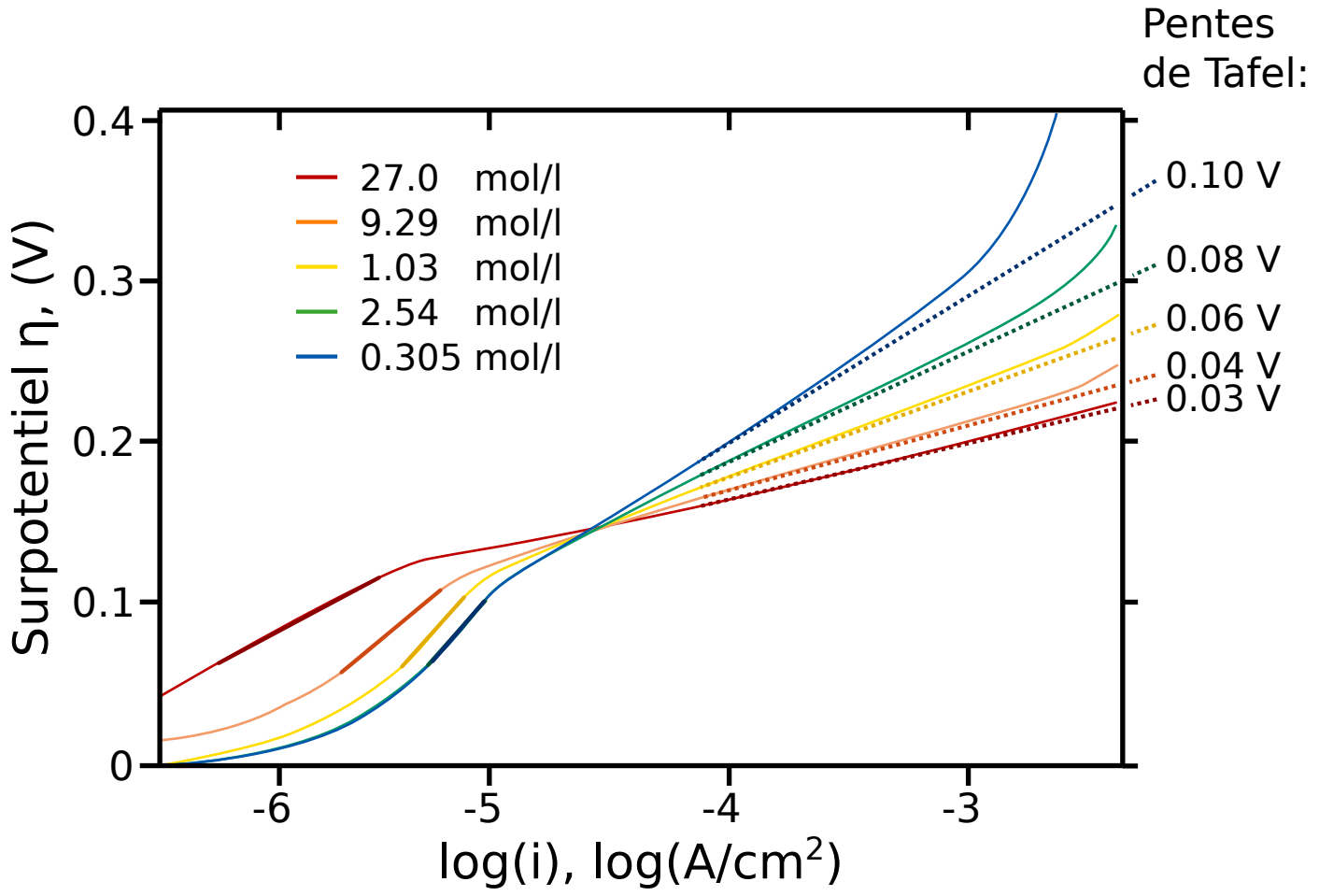

Figure 3.19 Courbes de polarisation anodique du Ge type p ($\rho = 10 m\Omega \cdot cm$) dans des solutions d'HF à différentes concentrations [8]

Cette étape n'implique pas d'HF ou d'ions de Fluor. Ceci est en contradiction avec une conclusion de E. Garralaga Rojas [39] basée sur des observations de dégagement d'hydrogène lors de la gravure, dont la dissolution bivalente ne prend place que sous des densités de courant supérieures à 7.5 $mA/cm^2$. Sachant la pente de Tafel (figure 3.19, section 1) $b = 0.10...0.12$ V et en supposant $\alpha = 0.5$, selon (3.13) $n \approx 1$.

Si le sur-potentiel est supérieur à 0.15 V, lors de la dissolution anodique, le Ge est oxydée au Ge(IV). Dans des solutions d'HF déci-molaires, $b = 0.10..0.12$ V, $n \approx 1$, $\alpha = 0.5$, l'ordre partiel de la réaction par rapport à l'HF est $\approx 0$, l'étape cinématique limitant correspond à celle du schéma de Beck et Gerischer (3.6d)

$$=\mathrm{Ge_2OOH^-} + \mathrm{e^+} \longrightarrow -\mathrm{GeOOH} \tag{3.15}$$

Lorsque la concentration d'HF augmente, les pentes des courbes anodiques diminuent ce qui correspond à une augmentation du nombre de charges dans l'étape cinématique déterminante. Autrement dit, quelques réactions s'associent dans cette étape. Dans les solutions d'HF dont la concentration est supérieure à 25 mol/l ($HF/H_2O \geq 45\%$), lorsque l'ordre partiel de la réaction par rapport à l'HF augmente jusqu'au 6 à 7, le processus est décrit par la réaction suivante :

$$\equiv \mathrm{Ge}\cdot + 6\mathrm{HF} + 4\mathrm{e}^+ \longrightarrow \mathrm{H_2GeF_6} + 4\mathrm{H}^+ \tag{3.16}$$

Seulement dans ce cas extrême la participation d'HF (plus précisément d'ions $HF^{2-}$ et $F^-$ au lieu d'ions $OH^-$ et $O_2^-$) dans les premières étapes de l'oxydation anodique du Ge est possible [8]. Le champ électrique créé par des ions $F^-$, dont le rayon est de 0.133 nm, est 1.6 fois plus faible que celui créé par des ions $O_2^-$ dont le rayon est de 0.146 nm [101].

Au lieu de la réaction 3.16, le processus peut comprendre deux étapes : une phase initiale, impliquant d'ions $OH^-$ et $O_2^-$ (3.17a) et une phase terminale se déroulant avec le HF (3.17b) lorsque les constantes de ces deux réactions sont similaires.

$$\equiv \mathrm{Ge}\cdot + 2\mathrm{H_2O} + 2\mathrm{e}^+ \longrightarrow =\mathrm{GeOOH}^- + 3\mathrm{H}^+ \tag{3.17a}$$

$$=\mathrm{GeOOH}^- + 6\mathrm{HF} + 2\mathrm{e}^+ \longrightarrow \mathrm{H_2GeF_6} + 2\mathrm{H_2O} + \mathrm{H}^+ \tag{3.17b}$$

La formation du Ge mesoporeux n'est possible que dans des solutions d'HF concentrées (>40%). La diminution de la concentration mène à une forte amorphisation de la surface. Dans des solutions d'HF moins concentrées (<20%), la surface est gravée de manière uniforme, sans formation de Ge poreux. Ainsi, la réaction 3.16 impliquant le fluor est considérée comme une clé dans la formation du Ge mésoporeux.

Finalement, l'équation (3.18) résume les réactions sur le Ge lors du régime anodique en fonction des conditions du procédé. Ce schéma montre les étapes clés de chaque réaction. Les étapes déterminantes de vitesse sont indiquées comme «r.d.»

$$\begin{cases} =\mathrm{Ge_2}: \xrightleftharpoons{+2\mathrm{e}^+} =\mathrm{Ge_2OOH}^- \xrightarrow[\mathrm{r.d.}]{+2\mathrm{e}^+} -\mathrm{Ge_2O_2^-} \longrightarrow \mathrm{Ge_2O_2};\ \eta_{<0.15\mathrm{V}};\ \mathrm{HF} \\ =\mathrm{Ge}: \begin{cases} \xrightleftharpoons{+2\mathrm{e}^+} =\mathrm{GeOOH}^- \begin{cases} \xrightarrow[\mathrm{r.d.}]{+\mathrm{e}^+} -\mathrm{GeOOH} \xrightarrow[\mathrm{r.d.}]{+\mathrm{e}^+} \mathrm{H_2GeO_3};\ \eta_{>0.15\mathrm{V}};\ \mathrm{HF}_{\leq 5\%} \\ \xrightarrow[\mathrm{r.d.}]{+2\mathrm{e}^+} \mathrm{H_2GeF_6};\ \eta_{\geq 0.15\mathrm{V}};\ \mathrm{HF}_{\geq 45\%} \end{cases} \\ \xrightarrow[\mathrm{r.d.}]{+4\mathrm{e}^+} \mathrm{H_2GeF_6};\ \eta_{>0.15\mathrm{V}};\ \mathrm{HF}_{\geq 45\%} \end{cases} \end{cases} \tag{3.18}$$

### 3.3.3 Régime cathodique

La réaction initiale lorsqu'une électrode de Ge est commutée de l'anode à la cathode, c'est à dire lorsque le courant est inversé dans la cellule électrolytique, est la réduction d'hydroxyde ou d'oxyde superficiel et la formation d'une couche d'hydrure, elle s'écrit comme suit [111] :

$$=\mathrm{Ge(OH)_2} + 4\mathrm{H^+} + 4\mathrm{e^-} \rightleftharpoons =\mathrm{GeH_2} + \mathrm{H_2O} \tag{3.19}$$

Il a été démontré que la réaction (3.19) comprend forcement un atome de surface radicalaire comme intermédiaire, qui se traduit comme suit [43] :

$$=\mathrm{Ge(OH)_2} + 2\mathrm{H^+} + 2\mathrm{e^-} \rightleftharpoons =\mathrm{Ge\colon} + 2\mathrm{H_2O} \tag{3.20a}$$

$$=\mathrm{Ge\colon} + 2\mathrm{H^+} \rightleftharpoons =\mathrm{GeH_2} + 2\mathrm{e^+} \tag{3.20b}$$

La réaction (3.20b) se déroule avec une consommation d'électrons de la bande de conduction du Ge lorsque le potentiel cathodique est faible ou avec une injection de trous dans la bande de valence du Ge lorsque le potentiel est élevé [81]. Une réduction cathodique du Ge nécessite toujours environ $4 \cdot 10^{-4}$ C/cm$^2$ de charge (soit 4 électrons par atome) quelle que soit la densité de courant cathodique ou le temps et le courant de pré-traitement anodique [111].

Le processus (3.20) est réversible lorsqu'un potentiel d'électrode est maintenu suffisamment faible ($< -0.2$ V vs SCE dans une solution aqueuses de $HClO_4$) — limite supérieure [76]. La limite inférieure est le potentiel de déclenchement de la dissolution cathodique $\varphi_{NPD}$ (angl. « Negative Potential Dissolution », NPD) suite à une protonation des liaisons covalentes entre des atomes superficiels et des atomes de volume. Le NPD implique un trou injecté à l'étape (3.20b), en créant la dissociation d'une liaison covalente. Une valeur de ce potentiel rapportée pour le Si est de $-1.5$ V [108] à $-5$ V [107] pour des solutions alcalines. Un phénomène de dissolution du Ge sous des courants négatifs élevés a été observé dans l'eau lourde [84] ainsi que dans une solution aqueuse de HCl [23]. L'énergie de dissociation d'une liaison covalente de Ge est 1.24 fois plus faible que celle du Si. Le potentiel $\varphi_{NPD}$ de Ge dans une solution d'HF est estimé à quelque volts. Cependant, aucune étude voltamétrique n'a été faite à ce jour.

Une fois que la surface de Ge est complètement hydrogénée, l'injection des électrons ensuite, mène à une réduction d'hydrogène atomique $H^+$ avec un dégagement d'hydrogène moléculaire $H_2$ (équation 3.21).

$$2H^+ + 2e^- \longrightarrow H_2 \tag{3.21}$$

En réalité, la réduction d'hydrogène se déroule en deux étapes via des radicaux superficiels [81], [45] :

$$=\text{Ge:} + 2H_3O^+ + 2e^- \longrightarrow =GeH_2 + 2H_2O \tag{3.22a}$$

$$=GeH_2 + 2H_3O^+ + 2e^- \longrightarrow =\text{Ge:} + 2H_2O + 2H_2 \uparrow \tag{3.22b}$$

Lorsque le Ge devient anode immédiatement après avoir été cathode, seulement deux électrons par atome de surface de Ge sont requis dans le processus d'oxydation initial. Ce résultat peut être expliqué si l'on suppose que les atomes d'hydrogène de l'hydrure de surface préfèrent se combiner chimiquement en hydrogène moléculaire ($H_2$) plutôt que d'être oxydés lorsque le courant est inversé [111] :

$$=GeH_2 \longrightarrow =\text{Ge:} + H_2 \tag{3.23a}$$

$$=\text{Ge:} + 2OH^- \longrightarrow Ge(OH)_2 + 2e^- \tag{3.23b}$$

Lors de l'application des pulsations carrées, le dégagement d'hydrogène est alors inévitable. L'effet peut être inhibée par commutation lente entre les régimes cathodique et anodique, ce qui augmentera, par ailleurs, la durée totale du procédé.

### 3.3.4 Analyse quantitative du procédé de porosification de Ge

Afin de déterminer les réactions chimiques qui se déroulent sur le Ge lors d'une porosification, l'étude du changement du potentiel anodique pendant la gravure $V(t)$ a été faite (Figures 3.20 et 3.21). Cette analyse est inspirée par la voltamétrie cyclique du Ge faite par D. R. Turner [111]. Lorsque le courant est appliqué sous forme de pulsations carrées de courant, une différence de potentiels entre le Ge et la pseudo-électrode de référence en Pt a été enregistrée. À noter que le niveau zéro correspond à la tension mesurée avant le début du procédé. Les potentiels tracés en échelle « SCE » (Électrode au calomel saturée en KCl) peuvent être décalés.

Des points caractéristiques peuvent être identifiés sur le graphique 3.21 :

I Croissance rapide du potentiel après le changement de la polarisation ;

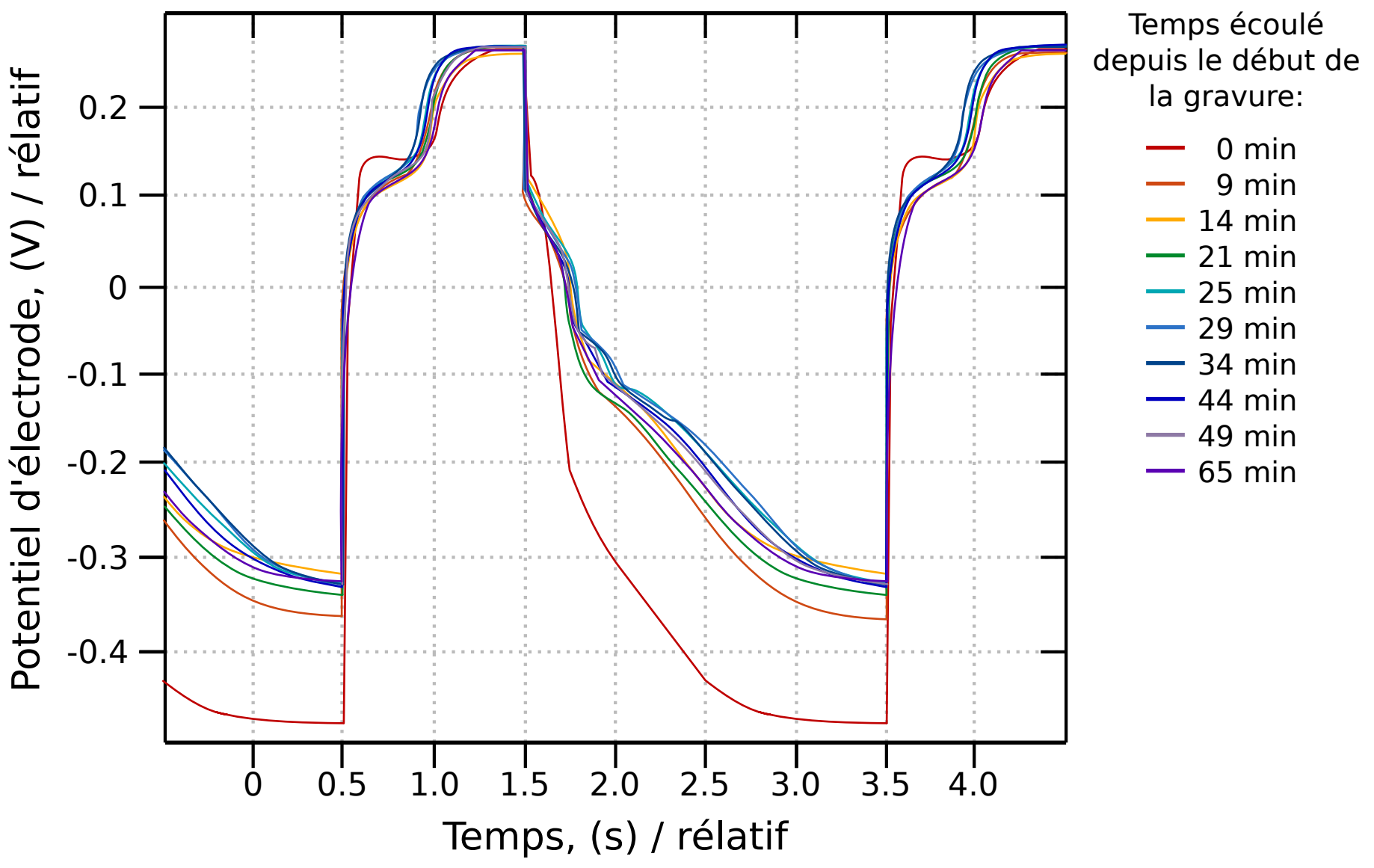

Figure 3.20 V(t) enregistré lors d'une gravure bipolaire dans une solution de $HF_{49\%}$ avec une amplitude de courant 2.1 mA/cm$^2$ et un rapport de cycle « 1"/2" ». La couleur d'une courbe chiffre le temps écoulé depuis le début de la gravure

II Dépassement de potentiel (angl. «Overshoot in potential ») au-dessus de la valeur normale qui a lieu lors de germination de premières pores sur la surface polie de Ge ; observable lors des 3 premiers cycles de gravure ;

III Droite de croissance + plateau : dissolution bivalente ;

IV Saut du potentiel, droite de croissance, plateau : probablement, une réaction de dissolution tétravalente déclenchée lorsque la tension dépasse le seuil ;

V Saut de potentiel suite à une commutation du courant : le taux de la dissolution tétravalente est limité par la vitesse de transport de charge à travers la ZCE.

VI Queue d'impulsion anodique due à une reconstruction de la double couche électrique à l'interface.

VII Hydrogénation de la surface de Ge.

VIII Plateau : décharge d'ions d'hydrogène et dégagement d'hydrogène gazeux.

La quantité de charge consommée pour hydrogéner la surface de Ge est suivante :

$$Q_c = J_c S_0 t_c \tag{3.24}$$

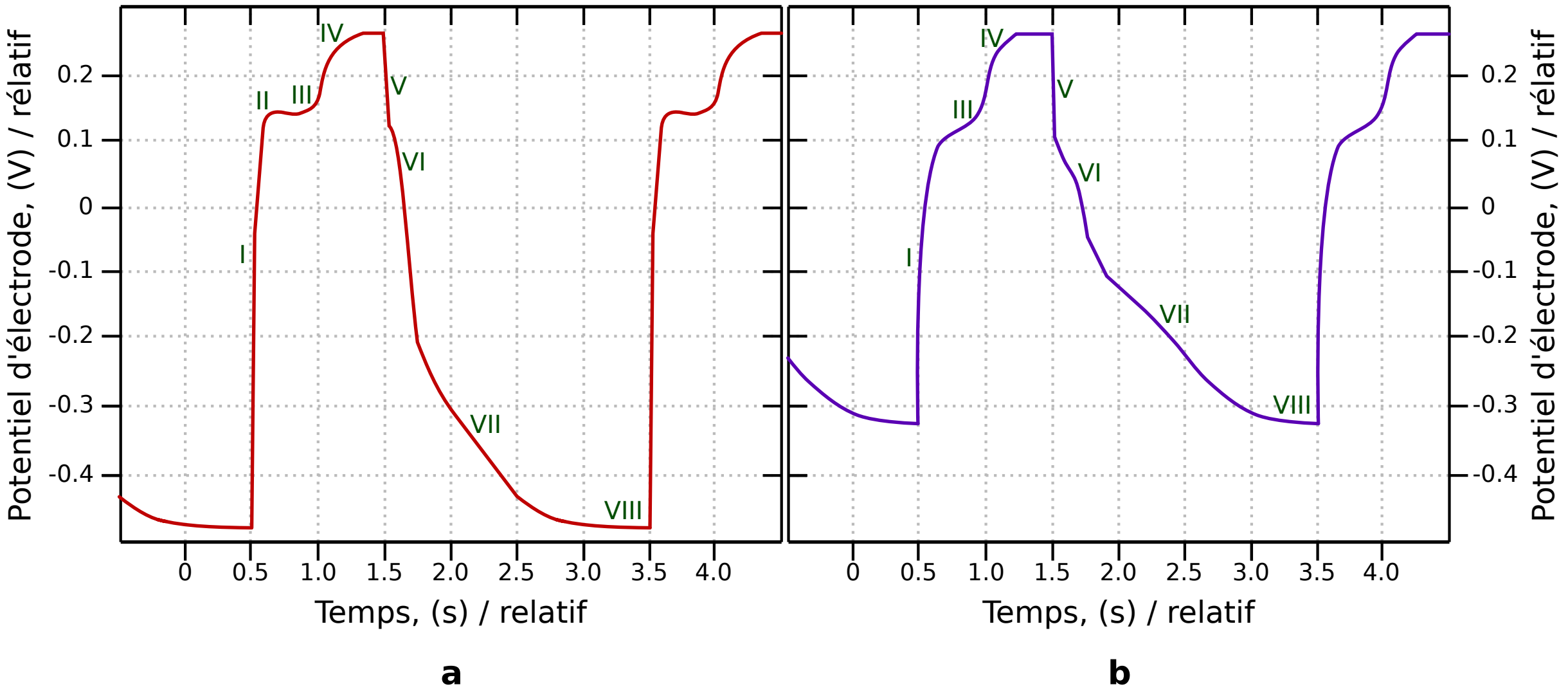

Figure 3.21 Enregistrements de V(t) au début (a) et à la fin (b) de la gravure. Les sections qui caractérisent la cinétique du procédé sont marquées avec des chiffres romains.

où $J_c$ — l'amplitude de la densité de courant cathodique, $S_0$ — la surface d'un échantillon porosifié,
$t_c$ — la durée de l'étape VII, soit le temps nécessaire pour passiver une surface de Ge avec une mono-couche d'hydrogène.

Afin de diminuer la durée du procédé, la durée des pulsations cathodiques peut être diminuée en maintenant la quantité de charge injectée (3.24). Le potentiel cathodique, cependant, doit être maintenu inférieur à $\varphi_{NPD}$ (Sections 3.2.4, 3.3.3).

Sachant $|J_c| = 2$ mA/cm$^2, S_0 \approx 1$ cm$^2, t_c = 0.75$–$1.1$ s, la surface hydrogénée interne passivée lors d'une impulsion cathodique est la suivante :

$$S_p = \frac{Q_c\ [\mathrm{C}]}{2 \cdot 10^{-4}\ [\mathrm{C} \cdot \mathrm{cm}^{-2}]} = 3.75\text{–}5.5\ \mathrm{cm}^2 \qquad (3.25)$$

où $2 \cdot 10^{-4}$ C — la quantité de charge nécessaire pour réduire 1 cm$^2$ de la surface de Ge [111].

La couche poreuse obtenue sous les conditions données possède une morphologie type « éponge » avec $D_p \approx 8.5$ nm, $D_{ip} \approx 7.5$ nm et une distribution des pores par taille très étroite (le matériau est composé de cavités séparées par des parois fines de 1–2 nm). Lorsque, l'épaisseur de la couche poreuse est de 1 µm, la surface totale interne est estimée au niveau de $S_t = 225$ cm$^2$.

Donc, $S_p \ll S_t$. À première vue, le résultat est surprenant :
Il n'y pas de besoin de passiver toute la surface interne du Ge poreux pour obtenir une couche uniforme et ne portant presque pas de signes d'une dissolution de la couche superficielle. Cependant, cela explique qu'il n'y a aucune différence entre un échantillon fabriqué par *BEE* avec $J_-$ constante et un autre fabriqué avec $J_-$ en rampe.La surface $S_p$ correspond à $\approx 1.5$ rangées des pores, probablement une dernière génération de pores nucléés et + les hémisphères inférieures d'une génération précédente (Figure 3.22).

La largeur de la ZCE $d_{ZCE}$ dans un semi-conducteur en contact avec un électrolyte peut être déterminée comme suit [80] :

$$d_{ZCE} = 2L_{D,eff}\left(\frac{e\varphi_{SC}}{kT} - 1\right)^{1/2} \tag{3.26a}$$

$$L_{D,eff} = \left(\frac{\varepsilon_r\varepsilon_0 kT}{2p_0e^2}\right)^{1/2} \tag{3.26b}$$

où $L_{D,eff}$ est la longueur de Debye effective ; $\varepsilon_r, \varepsilon_0$ — les permittivités diélectriques ; $k$ — la constante de Boltzmann ; $T$ — la température ; $p_0$ — la concentration de trous ; $e$ — la charge unitaire ; $\varphi_{SC}$ — la chute de tension aux bornes de la ZCE du semi-conducteur. En termes numériques :

$$d_{ZCE}\ [\text{nm}] = 1.33 \cdot (V[\text{mV}] - 25.8)^{1/2} \tag{3.27}$$

La ZCE dans le Ge sous une polarisation négative occupe tout l'espace entre les pores, ce qui bloque la propagation du courant injecté. Un allongement de $t_{off}$ se traduirait par un dégagement d'hydrogène et non pas par une augmentation de la surface passivée $S_p$. Sous une polarisation positive, la largeur de la ZCE est beaucoup plus petite, ainsi des charges libres existantes dans les cristallites mèneront à la dissolution de la couche de Ge poreux déjà formée. Néanmoins, les résultats expérimentaux indiquent qu'une dissolution du Ge a lieu principalement au front de propagation des pores.

### 3.3.5 Modèles qualitatifs

Les résultats expérimentaux sur la formation des différentes morphologies du Ge poreux sont expliqués ci-dessous avec des modèles qualitatifs simples.

Les processus électrochimiques sont résumés sur la Figure 3.23. La surface d'un pore (en bleu) bascule entre un état hydrogéné et un état hydroxylé sans perte de masse du Ge grâce aux réactions réversibles. Une dissolution anodique doit se terminer avant la dernière étape d'oxydation du $=GeOOH^{-}$ irréversible. La durée de vie de la passivation $\tau_p$ se réfère alors à la durée du processus de transformation de $=GeH_2$ en $=GeOOH^{-}$. Lorsque la réaction d'hydroxylation est rapide, la vitesse du processus est limitée par le transport de charge et augmente légèrement avec la concentration d'HF. Le $\tau_p$ dépend alors du dégré de passivation et du courant d'anodisation, ce qui confirme l'hypothèse de la section 3.1. Une partie de la surface d'un pore où le champ électrique est plus élevé est dissoute directement au $H_2GeF_6$. La nature de cette inhomogénéité du champ électrique détermine la morphologie du Ge poreux résultant.

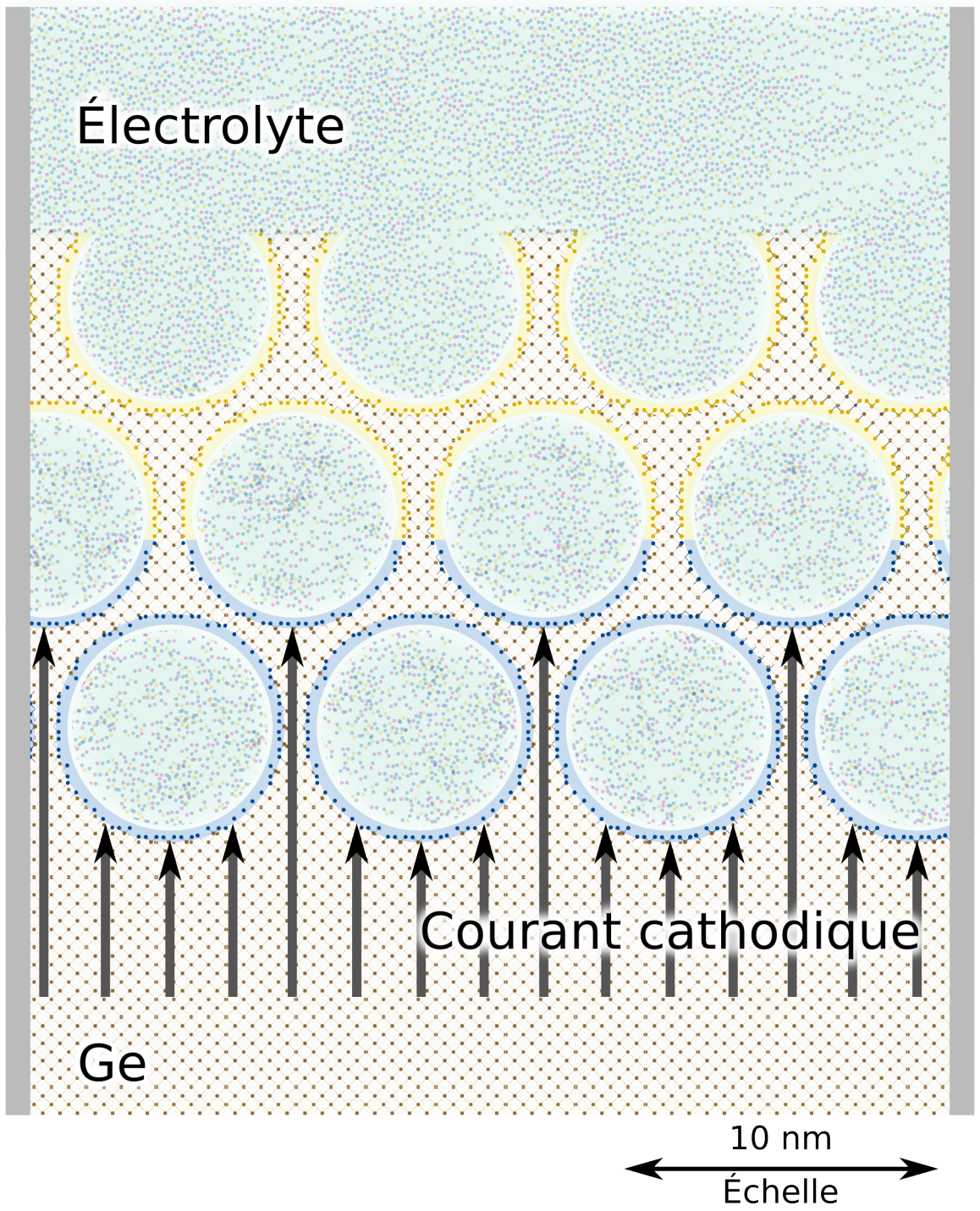

Figure 3.22 Illustration de la passivation d'une surface de Ge poreux par la charge négative injectée pendant 1 pulsation. Le fond bleu surligne l'hydrure de Ge, le fond jaune — l'(hydr)oxyde de Ge. Les atomes et les liaisons covalentes de Ge sont marrons, le Fluor — rose. Les flèches indiquent le sens du mouvement des électrons (qui est opposé à la direction du courant).

**Éponge** Les couches en éponge sont formées lorsque toute la surface d'un pore est hydrogénée (passivation complète). Des champs locaux, crées par des défaut du réseau ou des impuretés ionisées, sont plus forts qu'un champ électrique externe. Ce qui conduit à la germination des nouveaux pores. étant plus fortes qu'un champ électrique externe forment un germe du nouveau pore, Grâce au petit rayon de courbure de la surface sur l'apex des pores par exemple, le courant se concentre sur ce germe en déclenchant la réaction de la dissolution de Ge à l'état tétravalent. Ainsi la gravure s'effectue de manière inhomogène.

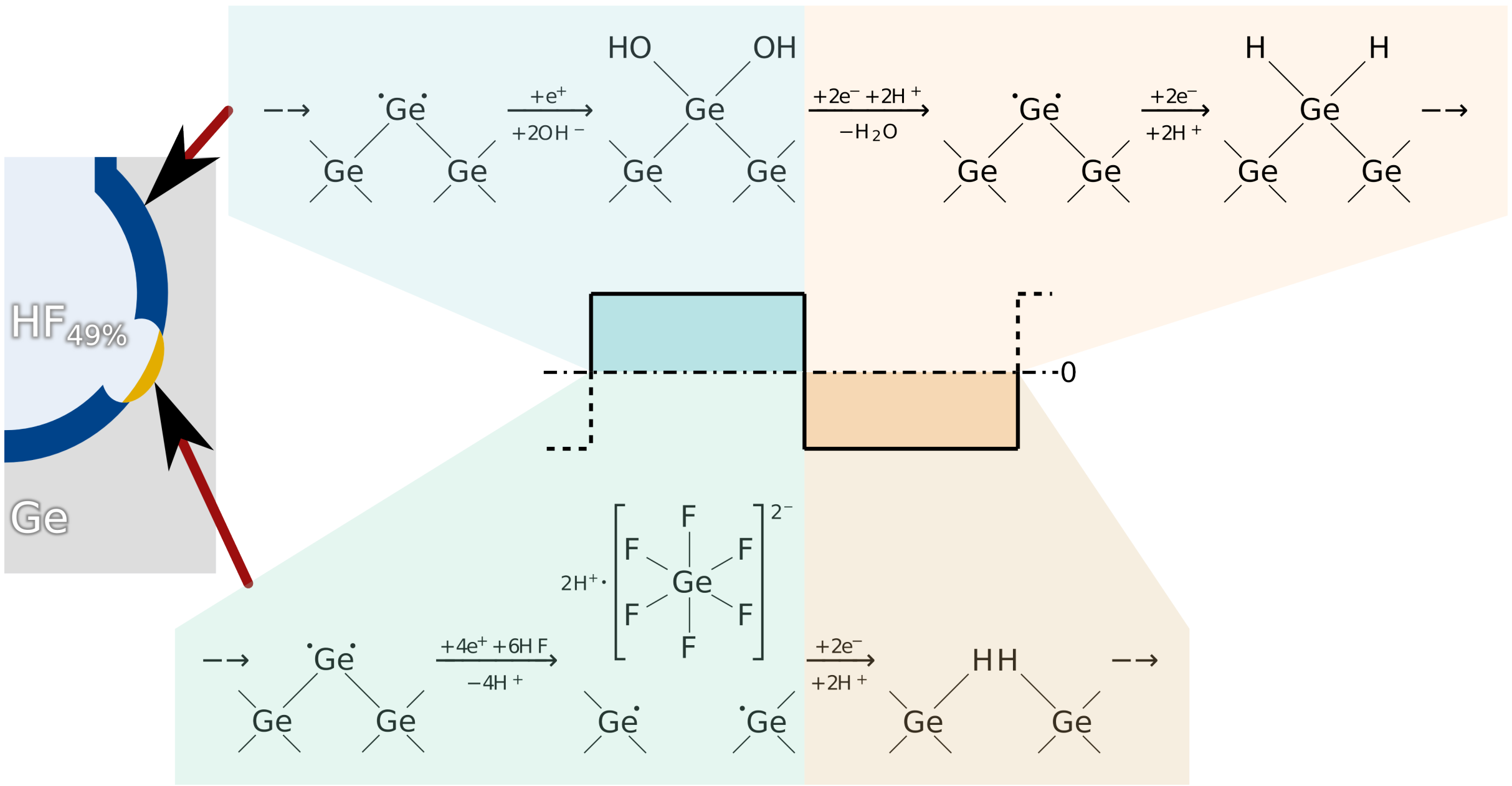

Figure 3.23 Réactions chimiques ayant lieu sur la surface de Ge pendant une gravure BEE lorsque la distribution de courant n'est pas uniforme.

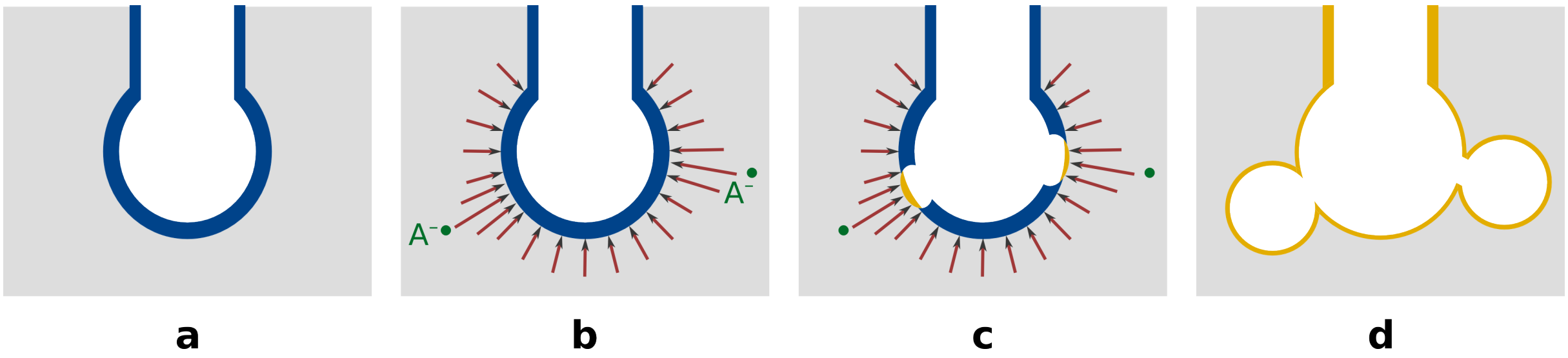

Figure 3.24 Schéma de propagation des pores dans le cas d'une passivation complète et un faible courant. Le fond bleu surligne l'hydrure de Ge, le fond jaune — l'(hydr)oxyde de Ge.
(a) État de départ : commutation du régime cathodique au régime anodique ;
(b) Application d'un champ électrique externe et sa redistribution due aux défauts (« A ») ;
(c) Gravure du Ge inhomogène suivant la distribution du champ électrique ;
(d) Fin du cycle.

**Colonnes.** Formation de couches de Ge poreux en colonnes est, en mode de passivation complète, possible lorsque l'amplitude du courant est relativement élevée. Le champ électrique est concentré sur l'apex d'un pore grâce à son petit rayon de courbure (Figure 3.25).

En faisant une analogie avec le Si poreux, le transfert des charges se fait par un claquage de la ZCE. La gravure résulte en pores cylindriques suivant le champ électrique en respectant une direction <100>, <111> ou <113>.

Il est important de noter que la valeur du courant de seuil dépend des propriétés du substrat de Ge. Comme il a été mentionné dans la section 3.2, dans le cas de substrats Umicore, une formation de colonnes a lieu sous des densités de courant supérieures à 2 mA/cm$^2$. Dans le cas de substrats AXT (avec la concertation des défauts élevée), des couches en éponge ont été obtenues sous des densités de courant allant jusqu'au 5 mA/cm$^2$. Lorsque la résistivité électrique des deux types de substrats est similaire, ce phénomène est lié à une concentration plus importante des défauts cristallins qui créent des champs locaux.

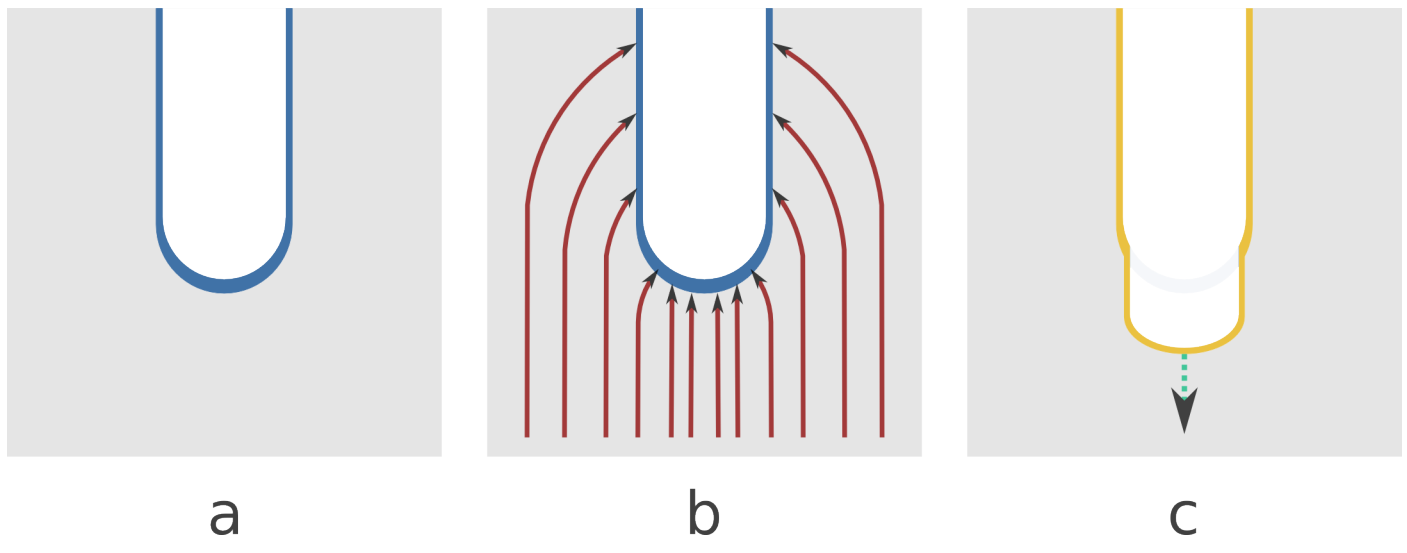

Figure 3.25 Schémas de propagation des pores dans le cas d'une passivation complète et un courant élevé. Le fond bleu surligne l'hydrure de Ge, le fond jaune — le (hydr)oxyde de Ge.
(a) État de départ : commutation du régime cathodique au régime anodique ;
(b) Application d'un champ électrique externe et la concentration des lignes de champ à l'apex d'un pore.
(c) Gravure de Ge par claquage de ZCE à l'apex jusqu'à la fin du cycle

**Arbres.** Lorsqu'une formation des pores en arbres obéit aux mêmes principes qu'une formation des colonnes, des branches latérales apparaissent dues à une passivation insuffisante de la surface latérale du pore. La distance entre les branches correspond à une largeur de ZCE de Ge.

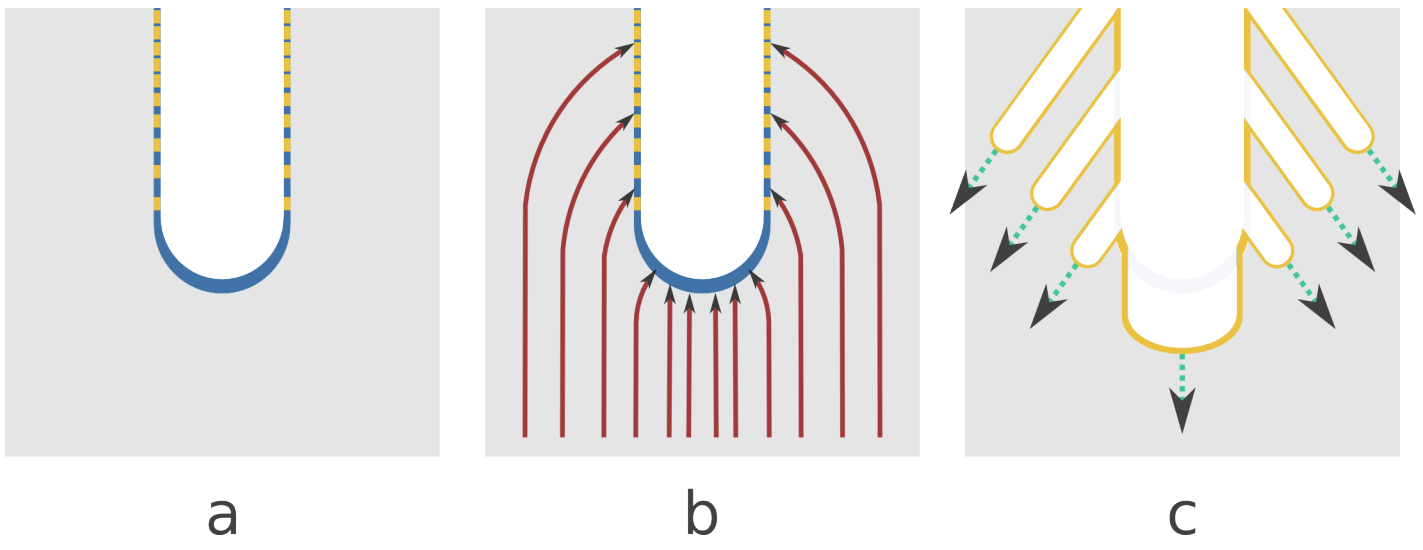

Figure 3.26 Schéma de propagation des pores dans le cas d'une passivation partielle et un courant élevé. Le fond bleu surligne l'hydrure de Ge, le fond jaune — l'(hydr)oxyde de Ge.
(a) État de départ : commutation du régime cathodique au régime anodique ;
(b) Application d'un champ électrique externe et sa concentration à l'apex d'un pore.
(c) Gravure de Ge par claquage de la ZCE à l'apex jusqu'à la fin du cycle

Les branches sont formées par les étapes.

1. D'abord, les premières branches se forment.
2. Ensuite la zone de charge d'espace apparait toute autour de ces pores.
3. Les pores suivants alors peuvent se germinale uniquement à la distance supérieure à la ZCE, car leur germination nécessite la présence des porteurs de charges libres.

Les pores se propagent alors de manière qu'il n'y ait pas de porteurs de charge libres entre les branches. Consécutivement, la croissance des branches se fait par des apex uniquement. La formation d'un nouveau pore sur le tronc principal n'est possible qu'à distance minimale $L_{ZCO}$ par rapport à un pore précédent. Une propagation des branches de deux troncs voisins s'arrête lorsque l'espacement entre leurs apex devient $\leq L_{ZCO}$. Une auto-organisation de la structure est observée. Le choix de la direction de propagation <111> ou <113> (morphologies « sapin » et « arêtes de poisson » respectivement) en fonction du degré de passivation est un phénomène encore à expliquer.

### 3.3.6 Algorithme pratique pour obtenir une couche de Ge poreux avec des paramètres prédéfinis

1. Choisir les paramètres de la couche poreuse désirée : morphologie, porosité, épaisseur.
2. Déterminer l'amplitude d'une impulsion anodique $J_+$ à partir de la Figure 3.7.
3. Si la couche doit être épaisse et si le changement de morphologie liée au temps est indésirable, choisir la rampe de courant (section 3.2.2, discussion de la figure 3.10).
4. Fixer l'amplitude d'une impulsion cathodique $J_- = -J_+$.
5. En fonction de la morphologie choisie, déterminer le régime de passivation (complète/partielle) et donc la durée de l'impulsion cathodique $t_{off}$ qui lui correspond. S'il n'y a pas de données empiriques pour la densité de courant donnée, fixer $t_{off} = 2$ s comme une valeur de départ.
6. Sachant que la valence de la réaction de dissolution anodique sous les conditions données $n = 2..4$, fixer $t_{on} = \frac{n}{4} \cdot t_{off}$.
   La valeur de $n$ est un paramètre empirique. Une position du point de fonctionnement sur le graphique 3.19 peut en donner l'idée.
7. Identifier la section VII sur la courbe $V(t)$ (figure 3.21) mesurée lors d'une application du premier couple d'impulsions $J_+; J_-$. Déterminer $t_c$.
8. Redéfinir le $t_{off} \leq t_c$ afin d'assurer le régime de passivation nécessaire.
9. Lors d'un des prochains cycles, augmenter l'amplitude $J_-$ restant en-deça de la tension de déclenchement de *NPD*. Diminuer le $t_{off}$ proportionnellement.
   Il y a très peu de données sur *NPD* du Ge dans des électrolytes à base d'HF. Dans le cas d'une solution d'HF concentrée, $-1.5$ V est une valeur sûre.
10. Réajuster le $t_{on}$ (étape 6).

## 3.4 Ingénierie des structures de Ge mesoporeux

### 3.4.1 Double-couches

Contrairement au procédé proposé par E. Garralaga Rojas et al., nous n'observons pas d'augmentation de la porosité de la partie inférieure du PGe lorsqu'une gravure est prolongée. Une couche uniforme se forme jusqu'à une commutation de morphologie. Lorsque la porosification est faite en mode de passivation complète, la morphologie de la couche principale est une éponge. La couche poreuse en fils qui se forment par la suite a une porosité plus faible que l'éponge.

L'autre type de structure double-couche est un empilement de deux couches avec la même morphologie mais de porosités différentes.

L'approche dite « classique » de réaliser une structure à double porosité consiste à moduler la porosité d'une couche poreuse pendant sa croissance par changement de la densité de courant anodique. La gravure commence avec une faible densité de courant et se termine avec une densité de courant plus élevée. Rappelons, que dans le cas du Si poreux, la porosité est contrôlé directement par l'amplitude du courant anodique.

L'effet de l'influence de la porosité de la couche sur restructuration du Ge poreux durant le recuit sera discuté en détailles dans la section 5.1. Dans les procédés de transfert de couche mince à base du Si poreux, la porosité de la couche inférieure doit être de 70–80% (Section 1.2.1)

Bien que la porosité du Ge poreux soit aussi ajustable, l'adaptation de cette méthode rencontre quelques limitations :

- Il apparaît assez difficile d'augmenter la taille des pores en appliquant un courant élevé pour réaliser une couche poreuse avec une porosité supérieure à 50%. La densité de pores et alors la porosité tendent à saturer.
- La diminution de la densité de courant dans le but de diminuer la porosité augmente notablement la durée du procédé. La vitesse de formation d'une couche avec la porosité inférieure à 15% est comparable à la vitesse de propagation de la couche amorphe superficielle.
- Les parois des pores sont couvertes par l'oxyde qui bouche ainsi les pores. L'oxyde étant instable, il peut disparaître au cours du traitement thermique ou bien aussi après un rinçage dans l'eau. De plus, ces deux procédures risquent de modifier la structure des deux couches.

Afin d'augmenter la porosité de la couche de séparation, la durée des impulsions anodiques a été augmentée au-delà du temps de persistance de la couche de passivation. Comme on peut le voir sur la Figure 3.27, suite à la destruction des parois entre les pores, des pores voisins se conjuguent ce qui entraîne une augmentation de la porosité de la couche.

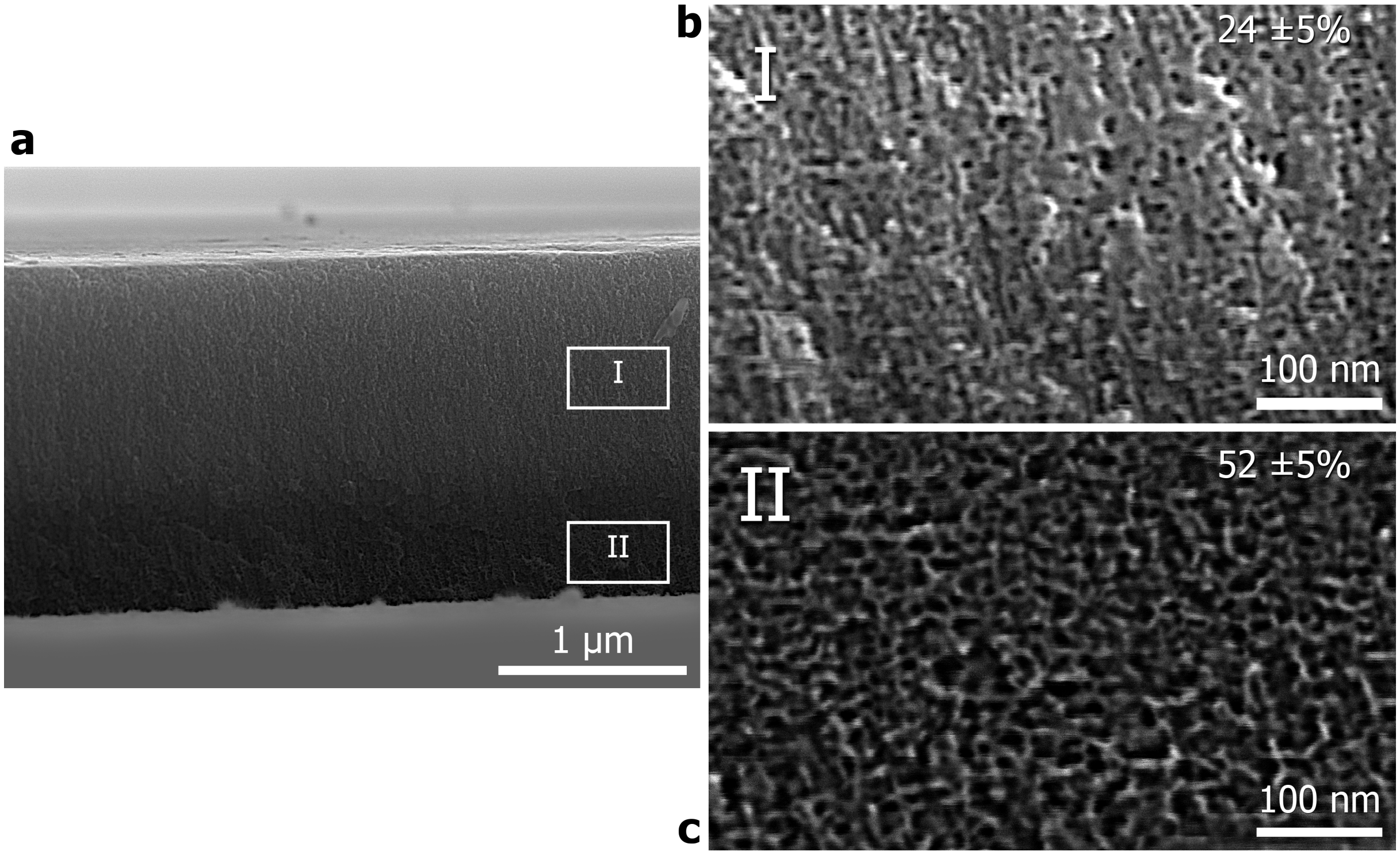

Figure 3.27 (a) Image MEB de la structure mésoporeuse de Ge dont la porosité des couches I et II est ≈ 24% et ≈ 52% respectivement ; (b) et (c) — grandissements des régions I et II de l'image (a) respectivement.

### 3.4.2 Nouvelle morphologie

La compréhension actuelle des mécanismes de formation du Ge mésoporeux nous a permis de concevoir la nouvelle morphologie présentée sur la Figure 3.28. La structure complexe constituée de colonnes verticales se propageant dans la direction <100> et ayant de nombreuses branches latérales a été obtenue en alternant une impulsion $2.1/-2.1$ mA/cm$^2$ et trois impulsions $0.5/-0.5$ mA/cm$^2$ avec le couple de gravure bipolaire fixe 1 s/2 s. En effet, une impulsion d'amplitude élevée déclenche la formation de colonnes, alors que trois impulsions consécutives de faible amplitude favorisent le développement des branches latérales avec une morphologie de type éponge, en bon accord avec la théorie de formation.

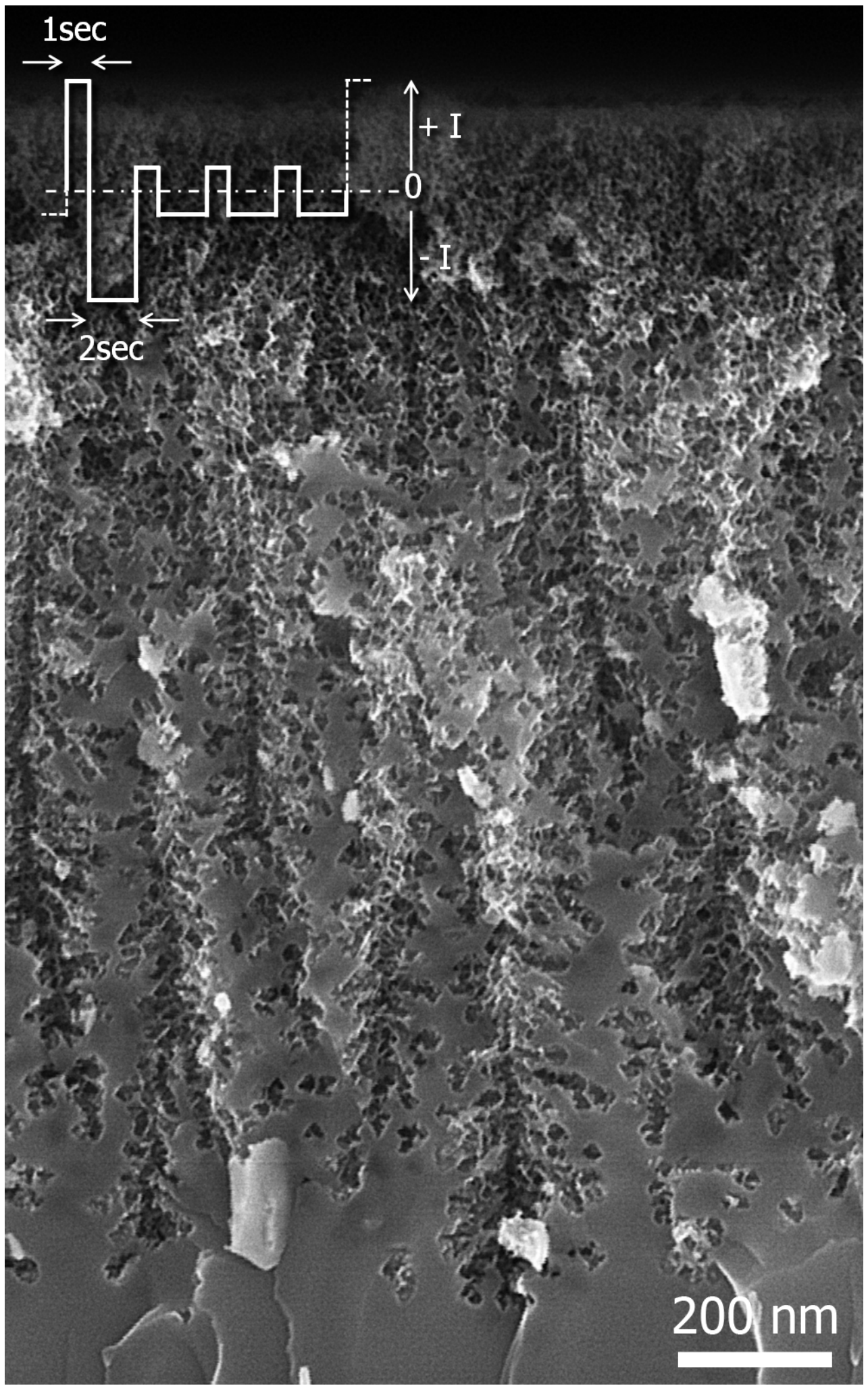

Figure 3.28 Image MEB en coupe transversale de la structure complexe réalisée en alternant une impulsion $2.1/-2.1$ mA/cm$^2$ et trois impulsions $0.5/-0.5$ mA/cm$^2$ avec le couple de gravure bipolaire fixe 1 s/2 s

## 3.5 Conclusion du chapitre

- La gravure bipolaire permet de fabriquer des couches de Ge mesoporeux.
- La durée d'impulsion anodique doit être comparable avec le temps de persistance de la couche de passivation.
- L'injection d'une charge négative de 2–4 mC/cm$^2$ assure une passivation complète de la surface des pores.
- Des couches poreuses avec la morphologie de pores de type colonnes, arbres, dendrites, arête de poisson et éponge ont été réalisées.
- La morphologie de la couche de Ge poreux dépend principalement des facteurs suivants :
  - densité de courant anodique ;
  - degré de passivation de la surface ;
  - diffusion des anions de fluor en profondeur de pores ;
  - diffusion de la charge électrique en profondeur de la couche poreuse
- La formation de structures multi-couches avec différentes morphologies des couches est observée lorsque la durée du procédé est longue.
- L'alternance de morphologies s'arrête sur des pores en fils quelles que soit les conditions de départ.
- La surface de l'échantillon ne se déplace pas lors de la porosification. Cependant, une amorphisation graduelle de la surface externe indique la présence d'une dissolution chimique. L'effet peut être partiellement inhibé en mettant le montage dans l'obscurité.
- L'épaisseur des couches peut être contrôlée dans la gamme de 100 nm–5.5 µm.
- Des couches poreuses d'épaisseur supérieure à 5 µm se détériorent souvent lors de séchage dû aux contraintes mécaniques très importantes.
- La porosité d'une couche en éponge est ajustable dans la gamme de 15–50% par changement de la densité de courant anodique.
- La taille des pores est dans la gamme de 4–10 nm avec une répartition volumique homogène.

- La compréhension actuelle des mécanismes de formation du Ge mésoporeux nous a permis de réaliser de nouvelles morphologies.

Le chapitre suivant sera consacré à l'analyse des propriétés structurales de Ge.

# CHAPITRE 4

# Structure et chimie des couches du germanium mesoporeux à l'échelle nanométrique

Dans ce chapitre nous présenterons l'analyse structurale détaillée par MET, la diffraction des électrons (SAED) et la spectroscopie µ-Raman des couches de Ge mésoporeux de deux types de morphologies « l'éponge » et « le sapin». La composition chimique de ces couches est étudiée par spectroscopies RBS et FTIR. L'ensemble de ces informations nous a permis de proposer une modèle de la structure des nanoparticules individuelles du Ge constituant les couches poreuses pour les deux types de morphologies. Ce modèle complète la théorie des mécanismes de nanostructuration du Ge décrite dans le chapitre 3.

## 4.1 Analyse structurale

### 4.1.1 TEM et SAED

La restriction majeure du MEB est liée à sa limite de résolution qui est autour de 2.5 nm. Il est impossible d'accéder à l'information concernant la structure fine des nanopores ou des nanoparticules constituant les couches du Ge mesoporeux. La microscopie TEM fournit une information pointue concernant la structure et la cristallinité à l'échelle nano de la zone observée de l'échantillon. Bien que cette technique est plus précise que le MEB, elle nécessite une préparation longue et minutieuse des échantillons ce qui limite son applicabilité de façon routinière. De plus, contrairement au MEB, cette technique est destructive.

L'image est complétée par l'analyse de diffraction du faisceau d'électrons sur des échantillons (SAED, angl. *Selected area electron diffraction*). Étant une technique d'analyse locale, la SAED permet d'extraire l'information concernant l'état cristallin de l'échantillon (ou d'une partie de celui-ci), de déterminer le paramètre de maille cristallin et donc le degré de contraintes dans la structure.

#### Analyse des structures de type « Éponge »

En tenant compte de l'isotropie de la structure de type « éponge », son orientation par rapport au faisceau d'électrons lors des observations TEM n'est pas si critique comme dans le cas d'une structure poreuse avec des pores cristallographiques. Suite à quoi, la préparation de ce type d'échantillons consiste à transférer une partie de la couche poreuse sur une grille de TEM sous forme des flocons poreux (Figure 4.1 (a)). Comme nous pouvons le constater, la structure mésoporeuse est constituée de nombreuses nanoparticules quasi-sphériques interconnectées entre elles (Figure 4.1 (b)), ce qui est très caractéristique d'autres semi-conducteurs poreux, tels que le Si poreux, par exemple. Il est à noter néanmoins que le degré d'interconnections reste faible : les nanoparticules voisines sont interconnectées par de petites isthmes de 2.5–4 nm larges (Figure 4.1 (d)).

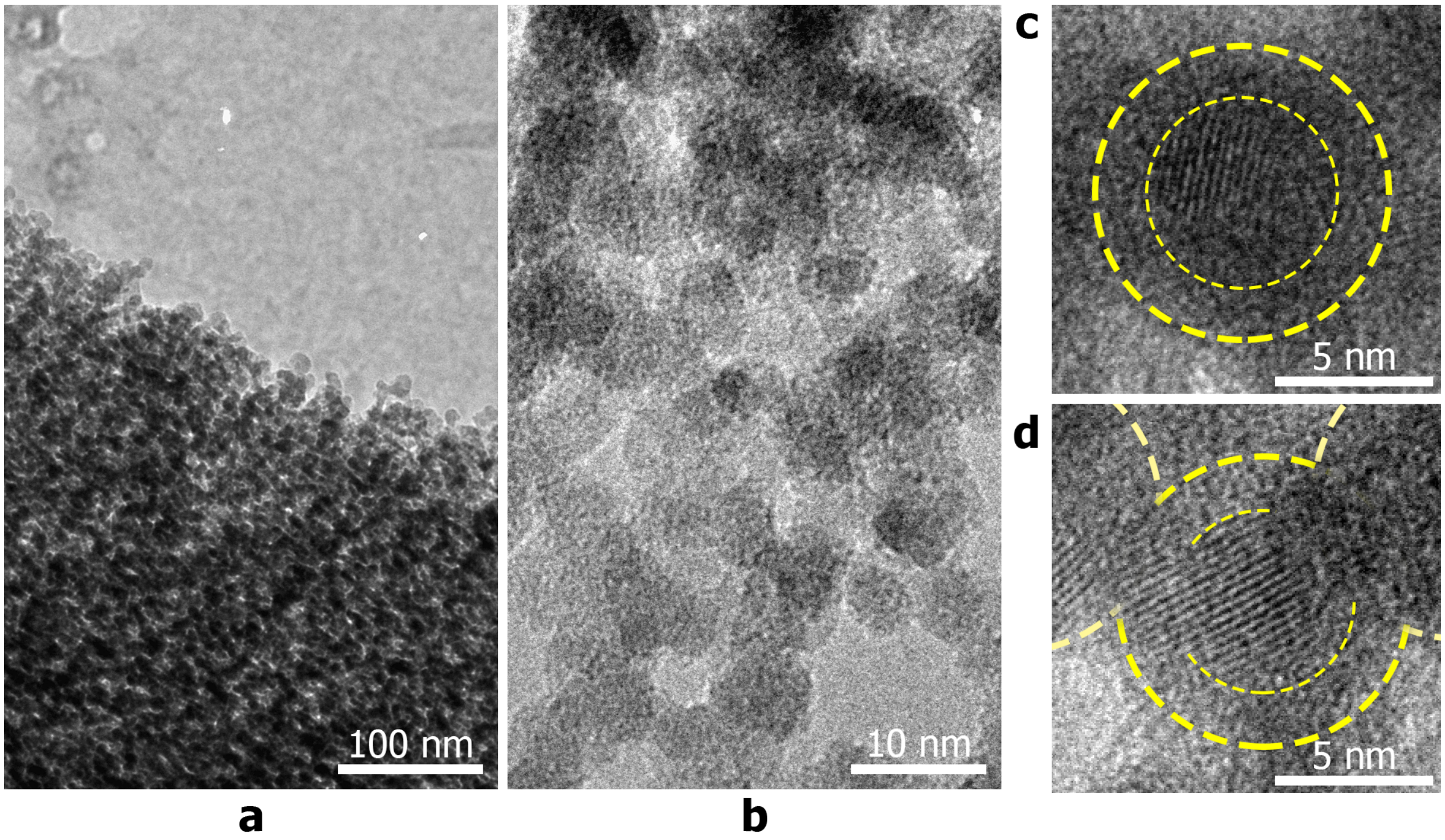

Figure 4.1 Images TEM du champ clair d'un flocon du Ge mesoporeux en éponge ($HF_{49\%}$, 2 $mA/cm^2$, 1 s/3 s, 2 h) à faible(a), moyen(b) et fort(c) grossissement. Des cristallites sont cerclés avec une ligne jaune.

La nanostructure des nanoparticules individuelles de Ge formant la couche mésoporeuse est présentée aux Figures 4.1 (c) et (d). Comme nous pouvons le voir chaque nanoparticule est composée d'un noyau cristallin entouré d'une coquille amorphe. En utilisant les autres techniques de caractérisation (comme il sera présenté dans les paragraphes suivants), sa composition chimique a été trouvée être le Ge amorphe oxydé. Les lignes pointillées, utilisées comme guide visuel indiquent les limites du noyau et de sa coquille. Le diamètre moyen du cœur cristallin mesuré à partir des images TEM (c) et (d) est de 6–7 nm ; l'épaisseur moyenne de la coquille est de 1–1.5 nm, ce qui indique une nanoparticule de Ge mesurant 8.5–9.5 nm en moyenne. Il est à noter que l'épaisseur de la coquille reste quasi-constante quelle que soit la taille du noyau cristallin.

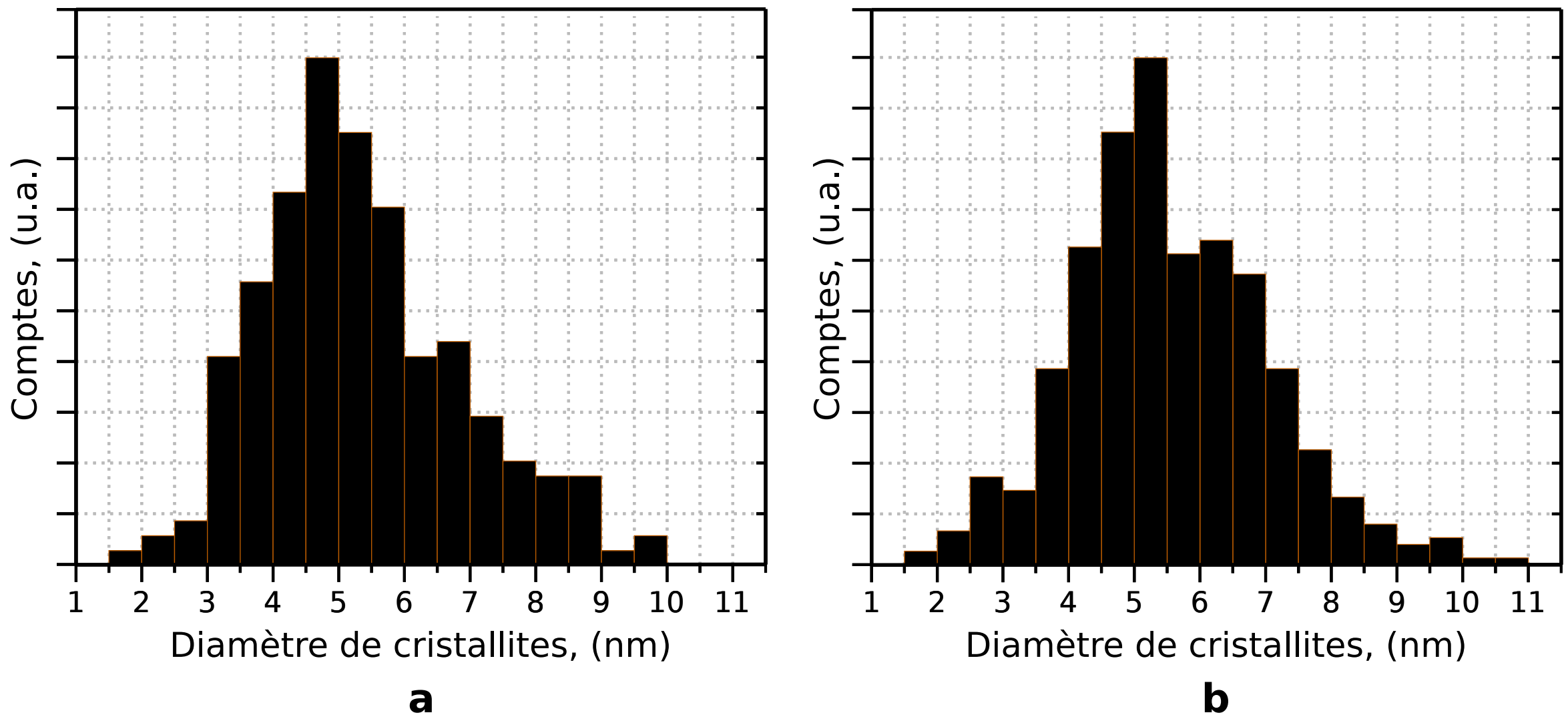

Figure 4.2 Distribution de cristallites par taille de la couche de PGe en éponge ($HF_{49\%}$, 2 mA/cm$^2$, 1 s/3 s, 2 h), calculé à l'aide d'observations TEM(a) et MEB(b)

Les distributions en taille des cristallites mesurées avec le logiciel Gatan Digital Micrograph™ sur la base des images MET et MEB, sont comparées sur les Figures 4.2 (a) et (b), respectivement. Les deux distributions sont cohérentes. En effet, le type de chaque distribution est une gaussienne dont le pic est centré sur 5 nm. L'apparition du deuxième pic large autour de 6.5 nm est due au fait que certains cristallites sont connectés par des isthmes dont la largeur est comparable à celle des cristallites voisins, qui sont désormais considérés comme un seul. Le diamètre moyen des cristallites estimé par le TEM est $5.1 \pm 1.2$ nm.

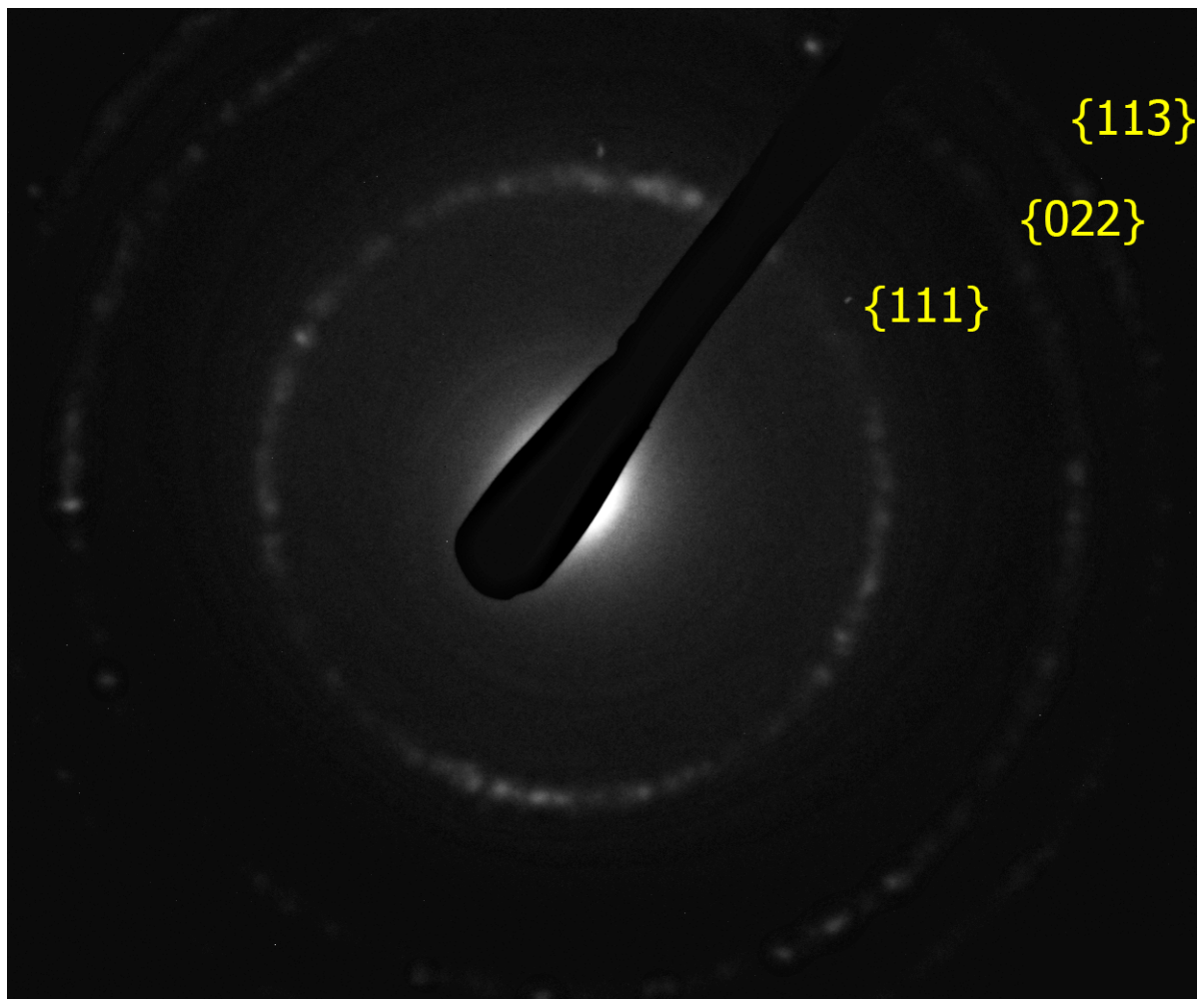

Figure 4.3 Cliché de diffraction d'un flocon de la couche de PGe en éponge ($HF_{49\%}$, 2 mA/cm$^2$, 1 s/3 s, 2 h)

Le cliché de diffraction de la couche du PGe en éponge est présenté à la Figure 4.3. L'apparition de plusieurs points de diffraction organisés en anneaux indique la nature polycristalline de l'échantillon. La polycristallinité du PGe du type éponge peut être expliquée par l'existence d'une coquille amorphe oxydée sur la surface des nanocristallites de Ge. En effet, la contrainte mécanique importante provenant de cette couche amorphe conjuguée à la fragilité de la structure du Ge poreux, provoquée par le faible niveau des interconnections entre les nanoparticules de Ge, entrainent une désorientation des nanoparticules du Ge l'une par rapport à l'autre. Ainsi, par cette désorientation, la relaxation de la contrainte totale a lieu.

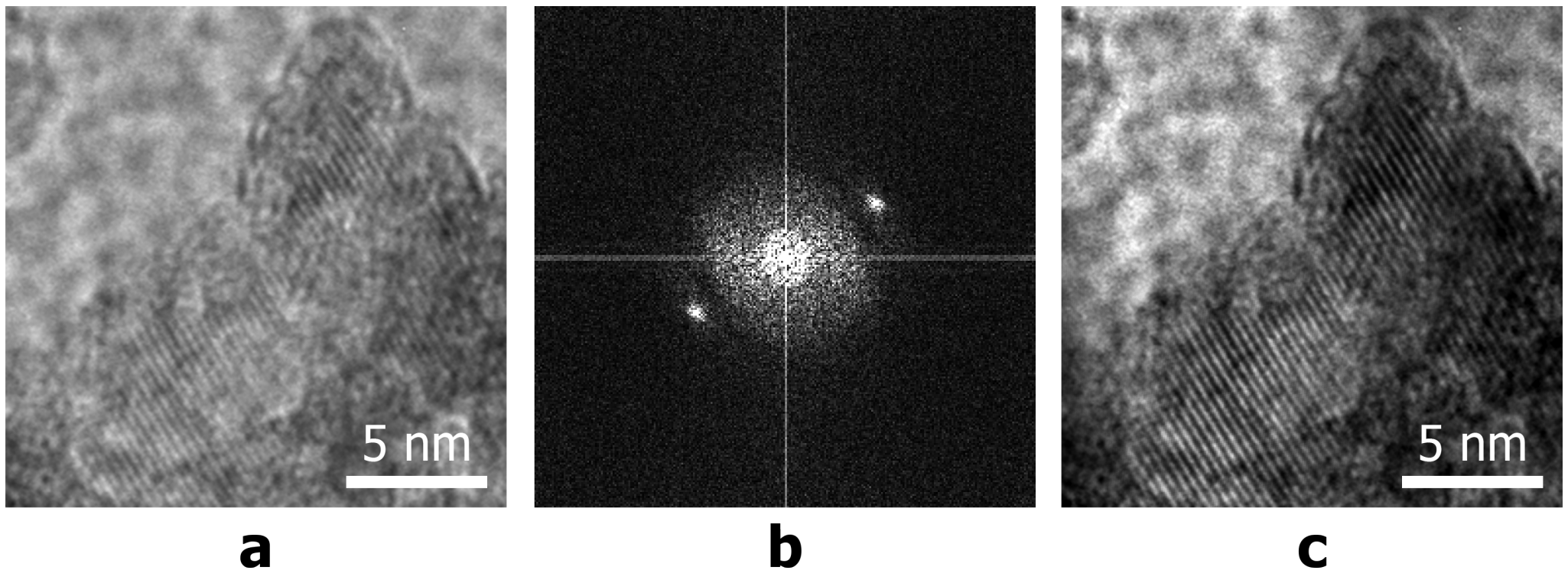

Figure 4.4 (a) Image TEM de haute résolution d'un cristallite de Ge mesoporeux en éponge ; (b) transformation de Fourier de (a) ; (c) image reconstruite par une transformation inverse de Fourier de (b) après une filtration du bruit et amplification des composants de basse fréquence.

Une autre image TEM de haute résolution du PGe en éponge illustrant la structure des nano-cristallites est montrée sur la Figure 4.4 (a). Comme nous pouvons le voir, la couche amorphe est présente partout sur la surface interne du PGe. Afin d'estimer la contrainte mécanique existante dans la structure poreuse, une transformation de Fourier de l'image TEM à haute résolution d'un cristalline a été effectuée (Figure 4.4 (b)). La tache principale et les deux taches symétriques satellites correspondent à une structure périodique — les rangées des atomes. La fréquence de cette structure périodique est $3.4 \pm 0.1$ Å. Les rangées atomiques sont alors identifiées comme les plans (111) dont l'espacement dans un cristal volumique est $5.646/\sqrt{3} = 3.259$ Å. Ainsi, la déformation de la structure est estimée à 3%. Cependant, cette dernière valeur est à considérer avec prudence, parce que le valeur de référence tombe dans la limite inférieure du domaine d'incertitude. Finalement, une image de cristallite améliorée (Figure 4.4 (c)) peut être reconstruite par multiplication de l'image 4.4 (a) avec le résultat de transformation de Fourier inverse de 4.4 (b), après une filtration des composants de haute fréquence.

### Analyse des structures de type « Sapin »

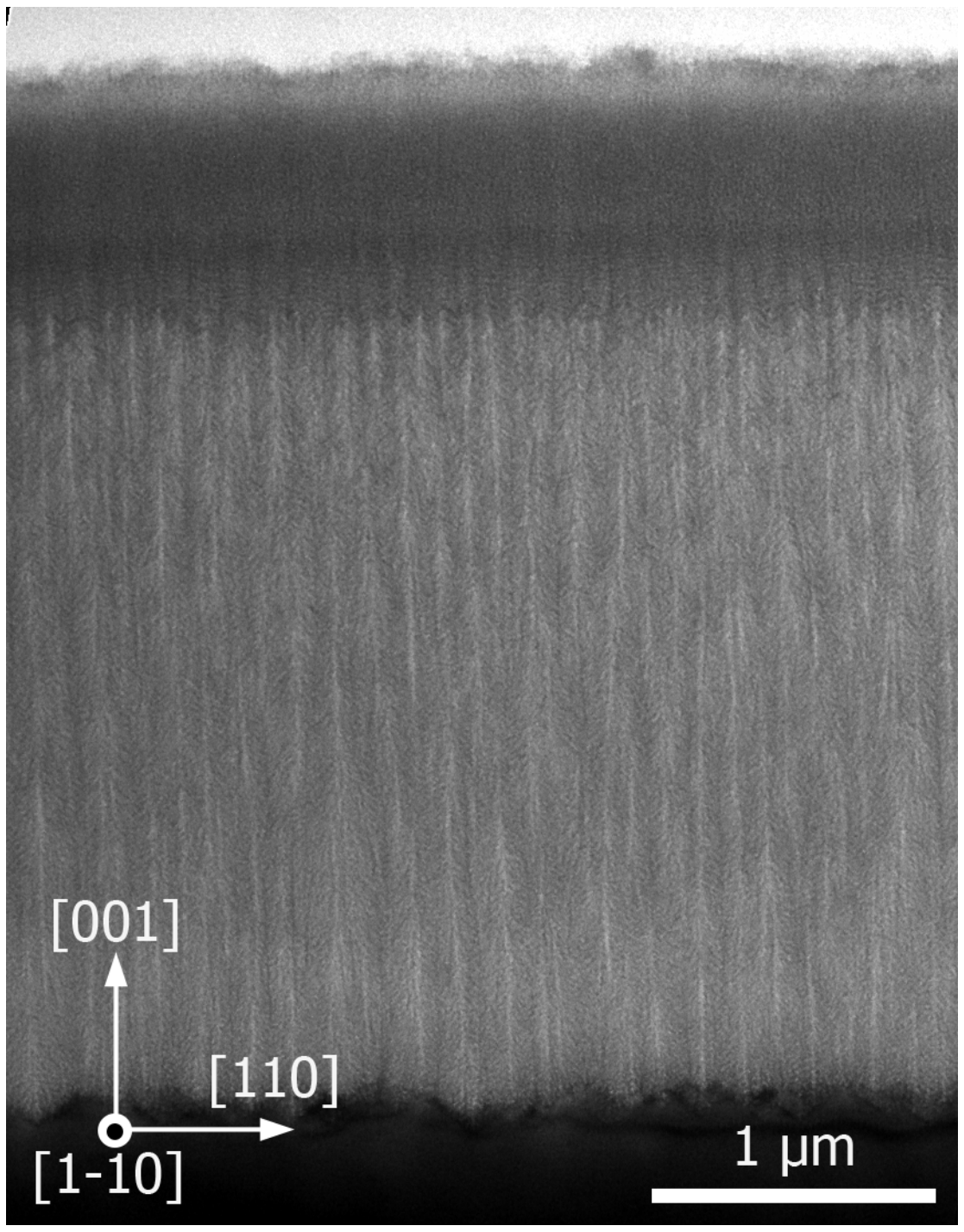

Figure 4.5 Image TEM du champ clair en coupe transversale de la couche de Ge mesoporeux en « arbres » ($HF_{49\%}$, 2 $mA/cm^2$, 0.5 s/0.25 s, 2 h) à faible grandissement

L'image TEM à faible grossissement (Figure 4.5) de la couche de PGe en « arbres » ressemble beaucoup à celle obtenue par le MEB (Figure 3.11) (les pores sont imagés clairs et le Ge cristallin est noir). Des lignes claires représentent le réseau de pores primaires parallèle à la direction [001] avec une périodicité de l'ordre de 100–150 nm. Plusieurs couches sont visibles : une première couche poreuse fortement amorphisée de 500–600 nm, une seconde couche poreuse plus cristalline de 3000 nm avec une couche intermédiaire entre ces deux première couches de 300 nm. Les observations MEB avaient laissé un doute, à savoir si la couche supérieure consiste en une couche superficielle amorphe ou cristalline fortement oxydée. Ici, la couche amorphe est beaucoup mieux définie par rapport à celle des images MEB ; sa nature amorphe est aussi confirmée par la prise du cliché de diffraction (Figure 4.6).

La Figure 4.7 (a) démontre une vue plus détaillée sur l'interface entre le Ge mesoporeux en arbres et le Ge volumique. L'alignement du système permet de confirmer la géométrie de la structure : Les pores secondaires se propagent selon les directions de type [11-1]. Leur périodicité est de l'ordre de 10-15 nm. Nous constatons que le front de gravure n'est pas plan comme dans le cas du silicium mais rugueux avec apparition de facettes parallèles à

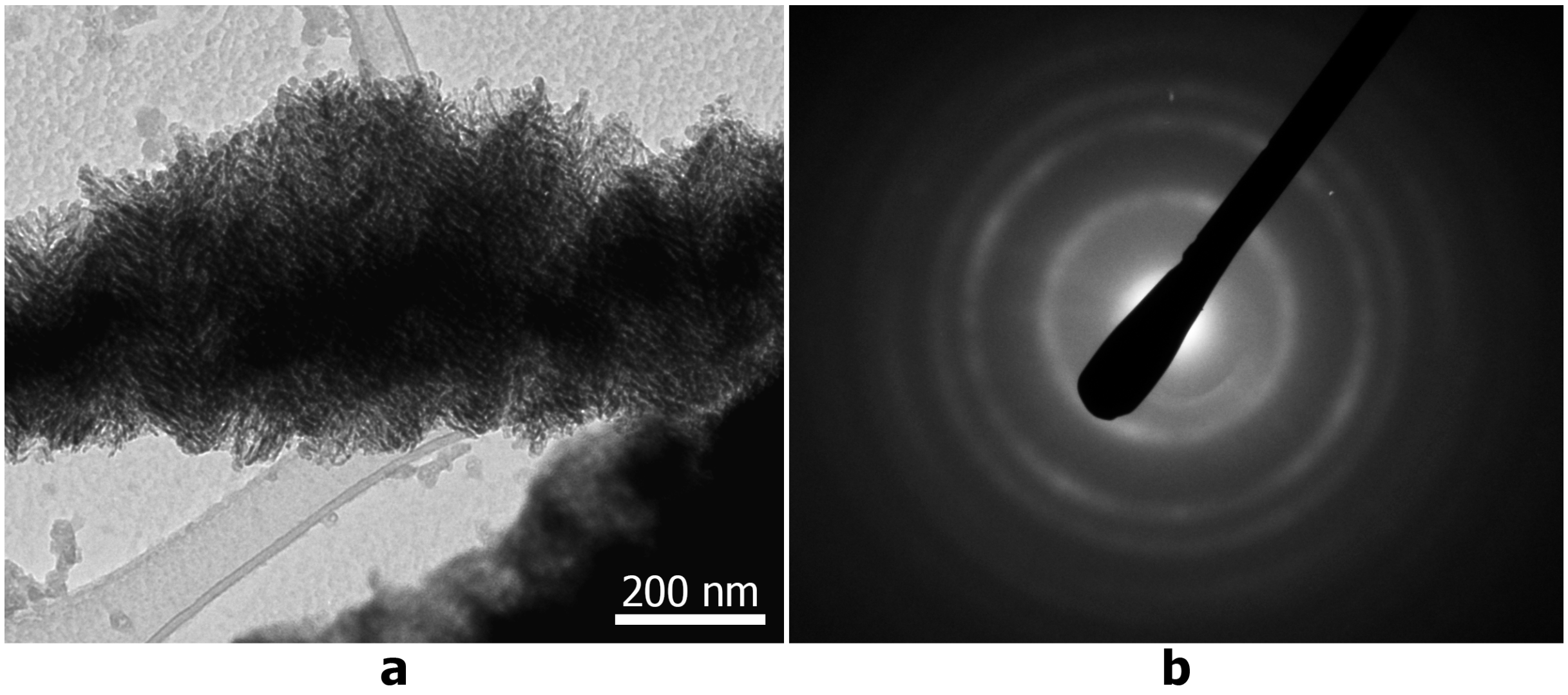

Figure 4.6 (a) Image TEM à faible grandissement d'un flocon du Ge mesoporeux en arbres, posé directement sur le grille de TEM ;
(b) cliché de diffraction du flocon poreux montré sur (a), soit une superposition des anneaux (représentation de la partie cristalline) et une tache floue et continue (représentation de la couche amorphe superficielle)

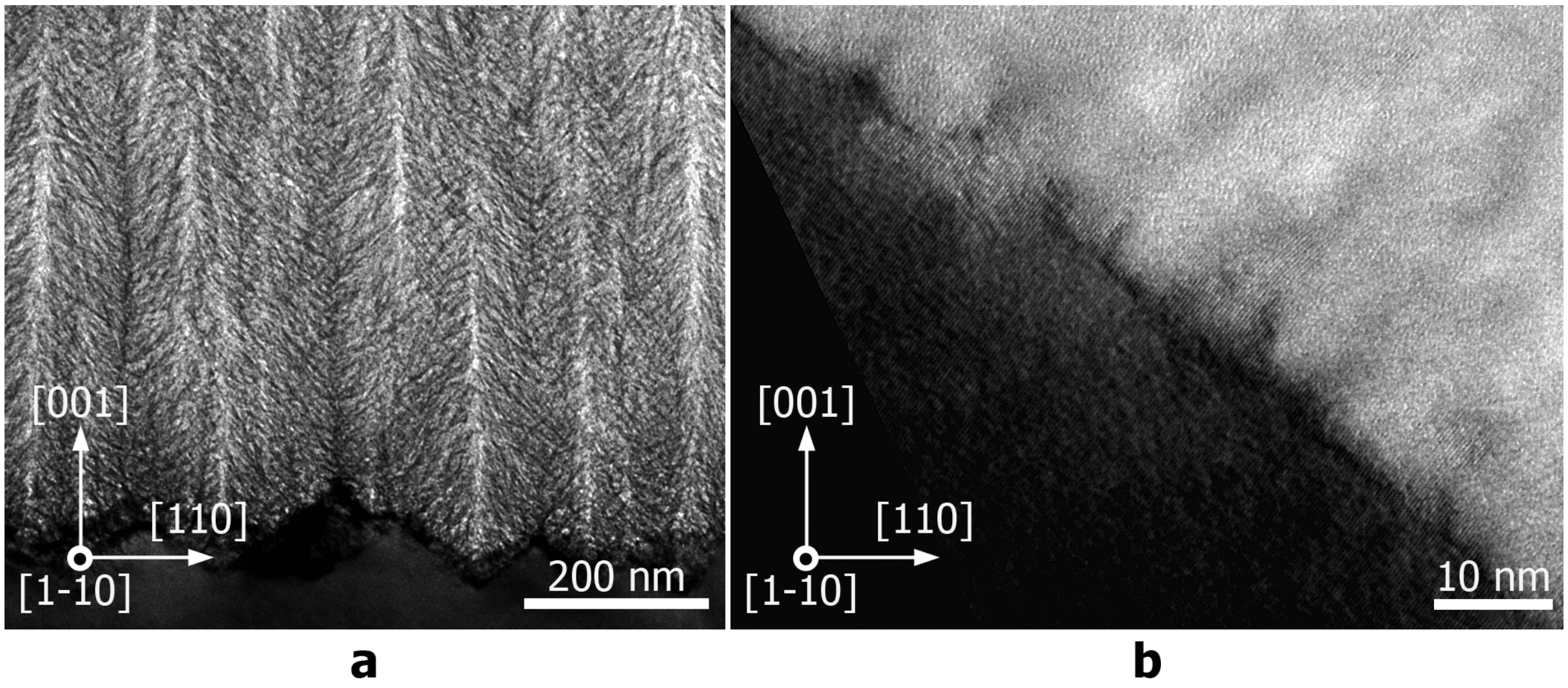

Figure 4.7 Image TEM du champ clair en coupe transversale de l'interface entre la couche de Ge mesoporeux en « arbres » ($HF_{49\%}$, 2 mA/$cm^2$, 0.5 s/0.25 s, 2 h) et le Ge volumique à moyen (a) et à fort (b) grandissement

la famille de plans compacts {111}. La zone interfaciale entre les deux couches correspond en fait à la superposition vue par le faisceau de la couche poreuse et du substrat. On peut en déduire la forme présente autour d'une pore primaire comme étant une obélisque dont la taille de la base carrée est limitée par la présence des autres pores primaires avec une « tête » pyramidale à base carrée dont les facettes sont des plans de type [101]. Les côtés de la base carrée sont parallèles aux directions de type [100]. Les facettes orientées selon {111} résultent de l'intersection des faces [101] de la pyramide avec la lame mince taillée dans le plan [1-10].

Une image de front de gravure à très haute résolution (dont les atomes de Ge peuvent être distingués) est montrée sur la Figure 4.7 (b) On peut noter la continuité des rangées atomiques à l'interface du Ge volumique et le squelette cristallin de la couche de PGe. Les branches latérales des pores ($\approx 10\ nm$ larges) suivent les directions <111> mais ne sont pas parfaitement droites. Leurs croisements et ramifications conduisent au fait que tout espacement entre deux branches n'est pas occupé par un seul cristallite. Les cristallites ainsi formés sont larges de 4–5 nm et longs de 10–20 nm, ce qui est une précision importante par rapport aux valeurs mesurées par le MEB.

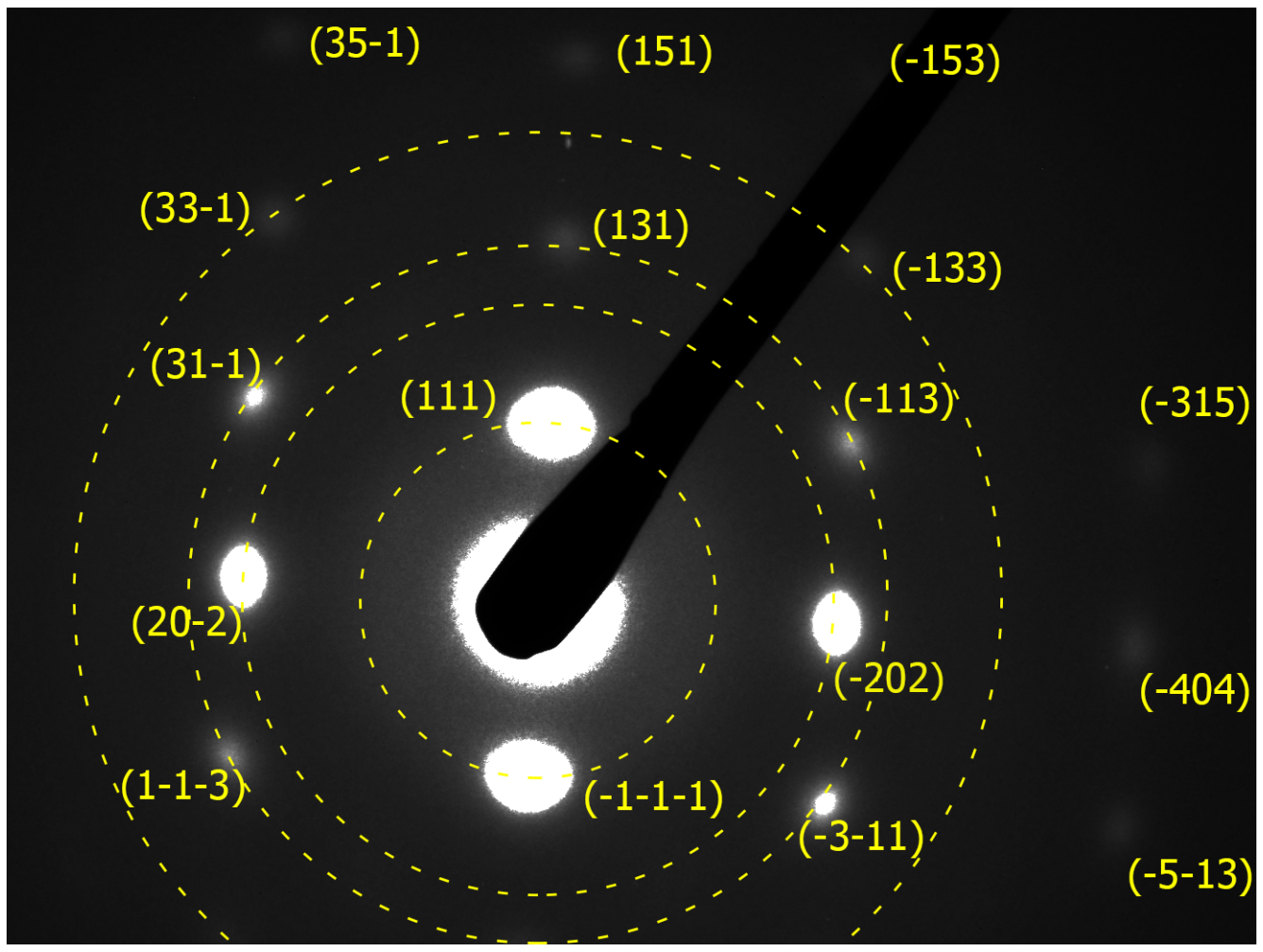

Figure 4.8 Cliché de diffraction d'un cristallite de la couche de PGe en arbre ($HF_{49\%}$, 2 $mA/cm^2$, 0.5 s/0.25 s, 2 h)

La Figure 4.8 présente une cliché de diffraction d'une couche de PGe en arbre. Le cliché est similaire à celui d'un mono-cristal de Ge. Les positions des taches correspondent à ceux du Tableau 2.4, ce qui permet de conclure à une préservation de l'ordre cristalline à grande distance dans la couche poreuse. Malheureusement, la grandeur importante des taches ne permet pas de déterminer les distances interplanétaires afin d'analyser les contraintes dans la structure.

### 4.1.2 µ-Raman

Étant très sensible à l'ordre à courte distance, la spectroscopie µ-Raman est une méthode complémentaire à la HR-TEM permettant de caractériser de façon non-destructive l'état structural des solides.

#### Ge monocristallin

Le spectre Raman du Ge monocristallin (100) est présenté à la Figure 4.9. Dans la configuration des mesures en rétrodiffusion, ce spectre correspond aux vibrations des phonons LO (optique longitudinale). Il est connu que le profil d'une raie de vibration d'un solide est une fonction de Lorentz symétrique autour de la fréquence de vibration $\omega_0$, qui se situe vers le nombre d'onde de 300 cm$^{-1}$ pour le Ge monocristallin (équation 4.1).

$$I(w) \propto \frac{1}{(w - w_0)^2 + (\Gamma_0/2)^2} \tag{4.1}$$

où $\Gamma_0$ est le paramètre d'élargissement.
Cet élargissement est principalement dû à l'amortissement naturel des phonons. Pour nos conditions expérimentales, la largeur à mi-hauteur $\Gamma_0$ du Ge monocristallin obtenue à partir du profil simulé (trait plein sur la Figure 4.9), est de 5 cm$^{-1}$.

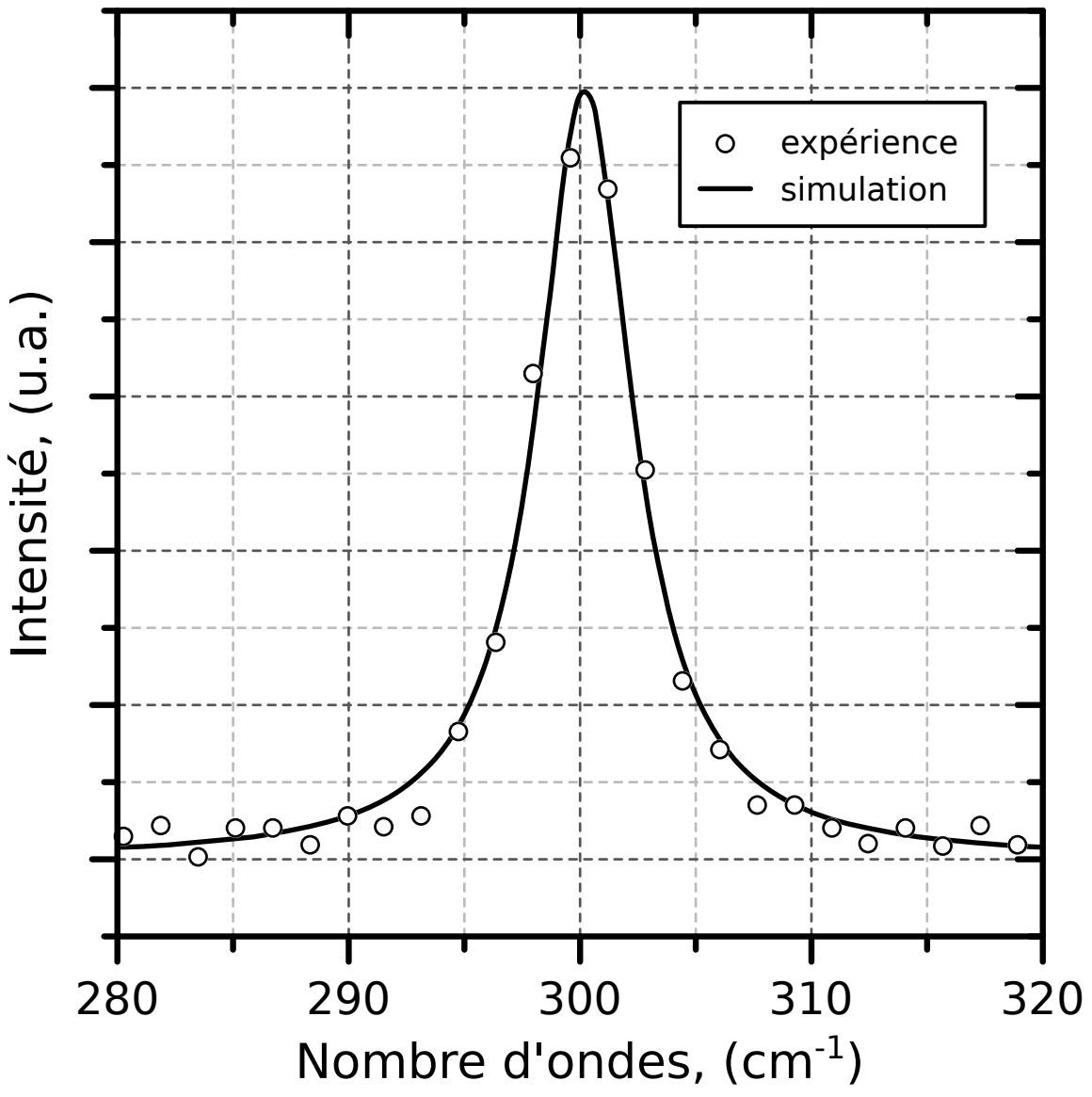

Figure 4.9 Spectre de diffusion Raman du Ge monocristallin

### Ge poreux de type « éponge »

Le spectre Raman de la couche de Ge poreux de type « éponge » est présenté à la Figure 4.10. La position du pic Raman du Ge monocristallin est également reportée sur la même figure par la ligne pointillée. On observe une claire distinction de ce spectre par rapport à celui-ci du Ge monocristallin présenté sur la Figure 4.9. Il est fortement asymétrique, plus large et la position de son pic principal est décalée vers les nombres d'ondes plus courts. Afin d'extraire l'information quantitative, le modèle du confinement des phonons dans les nanocristallites sphériques a été appliqué. En effet, comme il a été présenté dans le paragraphe précédent, la structure du Ge mesoporeux de type « éponge » est constituée de nombreuses nanoparticules quasi-sphériques interconnectées entre elles. Dans ce cas, selon le modèle du confinement des phonons (section 2.2.3), l'intensité de diffusion Raman est décrite par une superposition continue des courbes lorentziennes centrées sur le nombre d'ondes de la courbe de dispersion des phonons $\omega(q)$ et pondérées par le facteur $|C(\vec{q_0}, \vec{q})|^2$ (Equation 2.12).

La courbe simulée est présentée sur la Figure 4.10 (b) (courbe bleue). Dans cette simulation on utilise la distribution des particules par taille de type gaussienne qui a été mesurée préalablement par le TEM. Comme nous pouvons le voir, seule la position du pic principal et la partie des hautes nombres d'ondes ($> 290$ cm$^{-1}$) sont bien ajustées. Il est important de rappeler, que ce modèle décrit uniquement la phase cristalline du matériau. L'apparition de l'épaulement autours de 280 cm$^{-1}$ ne peut pas être décrite par ce modèle. Afin d'analyser cette bande de diffusion, le spectre simulé (courbe bleue) a été soustraite du spectre initial (courbe noire). Le spectre résultant (courbe rouge), dont le centre est situé autour de 280 cm$^{-1}$, est également asymétriquement allongé vers le rouge. Cette bande a été attribuée à la phase du Ge amorphe (a-Ge).

Le choix de la fonction d'ajustement pour le Ge amorphe n'est pas si trivial comme dans le cas du Si amorphe, pour lequel le spectre est modélisé par la courbe de type gaussien centrée sur la fréquence du mode TO [11, 55]. Le spectre du Ge amorphe est plus complexe. Il est composé de la bande TO centrée sur le nombre d'ondes de 280 cm$^{-1}$ avec une largeur à mi-hauteur de $\approx 30$ cm$^{-1}$ [102, 103]. La bande LO, dont le pic principal est situé autour de 240 cm$^{-1}$, est contigüe à gauche, ce qui résulte en un spectre final large et asymétrique pour le Ge amorphe [117]. Les positions des bandes TO et LO varient dans la gamme de $\pm 10$ cm$^{-1}$ en fonction de la méthode utilisée pour la formation de la phase amorphe. Dans notre cas, le choix du a-Ge de type électrolytique [117] est le plus adapté. Son spectre peut être déconvolué en utilisant deux courbes de type gaussien, centrées autour de 279 cm$^{-1}$ et 246 cm$^{-1}$ (Figure 12). Il est important de noter que la large bande autour de 246 cm$^{-1}$

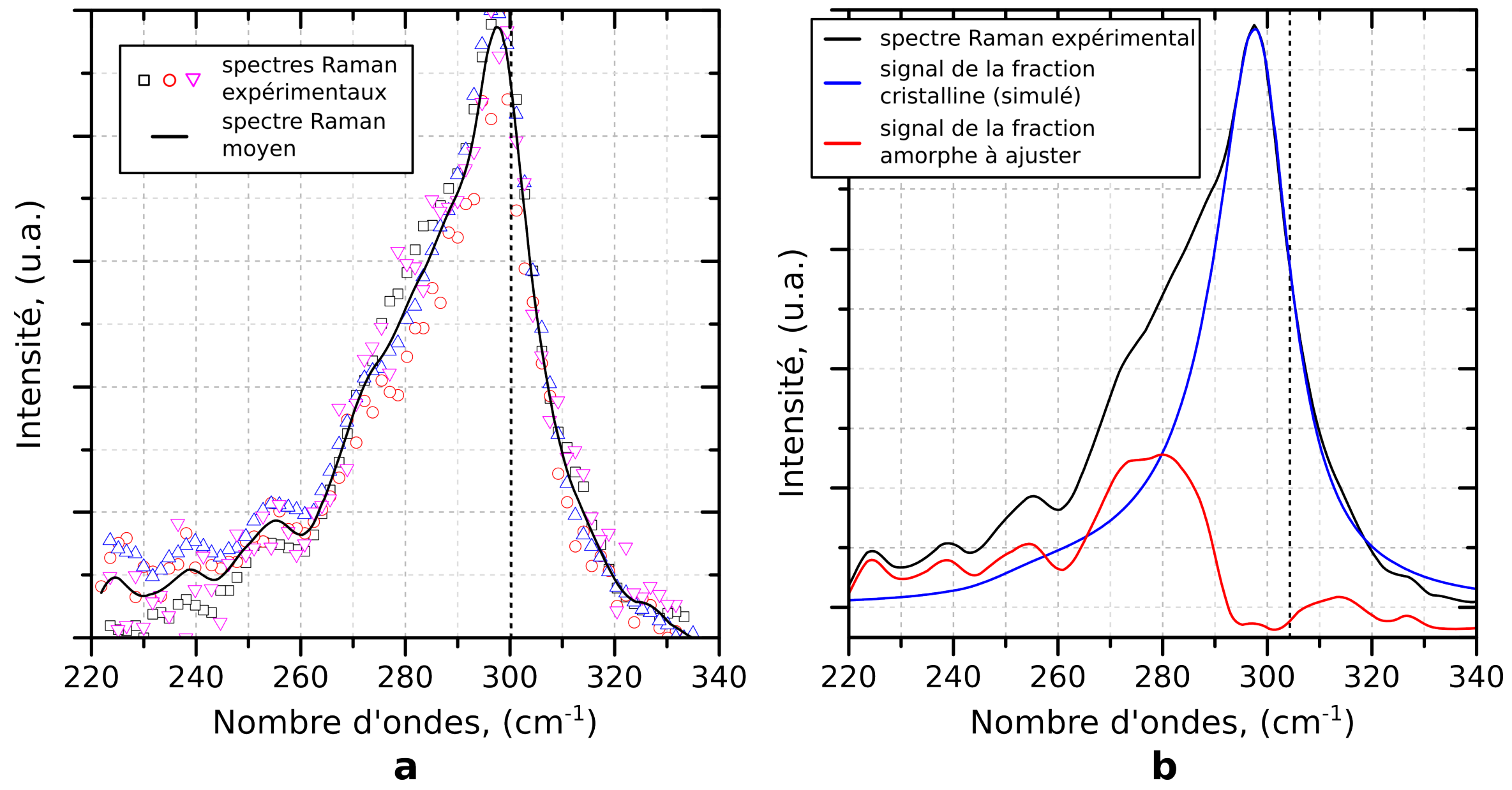

Figure 4.10 Spectre de diffusion Raman de la couche de PGe en éponge ($HF_{49\%}$, 2 mA/cm$^2$, 1 s/1 s, 2 h) (a) et sa modélisation selon le modèle de confinement des phonons (b)

peut être également attribuée au sous-oxyde du Ge ($GeO_x$) [102]. En tenant compte de la superposition des signaux du mode TO du Ge amorphe et du $GeO_x$, il parait assez difficile de les distinguer en utilisant uniquement la technique de la spectroscopie Raman.

L'existence de deux phases de Ge (amorphe et cristallin), révélées par la technique de la diffusion Raman, se trouve en cohérence avec les observations TEM. Le diamètre moyen du noyau cristallin obtenu à partir du modèle du confinement des phonons est de $2.65 \pm 0.7$ nm, ce qui est en bon accord avec la valeur mesurée directement par HRTEM $3.1 \pm 0.5$ nm

Il est à noter que le paramètre d'élargissement $\Gamma_0$ a été également ajusté dans la modélisation du spectre de la phase cristalline. La valeur trouvée est égale à 14, ce qui est environ trois fois supérieurs à la valeur de $\Gamma_0$ du Ge monocristallin. En effet, en tenant compte du fait que ce paramètre décrit l'amortissement des phonons, sa dépendance face à la taille des nanostructures est évidente : plus la taille des nanoparticules est petite, plus l'amortissement est important et, par conséquence, plus la valeur du $\Gamma_0$ est importante.

La technique de spectroscopie µ-Raman permet également de quantifier les fractions des phases amorphes et cristallines présentent dans l'échantillon. En intégrant l'aire sous la courbe attribuée au Ge amorphe (zones hachurées rouge et verte), il a été trouvé que 33% du volume de l'échantillon est amorphe. Le reste 67% est cristallin.

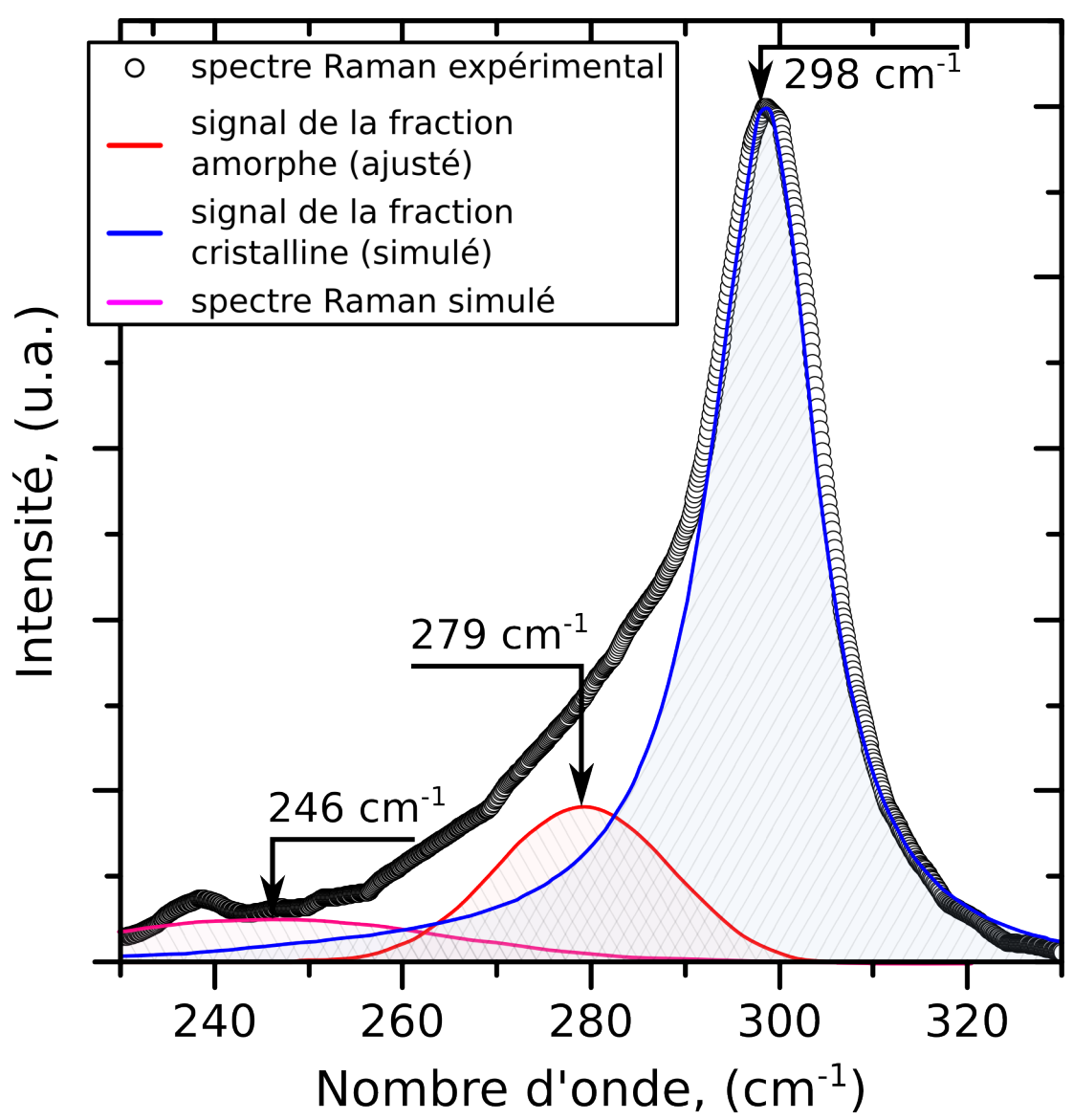

Figure 4.11 Déconvolution du spectre Raman de la couche du PGe en éponge ($HF_{49\%}$, 2 mA/cm$^2$, 1 s/1 s, 2 h). Les zones hachurées bleue et rouge correspondent aux signaux des cristallites et de la fraction amorphe, respectivement. La ligne en pointillées indique la position du pic Raman du Ge cristallin.

### Ge poreux de type « sapin »

Le modèle de confinement de phonons est aussi applicable aux structures qui sont constituées des nanoparticules oblongues [1] (ce qui est le cas des couches de Ge mesoporeux de type « sapin »).

Le spectre Raman de la couche du Ge poreux de type sapin (les points expérimentaux et la courbe moyenne) est présenté à la Figure 4.12 (a). En comparant ce spectre avec celui-ci du PGe de type « éponge », deux particularités peuvent être remarquées. Tout d'abord le spectre du PGe de type « sapin » est plus large. Deuxièmement, la position de son pic principal est encore décalée plus vers les énergies plus faibles. L'influence simultanée de la taille des nanoparticules du Ge constituant cette couche et la fraction du Ge amorphe dans cet échantillon peuvent expliquer les deux effets observés.

La déconvolution de ce spectre selon le protocole développé pour le PGe de type « éponge » a été effectuée (Figure 4.12 (a)). Comme nous pouvons le voir, la position du pic de la phase cristalline (zone hachurée bleue) est autour de 297 cm$^{-1}$ ce qui correspond à la taille moyenne du noyau cristallin de $2.0 \pm 0.4$ nm. De plus, la valeur de $\Gamma_0$ a été trouvée être égale à 27, ce qui indique également que la taille du noyau cristallin du PGe de type « sapin » est plus petite par rapport à celle du PGe de type « l'éponge» . En ce qui concerne la fraction du Ge amorphe (zone hachurée rouge) dans cette couche, elle est aussi plus importante par rapport à celle du PGe de type éponge. Elle constitue environs 60% du volume totale. Il est à noter également que la phase résiduelle amorphe (courbe rouge) est asymétrique, comme dans le cas du PGe de type « éponge » et peut être ajustée par deux courbes de type gaussien, centrées à 278 $cm^{-1}$ et 250 $cm^{-1}$.

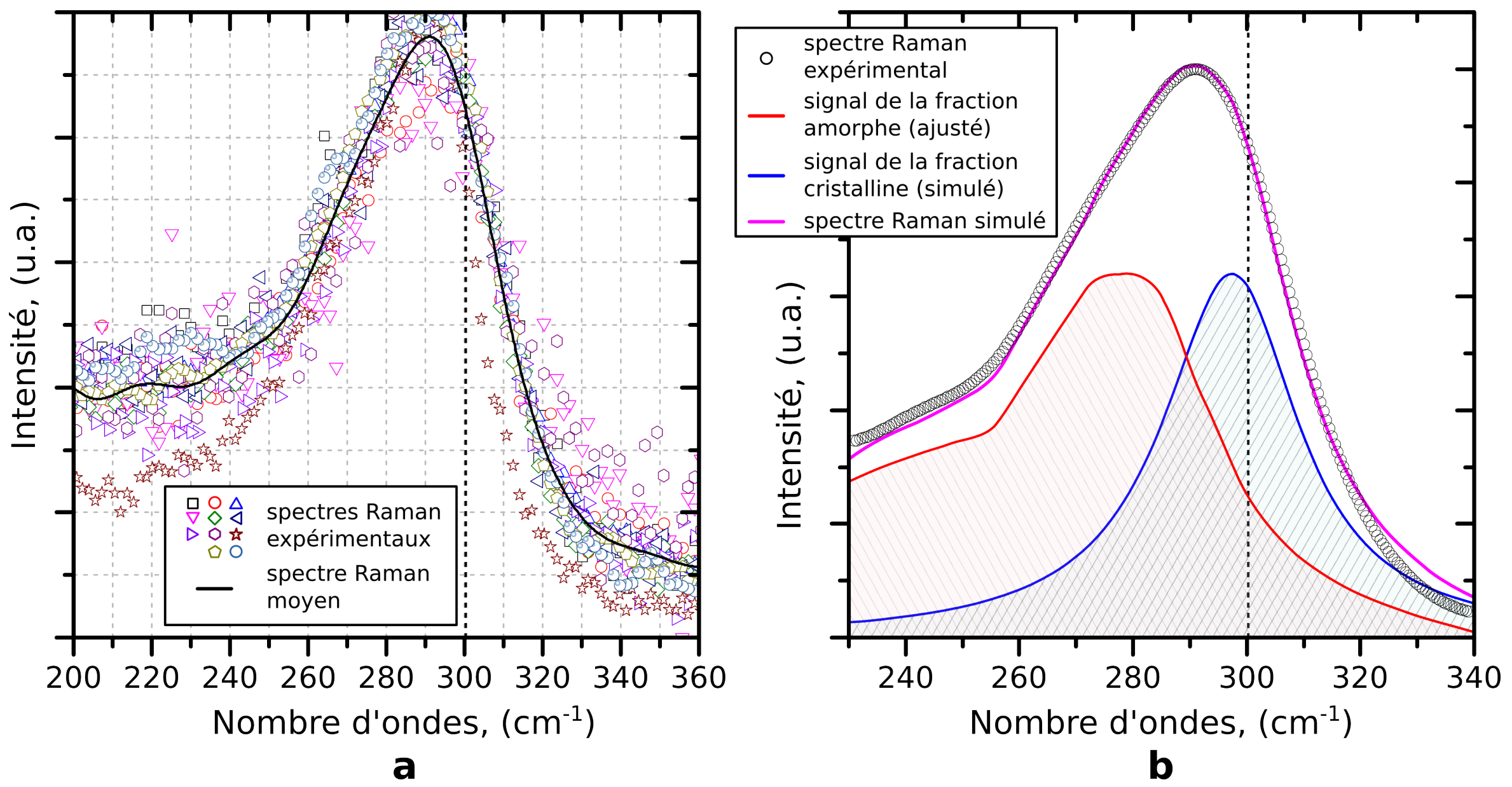

Figure 4.12 Spectre de diffusion Raman de la couche de PGe de type « arbres » ($HF_{49\%}$, 2 $mA/cm^2$, 0.5 s/0.25 s, 2 h) (a) et sa modélisation selon le modèle de confinement des phonons (b). Les zones hachurées bleue et rouge correspondent aux signaux des cristallites et de la fraction amorphe, respectivement. La ligne en pointillés indique la position du pic Raman du Ge cristallin.

## Discussion

Deux mécanismes d'apparition de la couche de Ge amorphe sur la surface interne du PGe peuvent être suggérés. Le premier est lié à l'effet de la reconstruction de la surface des nanoparticules du Ge induite par le désordre [47, 102]. L'existence de la coquille amorphe sur la surface des nanocristaux de Ge libres a été observée par Williamson et al [119] qui ont corrélé les résultats de photoémission de ces nanocristaux avec des résultats de calculs théoriques. Ils ont démontré que la densité des états observée dans la bande de valence peut être expliquée par une structure de type diamant désorientée qui se forme par la reconstruction des liaisons pendantes de la surface. En outre, l'apparition de la bande amorphe dans les spectres Raman des nanocristaux libres de Ge, Si et GaP formés par la technique d'évaporation, a été récemment observée [47]. Sa nature a été attribuée à une coquille amorphe entourant le noyau cristallin [47]. De plus, les spectres Raman de ces nanocristaux intégrés dans une matrice et investigués par les mêmes auteurs [37, 38] ne présentaient pas la bande liée avec la phase amorphe. Cette différence a été attribuée aux conditions de vibration différentes des atomes près de la surface dans ces deux cas. En effet, la présence de la matrice stabilise les atomes de la surface et réduit leur contribution à l'intensité Raman. Le deuxième mécanisme supplémentaire qui peut être responsable de la formation de la phase amorphe est la dissolution chimique du Ge poreux dans les électrolytes à base d'HF. Ce phénomène a été discuté dans le Chapitre 3.

## 4.2 Analyse chimique

### 4.2.1 FTIR

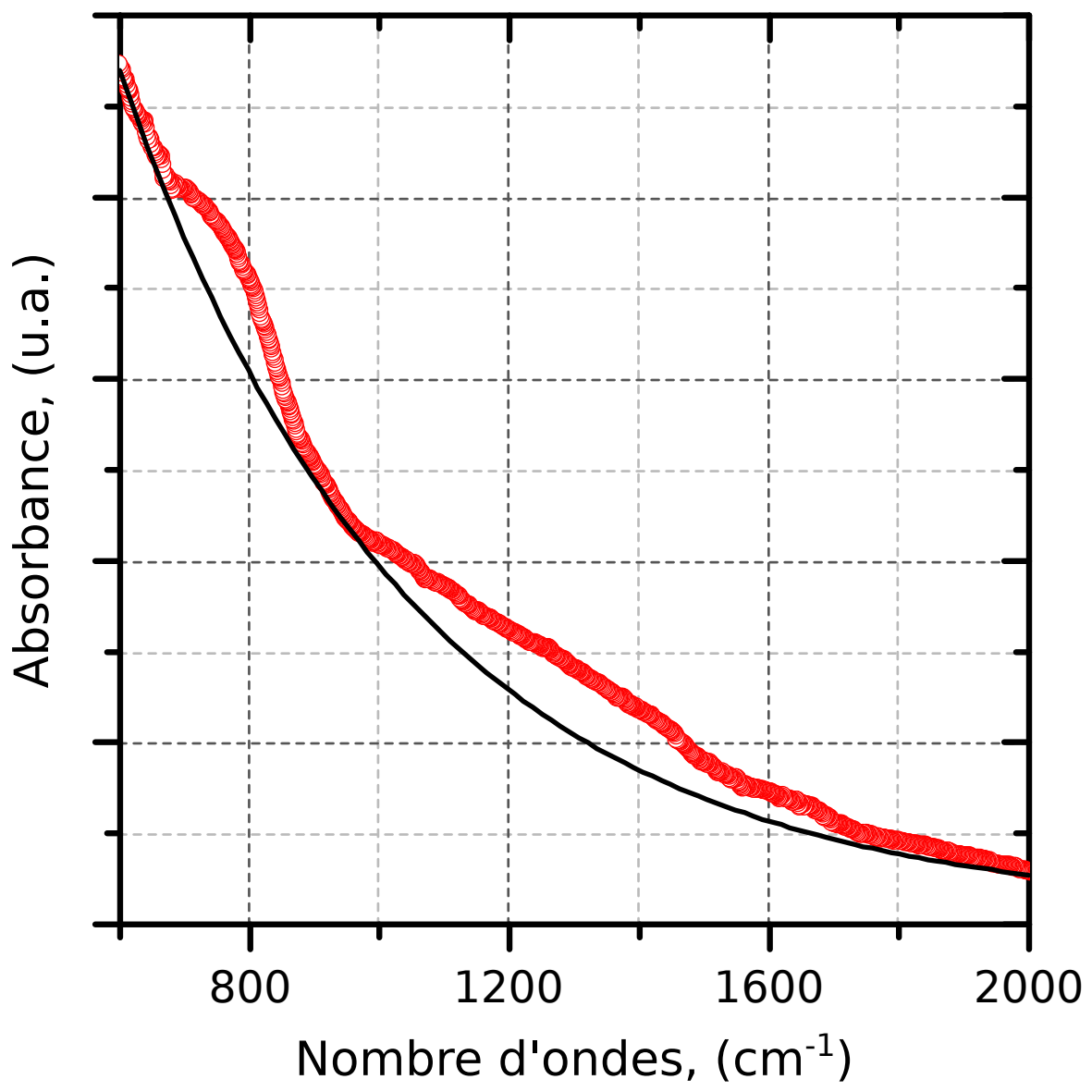

Figure 4.13 Modélisation du spectre expérimental d'absorbance de la couche du PGe (trait rouge) par le modèle de Drude [110]

La présence du sous-oxyde de Ge dans les couches poreuses de deux morphologies a déjà été suggérée d'après les analyses des spectres Raman (bande à 246 $cm^{-1}$). Afin de confirmer ou d'infirmer cette hypothèse l'analyse FTIR a été effectuée. Cette analyse permet de décrire qualitativement des groupes chimiques présentes sur la surface interne du PGe. Les mesures ont été réalisées en mode de mesure de réflexion totale atténuée (ATR). Le spectre typique d'absorption IR expérimental de la couche de PGe est indiqué sur la Figure 4.13 (trait rouge). La forte augmentation quadratique de l'absorbance lorsque la fréquence diminue est attribuée à l'absorption IR par des porteurs de charge libres (trous). Afin de révéler les caractéristiques de vibration des liaisons chimiques, ce signal superposé a été modélisé par le modèle de Drude selon l'approche publié par Timoshenko et al. [110]. Cette modélisation consiste d'abord à ajuster le spectre par la fonction quadratique et ensuite, à soustraire la courbe modélisée du spectre expérimental IR initial. Les spectres résultants qui correspondent à l'absorption des liaisons chimiques des couches poreuses de deux types de morphologies : (a) sapin et (b) éponge sont présenté à la Figure 4.14. Comme nous pouvons le voir dans les deux cas, la bande la plus intense du spectre se situe à 780 $cm^{-1}$ et peut être attribuée au mode d'étirement Ge−O du sous-oxyde superficiel de Ge ($GeO_x$, $x < 2$)[93]. Ainsi, l'hypothèse de la présence de la phase oxydée dans les couches du Ge poreux se confirme. De plus, une bande de faible intensité située entre 830 et 950 $cm^{-1}$ présente dans chaque morphologie, correspond au dioxyde de Ge : $GeO_2$. Les bandes d'espèces $GeH_x$ (l'étirement à 1900–2100 $cm^{-1}$ [15]) ne sont pas clairement observées, en raison de leur oxydation rapide à l'ambiante. L'interprétation de la large bande centrée à 1280 $cm^{-1}$ n'est pas tout à fait claire. Probablement, qu'elle peut être liée à certaines harmoniques et/ou aux bandes de composition de sous-oxyde $GeO_x$. Le petit épaulement à 1640 $cm^{-1}$ correspond aux vibrations de cisaillement de molécules d'eau adsorbées physiquement.

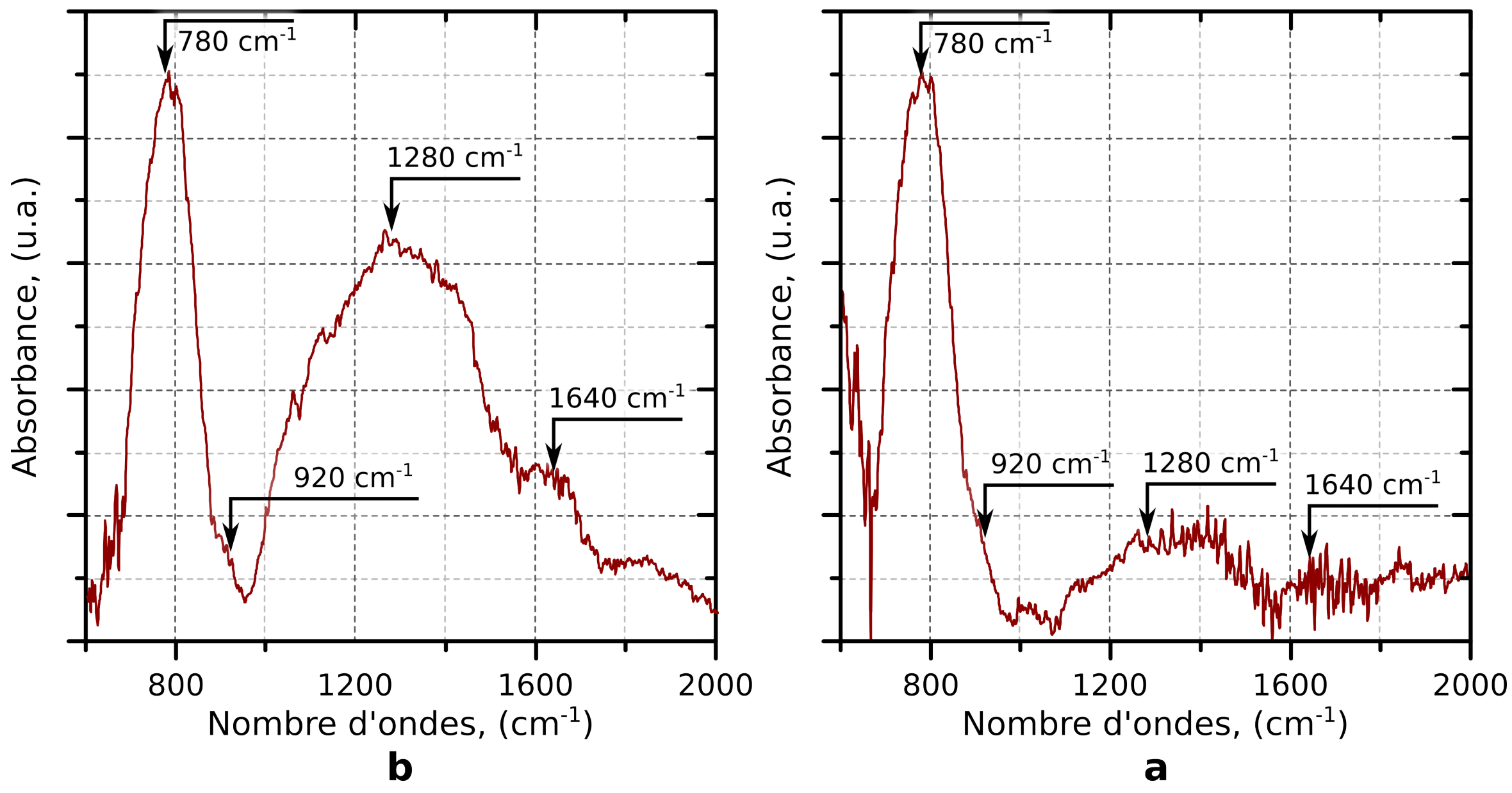

Figure 4.14 Spectres FTIR en mode ATR d'absorbance des couches du Ge mésoporeux avec une morphologie de type « sapin » ($HF_{49\%}$, 2 $mA/cm^2$, $1s/0.5s$, 2 h) (a) et de type « éponge » ($HF_{49\%}$, 0.5 $mA/cm^2$, 1 s/2 s, 2 h) (b).

### 4.2.2 RBS

Étant donné qu'une partie importante du PGe est oxydée, pour quantifier la composition chimique générale et, en particulier, pour déterminer la proportion des atomes de germanium et de d'oxygène dans la couche du PGe, des expériences RBS ont été réalisées.

La Figure 4.15 montre un spectre RBS de la couche mésoporeuse de Ge en éponge sur le substrat de Ge. Pour expliquer la forme du spectre RBS, considérons pas par pas une diffusion d'ions dans une couche de PGe. En supposant, que la surface libre de PGe est oxydée, une couche de PGe peut être approximée par un mélange d'atomes de germanium et d'oxygène, dont la densité atomique est plus faible par rapport au matériau cristallin.

Le seuil du signal RBS se situe à une énergie $\approx$ 1600 keV (point I). À cette énergie les ions interagissent avec les atomes de Ge de la surface sans pénétrer dans l'échantillon : ils perdent leur énergie par le facteur cinématique $K_{\mathrm{Ge}} \approx 0.8$. L'autre partie d'ions est rétrodiffusée après une décélération dans une couche de matière. La probabilité de rétrodiffusion est inversement proportionnelle à l'énergie des ions : elle peut être modélisée par une fonction hyperbolique. Cependant, la section efficace d'arrêt d'ions par des atomes de O est plus petite que celle de Ge. Ainsi, moins d'ions sont rétrodiffusés dans la partie poreuse de l'échantillon (région I-II) par rapport à une couche de même épaisseur de substrat (région Ia-II). La déviation et la largeur du saut de signal RBS observable entre les énergies de 1100 à 1200 keV est une indication de nature rugueuse de l'interface entre le substrat et la région poreuse et/ou une inhomogénéité latérale de l'épaisseur de la couche de PGe. Finalement, la région II-V est le signal des atomes de Ge du substrat. Il convient de noter, qu'une petite fraction des ions est rétrodiffusée dans la couche de PGe par des atomes d'oxygène. En effet, étant un atome plus léger que le Ge, l'oxygène a un facteur cinématique encore plus petit — $K_O \approx 0.32$. Le signal associé à de tels ions peut être retrouvé dans la plage d'énergies inférieures à 720 keV (région III-IV).

Le spectre RBS a été traité par le logiciel de simulation SIMNRA [78]. La composition de la couche poreuse qui correspond a une telle courbe est $\mathrm{Ge}_y\mathrm{O}_{1-y} = \mathrm{Ge}_{51\%}\mathrm{O}_{49\%}$ avec la densité atomique $N = 4220 \cdot 10^{15}$ atomes/cm$^2$. Sachant l'épaisseur de la couche poreuse $L = 1558$ nm, les fraction volumiques de Ge, $\mathrm{GeO}_2$, ainsi que la porosité de la couche peuvent être évaluées. Il convient de noter, que dans cette modélisation, la rugosité à l'interface Ge-PGe a été ajoutée. La valeur de la rugosité estimée de 150 nm est comparable avec les observations microscopiques. Une estimation grossière sur la porosité de la couche peut être faite :

$$P = 1 - \frac{N/N_A}{L\rho(\mathrm{cGe})/M(\mathrm{Ge})} \tag{4.2}$$

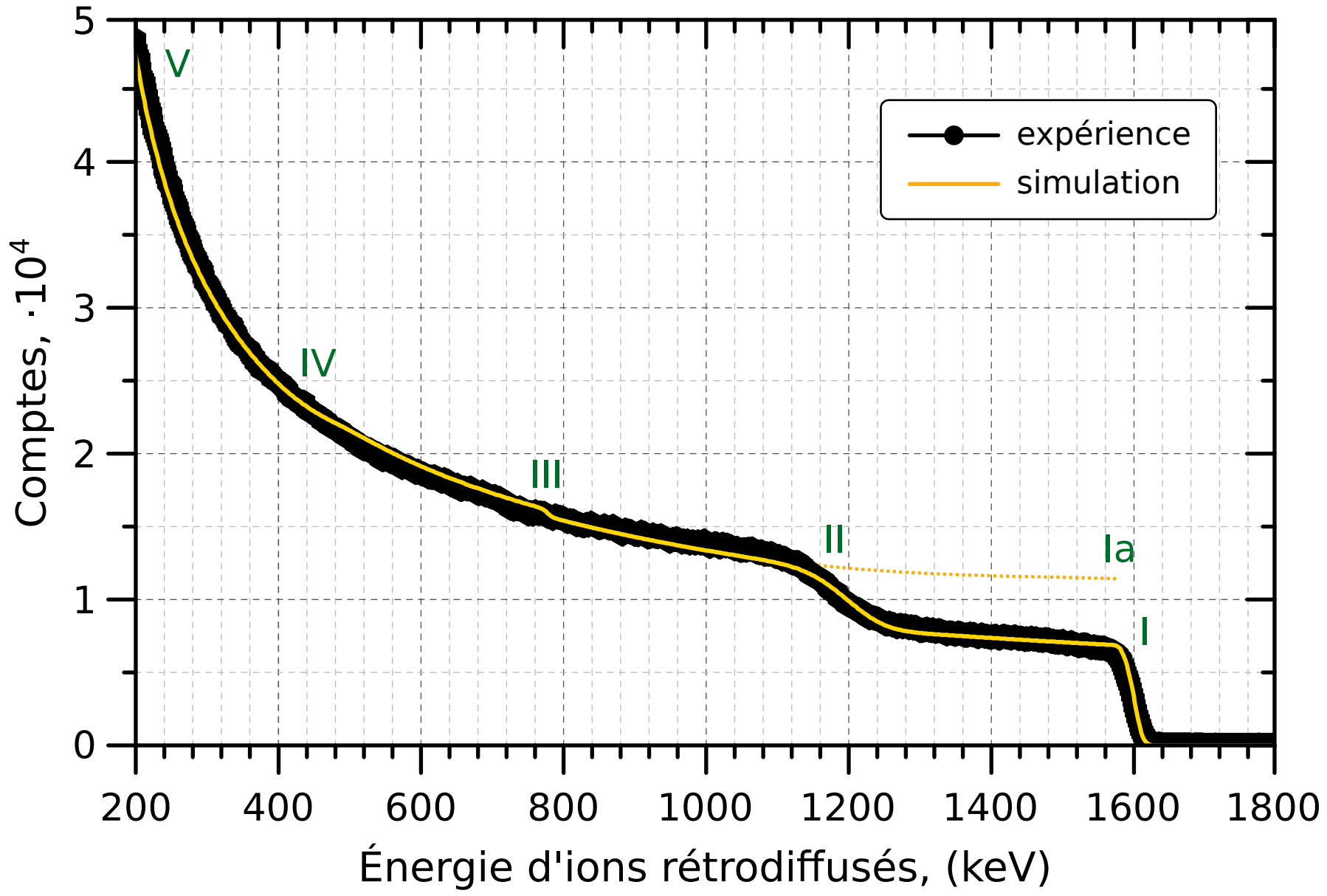

Figure 4.15 Spectre RBS de la couche de PGe en éponge ($HF_{49\%}$, 2 mA/cm$^2$, 1 s/3 s, 2 h). La ligne pointillée trace le spectre du Ge monocristallin sans couche de PGe.

où $\rho(Ge) = 5.323$ g/cm$^3$ est la masse volumique du Ge cristallin ; $M(\mathrm{Ge})$ — la masse molaire du Ge ; $N_A$ — la constante d'Avogadro.

Pour la couche donnée, la porosité est estimée au niveau de 39%.

Déterminons maintenant la composition de la couche poreuse plus précisément.

Établissons une relation entre la quantité d'atomes et le volume :

$$\frac{V(\mathrm{cGe})\rho(\mathrm{Ge})}{M(\mathrm{Ge})} = \frac{N(\mathrm{cGe})}{N_a} \tag{4.3a}$$

$$\frac{V(\mathrm{GeO}_x)\rho(\mathrm{GeO}_x)}{M(\mathrm{GeO}_x)} = \frac{N(\mathrm{GeO}_x)}{N_a} \tag{4.3b}$$

où $V(\mathrm{cGe})$ et $V(\mathrm{GeO}_x)$ sont les volumes de Ge cristallin et d'oxyde de Ge dans la couche poreuse, respectivement ; $N(\mathrm{cGe})$ et $N(\mathrm{GeO}_x)$ — le nombre d'entités dans la couche poreuse ; $M(\mathrm{GeO}_x)$ — la masse molaire d'une entité d'oxyde de Ge ;

L'analyse RBS ne fournit pas un rapport des entités $N(\mathrm{GeO}_x)/N(\mathrm{cGe})$ mais les fractions d'atomes $N(\mathrm{Ge})/N = y$ et $N(\mathrm{O})/N = 1-y$. Les atomes de Ge sont dans le noyau cristallin, mais aussi dans l'oxyde :

$$N(\mathrm{Ge}) = yN = N(\mathrm{cGe}) + \frac{1}{1+x} N(\mathrm{GeO}_x) \tag{4.4a}$$

$$N(\mathrm{O}) = (1-y)N = \frac{x}{1+x} N(\mathrm{GeO}_x) \tag{4.4b}$$

d'où :

$$N(\mathrm{cGe}) = N \frac{y(x+1)-1}{x} \tag{4.5}$$

Depuis les équations (4.3) et (4.4), une fraction volumique $\Upsilon^3$ du noyau par rapport au volume d'un cristallite peut être calculée :

$$\frac{V(\mathrm{cGe})}{V(\mathrm{cGe}) + V(\mathrm{GeO}_x)} = 1 - \left[ 1 + \frac{\rho(\mathrm{GeO}_x)}{\rho(\mathrm{Ge})} \frac{M(\mathrm{Ge})}{M(\mathrm{Ge}) + xM(\mathrm{O})} \left( \frac{y(x+1)-1}{(1+x)(1-y)} \right) \right]^{-1} = \Upsilon^3 \tag{4.6}$$

Dans une approximation de cristallites quasi-sphériques, dont le diamètre est $D$, le diamètre du noyau cristallin $d = \Upsilon D$. Si les cristallites sont couverts par le dioxyde de Ge parfait ($x = 2$, $\rho(\mathrm{GeO}_2) = 4.228$ g/cm$^3$), $d = 0.55D$.

En réalité, le coefficient stœchiométrique $x$ n'est pas forcément égal à 2. La densité de Ge amorphe est la même, que celle de Ge cristallin [69]. La masse volumique de $\mathrm{GeO}_x$ peut être alors approximée comme suit :

$$\rho(\mathrm{GeO}_x) = \rho(\mathrm{GeO}_2) + [\rho(\mathrm{Ge}) - \rho(\mathrm{GeO}_2)] \frac{2-x}{2} \tag{4.7}$$

Par exemple, pour $x = 1.5$, $d = 0.5D$. Cependant, la fonction $\Upsilon(x)$ décroît rapidement lorsque $x$ diminue et au-delà de $x = 1.25$, elle tombe dans des valeurs peu réalistes pour une valeur $x = 0.51$ donnée.

Ainsi, pour une couche poreuse dont le diamètre moyen de cristallites est $5.1 \pm 1.2$ nm, le le modèle donne une valeur pour le diamètre du noyau de $2.8 \pm 0.7$ nm et pour l'épaisseur de la coquille amorphe de $1.2 \pm 0.4$ nm, ce qui est en cohérence avec les valeurs mesurées par le TEM. Pour des cristallites plus grands, comme exposés sur les Figures 4.1 (c) et (d), les valeurs du diamètre de noyau et de l'épaisseur de la coquille attendues sont 5.0 nm et $\approx 2$ nm, respectivement. En réalité, la coquille est plus fine — 1.7 nm ; elle n'est pas, alors, en relation directe avec la taille du cristallite.

Il est important de noter, que l'analyse RBS des couches de PGe en éponge réalisée sous la même densité de courant mais avec différents rapports de cycles, donne les mêmes résultats. Pour la couche réalisée avec le cycle 1 s/1 s, $y = 0.52$ et $\Upsilon = 0.56$ tandis que pour la couche réalisée avec le cycle 1 s/1 s, $y = 0.515$ et $\Upsilon = 0.55$, ce qui confirme la similitude de toutes structures poreuses réalisés par le methode BEE avec $t_{off}$ supérieure à 1 s.

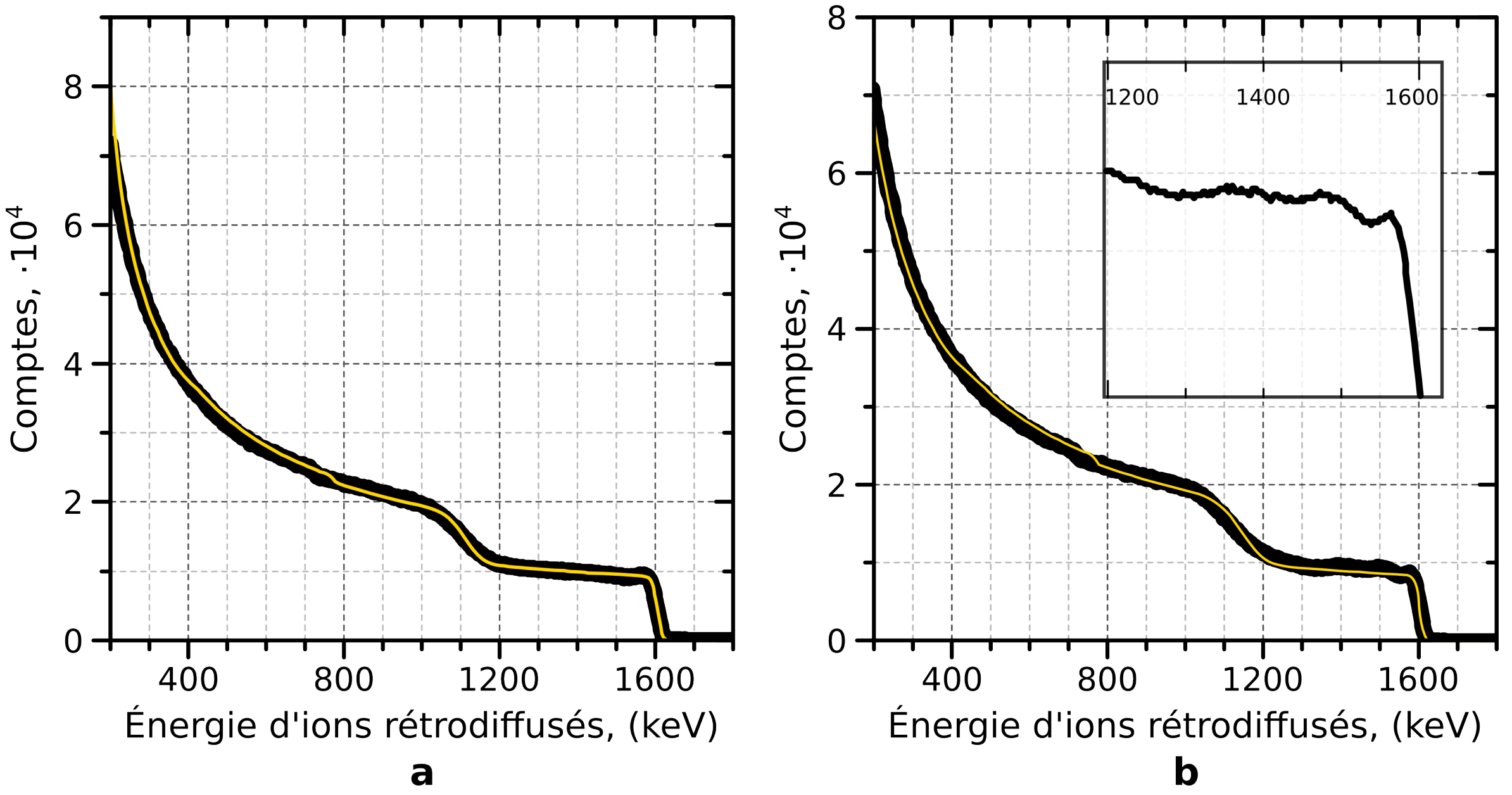

Figure 4.16 Spectres RBS de la couches de PGe de type « arbres » ($HF_{49\%}$, 2 mA/cm$^2$, 0.5 s/0.25 s, 2 h) (a) et de la structure « arête de poisson - arbres » ($HF_{49\%}$, 2 mA/cm$^2$, 1 s/0.5 s, 2 h) (b). Le zoom de la région qui correspond à la couche poreuse est donné dans l'encadre de la Figure b

L'analyse de la couche avec la morphologie de type « sapin » révèle les mêmes tendances générales (Figure 4.16 (a)). Notons cependant la teneur en oxygène plus élevée ($y = 0.436$, $\Upsilon = 0.45$) et la largeur de la région de transition entre le signal de la couche poreuse et le signal du substrat plus importante.

Finalement, le spectre d'empaillement des couches de différentes morphologies est présenté sur la Figure 4.16 (b). La différence en porosité et en teneur d'oxygène entre le PGe de type « arbre » et le PGe de type « arête de poisson » fait apparaitre des oscillations sur la partie du spectre qui correspond à la couche poreuse. Sachant les épaisseurs des couches et en mesurant le rapport des amplitudes de signaux de ces deux couches, la teneur en oxygène dans la couche de type « arête de poisson » est estimée au niveau de 60%. Il convient de préciser que la possibilité d'évolution de la teneur en oxygène en profondeur n'est pas exclue. Toutefois, dans les conditions d'analyse actuelles, elle ne peut pas être extraite du spectre RBS en raison d'une faible section de rétrodiffusion d'atomes d'oxygène et donc un faible signal de ces atomes (Figure 4.15, région III-IV).

## 4.3 Conclusions du chapitre

- Les structures en éponge consistent en des nano-particules quasi-sphériques, avec un diamètre moyen $5 \pm 1$ nm ;
- Dans des structures en arbre les nano-particules sont oblongs, avec une largeur 4–5 nm et la longueur 10–20 nm ;
- Les cristallites, étant connectés par des petites isthmes de la largeur de 2.5–4 nm, peuvent être considérés isolés.
- Les nano-particules sont constitués d'un noyau cristalline et d'une coquille amorphe de 1–1.5 nm d'épaisseur ;
- Une « éponge », se comporte plutôt comme un poly-cristal tandis que les structures de type « arbre » préservent la structure cristalline du Ge lors de sa porosification.
- le teneur en oxygène dans les couches de type « éponge », « arbres » et « arêtes de poisson » est de 49%, 54% et 60%, respectivement.
- La porosité est estimée à 39%, mais le fait que les pores soient bouchés par l'oxyde superficiel se traduit par une porosité apparente $\approx$ 20%
- Le facteur stœchiométrique $x$ de la composition du sous-oxyde de Ge est dans la gamme de $1.25 < x < 2$
- La couche amorphe superficielle est un sous-oxyde de Ge ($GeO_x$, $x < 2$). Aucune trace de dioxyde de Ge ($GeO_2$) n'est détectée.
- L'amorphisation de la surface des cristallites est probablement due à une reconstruction de la surface des cristallites ayant des liaisons pendantes.

# CHAPITRE 5

# Traitement de recuit du Ge poreux

## 5.1 Recuit des monocouches du Ge poreux

Un traitement thermique à des températures élevées est une partie intégrante de nombreux processus technologiques où le PGe peut être impliqué. Par exemple, lors du procédé de transfert de couches minces [97], une couche de Ge poreux est exposé à des températures jusqu'à 600–650°C pendant l'étape d'épitaxie. Bien que plusieurs travaux soulignent des changements structurels induits par le recuit du Ge poreux chimiquement gravé [57, 58], un seul article se réfère au recuit et à la réorganisation du Ge mésoporeux réalisé par gravure électrochimique[40].

Dans ce chapitre, les résultats d'analyse approfondie du comportement structural de Ge poreux au cours de son recuit sont présentés. En particulier, les influences de : (i) la température de recuit, (ii) l'hybridation temps et (iii) l'épaisseur de la couche de PGe initial sur la réorganisation induite par la chaleur des couches de PGE sont décrites en détail. Les résultats obtenus sont discutés par rapport à la théorie Lifshitz-Slyozov-Wagner (LSW) [42, 68, 113].

En général, le recuit dénote un traitement thermique des matériaux poreux ou en poudre, en dessous du point de fusion, qui est associé à des changements structuraux et/ou morphologiques [90]. Dans le cas du Si poreux, par exemple, les changements morphologiques deviennent visibles à des températures aussi basses que 350°C [90] ce qui est bien en dessous du point de fusion de Si volumique (1412°C) [67]. La force motrice de cette réorganisation induite thermiquement est liée à une minimisation de l'énergie de surface du Si poreux liée à sa surface interne hautement développée [48, 65, 82, 83, 90, 91, 96]. De même que pour le Si poreux, le Ge poreux est attendue de montrer la même tendance de réorganisation structurelle au cours du processus de recuit en dessous de son point de fusion. D'après la théorie du frittage classique mise au point pour les matériaux poreux [42], l'activation thermique stimule les atomes à occuper des positions d'énergie plus favorables, ce qui réduit le rapport surface/volume. La température et la durée du recuit jouent un rôle crucial [42] dans un tel processus. Ici, nous examinons indépendamment l'influence de chacun de ces paramètres sur la réorganisation des couches de Ge poreux. Dans toutes les expériences, le recuit a été conduit dans la chambre sous vide.

### 5.1.1 Influence de la température

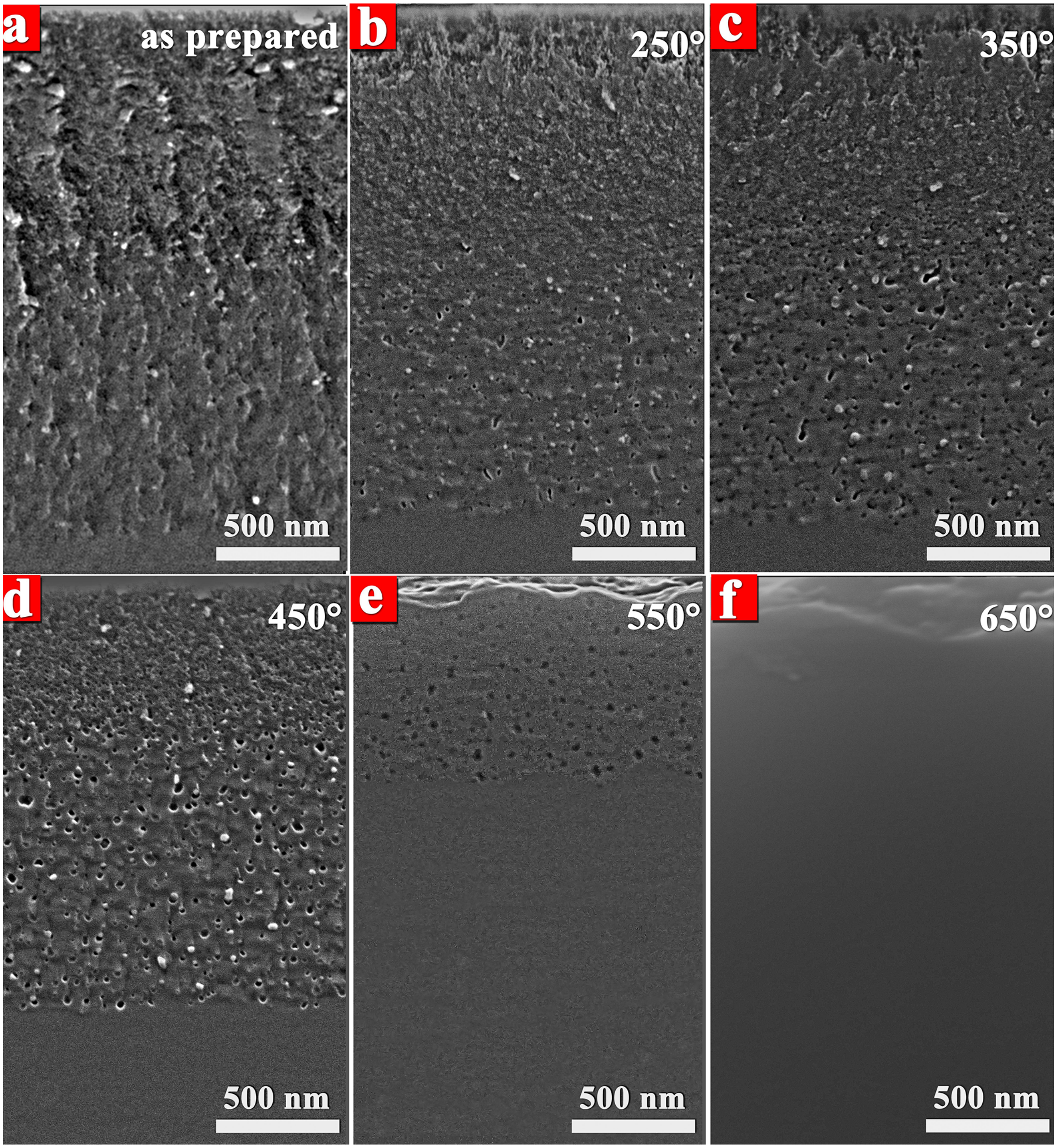

Figure 5.1 Images MEB en coupe transversale de la couche de Ge poreux (a) avant et (b-f) après un recuit à différentes températures pendant 30 min.

Dans la première série d'expériences, la durée de recuit a été fixée à 30 min et la température de recuit a varié entre 250°C et 650°C. La Figure 5.1 (a) présente la couche de Ge poreux en éponge d'épaisseur 2.1 µm telle que préparée. Un réseau de pores fortement interconnectés sans direction de croissance clairement définie peut être observée. Le diamètre moyen des pores se trouve être dans la gamme 5–8 nm. Les images MEB de cet échantillon après un recuit à cinq températures différentes sont représentées dans les Figures 5.1 (b-f). Même à une température aussi basse que 250℃ (Figure 5.1 (b)), la structure de Ge poreux est densifiée et des cavités commencent à apparaitre. Celles qui sont proches de l'interface Ge poreux / Ge volumique sont bien séparées les unes des autres et présentent deux formes principales : sphérique et allongée. La taille des cavités augmente progressivement en approchant de la surface. La même tendance est observée pour les échantillons recuits à 350℃ (Figure 5.1 (c)) et 450℃ (Figure 5.1 (d)). Un changement important dans l'évolution de la nanostructure du Ge poreux apparaît à la température de recuit de 550℃ (Figure 5.1 (e)). Une implosion de pores bien prononcée peut être observée à cette température. Lorsque des cavités implosent, la couche poreuse rétrécie (tandis que l'épaisseur totale du substrat reste inchangée), jusqu'à sa transformation complète en couche cristalline. A partir de 650℃ (Figure 5.1 (f)),toutes les cavités disparaissent. Une augmentation du diamètre moyen de cavités peut être notée lors de l'élévation de la température de recuit.

### Forme des pores

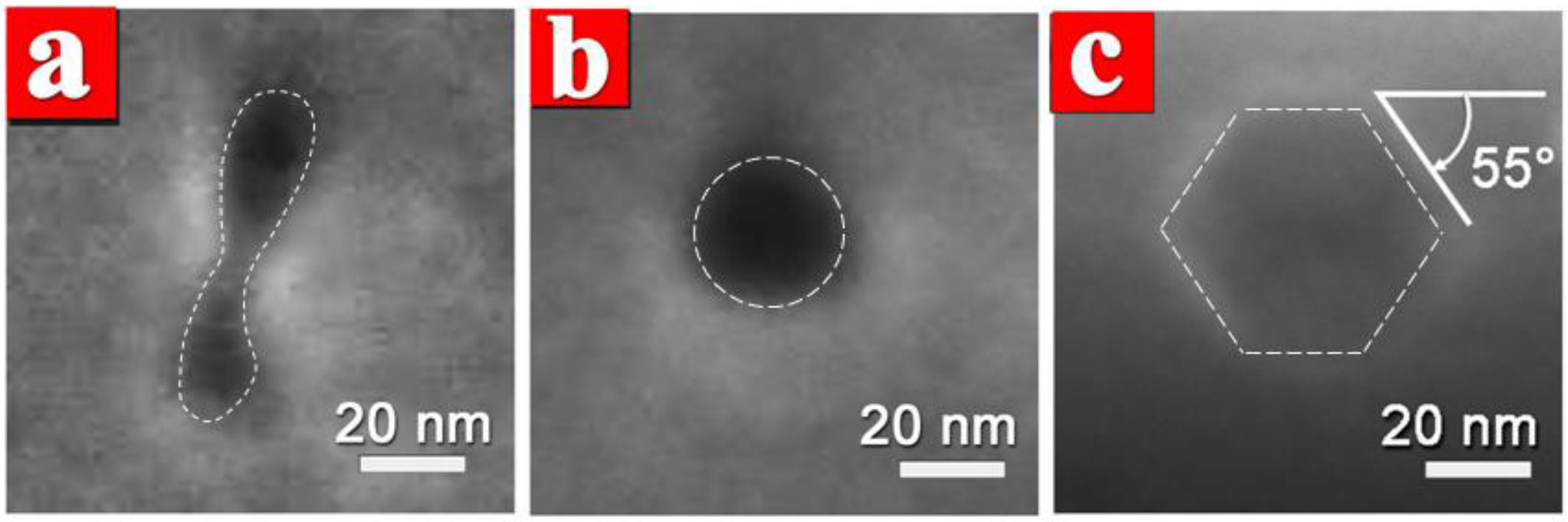

Figure 5.2 Différents types de formes de cavités observés pendant le recuit. Au début du traitement thermique (a) des cavités allongées formées par la coalescence de deux ou plusieurs pores voisins, et (b) des cavités sphériques peuvent être observées. Plus tard, (c) des cavités facettées commencent à dominer. L'angle mesuré entre les deux orientations les plus importantes est d'environ 55°

Trois formes typiques de nano-cavités qui dominent à différentes étapes de frittage du Ge poreux peuvent être distinguées (Figure 5.2). Au début du recuit (Figure 5.1 (b)), lorsque les distances entre les pores sont relativement petites, le rapprochement des pores voisins par l'intermédiaire d'une diffusion à partir de la surface et de leur fusion consécutive mène à une formation de cavités allongées (Figure 5.2 (a)). Comme il a été déjà mentionné, le traitement thermique du Ge poreux engendre un réarrangement morphologique et une diminution de son énergie de surface. Pour un volume donné de cavité, la forme sphérique présente la surface spécifique la plus petite, ainsi que l'énergie de surface la plus faible (cas isotrope). Par conséquent, une augmentation de la température et du temps de recuit provoque l'évolution de la forme des cavités vers une géométrie sphérique (Figure 5.2 (b)). Étant donné que l'énergie de surface du Ge est anisotrope [109], l'énergie de surface d'une cavité d'un volume constant peut être encore réduite par la formation de facettes. En effet, à partir de 450°C, les facettes d'une cavité avec une orientation bien définie commencent à être bien apparente, comme il est montré sur la Figure 5.2 (c). L'angle mesuré entre les deux orientations les plus importantes est de $55 \pm 1°$ avec une incertitude due à un clivage non-idéal (Figure 5.2 (c)). Cet angle est en parfait accord avec la valeur de l'angle de 54.73° entre les facettes (100) et (111) du Ge cristallin. Les plans {111} et {100} observés possèdent l'énergie de surface la plus faible et, par conséquent, ils sont les plus stables [109].

### Caractérisation structurale par spectroscopie μ-Raman

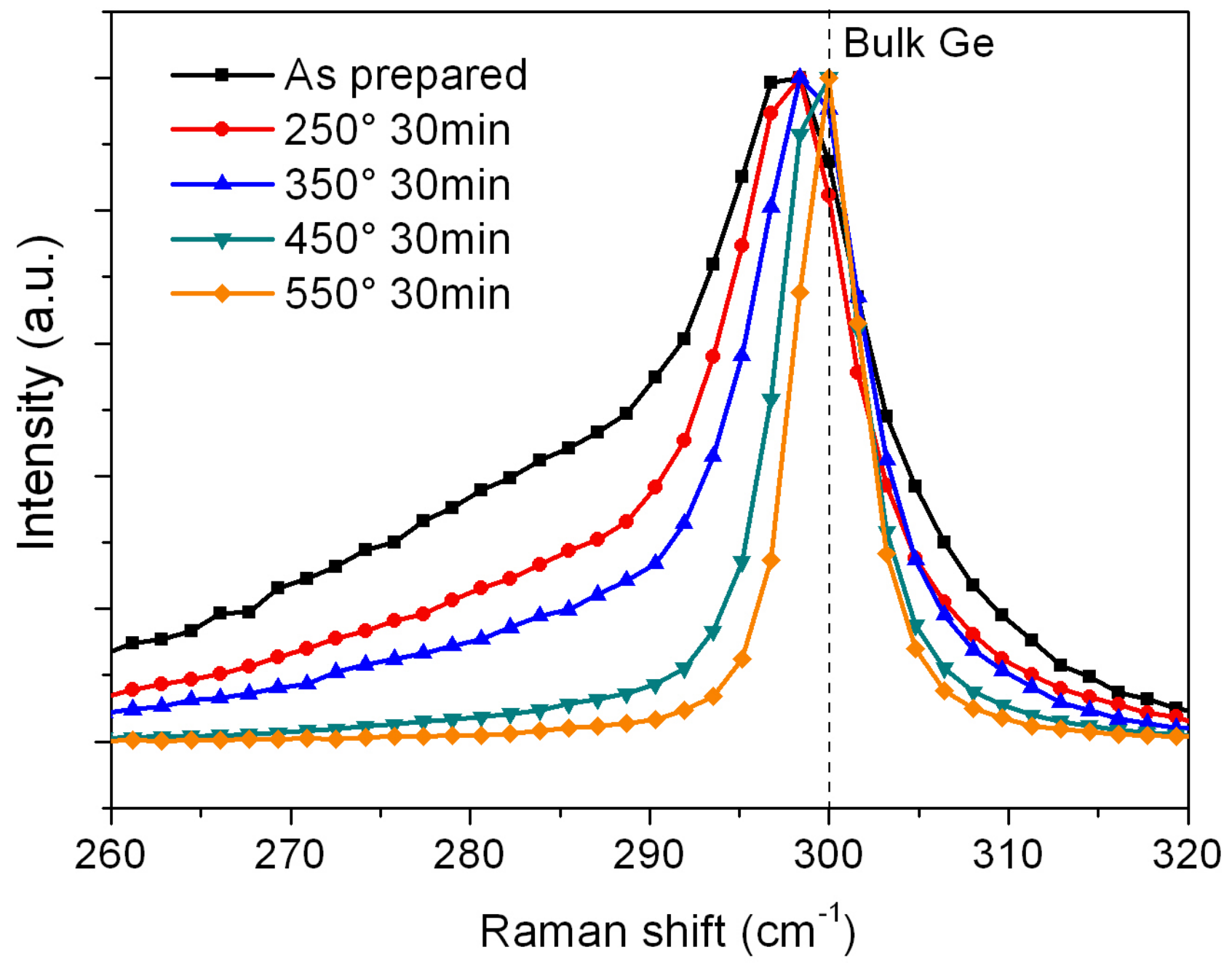

Figure 5.3 Spectres Raman correspondant aux couches de PGe recuites et présentées sur la figure 5.1. Les spectres Raman d'une couche de PGe telle que préparée et du Ge volumique sont aussi montrés à titre comparatif.

La spectroscopie de diffusion Raman est connue pour être un excellent outil optique non destructif pour étudier des propriétés structurelles des nanomatériaux. En particulier, une morphologie à l'échelle nanométrique et/ou une désorientation cristalline partielle peuvent influer sensiblement sur les formes de spectres Raman. L'évolution des spectres Raman des couches de Ge poreux recuites à des différentes températures pendant 30 min (Figure 5.1 (b-f)) est présentée sur la Figure 5.3. Le spectre Raman du Ge volumique monocristallin (c-Ge) est également donnée à titre de comparaison. Comme on peut le voir, la largeur, la forme et les positions spectrales des pics Raman de PGe tel qu'il est préparé, diffèrent fortement de celles du c-Ge. En effet, les premiers sont plus larges, décalées vers le rouge et élargis asymétriquement vers les basses fréquences. Une large bande supplémentaire situé à 275 $cm^{-1}$ présente dans la couche du PGe tel qu'il est préparé, peut être associée à une petite fraction de phase amorphe [59].

La réorganisation induite par le recuit d'une matrice poreuse peut être également révélée sur les spectres Raman. En effet, l'augmentation de la température de recuit est accompagnée par les évènements suivants :

- une disparition progressive d'une asymétrie du pic Raman principal ($\approx 300$ $cm^{-1}$) ;
- une diminution de la largeur du pic ;
- un décalage continue de la position du maximum du pic vers celle de Ge volumique.

Notons également que la fraction de Ge amorphe disparait progressivement avec l'augmentation de la température. Le spectre Raman d'une couche de PGe recuit à 550℃ coïncide totalement avec le spectre du Ge monocristallin. Ceci indique la restauration de l'ordre cristallin à l'intérieur de la phase solide de la matrice de PGe recuit. Les observations MEB ainsi que les mesures Raman nous permettent de conclure sur la nature monocristalline des échantillons recuits à 650℃ pendant 30 min.

## 5.1.2 Influence du temps de recuit

### Reconstruction du volume

Le recuit des échantillons à une température constante pour différents temps allant de 5 min à quelques heures, conduit à une évolution de la morphologie similaire à celle assurée par la variation de la température de recuit. Comme on peut le voir sur la figure 5.4, un recuit à 650℃ de la couche de PGe pendant 5 et 10 min change sa morphologie. Les pores ouverts, qui forment une « éponge » avant le recuit (figure 5.4 (a)) se transforment en cavités isolées, sphériques ou à facettes (figure 5.4 (b)). Par la suite ces cavités disparaissent progressivement (Figure 5.4 (c)). Il convient de noter que, pour une épaisseur donnée de la couche PGe, plus la température de recuit est élevée, plus le temps de recuit nécessaire pour transformer complètement une structure poreuse en structure monocristalline, diminue.

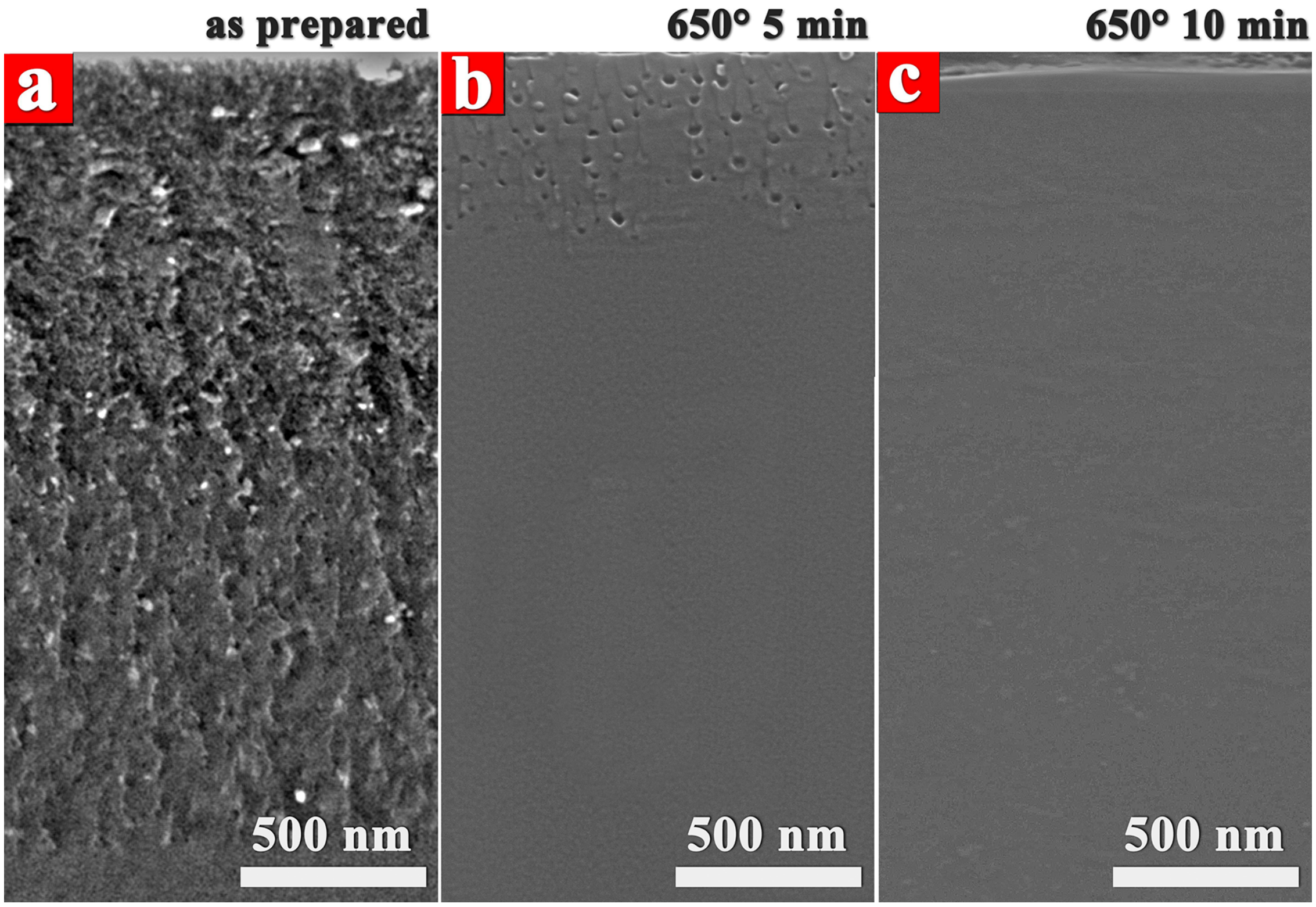

Figure 5.4 Images MEB en coupe transversale d'une couche de PGe avant (a) et après un recuit à 650℃ pendant 5 (b) et (c) 10 min.

La figure 5.5 présente l'évolution avec la température du taux de rétrécissement de la couche de PGe présentée sur la figure 5.1. Tandis que seulement 10 min est nécessaire pour imploser complètement les cavités dans la couche de PGe 2.1 µm d'épaisseur à 650℃ ; ce temps augmente jusqu'à 45 min à 550℃ et à plus de 2.5 h à 450℃.

Il convient également de souligner la dépendance cruciale de la durée d'implosion totale des pores avec l'épaisseur d'une couche poreuse. La Figure 5.6 (a)-présente une couche de PGe telle que préparée de 1 µm d'épaisseur. Des images MEB de cet échantillon après un recuit à 550℃ pendant 5 et 10 min sont présentées sur la figure 5.6 (b) et 5.6 (c), respectivement. Déjà après 5 min de recuit, un rétrécissement prononcé de la couche poreuse est observé. Une transformation complète du film de PGe en Ge monocristallin survient après 10 min de traitement thermique contrairement à la couche de PGe de 2.1 µm d'épaisseur qui nécessite 45 min. Ainsi, la durée totale de rétrécissement est fortement non-linéaire et dépend de l'épaisseur de la couche de PGe.

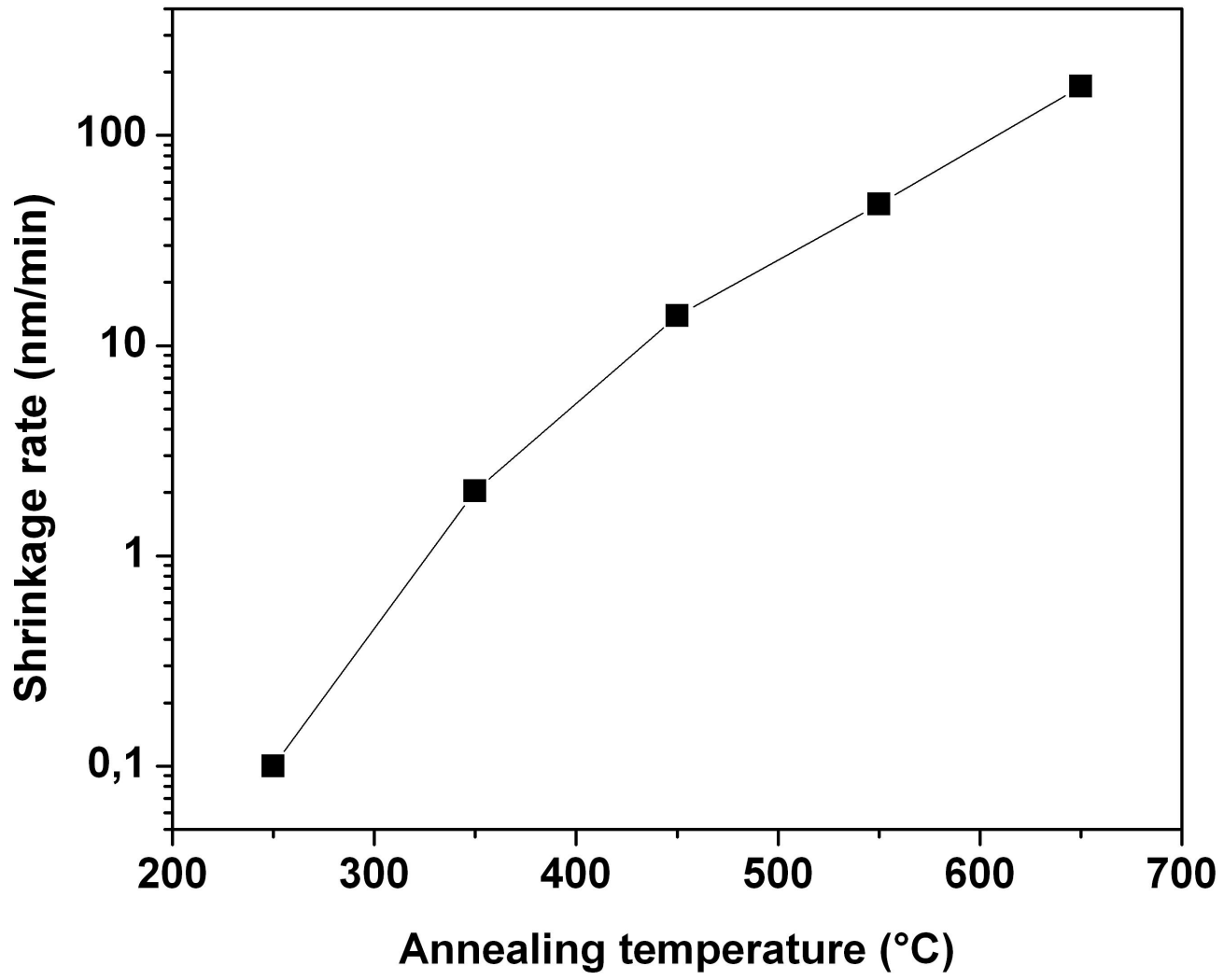

Figure 5.5 Taux de rétrécissement d'une couche de PGe en fonction de la température

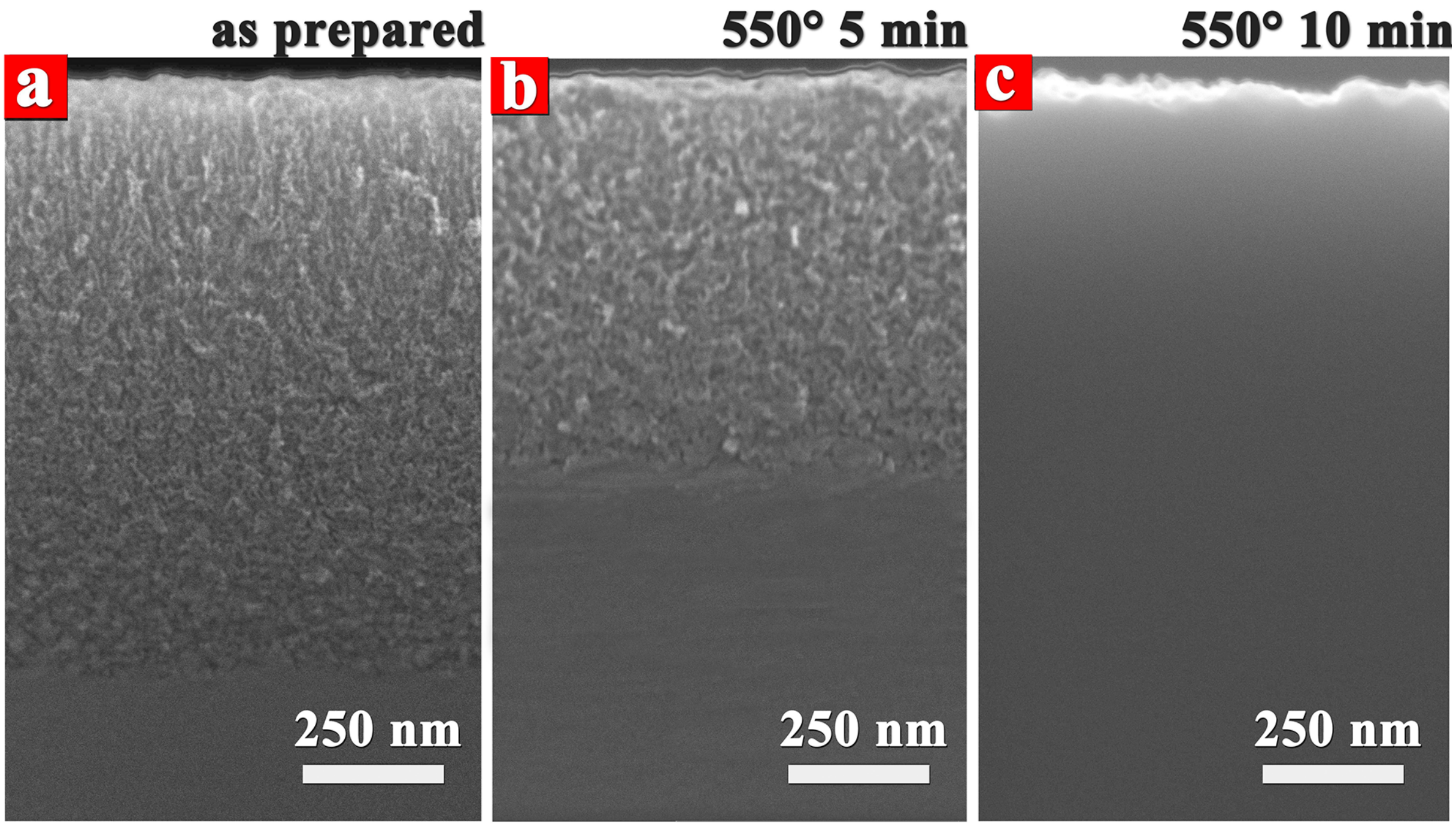

Figure 5.6 Images MEB en coupe transversale d'une couche mince de PGe (1 µm) avant (a) et après un recuit à 550°C pendant (b) 5 min et (c) 10 min.

### Reconstruction de la surface

L'évolution de la reconstruction de la surface d'une couche de PGe de 1 µm d'épaisseur, recuite à 550°C durant un temps de 5 à 45 min est montrée sur la figure 5.7 (a-f). Un réseau de pores ouverts fortement interconnectés et relativement homogènes peut être observé sur la surface de la couche de PGe telle que préparée (figure 5.7 (a)). Cependant, une formation de domaines non-poreux de quelques centaines de nanomètres a déjà lieu après seulement 5 min de recuit (figure 5.7 (b)). Ces domaines sont entourés par un réseau résiduel formé par des canaux de pores ouverts. L'exemple suivant (figure 5.7 (c)) illustre une transformation complète de la couche de PGe en Ge monocristallin avec une rugosité de surface considérablement augmentée.

Il convient de noter, que désormais le système cherchera toujours à se débarrasser de l'excès d'énergie de surface. En effet, les micro-hétérogénéités superficielles sont lissées avec le temps et les premières macro-cavités apparaissent (figure 5.7 (d)). Après 20 min de recuit (figure 5.7 (e)), on peut clairement observer une reconstruction de surface selon des orientations cristallographiques favorables pour lesquelles l'énergie de surface, et donc l'énergie totale du système, sera minimale. Tenant en compte le fait que les cavités ont tendance à former des espèces définies par des facettes $\{111\}$ et $\{100\}$, les plans d'une surface reconstruite sont supposés être des facettes de faible indice $\{111\}$ et $\{100\}$. Ce phénomène est encore plus prononcé pour les cas plus ultérieurs (voir l'image de la figure 5.7 (f)). Ainsi, du point de vue de la reconstruction de la surface, la rugosité macroscopique de la surface de PGe est renforcée avec le temps de recuit, de la même manière que dans le cas du Si poreux [64].

### 5.1.3 Discussion des résultats

#### Analyse qualitative

L'évolution de la morphologie à l'échelle micro et nano des couches de PGe décrite ci-dessus est une manifestation typique du phénomène de mûrissement d'Ostwald bien connu [26]. Ce phénomène peut être décrit quantitativement en termes de la théorie de Lifshitz-Slyozov-Wagner (LSW) [42, 68, 113], puisque le PGe tel qu'il est préparé peut être traité comme un cristal nano-creux sursaturé avec des lacunes. C'est l'une des théories assez bien établies décrivant la façon dont les lacunes évoluent pour atteindre un état d'équilibre final. Chaque cavité est considérée comme un massif des lacunes. Le terme « lacune » est introduit en plus du terme « cavité » pour distinguer les espèces mobiles (lacunes) qui participent au transfert de la matière et celles qui sont considérés momentanément immobiles (cavités).

La théorie LSW est basée sur les hypothèses suivantes :

- Dans le voisinage immédiat de chaque cavité, la répartition des lacunes est déterminée uniquement par le rayon de cette cavité et ne dépend pas des cavités environnantes.

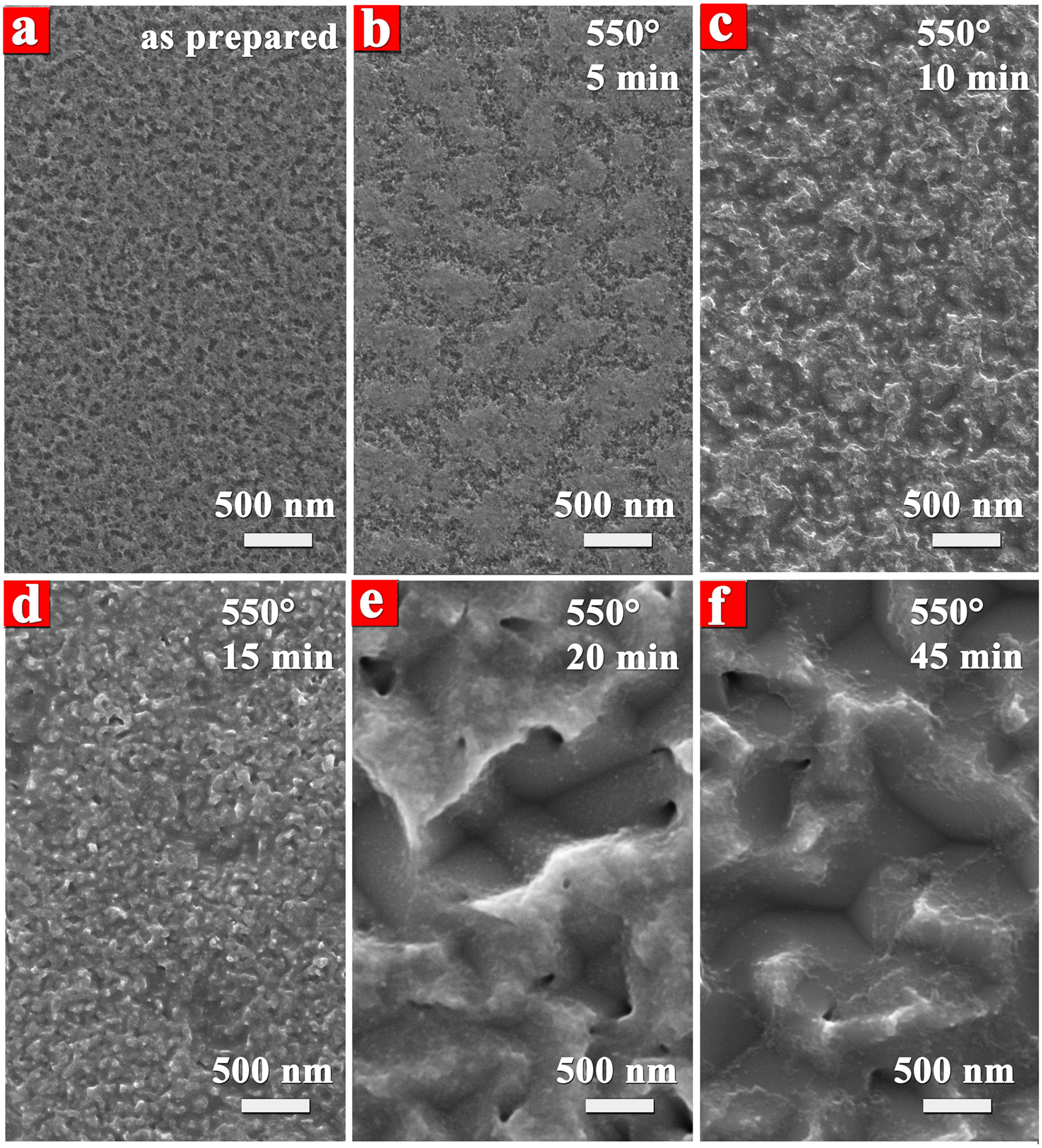

Figure 5.7 Images MEB en vue plane d'une couche mince de PGe (1 µm) avant (a) et après un recuit à 550°C pendant (b) 5 min, (c) 10 min, (d) 15 min, (e) 20 min et (f) 450 min.

- L'ensemble complet des cavités définit un niveau moyen de sursaturation de lacunes dans le réseau (champ de lacunes).

- Le sort de chacune des cavités est déterminée par son interaction avec la matière.

Le terme « champ de lacunes » est une abstraction. Avec son introduction le mouvement des lacunes dans le champ de lacunes est traité par analogie avec le cas des porteurs de charge dans le champ électrique effectif d'autres porteurs.

Sur cette base, les cavités peuvent être divisées en trois groupes principaux.

1. Le premier groupe comprend les grandes cavités, à proximité desquelles la concentration de lacunes (déterminée par la courbure de la surface), est moindre que celle du champ moyen. Dans ce cas, le flux de lacunes sera dirigé vers ces cavités qui vont grandir.

2. Le deuxième groupe est constitué de petites cavités dont la concentration de lacunes à leur périphérique est supérieure à celle du champ moyen. Le flux de lacunes sera dirigé loin de ces cavités et elles imploseront.

3. Pour les cavités du troisième groupe, la concentration de lacunes à proximité de leur surface est exactement la même que la concentration de lacunes dans le champ. Par conséquent, aucun flux de lacune ne se produit et le rayon d'une cavité (appelée rayon critique ($R_c$)) restera le même.

Cependant, avec le temps, la croissance de certaines cavités et l'implosion d'autres se traduira en une diminution de la surface totale interne du système. En conséquence, le niveau de la sursaturation de lacunes dans le réseau diminue et le rayon critique augmentera. Cela signifie que si au début certaines cavités appartiennent au premier groupe et grandissent ($t_1 : \ R > R_c(t)$), en raison de l'augmentation du rayon critique avec le temps, elles commencent à appartenir au deuxième groupe et, par conséquent, elles se rétrécissent ($t_2 > t1 : \ R < R_c(t)$).

Le destin des cavités situées près de la surface d'un cristal est différent de ceux situés dans sa profondeur, loin de la frontière. Loin de la surface, les cavités se partageront les lacunes et vont changer leur taille et leur forme de manière à réduire au maximum l'énergie totale de surface. D'autre part, près de la surface, les cavités peuvent transférer leurs lacunes non seulement aux grandes cavités voisines mais aussi à l'espace extérieure, qui est également situé à proximité immédiate. Ainsi, la surface de l'échantillon agit comme un puits de lacunes, près de laquelle les cavités implosent en créant une croûte non poreuse (région I Figure 5.8 (a)). Cette région est suivie par une couche dans laquelle la taille moyenne des cavités augmente progressivement avec la distance par rapport à la surface (région II Figure 5.8 (a)). La structure poreuse se termine en dessous par un ensemble de cavités qui ne « perçoit » pas du tout la surface (région III Figure 5.8 (a)).

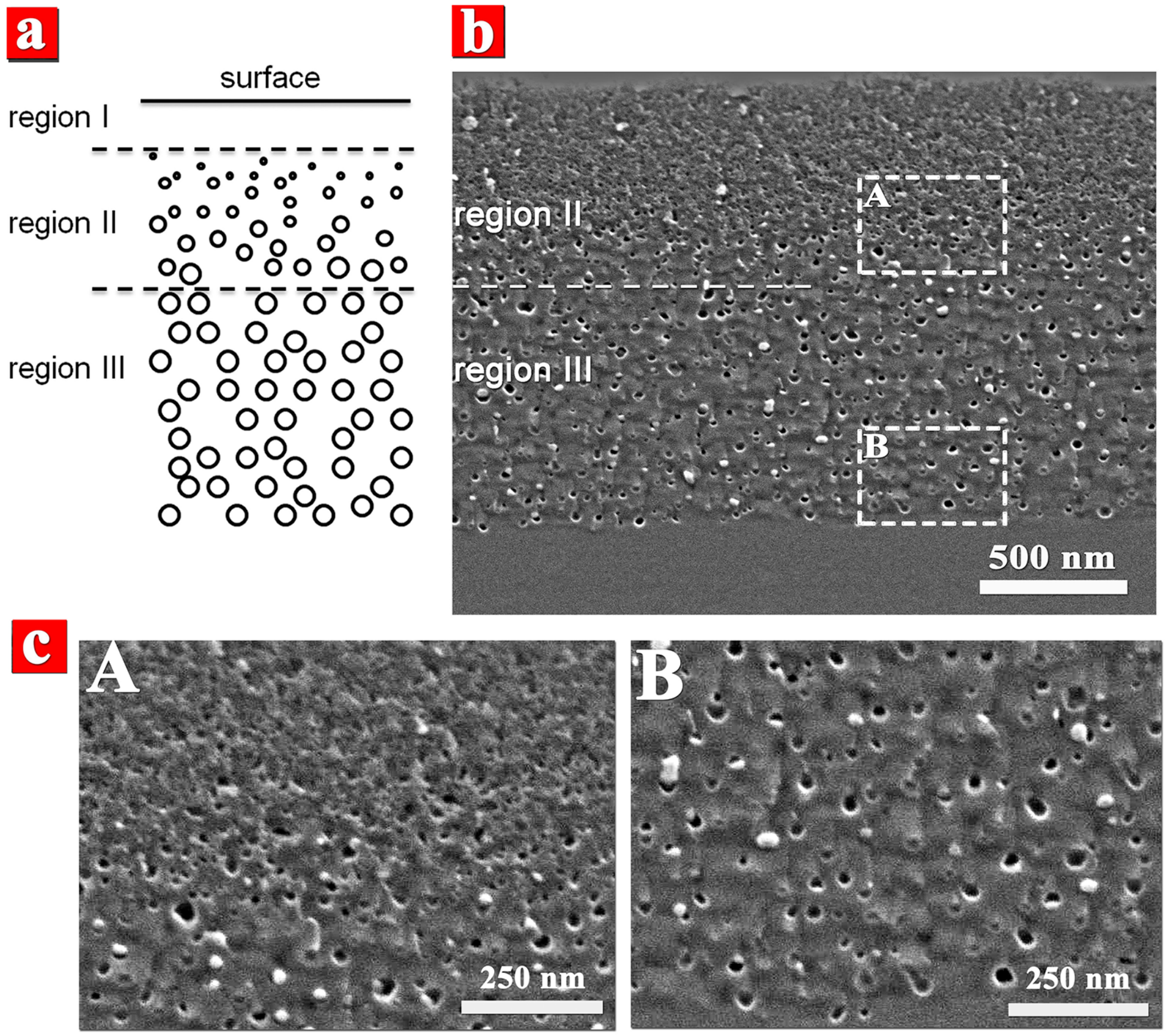

Figure 5.8 (a) Vue schématique d'une distribution de cavités à l'intérieur d'un matériau poreux fritté selon la théorie LSW ;
(b)une couche de PGe recuit à 450°C pendant 30 min. Les régions II et III sont indiquées.
(c) zooms détaillées A et B des régions II et III, respectivement.

La Figure 5.8 (b) présente une vue détaillée de la couche de PGe recuit à 450°C pendant 30 min. Sa nanostructure correspond bien au schéma de la Figure 5.8 (a). Deux régions spécifiques II et III peuvent être clairement identifiées. Des zooms pertinents A et B de chaque région sont présentés sur la Figure 5.8 (c). Comme on peut le voir, la région III se caractérise par une répartition homogène de la taille de cavités indiquant l'absence d'interaction entre les lacunes de cette région et la surface extérieure. D'autre part, la région II présente une diminution progressive de la taille des cavités vers la surface externe, qui est en bon accord avec les prédictions de la théorie LSW. Il convient de souligner que l'imagerie MEB de la croûte cristalline non-poreuse (région I) n'était pas possible pour cet échantillon.

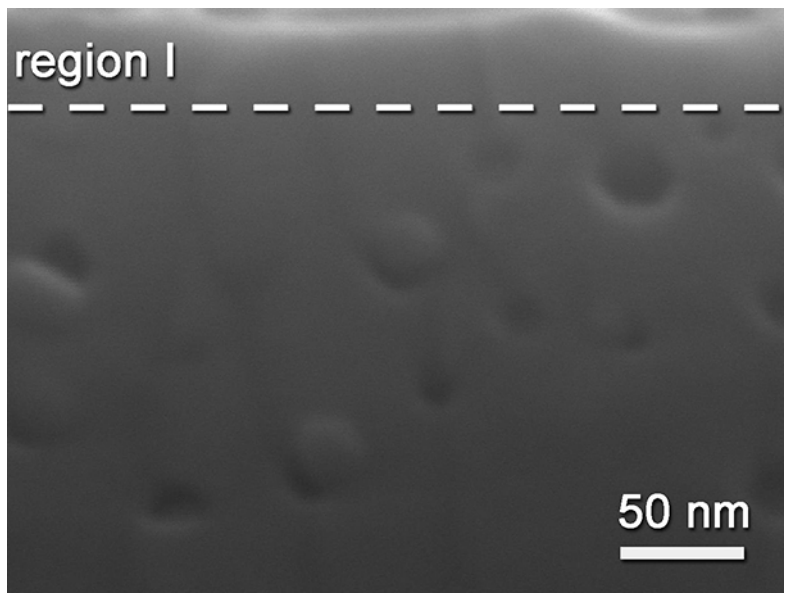

Figure 5.9 Formation d'une croûte de 35 nm d'épaisseur (région I) sur la surface d'une couche de PGe recuite à 650°C pendant 5 min.

A partir de 550°C, une croûte de surface bien visible apparait. La Figure 5.9 présente une région de 35 nm d'épaisseur formée à 650°C après de 5 min de recuit. La largeur de la croûte superficielle augmente de manière non linéaire avec le temps et la température de recuit suite à l'implosion de cavités de la région adjacente II. Plus la région I est épaisse, plus le temps qu'il faut pour une lacune d'atteindre la surface est grand, et la probabilité de dissolution de cette région dans une cavité proche (sans jamais atteindre la surface) est grande. Ainsi, le diamètre moyen des cavités augmente avec le temps de recuit. Les cavités se développent jusqu'à ce qu'elles atteignent la surface, puis elles disparaissent créant une rugosité de surface (Figure 5.7 (c)).

Une attention particulière devra être accordée aux couches minces de PGe. En effet, l'absence de grandes cavités dans la structure de PGe d'épaisseur 1 µm recuite à 550°C pendant 5 min (Figure 5.6 (b)) est constatée. Ainsi, pour ce type d'échantillons, seulement deux régions (I et II) peuvent être identifiées. Dans ce cas, toutes les cavités « perçoivent » la surface et, par conséquent, chacune d'elles va se réduire jusqu'à l'implosion. Ainsi, le taux de rétrécissement sera plus important que dans le cas d'échantillons de PGe relativement épais.

### Analyse quantitative

Comme dans le cas de nanostructures de Si poreux fritté [91], les couches de PGe recuites sont également caractérisées par une distribution log-normale de la taille des cavités :

$$\Lambda(R) = \frac{A}{\sqrt{2\pi}\sigma R} \exp\left(-\frac{1}{2}\left(\frac{\ln R/m}{\sigma}\right)^2\right) \tag{5.1}$$

avec $A$ — un facteur d'échelle, $R$ — le rayon d'une cavité, $m$ et $\sigma$ la médiane et la largeur de la distribution, respectivement.

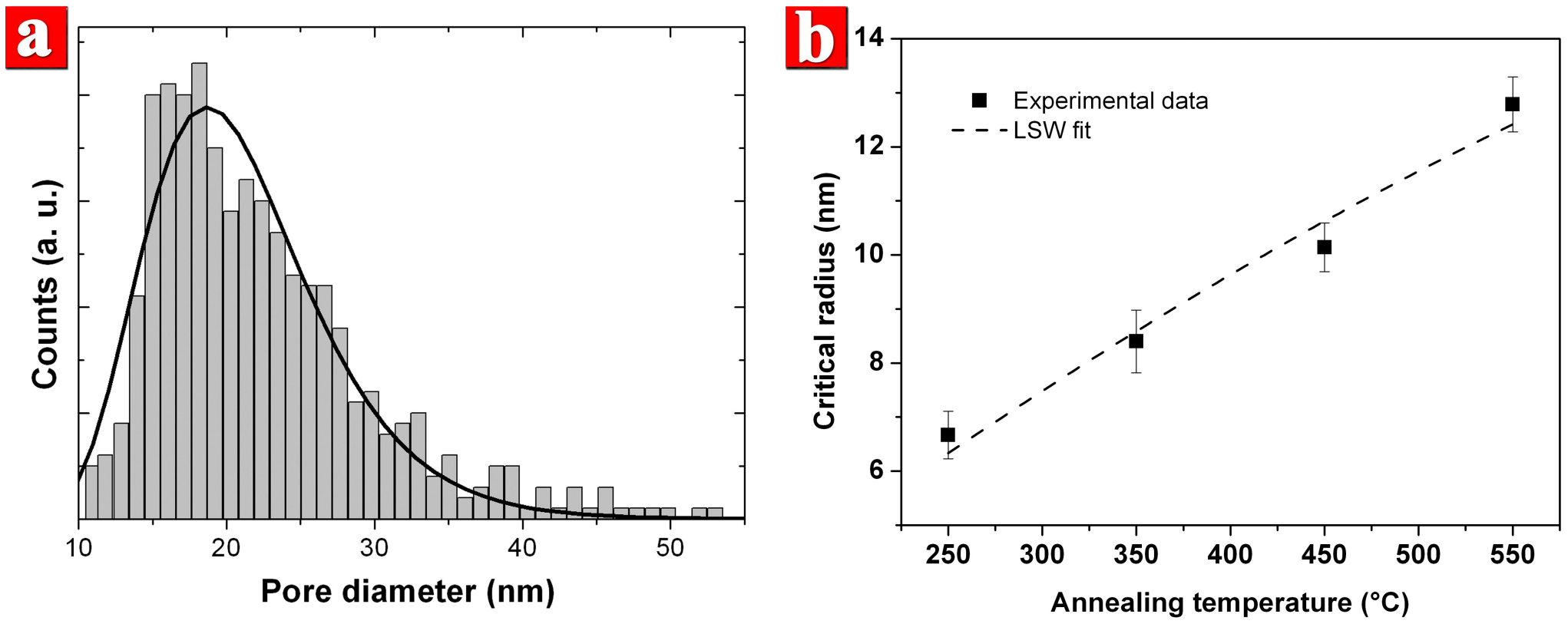

Figure 5.10 (a) Distribution typique de la taille des cavités ajustée avec une fonction log-normale (ligne continu).
(b) Le graphique d'évolution du rayon critique avec la température de recuit est ajusté à l'aide de l'équation 2

Une distribution typique de la taille des cavités ajustée avec une fonction log-normale est présentée sur la Figure 5.10 (a). La médiane de cette distribution, qui correspond au rayon critique d'une cavité, augmente avec la température de recuit tel qu'il est présenté sur la Figure 5.10 (b). Selon la théorie LSW, la dépendance du rayon critique de la température (T) et le temps (t) de recuit est donnée par la relation suivante :

$$R_c(t) \sim \left(\frac{4Dt}{9T}\right)^{1/3} \tag{5.2}$$

où $D$ est un coefficient de diffusion des lacunes défini par la Loi d'Arrhenius (5.3) :

$$D \sim \exp\left(-\frac{E_D}{k_B T}\right) \tag{5.3}$$

où $k_B$ est la constante de Bolzmann, $E_D$ est l'énergie d'activation thermique pour la diffusion d'une lacune dans le PGe.

La valeur de $E_D$ déterminée par ajustement des courbes expérimentales du PGe est de $0.306 \pm 0.032$ eV.

### 5.1.4 Conclusions du chapitre

- La dynamique de l'évolution d'une cavité à l'intérieur de la couche de PGe lors du recuit suit les prédictions de la théorie LSW.
- Les cavités ont tendance à former des espèce avec des facettes $\{111\}$ et $\{100\}$ afin de minimiser l'énergie de surface.
- La répartition des cavités par taille est log-normale.
- Des croûtes superficielles monocristallines non poreuses ont également été observées conformément aux prédictions de la théorie LSW.
- L'énergie d'activation du processus de maturation d'Ostwald induite thermiquement dans le PGe est de $306 \pm 0.032$ eV.
- Le temps du traitement thermique doit être suffisamment long pour former une croûte superficielle non-poreuse.
- En revanche, l'épaisseur de la couche poreuse initiale doit être suffisamment fine pour éviter la formation de grandes cavités, ce qui cause l'augmentation de la rugosité de la surface.
- La formation des facettes sur la surface du PGe recuit est observée comme la suite de la reconstruction de la surface. Ainsi, pour améliorer la rugosité de surface dans la perspective d'une épitaxie, un long temps de recuit doit être évitée.

# CHAPITRE 6

# Conclusions générales

La structuration à l'échelle nanométrique produit des modifications importantes des propriétés physiques des matériaux et des structures fonctionnelles. Cela ouvre de nouvelles possibilités d'applications innovantes des matériaux nanostructurés dans plusieurs domaines. Actuellement, les nanostructures de Ge font l'objet de nombreuses études fondamentales et appliquées.

Parmi les techniques de nanostructuration, l'anodisation électrochimique des semiconducteurs IV-IV et III-V est apparue comme une approche alternative à bas coût permettant l'obtention des nanomatériaux formant en volume un réseau de faible dimensions à température et sous atmosphère ambiante. Cependant, à ce jour et contrairement aux autres semiconducteurs IV-IV et III-V, le verrou technologique important de la gravure électrochimique du Ge est l'obtention de couches épaisses de Ge mésoporeux ayant une morphologie ajustable et contrôlable. Dans le cadre de cette thèse, nous avons levé ce verrou en appliquant la technique de gravure électrochimique bipolaire. Cette technique consiste à alterner périodiquement la polarisation de l'électrode de travail afin d'inhiber la destruction des pores formées par une gravure latérale. La formation des couches de Ge poreux avec une porosité ajustable dans la gamme 15–60% et des épaisseurs comprises entre 100 nm à 10 µm est démontrée. Des couches poreuses avec une morphologie de pores de type colonnes, sapin, dendrites, arêtes de poisson et éponge ont été réalisées. Il a été démontré que la morphologie du Ge poreux dépend principalement des facteurs suivants :

- la densité de courant anodique ;
- le degré de passivation de la surface ;
- la diffusion des anions du Fluor en profondeur de pores ;
- la diffusion de la charge électrique en profondeur de la couche poreuse

Il a été constaté que lorsque la durée du procédé est supérieure à 4h la formation des structures à multi-couches avec différentes morphologies a lieu. Quelles que soient les conditions de départ, les nanopores de type fils se forment à l'interface entre la couche poreuse et le Ge massif. Il est important de mentionner que le niveau de la surface de l'échantillon ne se déplace pas lors de la porosification, ce qui indique l'absence de dissolution chimique importante de la couche formée. Néanmoins, une amorphisation graduelle de la couche superficielle due à la dissolution chimique a été constatée. Cet effet peut être partiellement inhibé en réalisant l'anodisation sous obscurité.

Sur la base des résultats expérimentaux, un modèle électrochimique de formation des pores a été élaboré. Conformément à ce modèle, la surface d'un pore bascule entre un état hydrogéné et un état hydroxylé sans perte de masse du Ge grâce aux réactions réversibles.

Sous une polarisation cathodique la surface libre de Ge est hydrogénée. Il a été démontré que les réactions électrochimiques n'ont lieu qu'à l'interface Ge-Ge poreux. Afin de minimiser le dommage de la structure, il est important d'arrêter l'injection de la charge négative avant le déclenchement de la réaction de réduction de l'hydrogène moléculaire. Il est également important de noter que la durée des impulsions cathodiques (et alors la durée totale du procédé) peut être diminuée grâce à l'augmentation proportionnelle de l'amplitude des pulsations cathodiques.

Sous un régime anodique deux types de réactions ont lieu : la dissolution à l'état bivalent et la dissolution à l'état tétravalent. Une dissolution anodique doit se terminer avant la dernière étape d'oxydation du $=GeOOH^-$ irréversible. Lorsque la réaction d'hydroxylation est rapide, la vitesse du processus est limitée par le transport de charge et augmente légèrement avec la concentration de HF. Une partie de la surface d'un pore où le champ électrique est plus élevé est dissoute directement au $H_2GeF_6$. La nature de cette inhomogénéité du champ électrique détermine la morphologie du Ge poreux résultant.

La compréhension actuelle des mécanismes de formation du Ge mésoporeux nous a permis de concevoir de nouvelles morphologies du Ge poreux et de réaliser des empilements de couches de différentes porosités.

Afin de compléter le modèle électrochimique, nous avons analysé la structure des nanoparticules de Ge individuelles dans la couche de Ge poreux. Il a été démontré que les nanoparticules sont quasi-sphériques avec un diamètre moyen de $5 \pm 1$ nm et une distribution par taille de type gaussienne. Les particules sont composées du noyau cristallin de $\approx 3$ nm de diamètre qui est enveloppé par une coquille amorphe de 1–1.5 nm d'épaisseur. Cette coquille est partiellement oxydée avec une diminution du contenu en oxygène quand on se rapproche de l'interface noyau/coquille. Le facteur moyen stœchiométrique x de la composition du sous-oxyde $GeO_x$ est entre 1.25 et 1.5. L'apparition est liée à la reconstruction de la surface des cristallites lorsque la gravure électrochimique ou chimique créait des liaisons pendantes à la surface.

Finalement, nous avons analysé la reconstruction du volume et de la surface du Ge poreux induite par le traitement thermique qui est une partie intégrale du procédé de report de la cellule solaire du substrat de Ge. Les influences de la température, de la durée de recuit et de l'épaisseur des couches sur le temps d'implosion des pores et sur la structure cristalline des couches recuites ont été étudiées. Les résultats obtenus ont été discutés par rapport à la théorie Lifshitz-Slyozov-Wagner. Il a été trouvé, que le choix des conditions optimales de recuit est ambigu :

- Le temps du traitement thermique doit être suffisamment long pour former une croûte superficielle non-poreuse.
- Un recuit prolongé conduit à la formation des facettes et en conséquence à l'augmentation de la rugosité de la surface.
- L'épaisseur de la couche poreuse initiale doit être suffisamment fine pour éviter la formation de grandes cavités, ce qui cause l'augmentation de la rugosité de la surface.
- La couche poreuse supérieure doit être suffisamment épaisse pour éviter la migration des lacunes depuis la couche inférieure vers la surface.

Nous concluons que pour établir un compromis entre la qualité cristalline de la structure recuite (ce qui détermine ses propriétés électriques) et sa rugosité de surface (ce qui détermine la qualité des couches épitaxiales qui seront formées par la suite), l'épaisseur de la couche poreuse de faible porosité doit être de $\approx 1$ µm et le recuit doit être conduit sous une haute température ($\geq 650$°C) de manière rapide ($\approx 5$ min).

À l'issue de ce travail, différents champs d'exploration se sont ouverts. Les actions à mener afin de compléter le projet de report de couches minces d'une cellule solaire à triple jonctions seraient :

- Explorer les effets de la température et du dopage de substrat sur la morphologie de Ge poreux.
- Analyser la reconstruction des structures de Ge poreux à double porosité lors de recuit.
- Analyser des propriétés électriques (la résistivité, la mobilité et la durée de vie des porteurs de charge) des couches de Ge poreux recuites.
- Déposer la couche tampon du Ge sur la structure de Ge poreux recuite par l'épitaxie MBE. Optimiser les procédés de porosification et de recuit afin de réaliser la couche tampon de qualité monocristalline.

- Réaliser la croissance des couches de la cellule solaire par l'épitaxie CBE sur la couche tampon du Ge.
- Développer le procédé de détachement tenant compte de la procédure de métallisation à intégrer.

# ANNEXE A

# Publications et Communications à des congrès, symposiums

1. S. Tutashkonko, A. Boucherif, T. Nychyporuk, A. Kaminski-Cachopo, R. Arès, M. Lemiti, V. Aimez
   Mesoporous Germanium formed by bipolar electrochemical etching
   Electrochimica Acta 88 (2013) 256–262

2. S. Tutashkonko, A. Boucherif, T. Nychyporuk, A. Kaminski-Cachopo, R. Arès, V. Aimez, M. Lemiti
   Structural and morphological study of MesoporousGermanium layers formed by bipolar electrochemicaletching(oral)
   Pacific-Rim Meeting / Electrochemical Society, 7-12 October, Honolulu, Hawaii, USA

3. S. Tutashkonko, A. Boucherif, T. Nychyporuk, A. Kaminski-Cachopo, R. Arès, M. Lemiti, V. Aimez
   Porous Germanium for enhanced Ge substrate usage in multi-junction solar cells.(oral)
   Next Generation Solar 2012, 14-15 May 2012, Montréal, Canada.

4. A. Boucherif, S. Tutashkonko, T. Nychyporuk, A. Kaminski-Cachopo, M. Lemiti, V. Aimez, R. Arès
   Epitaxial growth of III-V semiconductors on porous germanium for layer transfer process of multijunction solar cells(oral).
   8th International Conference on Concentrating PV Systems, 16 - 18 April 2012, Toledo, Spain.

5. S. Tutashkonko , T. Nychyporuk, A. Kaminski-Cachopo, R. Arès, M. Lemiti, V.Aimez
   Germanium Thin Film Transfer Process via Nanoporous Sacrificial Layer for Concentrated Photovoltaics(oral).
   International Conference on Materials for Advanced Technologies – 2011, 26 June - 1 July 2011, Singapore.

6. S. Tutashkonko, A. Boucherof, T. Nychyporuk, A. Kaminski-Cachopo, R. Arès, M. Lemiti, V.Aimez
   Genanoporous layers for enhanced Ge substrate usage in CPV applications (poster)
   Photonics North 2011, 16-18 May 2011, Ottawa, Canada.

7. S. Tutashkonko, T. Nychyporuk, A. Kaminski-Cachopo, R. Arès, M. Lemiti, V.Aimez
   Germanium Thin Film Transfer Process via Nanoporous Sacrificial Layer for Concentrated Photovoltaics(poster).
   E-MRS 2011 Spring Meeting, 9-13 May 2011, Nice, France.

# LISTE DES RÉFÉRENCES